\documentclass[longauth]{aa}  

\usepackage{graphicx}
\usepackage{txfonts}
\usepackage[utf8]{inputenc}
\usepackage{multirow}
\usepackage{upquote} 
\usepackage[percent]{overpic} 
\usepackage{pdflscape} 

\usepackage[dvipsnames]{xcolor}
\usepackage{url}
\usepackage{hyperref}
\hypersetup{
    colorlinks=true,
    linkcolor=blue,
    urlcolor=blue,
    citecolor=blue
}

\usepackage{stackengine} 
\usepackage[maxfloats=256]{morefloats} 

\newcommand{\orcit}[1]{\protect\href{https://orcid.org/#1}{\protect\includegraphics[width=8pt]{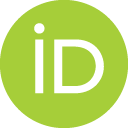}}}

\def\gaia{\textit{Gaia}\xspace}
\def\gdr3{\textit{Gaia}~DR3\xspace}
\def\g{$G$\xspace}
\def\bp{$G_{\rm BP}$\xspace}
\def\rp{$G_{\rm RP}$\xspace}
\def\bprp{\mbox{$G_{\rm BP}-G_{\rm RP}$}\xspace}
\def\bpg{\mbox{$G_{\rm BP}-G$}\xspace}
\def\grp{\mbox{$G-G_{\rm RP}$}\xspace}

\titlerunning{\gaia~DR3: All-sky classification of 12.4\,M variable sources} 

\begin{document} 

\title{\gaia Data Release 3}
\subtitle{All-sky classification of 12.4 million variable sources into 25 classes}

\author{Lorenzo~Rimoldini\inst{\ref{inst-ch-eco}}\fnmsep\thanks{\email{Lorenzo.Rimoldini@unige.ch}}\orcit{0000-0002-0306-585X}
\and
Berry~Holl\inst{\ref{inst-ch-eco},\ref{inst-ch-obs}}\orcit{0000-0001-6220-3266}
\and
Panagiotis~Gavras\inst{\ref{inst-es-rhea}}\orcit{0000-0002-4383-4836}
\and
Marc~Audard\inst{\ref{inst-ch-eco},\ref{inst-ch-obs}}\orcit{0000-0003-4721-034X}
\and
Joris~De~Ridder\inst{\ref{inst-be-ku}}\orcit{0000-0001-6726-2863}
\and
Nami~Mowlavi\inst{\ref{inst-ch-eco},\ref{inst-ch-obs}}
\and
Krzysztof~Nienartowicz\inst{\ref{inst-ch-sed}}\orcit{0000-0001-5415-0547}
\and
Gr\'{e}gory~Jevardat~de~Fombelle\inst{\ref{inst-ch-eco}}
\and
Isabelle~Lecoeur-Ta\"ibi\inst{\ref{inst-ch-eco}}\orcit{0000-0003-0029-8575}
\and
Lea~Karbevska\inst{\ref{inst-ch-eco},\ref{inst-fr}}
\and
Dafydd~W.~Evans\inst{\ref{inst-uk-ioa}}\orcit{0000-0002-6685-5998}
\and
P\'{e}ter~\'{A}brah\'{a}m\inst{\ref{inst-hu-obs},\ref{inst-hu-mta}}
\and
Maria~I.~Carnerero\inst{\ref{inst-it-to}}\orcit{0000-0001-5843-5515}
\and 
Gisella~Clementini\inst{\ref{inst-it-bo}}\orcit{0000-0001-9206-9723}
\and
Elisa~Distefano\inst{\ref{inst-it-ct1}}\orcit{0000-0002-2448-2513}
\and
Alessia~Garofalo\inst{\ref{inst-it-bo}}\orcit{0000-0002-5907-0375}
\and
Pedro~Garc\'{i}a-Lario\inst{\ref{inst-es-esa}}\orcit{0000-0003-4039-8212}
\and
Roy~Gomel\inst{\ref{inst-il-ta1}}
\and
Sergei~A.~Klioner\inst{\ref{inst-de-dre}}\orcit{0000-0003-4682-7831}
\and
Katarzyna~Kruszy{\'n}ska\inst{\ref{inst-pl-wa}}\orcit{0000-0002-2729-5369}
\and
Alessandro~C.~Lanzafame\inst{\ref{inst-it-ct1},\ref{inst-it-ct2}}\orcit{0000-0002-2697-3607}
\and
Thomas~Lebzelter\inst{\ref{inst-at-vie}}\orcit{0000-0002-0702-7551}
\and
G\'{a}bor~Marton\inst{\ref{inst-hu-obs}}\orcit{0000-0002-1326-1686}
\and
Tsevi~Mazeh\inst{\ref{inst-il-ta1}}\orcit{0000-0002-3569-3391}
\and
Roberto~Molinaro\inst{\ref{inst-it-na}}\orcit{0000-0003-3055-6002}
\and
Aviad~Panahi\inst{\ref{inst-il-ta1}}\orcit{0000-0001-5850-4373}
\and
Claudia~M.~Raiteri\inst{\ref{inst-it-to}}\orcit{0000-0003-1784-2784}
\and
Vincenzo~Ripepi\inst{\ref{inst-it-na}}\orcit{0000-0003-1801-426X}
\and
L\'{a}szl\'{o}~Szabados\inst{\ref{inst-hu-obs}}\orcit{0000-0002-2046-4131}
\and
David~Teyssier\inst{\ref{inst-es-tel}}\orcit{0000-0002-6261-5292}
\and
Michele~Trabucchi\inst{\ref{inst-ch-obs}}\orcit{0000-0002-1429-2388}
\and
{\L}ukasz~Wyrzykowski\inst{\ref{inst-pl-wa}}\orcit{0000-0002-9658-6151}
\and
Shay~Zucker\inst{\ref{inst-il-ta2}}\orcit{0000-0003-3173-3138}
\and
Laurent~Eyer\inst{\ref{inst-ch-obs}}\orcit{0000-0002-0182-8040}
}
\authorrunning{Rimoldini et al.}

\institute{Department of Astronomy, University of Geneva, Chemin d'Ecogia 16, 1290 Versoix, Switzerland\label{inst-ch-eco}
\and
Department of Astronomy, University of Geneva, Chemin Pegasi 51, 1290 Versoix, Switzerland\label{inst-ch-obs}
\and
RHEA for European Space Agency (ESA), Camino bajo del Castillo, s/n, Urbanizacion Villafranca del Castillo, Villanueva de la Cañada, 28692 Madrid, Spain\label{inst-es-rhea}
\and
Institute of Astronomy, KU Leuven, Celestijnenlaan 200D, 3001 Leuven, Belgium\label{inst-be-ku}
\and 
Sednai Sàrl, Rue des Marbriers 4, 1204 Genève, Switzerland\label{inst-ch-sed}
\and 
Universit\'{e} de Caen Normandie, C\^{o}te de Nacre, Boulevard Mar\'{e}chal Juin, 14032 Caen, France\label{inst-fr}
\and
Institute of Astronomy, University of Cambridge, Madingley Road, Cambridge CB3~0HA, United Kingdom\label{inst-uk-ioa}
\and
Konkoly Observatory, Research Centre for Astronomy and Earth Sciences, E\"{o}tv\"{o}s Lor\'{a}nd Research Network (ELKH), MTA Centre of Excellence, Konkoly Thege Mikl\'os \'ut 15-17, 1121 Budapest, Hungary\label{inst-hu-obs}
\and
ELTE E\"{o}tv\"{o}s Lor\'{a}nd University, Institute of Physics, P\'{a}zm\'{a}ny P\'{e}ter s\'{e}t\'{a}ny 1A, 1117 Budapest, Hungary\label{inst-hu-mta}
\and
INAF -- Osservatorio Astrofisico di Torino, Via Osservatorio 20, 10025 Pino Torinese, Italy\label{inst-it-to}
\and
INAF -- Osservatorio di Astrofisica e Scienza dello Spazio di Bologna, Via Gobetti 93/3, 40129 Bologna, Italy\label{inst-it-bo}
\and
INAF -- Osservatorio Astrofisico di Catania, Via S. Sofia 78, 95123 Catania, Italy\label{inst-it-ct1}
\and
European Space Agency (ESA), European Space Astronomy Centre (ESAC), Camino bajo del Castillo, s/n, Urbanizacion Villafranca del Castillo, Villanueva de la Cañada, 28692 Madrid, Spain\label{inst-es-esa}
\and
School of Physics and Astronomy, Tel Aviv University, Tel Aviv 6997801, Israel\label{inst-il-ta1}
\and
Lohrmann Observatory, Technische Universität Dresden, Mommsenstraße 13, 01062 Dresden, Germany\label{inst-de-dre}
\and
Astronomical Observatory, University of Warsaw, Al.~Ujazdowskie 4, 00-478 Warszawa, Poland\label{inst-pl-wa}
\and
Department of Physics and Astronomy, University of Catania, Via S. Sofia 64, 95123 Catania, Italy\label{inst-it-ct2}
\and
University of Vienna, Department of Astrophysics, Tuerkenschanz\-strasse 17, 1180 Vienna, Austria\label{inst-at-vie}
\and
INAF -- Osservatorio Astronomico di Capodimonte, Via Moiariello 16, 80131 Napoli, Italy\label{inst-it-na}
\and
Telespazio Vega UK Ltd for ESA/ESAC, Camino bajo del Castillo, s/n, Urbanizacion Villafranca del Castillo, Villanueva de la Ca\~{n}ada, 28692 Madrid, Spain\label{inst-es-tel}
\and
Porter School of the Environment and Earth Sciences, Tel Aviv University, Tel Aviv 6997801, Israel\label{inst-il-ta2}
}

\date{Received 30 November 2022 / Accepted 17 December 2022}

 
  \abstract 
   {\gdr3 contains 1.8~billion sources with \g-band photometry, 1.5~billion of which with \bp and \rp photometry, complemented by positions on the sky, parallax, and proper motion. The median number of field-of-view transits in the three photometric bands is between 40 and 44 measurements per~source and covers 34~months of data collection.}
   {We pursue a classification of Galactic and extra-galactic objects that are detected as variable by \gaia across the whole sky.}
   {Supervised machine learning (eXtreme Gradient Boosting and Random Forest) was employed to generate multi-class, binary, and meta-classifiers that classified variable objects with photometric time series in the \g, \bp, and \rp bands.}
   {Classification results comprise 12.4~million sources (selected from a much larger set of potential variable objects) and include about 9~million variable stars classified into 22 variability types in the Milky Way and nearby galaxies such as the Magellanic Clouds and Andromeda, plus thousands of supernova explosions in distant galaxies, 1~million active galactic nuclei, and almost 2.5~million galaxies. The identification of galaxies was made possible by the artificial variability of extended objects as detected by \gaia, so they were published in the \texttt{galaxy\_candidates} table of the \gdr3 archive, separate from the classifications of genuine variability (in the \texttt{vari\_classifier\_result} table). The latter contains 24 variability classes or class groups of periodic and non-periodic variables (pulsating, eclipsing, rotating, eruptive, cataclysmic, stochastic, and microlensing), with amplitudes from a few milli-magnitudes to several magnitudes.}
  {}


\maketitle

\section{Introduction\label{sec:introduction}}

Time-dependent brightness variations of celestial objects may be caused by different phenomena: intrinsic physical changes, such as pulsations, eruptions, and cataclysmic outbursts, or extrinsic reasons that depend on the direction of observation, such as eclipsing binaries, stars rotating with spots or with ellipsoidal shapes, and microlensing events, 
 as shown in fig.~1 of \citet{2019A&A...623A.110G}.
The detection of variability requires multi-epoch observations and, depending on the signal sampling, a certain set of classes can be identified.  \gaia's sparse sampling allows for the detection of periodic signals ranging from minutes to years and for medium to long-term non-periodic variability. The chance of detection of up to approximately six-week long transient phenomena, for example, crucially depends on the sampling at a given location in the sky \citep[see appendix~A in][]{2017arXiv170203295E}, which follows from the scanning law properties \citep{2016A&A...595A...1G}. Although the scanning law of \gaia was designed for astrometric goals, it allows for the identification of a broad variety of variability types, with different possible levels of completeness \citep{DR3-DPACP-162}.  

As the time span of \gaia data collection progressively increased from \gaia data release~1 (DR1) to DR2 and DR3 \citep[14, 22, and 34 months, respectively;][]{2016A&A...595A...2G,2018A&A...616A...1G,DR3-DPACP-185}, the classified variability types increased from Cepheids and RR\,Lyrae stars in a limited region of the sky \citep[in DR1;][]{2017arXiv170203295E}, to an all-sky\footnote{Due to the scanning law coverage, only from \gaia~DR2 onwards were sufficient epochs available at all locations on the sky, for example, see fig.~1 in \citet{2018A&A...618A..30H}.} classification of the DR1 classes plus long-period variables and $\delta$\,Scuti or SX\,Phoenicis stars \citep[in DR2;][]{2019A&A...625A..97R}, and 20 further variability classes in~DR3 (presented in this article and listed in Sect.~\ref{sssec:classes}). 
For brevity, we refer to Table~\ref{tab:training} for selected publications related to these classes, with representatives identified in various surveys. 

Machine learning is a practical tool to automate classification tasks that involve multiple known classes and a possibly high number of attributes to identify such classes and distinguish them from others \citep[e.g.\ see][]{2007A&A...475.1159D,2009A&A...494..739S,2010ApJ...713L.204B,2011ApJ...733...10R,2011MNRAS.414.2602D}. 
Herein, we present how a supervised classification was applied to \gaia~DR3 data to classify variable sources into two dozen classes (plus galaxies). 
In particular, we describe the details concerning the construction of the training set and of the classifiers, the verification of the results, and the generation of an overall classification score. Selection procedures, parameter distributions, and assessments of candidates are presented for each class.

Some of the classification results are further processed by specific object studies~(SOSs) dedicated to single classes, typically describing a subset of the most reliable candidates in detail. Such single-class processing modules are available in DR3 for active galactic nuclei \citep[AGNs][]{DR3-DPACP-167}, Cepheids \citep{DR3-DPACP-169}, compact companions \citep{DR3-DPACP-174}, eclipsing binaries \citep{DR3-DPACP-170}, long-period variables \citep{DR3-DPACP-171}, main-sequence oscillators \citep{DR3-DPACP-79}, planetary transits \citep{DR3-DPACP-181}, and RR\,Lyrae stars \citep{DR3-DPACP-168}. Other SOS modules, such as microlensing events \citep{DR3-DPACP-166}, short-timescale variables \citep[see sect.~10.12 of the \gdr3 documentation;][]{2022gdr3.reptE..10R}, and solar-like rotation modulation stars \citep{DR3-DPACP-173}, were executed independently of the classification results, as they relied on their own candidate selection. A summary of the variability results from all modules is presented in \citet{DR3-DPACP-162}.

This article is organised as follows. The classification input data are outlined in Sect.~\ref{sec:data}; the preparation, application, and verification of supervised learning procedures are described in Sect.~\ref{sec:method}; the results for each class are presented in Sect.~\ref{sec:results}; and conclusions are drawn in Sect.~\ref{sec:conclusions}.
Special training selections applied to a subset of classes are detailed in Appendix~\ref{app:training}; 
selected classification attributes are listed in Appendix~\ref{app:attributes}; 
additional class labels from the literature (among the false positive classes listed in Table~\ref{tab:results_details}) are defined in Appendix~\ref{app:FP_labels};
some examples of queries to facilitate the exploitation of classification results in the \gaia archive\footnote{\url{https://gea.esac.esa.int/archive/}} are provided in Appendix~\ref{app:queries}; 
and common diagrams for all classes, including a summary of trained and classified sources, an assessment of the results with respect to the literature, and sample light curves, are presented in Appendix~\ref{app:plots}. 
All table names in the \gaia archive that are mentioned in the text assume the prefix \texttt{gaiadr3} (as shown in Appendix~\ref{app:queries}).


\section{Data\label{sec:data}}

As part of the \gaia variability pipeline \citep{DR3-DPACP-162}, the general classification module received -- as input -- sources with photometric time series in the \g, \bp, and \rp bands \citep{2021A&A...649A...3R} that had at least five field-of-view~(FoV) measurements in the \g band, which were already identified as potential variable sources and characterised by basic statistics and periodicity parameters.
Before any computation, sources and associated epoch FoV~transits were processed by the chain of operators described in sect.~10.2.3 of the \gdr3 documentation \citep{2022gdr3.reptE..10R} and sect.~3.1 of \citet{DR3-DPACP-162}, which selected, transformed, and cleaned time series from spurious or doubtful observations. The balance between outlier removal and signal preservation favoured the latter, considering that some of the targeted variability types relied on a small number of outlier-like measurements (such as Algol-type eclipsing binaries and microlensing events). 
All time series and derived statistical numbers hereafter refer to these cleaned time series.
The median number of FoV~measurements in the three photometric bands is between 40 and 44 per~source \citep{DR3-DPACP-162}, within a time span of typically 900--1000~days in the \g band.

While the processing of \gaia (early)\,DR3 photometry included significant calibration improvements with respect to DR2 \citep{2021A&A...649A...3R}, some low-level uncalibrated systematic effects remained and their impact on epoch photometry are described in \citet{DR3-DPACP-142}.
Among instrumental effects, scan-angle dependent signals were induced mainly by asymmetric extended sources (such as barred spiral galaxies and tidally distorted stars) and multiple close pairs ($\lesssim 1\arcsec$) of point-like sources \citep{DR3-DPACP-164}. 
Although such signals helped the identification of galaxies from photometric variations, in general data artefacts might interfere with the correct identification of classes with genuine variability, especially those associated with low signal-to-noise ratios.

The classification of variables employed also astrometrically derived parameters such as parallax and proper motion \citep{2021A&A...649A...2L}. However, \gdr3 astrophysical parameters \citep{2022arXiv220606138A,DR3-AP1,DR3-AP3,DR3-AP2} could not be included as they were processed in parallel and became available after the results of the variability pipeline were finalised.

A subset of classified sources were analysed in more detail by subsequent SOS modules, typically focusing on specific classes, as mentioned in Sect.~\ref{sec:introduction}. The results of all variability modules were subject to additional source filtering before their ingestion into the public \gaia archive \citep{DR3-CU9}.
Statistical parameters of all the photometric time series published in \gdr3 are available in the \texttt{vari\_summary} table.


\section{Method\label{sec:method}}

For \gdr3, general classification relied on supervised machine learning, that is, training classifiers with sources of known variability types and applying the resulting models to classify sources of an unknown variability type. 
Known variables in the literature are cross-matched with \gaia sources, verified, selected, and characterised by attributes derived from the \gaia data. The use of both cross-match sources and (optimised) classification attributes for training was described in \citet{2019A&A...625A..97R} and it is not repeated herein.

An extensive cross-match of \gaia sources was compiled by \citet{DR3-DPACP-177}, which provided millions of variable objects from the literature and represented over 100~variability types. 
The robustness of the cross-match method, which included astrometric and photometric information in the identification of matches, and the verification of the genuineness of literature classifications ensured the reliability of training sources (critical to supervised classification) and of the validation of the results.


\subsection{Training set\label{ssec:training}}

Potential training sources from literature were vetted for each class to ensure the correct class membership. This was repeated for every catalogue that was deemed trustable for training the class under investigation.
The reliance of supervised classification on known objects makes it vulnerable to biases from the literature, for instance, related to their data acquisition and classification methods.
Thus, in addition to class verification, the cross-matched objects were probed in several dimensions to identify intrinsic biases, such as limited sky coverage or apparent magnitude range with respect to the ones of \gaia, in order to prevent (or minimise) the transfer of literature selection functions to the \gaia classifications.

\subsubsection{Published classes\label{sssec:classes}}

Since it was difficult to know a priori the full list of classes that could be identified in \gdr3, the verification of literature classifications and source selection for training purposes were performed for every variability type defined in \citet{DR3-DPACP-177}. However, only the actions on classes relevant to the published results are presented herein.

The published variability classes, corresponding acronyms, including the types trained within class groups or a (non-comprehensive) list of sub-types for some classes, are presented as follows:
\begin{enumerate}
  \item $\alpha^2$\,Canum Venaticorum (ACV) or (magnetic) chemically peculiar (MCP, CP) or rapidly oscillating Am- and Ap-type (ROAM and ROAP) or SX\,Arietis (SXARI) star (collectively denominated as ACV|CP|MCP|ROAM|ROAP|SXARI); 
  \item $\alpha$\,Cygni-type star (ACYG);
  \item AGN, from the perspective of variability, the general term AGN is favoured with respect to a quasar (or QSO), as brightness variability is caused by the activity of galactic nuclei, from processes in the accretion disc around a supermassive black hole (such as in Seyfert galaxies and QSOs), which can lead to the formation of relativistic plasma jets (identified as blazars when directed towards us);  
  \item $\beta$\,Cephei variable (BCEP);
  \item B-type emission line (BE) star or $\gamma$\,Cassiopeiae (GCAS) or S\,Doradus (SDOR) or Wolf-Rayet (WR) star (denoted as BE|GCAS|SDOR|WR);
  \item Cepheid (CEP), including anomalous Cepheid (ACEP), BL\,Herculis variable (BLHER, also known as CWB), W\,Virginis variable (CW), $\delta$~Cephei star (DCEP), RV\,Tauri-type star (RV), and generic type~II Cepheid (T2CEP);
  \item cataclysmic variable (CV), excluding supernova and symbiotic star (mentioned separately);
  \item $\delta$\,Scuti (DSCT) or $\gamma$\,Doradus (GDOR) or SX\,Phoenicis (SXPHE) star, including hybrid $\delta$\,Scuti\,+\,$\gamma$\,Doradus (DSCT+GDOR) stars (labelled as DSCT|GDOR|SXPHE);
  \item eclipsing binary (ECL), including Algol ($\beta$\,Persei) type (EA), $\beta$\,Lyrae type (EB), and W\,Ursae Majoris type (EW); 
  \item ellipsoidal variable (ELL); 
  \item star with exoplanet transits (EP); 
  \item long-period variable (LPV), including long secondary period variable (LSP), Mira ($o$\,Ceti) type (M), Mira or semi-regular variable (M|SR), OGLE small amplitude red giant (OSARG), small amplitude red giant (SARG), and semi-regular variable (SR) of sub-types SRA, SRB, SRC, SRD, and SRS; 
  \item microlensing event (MICROLENSING); 
  \item R\,Coronae Borealis variable (RCB); 
  \item RR\,Lyrae star (RR), including fundamental-mode (RRAB), first overtone (RRC), double-mode (RRD) RR\,Lyrae star (and anomalous double-mode, ARRD);
  \item RS\,Canum Venaticorum variable (RS); 
  \item short-timescale object (S); 
  \item subdwarf B~star (SDB) of type V1093\,Herculis (V1093HER) or V361\,Hydrae (V361HYA);
  \item supernova (SN); 
  \item solar-like star (SOLAR\_LIKE), including BY\,Draconis type (BY), rotating spotted star (ROT), BY and/or ROT star (BY|ROT), and flaring star (FLARES); 
  \item slowly pulsating B-type variable (SPB); 
  \item symbiotic variable star (SYST), including Z\,Andromedae type (ZAND); 
  \item variable white dwarf (WD), including a generic class (ZZ), objects of spectral type DA, such as ZZ\,Ceti (known as ZZA or DAV), DB, such as V777\,Herculis (known as ZZB, DBV, V777HER, GD358), or DO, such as GW\,Virginis pre-white dwarf (comprising ZZO, DOV, GWVIR, PG1159); extremely low mass and hot ZZ\,Ceti variables were labelled as ELM\_ZZA and HOT\_ZZA, respectively; 
  \item young stellar object (YSO), including dipper stars (DIP), eruptive YSOs such as FU\,Orionis type variables (FUOR), pulsating pre-main-sequence stars (PULS\_PMS), Herbig Ae or Be types (HAEBE), including UX\,Orionis stars (UXOR), and T\,Tauri star (TTS), among which classical (CTTS), weak-lined (WTTS), and late~G to early~K type pre-main-sequence (GTTS) stars. \label{itm:yso} 
\end{enumerate}
A brief definition of the 24 variability class labels is presented also in the  \texttt{vari\_classifier\_class\_definition} table of the \gaia archive.
These labels identify the best classification of a variable source in the field \texttt{best\_class\_name} of table \texttt{vari\_classifier\_result}.

In addition, galaxies (labelled as GALAXY) are classified not because they are intrinsically variable, but because they appear to be photometrically variable in the \gaia processing (see Sect.~\ref{sec:data}).
To prevent misinterpretation, galaxies are published exclusively in the extra-galactic \texttt{galaxy\_candidates} table, together with results on galaxies from other \gaia pipelines \citep{DR3-CU4,DR3-AP3}.

\subsubsection{Source selection\label{sssec:selection}}

For each class in \citet{DR3-DPACP-177}, two sets of sources were selected: one for training and one for testing purposes, generally with no sources in common, except for classes that were insufficiently represented. 
Among the trained types, sub-types, and possible combinations thereof that were part of the published classes (listed in Sect.~\ref{sssec:classes} and in Table~\ref{tab:training}), the ones associated with the following labels (in alphabetical order) were represented by less than about 500--600 (with a median of 86) sources, after the selections described in the subsequent paragraphs of this section: ACEP, ACV, ACYG, ARRD, BCEP, BY|ROT, CP, CTTS, DIP, DSCT+GDOR, ELM\_ZZA, EP, FLARES, FUOR, GTTS, GWVIR, HAEBE, HOT\_ZZA, MICROLENSING, PULS\_PMS, RCB, ROAM, ROAP, RV, SDOR, SN, SPB, SRA, SRC, SRD, SXARI, SXPHE, SYST, T2CEP, UXOR, V1093HER, V361HYA, V777HER, WR, WTTS, ZAND, ZZ, and ZZA.
In such cases, higher priority was given to the quality of results (by training with all known objects) rather than to their assessment, so the independence of training and test sets of poorly represented types was not pursued. 

For both training and test sets, up to six subsets of sources were created with different sample sizes (up to about 500 for the first one and then approximately 1000, 2000, ..., 5000, depending on the available number of sources per class), in order to have, for a given class, a set of the chosen size ready for use, while preserving the representation of the full magnitude range and sky coverage (see Sect.~\ref{sssec:trainingclasses}). 

General conditions and selection procedures were applied to all sources of known variables from the cross-match  with \gaia (with a few exceptions), as follows.
\begin{enumerate}
    \item At least five~FoV transit observations in the \g~band.
    \item Due to the importance of colour information, at least one observation in \bp or \rp was required (at this stage, independently of the trained attributes described in Sect.~\ref{sssec:attributes}).
    \item The distribution of angular distances of the cross-match was verified for each class and catalogue: potential spurious matches at significantly higher angular distances than most were removed, unless their correlation with parallax or median \g-band magnitude suggested nearby objects. Selections beyond simple angular distance thresholds are detailed in Appendix~\ref{app:training}.
    \item The minimum standard deviation versus median magnitude in the \g~band was generally set to the third quartile of the standard deviations of 1.6~million reference sources binned in 0.05\,mag intervals (see Appendix~\ref{app:plots}), except for a small number of classes such as planetary transits and $\gamma$\,Doradus stars.
    \item For catalogues whose distribution of sources in the sky had over- or under-represented regions due to, for example, limitations of ground-based observations or special targeted regions (and not because of the natural distribution of such objects), sources were sampled to prevent or minimise biases resulting from these limitations. 
    The same procedure was also applied to classes that were extremely rich with respect to the few thousand sources per class needed for training (not to overwhelm the identification of rare classes trained with only a few tens or hundreds of sources). In particular, the sky was subdivided into 3072 Hierarchical Equal Area iso-Latitude Pixelization \citep[HEALPix;][]{2002ASPC..281..107G} pixels (corresponding to a resolution of 4, with mean angular spacing of 3.6645\,deg)\footnote{\url{https://lambda.gsfc.nasa.gov/toolbox/pixelcoords.html}} and the $k$ nearest neighbours (with $k$ depending on catalogue) to each HEALPix centre were identified within a radius of 5\,degrees. 
    This radius allowed for sources to be drawn from nearby HEALPix as well, in order to avoid gaps or distributions reflecting single HEALPix boundaries. Sources were then randomly selected from such $k$~neighbours according to the desired downsampling and possible duplicated sources were removed.
    \item The steep increase in the number of objects towards faint magnitudes can make the bright end unnoticeable by classifiers, so sources were sampled in median \g magnitude to improve the detectability of objects at the bright end, while keeping the full range of represented magnitudes, for each class with more than 500 sources. 
    In particular, the magnitude distribution was binned in 100~intervals and unique sources were randomly sampled up to a maximum number per bin. The latter depended linearly on the magnitude  by a custom (positive) coefficient and offset per class. 
    These coefficient and offset were adjusted for each of the sub-sample sizes introduced earlier (for training and testing).
    This procedure made it possible to retain the brightest and the faintest objects, as well as an indication of the most represented magnitudes, with a supplementary option to keep full catalogues of special relevance before adding sources from other catalogues.
\end{enumerate} 

Additional per-class selections were applied in specific cases to ensure genuine class representation (see Appendix~\ref{app:training}).
The distribution of training sources in various diagrams is presented in Appendix~\ref{app:plots}, for each published class, as listed in Sect.~\ref{sssec:classes}.

\subsubsection{Training classes\label{sssec:trainingclasses}}

Commonly confused or physically related (sub-)types were combined as listed in the second column of Table~\ref{tab:training}, where the values in parentheses indicate an approximate number of representatives or an upper limit when fewer sources were available (see Sect.~\ref{sssec:selection}). 
In general, larger sample sizes were assigned to more frequent types, although not with a realistic  occurrence relative to other types (as known from the literature), otherwise (unweighted) decision-tree based classifiers might optimise models that neglected rare types. As some types overlapped in part with other ones (such as CP and MCP), some of the sets had sources in common, thus duplicated sources were removed. For one per~cent of the training set, some of the literature classifications were in conflict with other classes (represented in different class groups). In such cases, duplicated sources that were part of generic class groups (such as S and SOLAR\_LIKE) were removed in favour of specific classes and the remaining few hundred sources in conflict (none of which belonged to rare classes) were removed from the training set. Special objects that were wished not to be missed, such as class prototypes, were added to the training set for several types.
A similar procedure (except for the addition of special objects) was followed for the test set.

Classes that were trained, classified, but not published in \gdr3 (following the verification filters mentioned in Sect.~\ref{ssec:verification}), included blue large-amplitude pulsators \citep[BLAP,][]{2017NatAs...1E.166P}, FK\,Comae Berenices-type variables, heartbeat stars, high mass X-ray binaries, poorly studied irregular variables, post-common envelope binaries (or pre-cataclysmic variables), protoplanetary nebulae embedding yellow supergiant post-AGB stars, PV\,Telescopii-type variables, strong reflection (re-radiation) in close binary systems, UV\,Ceti stars, general sources with variable X-ray emission, ZZ\,Leporis stars, and a selection of constant objects from \textit{Hipparcos} \citep{1997ESASP1200.....E}, Optical Gravitational Lensing Experiment-IV \citep[OGLE-IV;][]{2012AcA....62..219S}, Sloan Digital Sky Survey \citep[SDSS;][]{2007AJ....134..973I}, Transiting Exoplanet Survey Satellite \citep[TESS;][]{2015JATIS...1a4003R}, and Zwicky Transient Facility \citep[ZTF;][]{2014htu..conf...27B}, from the cross-match of \citet{DR3-DPACP-177}; the class of constant sources was introduced to clean the classification of variables from false variability detections. 
Such omitted classes are not listed in Table~\ref{tab:training} to prevent false expectations.

\begin{table*}
\caption[Training catalogues]{Classification training classes (see Sect.~\ref{sssec:classes} for class label definitions), with the specification of their components (whose approximate representation is indicated in brackets when greater than~500; see Sects.~\ref{sssec:selection} and~\ref{sssec:trainingclasses} for details on the creation of class subsets), the number of training sources $N_{\mathrm{TRN}}$, and references. The class group ACV|CP|MCP|ROAM|ROAP|SXARI is abbreviated as ACV|CP|...|SXARI.\label{tab:training}} 
\centering                  
\begin{tabular}{llrl@{}}     
\hline\hline & \\[-2.0ex]                 
Class (group) & Type or sub-type \citep[see][]{DR3-DPACP-177} & {$N_{\mathrm{TRN}}$} & Reference \\
\hline \\[-1.5ex]
ACV|CP|...|SXARI & ACV, CP, MCP (2000), ROAM, ROAP, SXARI & 1572 & 1--10\\
ACYG & ACYG & 59 & 3, 10 \\
AGN & QSO (3000) & 3089 & 11, 12 \\
BCEP & BCEP & 173 & 3, 10, 13, 14 \\
BE|GCAS|SDOR|WR & BE (3000), GCAS (1000), SDOR, WR & 3546 & 3, 6, 9, 10, 15, 16 \\
CEP & ACEP, BLHER, CEP (1000), CW, DCEP (1000), & 4448 & 3, 6, 7, 10, 17--28 \\
 & RV, T2CEP &  &  \\
CV & CV (2000) & 1815 & 6, 10, 29--33 \\
DSCT|GDOR|SXPHE & DSCT, DSCT+GDOR, DSCT|SXPHE (1000), & 4259 & 3, 6, 7, 10, 18, 21, 25, 34--46   \\
                & GDOR (1000), SXPHE  &  &  \\
ECL & EA (2000), EB (2000), ECL (2000), EW & 6360 & 3, 6, 7, 10, 18, 21, 25, 35, 40, 47--52 \\
ELL & ELL (3000) & 2864 & 6, 10, 18, 25, 50, 52 \\
EP & EP & 66 & 53 \\
LPV & LPV, LSP, M, M|SR, OSARG, SARG, SR, SRA, & 5353 & 3, 6, 7, 9, 10, 18, 21, 25, 35, 38, 40, \\
    & SRB, SRC, SRD, SRS &  & 42, 47, 54--65 \\
MICROLENSING & MICROLENSING & 116 & 30, 66 \\
RCB & RCB & 69 & 6, 10, 30, 67 \\
RR & ARRD, RRAB (3000), RRC (2000), RRD (1000) & 6377 & 3, 6, 10, 17, 18, 21, 23, 25, 26, 35, \\
   &  &  & 38, 40, 44, 47, 68--85 \\
RS & RS (3000) & 2548 & 3, 10, 18, 35, 86  \\
S & S (2000) & 1965 & 10, 88, 89 \\
SDB & V1093HER, V361HYA & 62 & 6, 10  \\
SN & SN & 86 & 30  \\
SOLAR\_LIKE & BY (1000), BY|ROT, FLARES, & 2628 & 2, 3, 10, 38, 47, 90--105 \\
            & ROT (1000), SOLAR\_LIKE (1000) &  &  \\
SPB & SPB & 149 & 3, 10, 13, 106  \\
SYST & SYST, ZAND & 316 & 6, 10, 107 \\
WD & ELM\_ZZA, GWVIR, HOT\_ZZA, V777HER, ZZ, ZZA & 1075 & 6, 10, 108--119 \\
YSO & CTTS, DIP, FUOR, GTTS, HAEBE, PULS\_PMS, & 5148 & 10, 30, 120, 121 \\
    & TTS (1000), UXOR, WTTS, YSO (4000) &  &  \\
\hline & \\[-2ex]
GALAXY & GALAXY (3000) & 3116 & 17 (only galaxies), 122 \\
\hline                       
\end{tabular}
\tablebib{
(1)~\citet{2015A&A...581A.138B};
(2)~\citet{2019MNRAS.487.3523C};
(3)~\citet{1997ESASP1200.....E};
(4)~\citet{2019MNRAS.488...18H};
(5)~\citet{2018A&A...619A..98H};
(6)~\citet{2018MNRAS.477.3145J,2019MNRAS.486.1907J,2019MNRAS.485..961J};
(7)~\citet{2002AcA....52..397P};
(8)~\citet{2009A&A...498..961R};
(9)~\citet{2012ApJS..203...32R};
(10)~\citet[][VSX version 2019-11-12]{2006SASS...25...47W};
(11)~\citet{EDR3-DPACP-133};
(12)~\citet{2013yCat.1323....0M};
(13)~De~Cat \citep[catalogue COMP\_SPB\_BCEP\_DECAT\_PR in][]{DR3-DPACP-177};
(14)~\citet{2005ApJS..158..193S};
(15)~\citet{2002A&A...393..887M};
(16)~\citet{2005MNRAS.361.1055S};
(17)~\citet{2019A&A...622A..60C};
(18)~\citet{2014ApJS..213....9D};
(19)~M\,31 \citep[catalogue GAIA\_M31\_CEP\_GAIA\_2018 in][]{DR3-DPACP-177};
(20)~\citet{2009A&A...495..249M};
(21)~\citet{2013AJ....146..101P}; 
(22)~\citet{2011ApJS..193...26P};
(23)~\citet{2019A&A...625A..14R};
(24)~\citet{2008AcA....58..163S,2008AcA....58..293S,2010AcA....60...17S,2010AcA....60...91S,2011AcA....61..285S,2015AcA....65..297S,2020AcA....70..101S};
(25)~\citet{2012AcA....62..219S};
(26)~\citet{2017AcA....67..297S};
(27)~\citet{2018AcA....68..315U};
(28)~Zak \citep[catalogue GAIA\_CEP\_ZAK\_2018 in][]{DR3-DPACP-177};
(29)~\citet{2014MNRAS.441.1186D};
(30)~\gaia science alerts \citep[catalogue GAIA\_TRANSIENTS\_ALERTS\_2019 in][]{DR3-DPACP-177};
(31)~\citet{2015AcA....65..313M};
(32)~\citet{2003A&A...404..301R};
(33)~\citet{2011AJ....142..181S};
(34)~\citet{2015AJ....149...68B};
(35)~\citet{2020ApJS..249...18C}; 
(36)~De~Ridder \citep[catalogue GAIA\_DR2\_CLASS\_DSCT\_SXPHE\_SELECTION in][]{DR3-DPACP-177};
(37)~\citet{2007A&A...475.1159D,2011A&A...529A..89D};
(38)~\citet{2016AcA....66..197H}; 
(39)~\citet{2016MNRAS.458.2307K};
(40)~\citet{2009AcA....59...33P}; 
(41)~\citet{2010AcA....60....1P};
(42)~\citet{2019A&A...625A..97R}; 
(43)~\citet{2013A&A...550A.120S};
(44)~\citet{2012MNRAS.424.2528S}; 
(45)~\citet{2011A&A...534A.125U};
(46)~\citet{2015ApJS..218...27V};
(47)~\citet{2017MNRAS.469.3688D}; 
(48)~\citet{2016AJ....151...68K};
(49)~\citet{2013AcA....63..323P};
(50)~\citet{2016AcA....66..421P}; 
(51)~Rybizki \citep[catalogue GAIA\_ECL\_RYBIZKI\_2018 in][]{DR3-DPACP-177};
(52)~\citet{2016AcA....66..405S}; 
(53)~\citet[][updated list available at \url{https://www.astro.keele.ac.uk/jkt/tepcat/html-catalogue.html}]{2011MNRAS.417.2166S};
(54)~\citet{2012A&A...548A..79A};
(55)~\citet{2001A&A...369..178B};
(56)~\citet{2019A&A...625A.151B};
(57)~\citet{2007A&A...473..143D};
(58)~\citet{2018AJ....156..241H};
(59)~\citet{2014A&A...566A..43K};
(60)~\citet{2019A&A...626A.112M};
(61)~\citet{2004ApJS..154..623M};
(62)~\citet{2009AcA....59..239S,2011AcA....61..217S,2013AcA....63...21S};
(63)~\citet{2011A&A...536A..60S};
(64)~\citet{2017JKAS...50..131S};
(65)~\citet{2004AJ....128.2965W};
(66)~Kruszy{\'n}ska \citep[catalogue COMP\_MICROLENSING\_GAIA in][]{DR3-DPACP-177};  
(67)~\citet{2009AcA....59..335S};
(68)~\citet{2014MNRAS.441.1230A};
(69)~\citet{2006MNRAS.372.1657B};
(70)~\citet{2013AJ....146...94B};
(71)~\citet{2016AJ....152..170B};
(72)~\citet{2006AJ....132.1014C,2008AJ....135.1459C};
(73)~\citet{2013ApJ...763...32D,2013ApJ...765..154D}; 
(74)~\citet{2006ApJ...653L.109D,2012ApJ...752...42D};
(75)~\citet{2013ApJ...767...62G};
(76)~Garofalo \citep[catalogue GAIA\_RRL\_GAROFALO\_SELECTION in][]{DR3-DPACP-177};
(77)~\citet{2006AJ....132.1202K};
(78)~\citet{2009ApJ...695L..83M,2012ApJ...756..121M};
(79)~\citet{2002AJ....124..949P,2003AJ....126.1381P};
(80)~\citet{2014ApJ...793..135S,2017AJ....153..204S};
(81)~\citet{2006ApJ...649L..83S};
(82)~\citet{2015A&A...573A.103S};
(83)~\citet{2009AcA....59....1S,2010AcA....60..165S,2011AcA....61....1S,2014AcA....64..177S,2016AcA....66..131S,2019AcA....69..321S};
(84)~\citet{2015MNRAS.446.2251T};
(85)~\citet{2009MNRAS.398.1757W};
(86)~\citet{2008MNRAS.389.1722E};
(88)~\citet{2018A&A...620A.197R};
(89)~\citet{2011AJ....142..160S};
(90)~\citet{2015ApJ...814...35C};
(91)~\citet{2013A&A...555A..63D};
(92)~Distefano \citep[catalogues GAIA\_ROT\_GAIA\_2017, GAIA\_BY\_DISTEFANO\_2019, KEPLER\_GAIA\_BY\_ROT\_DISTEFANO\_2020, and TESSGAIA\_BY\_ROT\_DISTEFANO\_2020 in][]{DR3-DPACP-177};
(93)~\citet{2006AJ....131.1044D};
(94)~\citet{2010MNRAS.408..475H};
(95)~\citet{2016ApJ...831...27H};
(96)~\citet{2018A&A...616A..16L};
(97)~\citet{2010A&A...520A..79M};
(98)~\citet{2007A&A...469..713M};
(99)~\citet{2010A&A...520A..15M,2011A&A...532A..10M};
(100)~\citet{2015A&A...583A..65R};
(101)~\citet{2013ApJS..209....5S};
(102)~\citet{2019MNRAS.487.4695S};
(103)~\citet{2011AJ....141...50W};
(104)~\citet{2015ApJ...798...92W};
(105)~\citet{2017ApJ...835...61Z};
(106)~\citet{2003A&A...404..689N};
(107)~\citet{2019ApJS..240...21A};
(108)~\citet{1999ApJ...516..887B};
(109)~\citet{2020A&A...638A..82B};
(110)~\citet{2011ApJ...733L..19D};
(111)~\citet{2010ApJ...720L.159D};
(112)~\citet{2020svos.conf...11E};
(113)~\citet{2005ApJ...631.1100G};
(114)~\citet{2012ApJ...750L..28H,2013MNRAS.436.3573H,2013ApJ...765..102H};
(115)~\citet{2014MNRAS.442.2278K};
(116)~\citet{2013MNRAS.432.1632K};
(117)~\citet{2009ApJ...690..560N};
(118)~\citet{2007ApJS..171..219Q};
(119)~\citet{2016ApJ...817...27W};
(120)~\citet{2020MNRAS.496.3257B};
(121)~\citet{2020IAUS..345..378V};
(122)~Krone-Martins \citep[catalogue GAIA\_GAL\_GAIA\_2018 in][]{DR3-DPACP-177}.
}
\end{table*}

\subsubsection{Attributes\label{sssec:attributes}}

Classification attributes were used to characterise the light variations and the general properties of sources. About 60 attributes were generated, including time series statistics, photometric colours, astrometric parameters, periodicity indicators, combinations of photometric and astrometric quantities, comparisons of statistics and correlations between the \bp and \rp bands, and stochastic model parameters for AGNs \citep{2011AJ....141...93B}. Their effectiveness in the identification of variability classes was tested with Random Forest classifiers \citep{Breiman.Random.Forest}, which assessed the attribute usefulness with `out-of-bag' objects (unused for training).  Such classifiers were used to select attributes by adding the most useful one to the selection iteratively, until the reduction of the total error rate was insignificant. About 10\,\% of training sources per class (or 20\,\% for classes with less than 100~training objects and that were not merged with other types in the training set) were used in the identification of optimal attributes. 
The choices of method for attribute selection and of training-set downsampling were due to the necessity of computational efficiency, given the limited time available.

The final list of attributes and their definitions are presented in Appendix~\ref{app:attributes}. 
The adaptation of \citet{2011AJ....141...93B} parameters (\texttt{qso\_variability} and \texttt{non\_qso\_variability}) to the \gaia data is described in \citet{DR3-DPACP-167}.

The predictive power of classification attributes was limited by the unaccounted reddening of colours, extinguished magnitudes, and no priors on parallax (as a function of celestial location), in addition to the effects of residual outliers, artificial signals, and other photometric imperfections. 
 

\subsection{Classifiers\label{ssec:classifier}}

The numbers of classified variable sources and related classes are significantly higher in \gdr3 with respect to the previous data release. 
The richness of variability types was a major challenge and it was addressed with a multitude of classifiers, which provided different perspectives on the peculiarities of each type of variable object. 

Classifiers were trained with two machine learning algorithms, as implemented in the H2O platform \citep{h2o_platform}: Distributed Random Forest \citep[DRF;][]{Breiman.Random.Forest} and eXtreme Gradient Boosting \citep[XGBoost;][]{Chen:2016:XST:2939672.2939785}. Both methods had their pros and cons; DRF results seemed more robust with realistic posterior probability distributions, while XGBoost was more effective in the identification of rare and subtle classes (such as planetary transits). 

According to the attribute selection (Sect.~\ref{sssec:attributes}), optimal results were achieved with 300~trees and with the square root of the number of attributes as the number of randomly sampled attributes to test at each split of any DRF tree. The same number of trees was used for XGBoost too.

The following types of classifiers were trained:
\begin{enumerate}
    \item multi-class DRF and XGBoost classifiers with all classes at the same level (two versions, with minor updates to some class groups),
    \item binary DRF (unweighted and weighted) and XGBoost classifiers of one class versus all other classes,
    \item meta-classifiers that aimed to combine the best per-class results of the DRF and XGBoost methods, applied to both multi-class and binary classifiers,
    \item a two-stage classifier for main-sequence pulsating OBAF-type stars \citep{DR3-DPACP-79},
\end{enumerate}
which lead to over 100~classifiers in total, with up to 12~classifiers per class.
For some of the least represented classes, there were classifiers that could not recover any training sources, so they were excluded.
Also, results from all the available classifiers were not necessarily used for each class.

In the case of binary classifiers, DRF was more sensitive to the class imbalance than XGBoost, especially for classes that were two orders of magnitude less numerous than the rest of the training set. However, higher weights could be associated with the less represented classes, so DRF was executed also by weighting the targeted class such that the latter became as relevant as all other classes grouped into one, that is, equivalent to a binary classifier with two perfectly balanced classes.

In the two-stage classifier for main-sequence pulsating OBAF-type stars, the first stage separated 
ACV, BCEP, BE, CP, DSCT, GCAS, GDOR, MCP, PULS\_PMS, SPB, SXARI, and SXPHE stars (including variability types that may be confused with the pulsating OBAF-type stars, when using only \gaia data) from all other types. In the second stage, the first group of types was split into the targeted ones (BCEP, DSCT, GDOR, SXPHE, and SPB) versus the others.

As a result, for each source, only the combined solution of all these different classifiers is recorded in the \texttt{best\_class\_name} and \texttt{best\_class\_score} fields of the \texttt{vari\_classifier\_result} table (as explained in Sect.~\ref{ssec:verification}), and the \texttt{vari\_classifier\_definition} table lists only a single `combined’ classifier instead of the over 100~classifiers that were used to compile the classifications of all sources.

As a side note, the training set for the preceding module of general variability detection \citep[see sect.~10.2.3 of the \gdr3 documentation;][]{2022gdr3.reptE..10R} included the training set described in Sects.~\ref{sssec:selection} and~\ref{sssec:trainingclasses} (except for the constant objects) as a single `variable' class, for the targeted variable sources for \gdr3. In addition, a similar number of sources was selected from the 75\,\% least variable ones among 1.6~million reference sources binned in 0.05\,mag intervals (see Appendix~\ref{app:plots}), as the other class. A binary XGBoost classifier then identified variable objects according to a classifier probability greater than~0.5 (not published in \gdr3), which were subsequently processed by the over~100 classifiers of variability types.


\subsection{Verification and filtering of results\label{ssec:verification}}

After the automated execution of the classifiers (Sect.~\ref{ssec:classifier}), trained with the selected sources (Sect.~\ref{sssec:selection}) and attributes (Sect.~\ref{sssec:attributes}), the results of all classifiers for a given class were verified, alleviated from contaminants, and assessed. The classes whose value was deemed sufficient to be published (separately or merged with other classes) are listed in Sect.~\ref{sssec:classes}, while unconvincing results of other classes were excluded (Sect.~\ref{sssec:trainingclasses}).

Depending on class, different selective conditions were applied to the classifier results, such as minimum thresholds on the posterior probability to belong to a given class, adjustments of the minimum variability level, colour and/or magnitude cuts, conditions on astrometric and time series parameters, and limitations on environment crowding, among other restrictions, often guided by the distributions of known objects (among the classified ones) and of the candidates. The list of such filters and thresholds are presented for each class in Sect.~\ref{sec:results}, to help the interpretation of the results where they are described.  Exceptionally, following the feedback from SOS modules, a small fraction of sources (of particular interest) might override the general selection conditions (such as the special objects mentioned in Sect.~\ref{sssec:trainingclasses} or bona~fide sources with peculiar behaviour), provided they were classified correctly. 

Additional sources were removed following a further post-processing verification, typically involving suspect features for a very small fraction of candidates. 


\subsection{Classification score\label{ssec:class_score}}

After the verification filters, the remaining candidates were assigned a classification score, which was derived from the posterior probabilities of the classifiers used to identify the sources of a given class. 
Classifier posterior probabilities were not calibrated, so their values were not directly comparable, because of the use of different methods and classifier structures. 
In order to treat the posterior probabilities of all classifiers on the same footing for a given class, they were converted into normalised ranks (for each classifier), which could then be compared across different classifiers.

For each classifier, this normalised rank was computed by sorting posterior probabilities of a certain class in ascending order. The source rank depended on its location in this ordered list and identical probabilities corresponded to identical ranks. Denoting the number of different probabilities (that is, the number of sources with posterior probabilities of a given class minus that of sources with duplicated probability values for that class) by $N$, the possible score values were represented by the ranks of the probabilities, from the first to the $N$-th, normalised by $N$: \{1, 2, ..., $N$\}\,$/$\,$N$ (some of which were repeated in presence of duplicated probabilities). Thus, for each classifier, the score associated with a source of a given class ranged from a minimum value greater than zero to a maximum of one. 
The median of such normalised ranks from different classifiers, for a given source and class (\texttt{best\_class\_name}), was stored as \texttt{best\_class\_score} in the \texttt{vari\_classifier\_result} table. 
Exceptionally, for BCEP and WD, the normalised ranks from probabilities of the general variability detection classifier (Sect.~\ref{ssec:classifier}) were included in the score evaluation, when higher variability implied higher reliability.

Occasional gaps in the distribution of the classification score of some classes were created by the removal of sources associated with multiple classes (Sect.~\ref{ssec:overlaps}) and also of some questionable candidates, following the  post-processing verification.  The average and extreme values of the classification scores are presented in Table~\ref{tab:results} for each class.


\subsection{Multi-class source reduction\label{ssec:overlaps}}

After the verification filters (Sect.~\ref{ssec:verification}) and assignment of classification scores (Sect.~\ref{ssec:class_score}), the combination of all per-class candidates revealed about 110~thousand sources (less than 1\,\% of all classified sources) associated with multiple (usually two) classes. The vast majority of them were not due to genuine concurrent variability types (as it may happen, for example, in an eclipsing binary system with intrinsically variable components, perhaps with transiting exoplanets too). 
Multiple variability was not pursued in \gdr3, so a single class per source was enforced.

The following set of rules was applied to multi-class sources in order to reduce them to a single class per source.
\begin{enumerate}
    \item The least numerous classes would be unfairly reduced if their candidates were left to compete with those of the most numerous classes, so rare classes of ACYG, BCEP, BE|GCAS|SDOR|WR, EP, MICROLENSING, RCB, SDB, SN, SPB, SYST, and WD were safeguarded against any other alternative class. 
    \item\label{item:score-xm} For 1290~sources that remained with multiple classes after the application of all the other rules, the class corresponding to the highest classification score was kept. 
    \item Candidates of some classes were considered dispensable because they were byproducts of classification (such as galaxies), or related to SOS modules that did not rely on classification and were expected to provide higher quality candidates than a general classification (with expert procedures dedicated to a single class). Such classes included S, GALAXY, SOLAR\_LIKE, and RS, ordered from the most to the least dispensable class. 
    Dispensable classes were dropped from multi-class sources, provided at least one non-dispensable class per source remained. For sources with all dispensable multi-classes, only the least dispensable one was kept (following the above mentioned list ordered by decreasing dispensability).
    \item For the special objects (Sect.~\ref{sssec:trainingclasses}) with multiple classes, only the known class was kept, when available among the classifications, otherwise the standard treatment of other multi-class sources was followed.
\item For multi-class sources that were present also in at least one SOS module, different scenarios were possible: 
  \begin{enumerate}
        \item the classified classes included at least one class for which a corresponding SOS module did not exist or whose source was absent from the corresponding SOS module: 
        \begin{enumerate}
           \item all classes with the source absent from the corresponding SOS module, provided the latter existed, were removed
       (e.g.\ for a source classified as BE+AGN and present in the upper main-sequence oscillator SOS module but not in the AGN SOS one, the AGN class was removed);
           \item the classes with the source present in the corresponding SOS modules were kept and all other classes were removed (e.g.\ for a source classified as BE+AGN+RR and present only in the AGN SOS module, the BE and RR classes were removed); 
        \end{enumerate}         
        \item all classes had the source present in the SOS modules: classes unmatched by the corresponding SOS classes were removed, as long as at least one of the other classes matched an SOS class, otherwise item~(\ref{item:sos-similarity}) was applied;
        \item\label{item:sos-similarity} for all classes that had the source present in SOS modules but that did not match any of the SOS classes, the class matching criterion was extended to include the similarity of such classes with the following SOS modules, for partial validation:
        \begin{enumerate}
          \item BE|GCAS|SDOR|WR and GALAXY sources in the AGN SOS module,
          \item CEP sources in the RR\,Lyrae and the long-period variable SOS modules,
          \item ECL sources in the RR\,Lyrae SOS module,
          \item ELL and EP sources in the eclipsing binary SOS module,
          \item RR sources in the Cepheid or eclipsing binary SOS modules,
          \item S in the eclipsing binary or the RR\,Lyrae SOS modules,
          \item YSO in the rotational modulation SOS module;
    \end{enumerate}
    classes not similar to those of SOS modules were removed, as long as at least one of the other classes was similar to an SOS class, otherwise item~(\ref{item:score-xm}) was applied.
  \end{enumerate}
  The existence of SOS modules matching some of the multiple classes, but without the multi-class sources present in any of them, added no information to the selection rules (bona~fide candidates can be excluded from the corresponding SOS modules for various reasons, such as insufficient sampling for reliable model results).
\end{enumerate}


\subsection{Classification versus SOS modules\label{ssec:class-sos}}

Although SOS modules provided purer class samples than those from classification, the latter did not necessarily represent a superset of the corresponding SOS modules. 
As shown in Table~\ref{tab:results}, in addition to the `extra' sources in classification and the ones in `common' with respect to the SOS modules, some sources were classified as `other' classes \citep[see fig.~7 of][]{DR3-DPACP-162}, different from the expected SOS modules, and others were `missed' in the published classification results. The reasons for the `other' and `missed' classifications are listed in the following.
\begin{enumerate}
    \item Some SOS modules did not depend on classification but relied on special variability detection or extraction of candidates (as for microlensing events, short timescales, and solar-like rotation modulation).
    \item For the SOS modules that depended on classification:
    \begin{enumerate}
        \item they received input candidates before the multi-class source treatment described in Sect.~\ref{ssec:overlaps}, so sources could be assigned to `other' classes as a consequence of enforcing a single class per source;
        \item some of these SOS modules included classification candidates from multiple classes, due to similar features (such as long-period variables and symbiotic stars);
        \item given the advanced class-specific SOS verification, permissive classifier probability thresholds or their combination with other parameters allowed for a larger initial set of candidates (before verification cuts) than the one considered for classification, for a given class.
    \end{enumerate}
    \item The results of several SOS modules were not mutually exclusive \citep[see fig.~6 of][]{DR3-DPACP-162}, while only one class per source was required by classification.
\end{enumerate}


\section{Results\label{sec:results}}

From the about 400 million sources identified as potentially variable by the preceding general variability detection module (Sect.~\ref{ssec:classifier}), 12\,428\,245 objects were selected for the \gdr3 classification results. Among them, 9\,976\,881 variable sources, classified into 24 groups of variability types, and 2\,451\,364 galaxies were published in the \texttt{vari\_classifier\_result} and \texttt{galaxy\_candidates} tables, respectively. The latter includes also results from other coordination units\footnote{\url{https://www.cosmos.esa.int/web/gaia/coordination-units}} of the \gaia Data Processing and Analysis Consortium (DPAC), as described in \citet{DR3-DPACP-101}. The exclusion of galaxies from the variability result tables was motivated by their spurious light curve variability \citep{DR3-DPACP-164}. For the same reason, galaxy light curves were not published, except for the ones that were automatically included in the \gaia Andromeda Photometric Survey \citep{DR3-DPACP-142}.

The classified variable sources are identified by \texttt{source\_id}, they are assigned class labels (\texttt{best\_class\_name}), and they are associated with classification scores (\texttt{best\_class\_score}) in the \texttt{vari\_classifier\_result} table.
The galaxies whose classification was based on photometric time series are labelled as GALAXY in the field \texttt{vari\_best\_class\_name}, with the classification score stored in the field  \texttt{vari\_best\_class\_score} of the \texttt{galaxy\_candidates} table (see Appendix~\ref{app:queries}).

Figure~\ref{fig:histo} depicts the number of classified sources per class group, from the most to the least numerous class (galaxies and R\,Coronae Borealis stars, respectively).
Besides common and rare classes, certain class groups such as ACV|CP|MCP|ROAM|ROAP|SXARI and S were published for exploratory use to the benefit of the community.

\begin{figure}
\centering
 \includegraphics[width=\hsize]{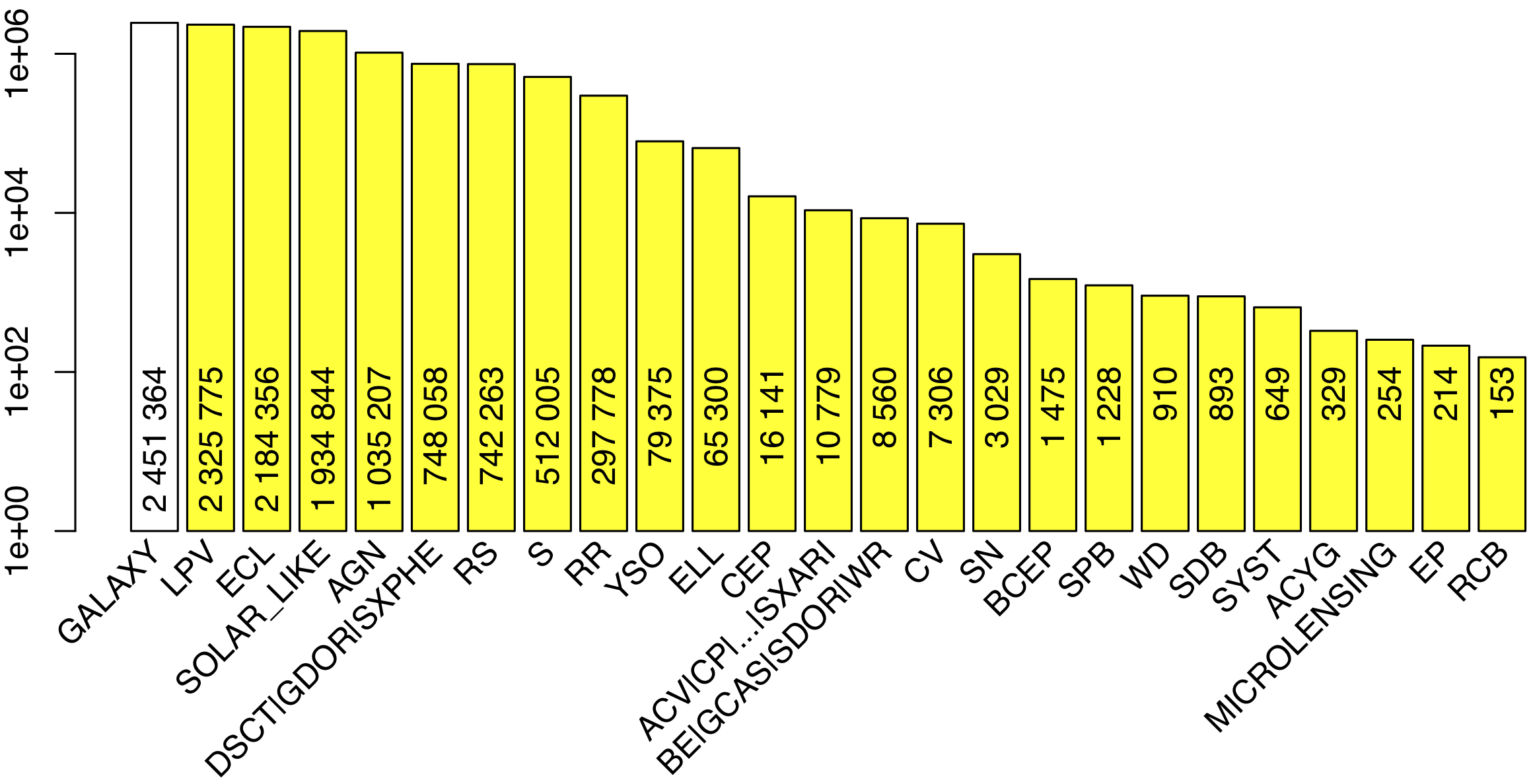}
 \caption{Number of sources per class, sorted in decreasing order. The bars shaded in yellow correspond to variability types published in the \texttt{vari\_classifier\_result} table, while galaxies, identified by their artificial photometric variations in \gaia, are published exclusively in \texttt{galaxy\_candidates}. The ACV|CP|MCP|ROAM|ROAP|SXARI class group is abbreviated as ACV|CP|...|SXARI.}
 \label{fig:histo}
\end{figure}

Table~\ref{tab:results} summarises the classification results for each class: the source counts, the distribution of classification scores (Sect.~\ref{ssec:class_score}), completeness (the ratio of identified to all known sources of a given class), contamination (the fraction of contaminants among the classifications of a given class, i.e.\ 1$-$\,purity), the $F_1$ score (the harmonic mean of completeness and purity), the maximum $F_1$ value and the corresponding  minimum classification score (for an optimal balance between completeness and contamination), and a comparison with SOS modules (explained in Sect.~\ref{ssec:class-sos}). 
Sample ADQL queries in Appendix~\ref{app:queries} include indications of how to retrieve all the candidates of a given class (from classification and SOS modules) and how to reproduce the comparison of results from classification versus SOS modules in Table~\ref{tab:results}. 

Completeness and contamination rates are computed for each class globally, with no restriction on sky location, magnitude, amplitude, signal-to-noise, period range, or other parameters, thus they might be more conservative than the detailed estimates presented in the papers describing the SOS module results. Typically, such rates depend on surveys from the literature in optical bands (or with detectable counterparts in \gaia), which are employed as reference. For this study, we take advantage of the catalogues of variable objects compiled by \citet{DR3-DPACP-177}, in particular considering a subset of reliable catalogues (flagged in their \texttt{selection} field). 
More details on the class composition of the true and unknown positives, in addition to the false positives and negatives, are presented in Table~\ref{tab:results_details}.

\begin{figure}[t]
\centering
 \includegraphics[width=\hsize]{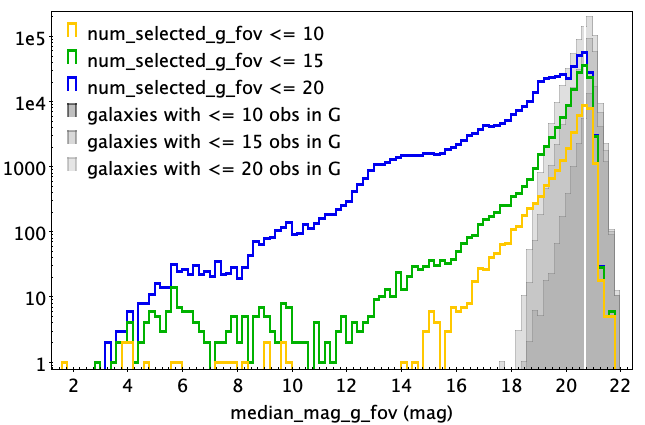}
 \caption{Number of the least-sampled sources in 0.2\,mag intervals as a function of the median \g-band magnitude. The number of variable sources (in the \texttt{vari\_classifier\_result} table of the \gaia archive) is colour coded by the maximum number of selected FoV~observations in the \g~band (\texttt{num\_selected\_g\_fov} up to 10, 15, and 20) as shown in the legend (with orange, green, and blue colours, respectively). The bars shaded in grey refer to the same conditions but for the galaxies identified by their artificial variability (published in the \texttt{galaxy\_candidates} table), including unpublished values of \texttt{num\_selected\_g\_fov} for galaxies outside the \gaia Andromeda Photometric Survey. The white vertical line marks the median \g~magnitude of 20.7. The fields \texttt{num\_selected\_g\_fov} and \texttt{median\_mag\_g\_fov} are published in the \texttt{vari\_summary} table.}
 \label{fig:leastsampled_histo}
\end{figure}

\begin{figure}[t]
\centering
 \includegraphics[width=\hsize]{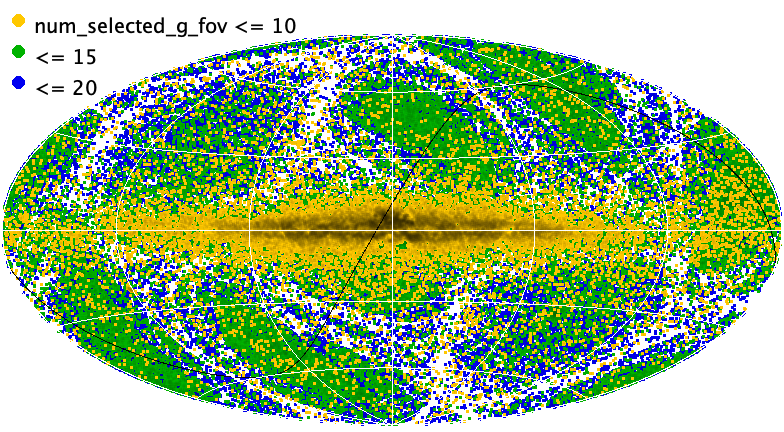}
 \caption{Sky map of the least-sampled sources in Galactic coordinates (white grid), colour coded as in Fig.~\ref{fig:leastsampled_histo} for variable sources with \texttt{num\_selected\_g\_fov} up to 10, 15, and 20. Galactic longitude is zero at the centre and increases towards the left. The thin line in black denotes the Ecliptic.}
 \label{fig:leastsampled_sky}
\end{figure}

For an overview on the identified classes of the least sampled sources, Fig.~\ref{fig:leastsampled_histo} shows the occurrence of classified objects as a function of median \g-band magnitude for sources with up to 10, 15, and 20~\g-band observations. Variable objects and galaxies are shown separately and in both cases the least sampled sources are the most numerous at the faint end, as a consequence of \gaia's magnitude limit, with thicker tails of the magnitude distribution towards bright magnitudes from objects with more observations.
Similarly, the distributions in the sky of variable classifications with up to 10, 15, and 20~\g-band observations are presented in Fig.~\ref{fig:leastsampled_sky}. The gaps created by sources with more measurements (following the Gaia scanning law) split the sky distributions in several regions: variables with up to 15 and 20~\g-band observations tend to be more numerous in regions intersected by the Ecliptic, while sources with with up to 10~measurements in the \g~band concentrate towards the Galactic plane. Although galaxies are not shown in Fig.~\ref{fig:leastsampled_sky}, most of them are located in the previously mentioned regions except for the Galactic plane, for all samples (including the galaxies with up to 10~observations).
The top-five most common classes among the classified sources with the lowest number of observations are listed as follows (with the number of sources satisfying the sampling conditions indicated in parentheses):
\begin{enumerate}
    \item GALAXY (43\,428), LPV (30\,203), SN (2515), CV (818), and AGN (579), for sources with up to 10~\g-band FoV observations;
    \item GALAXY (273\,603), LPV (104\,133), AGN (20\,846), RR (4295), and SN (2833), for sources with up to 15~\g-band FoV observations;
    \item GALAXY (643\,935), LPV (170\,681), ECL (104\,043), AGN (96\,760), and RR (22\,005), for sources with up to 20~\g-band FoV observations.
\end{enumerate}

\begin{landscape}
\begin{table}  
\caption[Classification result statistics]{Statistics of classification results for each class (source counts, classification and $F_1$ scores, completeness and contamination rates) and their distribution with respect to the corresponding SOS modules (sources in `common' between the ones classified as a given class and the corresponding SOS module, `extra' sources in classification, sources classified as `other' classes or `missed' by classification). The class group ACV|CP|MCP|ROAM|ROAP|SXARI was abbreviated as ACV|CP|...|SXARI. Dashes indicate cases when columns are not applicable (absence of SOS modules). 
In total, 9\,976\,881 sources from 24 variability types (or type groups) and 2\,451\,364 galaxies are published in the \texttt{vari\_classifier\_result} and \texttt{galaxy\_candidates} tables, respectively. Estimates of completeness and contamination rates generally exclude the training sources of the corresponding class, with the exception of the class labels followed by an asterisk. For the explanation of the values within parentheses, see the sub-sections on the relevant classes in Sect.~\ref{sec:results}. See Table~\ref{tab:results_details} for further details on completeness and contamination.\label{tab:results}} 
\centering                  
\begin{tabular}{@{}lrrrrrrcccrrrrc@{}}     
\hline\hline & \\[-2.0ex]                 
\multirow{2}{*}{Class} & \multirow{2}{*}{$N_{\mathrm{CLS}}$} & \multirow{2}{*}{$N_{\mathrm{SOS}}$} & \multicolumn{4}{c}{Classification versus SOS module} &
   \multicolumn{3}{c}{Classification score} & Compl. & Cont. & $F_1$ & Max $F_1$ & Min score\\
 & & & Common & Extra & Other & Missed & Min & Avg & Max & (\%) & (\%) & (\%) & (\%) & (at max $F_1$) \\  
\hline & \\[-1.5ex]
ACV|CP|...|SXARI (1) & 10\,779 & $<$ 54\,476 & 0 & $<$ 10\,779 & 54\,476 & 0 & 0.00 & 0.54 & 1.00 & 13.0 & 83.4 & 14.5 & 14.6 & 0.04 \\
ACYG (1,*) & 329 & $<$ 54\,476 & 5 & $<$ 324 & 54\,471 & 0 & 0.07 & 0.54 & 0.99 & $<$89.1 & $>$44.7 & $<$68.3 & $<$81.2 & 0.63 \\
AGN & 1\,035\,207 & 872\,228 & 872\,181 & 163\,026 & 5 & 42 & 0.00 & 0.54 & 1.00 & 50.6 & $>$0.1 & $<$67.1 & $<$67.1 & 0.00 \\
BCEP (1,*) & 1\,475 & $<$ 54\,476 & 174 & 1\,301 & 54\,302 & 0 & 0.02 & 0.47 & 0.99 & $<$78.8 & $>$44.9 & $<$64.9 & $<$71.8 & 0.51 \\
BE|GCAS|SDOR|WR (1) & 8\,560 & $<$ 54\,476 & 223 & 8\,337 & 54\,253 & 0 & 0.01 & 0.72 & 1.00 & 3.9 & 65.3 & 7.0 & 7.0 & 0.01 \\
CEP & 16\,141 & 15\,021 & 14\,987 & 1\,154 & 34 & 0 & 0.00 & 0.48 & 1.00 & 75.2 & 4.5 & 84.1 & 84.1 & 0.00 \\
CV & 7\,306 & -- & -- & -- & -- & -- & 0.01 & 0.59 & 1.00 & 15.0 & 47.3 & 23.4 & 23.7 & 0.20 \\
DSCT|GDOR|SXPHE (1) & 748\,058 & $<$ 54\,476 & 53\,640 & 694\,418 & 836 & 0 & 0.00 & 0.49 & 1.00 & 44.6 & 43.2 & 50.0 & 51.0 & 0.15 \\
ECL & 2\,184\,356 & 2\,184\,477 & 2\,184\,337 & 19 & 60 & 80 & 0.00 & 0.54 & 1.00 & 47.8 & 4.7 & 63.7 & 63.7 & 0.00 \\
ELL (2) & 65\,300 & 6\,306 & 6\,303 & 58\,997 & 3 & 0 & 0.00 & 0.48 & 1.00 & 8.9 & 56.5 & 14.7 & 15.2 & 0.11 \\
EP (3) & 214 & 214 & 214 & 0 & 0 & 0 & 1.00 & 1.00 & 1.00 & (10.2) & (4.7) & (18.4) & (18.4) & 0.00 \\
LPV & 2\,325\,775 & 1\,720\,588 & 1\,720\,066 & 605\,709 & 521 & 1 & 0.00 & 0.58 & 1.00 & 54.3 & 4.3 & 69.3 & 69.3 & 0.11 \\
MICROLENSING (4,*) & 254 & 363 & 187 & 67 & 11 & 165 & 0.00 & 0.43 & 0.96 & $<$56.6 & $>$7.2 & $<$70.3 & $<$71.1 & 0.22 \\
RCB (*) & 153 & -- & -- & -- & -- & -- & 0.03 & 0.48 & 0.97 & $<$68.6 & $>$52.9 & $<$55.8 & $<$63.2 & 0.30 \\
RR & 297\,778 & 271\,779 & 271\,576 & 26\,202 & 0 & 203 & 0.00 & 0.52 & 1.00 & 70.0 & 2.6 & 81.4 & 81.4 & 0.00 \\
RS & 742\,263 & -- & -- & -- & -- & -- & 0.00 & 0.41 & 1.00 & 30.0 & 73.3 & 28.2 & 28.2 & 0.00  \\
S (4) & 512\,005 & 471\,679 & 499 & 511\,506 & 285\,249 & 185\,931 & 0.00 & 0.59 & 1.00 & (30.5) & (95.2) & (8.3) & (17.8) & (0.87) \\
SDB (*) & 893 & -- & -- & -- & -- & -- & 0.17 & 0.73 & 0.99 & $<$76.1 & $>$28.2 & $<$73.9 & $<$73.9 & 0.00 \\
SN & 3\,029 & -- & -- & -- & -- & -- & 0.00 & 0.51 & 1.00 & 46.0 & 6.0 & 61.8 & 61.8 & 0.00 \\
SOLAR\_LIKE (4,5) & 1\,934\,844 & 474\,026 & 102\,573 & 1\,832\,271 & 21\,575 & 349\,878 & 0.00 & 0.39 & 1.00 & 16.7 & 6.3 & 28.4 & 28.4 & 0.00 \\
SPB (1,*) & 1\,228 & $<$ 54\,476 & 434 & 794 & 54\,042 & 0 & 0.02 & 0.49 & 1.00 & $<$56.4 & $>$76.1 & $<$33.6 & $<$41.8 & 0.50 \\
SYST (*) & 649 & -- & -- & -- & -- & -- & 0.00 & 0.50 & 1.00 & $<$81.7 & $>$51.3 & $<$61.0 & $<$61.8 & 0.08 \\
WD (*) & 910 & -- & -- & -- & -- & -- & 0.25 & 0.67 & 1.00 & $<$15.6 & $>$13.6 & $<$26.4 & $<$26.5 & 0.30 \\
YSO & 79\,375 & -- & -- & -- & -- & -- & 0.00 & 0.52 & 1.00 & 9.1 & $<$79.8 & 12.5 & 12.5 & 0.01  \\
\hline & \\[-2ex]
GALAXY (6) & 2\,451\,364 & -- & -- & -- & -- & -- & 0.00 & 0.74 & 1.00 & 55.5 & 0.2 & 71.4 & 71.4 & 0.00 \\
\hline & \\[-1.5ex]
Total & 12\,428\,245 &  &  &  &  &  &  &  &  &  &  &  &  &  \\
\hline                       
\end{tabular}
\tablefoot{(1) ACV|CP|MCP|ROAM|ROAP|SXARI, ACYG, BCEP, BE|GCAS|SDOR|WR, DSCT|GDOR|SXPHE, and SPB classifications were considered as input for the upper main-sequence oscillator SOS module, but the latter did not split its 54\,476 candidates into separate classes. (2) ELL classifications were used as input for the compact companion SOS module, but the latter did not target typical members of this class. (3) EP classifications were used as input for the planetary transit SOS module, but only SOS candidates were retained in the classification results. (4) MICROLENSING, S, and SOLAR\_LIKE classifications were not used as input for the respective SOS modules. (5) SOLAR\_LIKE classifications included more types (flaring stars) than the rotational modulation SOS module.  (6) GALAXY classifications were made available exclusively in the \texttt{galaxy\_candidates} table.  (*) Less than 12~sources classified as a given class were matched with the literature \citep{DR3-DPACP-177} after the removal of training set objects (aimed at reducing biases in the estimates of completeness and contamination rates). For these classes, training sources were included in the assessment of completeness and contamination (for upper or lower limits, respectively).}
\end{table}  
\end{landscape}

\begin{landscape}
\begin{table} 
\caption[Completeness and contamination details]{Completeness and contamination details of classification results for each class (the class group ACV|CP|MCP|ROAM|ROAP|SXARI was abbreviated as ACV|CP|...|SXARI). The literature compilation of \citet{DR3-DPACP-177} was adopted as reference by considering the 6.7~million sources in the catalogues flagged by the \texttt{selection} field and the types listed in \texttt{primary\_var\_type}. The literature variability types that are consistent with ours are indicated in the second column. Results with all classification scores are considered. Classes (see label definitions in Appendix~\ref{app:FP_labels}) and their representation (in parentheses) are listed for true and unknown positives, as well as false positives and negatives (with an indication of the most relevant literature for over 90\,\% of the missed classifications, in decreasing order). Source numbers exclude the training sources of the corresponding class, except for the classes followed by an asterisk (otherwise left  with $<$\,12 true positives). \label{tab:results_details}} 
\centering                  
\begin{tabular}{@{}lllrlll@{}}     
\hline\hline & \\[-2.0ex]                 
Class & Consistent type & True positive & Unknown  & False positive & False negative (FN) & Main FN \\
      &  & &  positive~~~ & (top-6 types)  &  & reference \\
\hline & \\[-1.5ex]
ACV|CP|...|SXARI & ACV, CP, MCP, & MCP(174), ACV(29),& 8\,914 & CST(786), DSCT(49), & MCP(1178), ACV(179), & 8, 10  \\
                 & ROAM, ROAP,  & SXARI(1) &  & EA(44), GCAS(34), & ROAP(8), SXARI(5) &   \\
                 & SXARI        &          &  & RAD\_VEL\_VAR(26),  &  &   \\
                 &              &          &  & ECL(17) &  &   \\
ACYG (*) & ACYG & ACYG(57) & 226 & GCAS(20),  MCP(7), & ACYG(7) & 10, 3  \\
         &      &          &     & CST(4), EA(4), &  &   \\
         &      &          &     & ECL(2),  ELL(2) &   &   \\
AGN & AGN, BLAZAR, & QSO(828\,728),  & $<$\,200\,081 & YSO(328),   & QSO(807\,955), AGN(4267), & 11, 123  \\
    & BLLAC, QSO & BLAZAR(2242),     &  & DSCT|SXPHE(275), & BLAZAR(121), BLLAC(9) &   \\
    &            & AGN(357), BLLAC(2) &  & RRAB(155), RRC(109),  &  &   \\
    &            &                    &  & SR(50), GALAXY(48) &  &   \\
BCEP (*) & BCEP & BCEP(145) & 1\,212 & GCAS(55), CST(29),  & BCEP(39) & 10, 3, 13, 14  \\
         &      &           &        & DSCT(11), ELL(3),  &  &   \\
         &      &           &        & RRC(3), SB(3)  &  &   \\
BE|GCAS|SDOR|WR & BE, GCAS,      & GCAS(78), BE(58), WR(1) & 6\,864 & EA(152), SR(20), & BE(2364), GCAS(982) & 124, 6, 16, 10, 15  \\
                & SDOR, WR       &                         &        & L(14), HMXB(10), & WR(13), SDOR(2) &   \\
                &                &                         &        & YSO(9), CST(7) &  &   \\
CEP & ACEP, BLHER,  & DCEP(10\,429), CW(240), & 1\,133 & RRAB(107), EW(73), & DCEP(2054), T2CEP(543), & 125, 26, 27, 6, \\
    & CEP, CW, DCEP,  & BLHER(210), T2CEP(115), &  & M|SR(73), RRC(51), & CW(364), BLHER(335), & 126, 17, 19, 10, \\
    & RV, T2CEP               & ACEP(64), CEP(27), RV(13) &  & RS(36), EB(31) & CEP(204), ACEP(123),RV(39) & 42, 7, 47, 35 \\
CV & CV & CV(223) & 5\,717 & RRAB(41), QSO(37), & CV(1264) & 10, 6, 32, 127, \\
   &    &         &        & PCEB(24), YSO(20), &  & 31, 33, 30  \\
   &    &         &        & EW(16), EA(12) &  &   \\
DSCT|GDOR|SXPHE & DSCT,           & DSCT(12\,294),       & 718\,199 & EW(3402),            & DSCT(12\,865),  & 35, 42, 41, 10,  \\
                & DSCT+GDOR,       & DSCT|SXPHE(2609),  &          & RAD\_VEL\_VAR(1813), & DSCT|SXPHE(4846), & 6, 43, 37  \\
                & DSCT|SXPHE,       & GDOR(50), SXPHE(4) &          & RS(1598), ROT(1052), & GDOR(809), DSCT+GDOR(25), &   \\
                & GDOR, SXPHE        &                    &          &  CST(602), RRC(525) & SXPHE(13) &   \\
ECL & EA, EB, ECL, EW & EW(300\,168), EA(144\,524), & 1\,634\,579 & RS(10\,791), RRC(4850), & EA(259\,020), EW(176\,469), & 52, 50, 10, 35, \\
    &                 & ECL(58\,115), EB(16\,449) &               & ROT(3747), BY(2057), & ECL(126\,766), EB(4116) & 6, 58  \\
    &                 &                           &               & ELL(1056), RRAB(983) &  &   \\
ELL & ELL & ELL(2211) & 59\,734 & EW(1519), EA(616) & ELL(22\,753) & 52, 10  \\
    &     &           &         & SR(295), RS(210), &  &   \\
    &     &           &         & OSARG(148), ECL(26) &  &   \\
EP & EP & EP(101)  & 47 & EA(3), CST(2) & EP(890) & 53, 10  \\
LPV & LPV, LSP, M, M|SR, & SR(107\,327), LPV(76\,081), & 1\,987\,257 & L(6653),  & OSARG(197\,336), SR(61\,361), & 62, 35, 10, 6  \\
    & OSARG, SARG,   & OSARG(71\,093),  &  & GALAXY(1364), & SRS(6173), LPV(2681),  &   \\
    & SR, SRA, SRB,  & M|SR(53\,568), M(11\,703),  &  & ECL(1034), EW(745), & M|SR(1723), M(1038),  &   \\
    & SRC, SRD, SRS & SRS(780), SRB(281),   &  & ELL(731), EA(728) & SRB(137), SRA(16), &   \\
    &               & LSP(9), SRA(1)                       &  &  & SRD(7), SRC(2), LSP(1) &   \\
\hline & \\  [-2.0ex]                    
\multicolumn{7}{r}{{(Continued on next page)}} 
\end{tabular}
\end{table}  
\end{landscape}

\setcounter{table}{2}
\begin{landscape}
\begin{table}  
\caption{(Continued)} 
\centering                 
\begin{tabular}{@{}lllrlll@{}}     
\hline\hline & \\[-2.0ex]                 
Class & Consistent type & True positive & Unknown  & False positive & False negative (FN) & Main FN \\
      &  & &  positive~~~ & (top-6 types)  &  & reference \\
\hline & \\[-1.5ex]
MICROLENSING (*) & MICROLENSING & MICROLENSING(64) & 185 & ECL(2), LPV(1), & MICROLENSING(49) & 66  \\
                 &              &  &  & M|SR(1), SR(1) &  &   \\
RCB (*) & RCB & RCB(48) & 51 & LPV(20),  M|SR(11), & RCB(22) & 6, 10, 67  \\
        &     &  &  & SR(8), YSO(5), &  &   \\
        &     &  &  & M(3), T2CEP(3) &  &   \\
RR & ARRD, RR, RRAB, & RRAB(145\,977),  & 84\,897 & EW(3270), EA(559), & RRAB(67\,666), & 42, 128, 10, 17,  \\
   & RRC, RRD        & RRC(52\,042),    &  & ECL(489), ROT(314), & RRC(17\,980), RRD(748), & 6, 129  \\
   &                 & RRD(3740), RR(4) &  & RS(165), ACEP(116) & RR(118), ARRD(36) &   \\
RS & RS & RS(23\,401) & 652\,943 & ROT(23\,176), BY(18\,737), & RS(54\,698) & 35, 10  \\
   &    &             &          & EW(9487), SR(5761), &  &   \\
   &    &             &          & EA(1421), DSCT(792) &  &   \\
S & BLAP, S & S(335) & 504\,195 & EW(4356), ECL(710), & S(760), BLAP(4) & 88, 10, 89, 130   \\
  &       &          &          & EA(647), GALAXY(244), &  &   \\
  &       &          &          & RRAB(168), CST(106) &  &   \\
SDB (*) & V1093HER, & V361HYA(42), & 822  & RAD\_VEL\_VAR(9), & V361HYA(15), & 6, 10  \\
        & V361HYA   & V1093HER(9) &  & SB(5), EA(3), CST(1), & V1093HER(1) &   \\
        &           &             &  & DSCT(1), ROT(1) &  &   \\
SN & SN & SN(373) & 2\,572 & CV(22), GALAXY(2) & SN(437) & 131, 30  \\
SOLAR\_LIKE & BY, BY|ROT, & ROT(35\,910), BY(15\,597), & 1\,873\,744 & RAD\_VEL\_VAR(1623), & ROT(164\,245), BY(88\,230), & 96, 35, 6, 105,  \\
            & FLARES, ROT, & SOLAR\_LIKE(4849), &  & CST(1290), EA(167), & SOLAR\_LIKE(27\,171), UV(973), & 10, 132, 100  \\
            & SOLAR\_LIKE, UV & FLARES(4), UV(4) &  & RS(148), EW(119), EP(93) & FLARES(387), BY|ROT(27) &   \\
SPB (*) & SPB & SPB(114)  & 751 & MCP(111), ACV(110), & SPB(88) & 10, 106, 3, 13  \\
        &     &           &  & CST(44), GCAS(42), &  &   \\
        &     &           &  & ELL(9), SB(9) &  &   \\
SYST (*) & SYST, ZAND & SYST(200),  & 199 & LPV(77), M|SR(58), & SYST(41), & 133, 107, 10  \\
         &            & ZAND(19)    &  & SR(48), OSARG(27), & ZAND(8) &   \\
         &            &             &  & L(6), M(5) &  &   \\
WD (*) & EHM\_ZZA, & ZZA(179), GWVIR(61), & 602 & CV(12), PCEB(5), & GWVIR(721), ZZA(366), & 112, 134 \\
   & ELM\_ZZA, GWVIR, & V777HER(21), &  & DSCT(4), EW(3), & V777HER(326), &   \\
   & HOT\_ZZA, & ZZ(3), ELM\_ZZA(2) &  & ROT(3), CST(2) & PRE\_ELM\_ZZA(10), &   \\
   & PRE\_ELM\_ZZA, &  &  &  & ELM\_ZZA(7), ZZ(7),  &   \\
   & V777HER, ZZ, ZZA &  &  &  & HOT\_ZZA(3)  &   \\
YSO & CTTS, DIP, FUOR,  & YSO(948), TTS(54), & 70\,131 & RS(1542), BY(1435), & YSO(9597), TTS(651), & 121, 6, 10  \\
    & GTTS, HAEBE, & CTTS(34),  WTTS(17), &  & ROT(487), EW(219), & CTTS(209), WTTS(129), &   \\
    & IMTTS, PULS\_PMS,& HAEBE(2),  &  & M|SR(143), SR(91) & HAEBE(15), UXOR(11), &   \\
    &  TTS,  UXOR, & GTTS(1), UXOR(1) &  &  & DIP(2), IMTTS(2),  &   \\
    & WTTS, YSO &  &  &  & PULS\_PMS(1) &   \\
\hline & \\[-2ex]
GALAXY & GALAXY & GALAXY(968\,265) & 1\,478\,607 & QSO(1314), AGN(570), & GALAXY(774\,829) & 122  \\
       &        &                  &             &  RRAB(156),  SN(106), &  &   \\
       &        &                  &             &  RRC(88), CST(19) &  &   \\
\hline                       
\end{tabular}
\tablebib{See the identifiers up to (122) at the bottom of Table~\ref{tab:training}, in addition to those beyond (122) as follows: 
(123)~\citet{2019arXiv191205614F}; 
(124)~\citet{2008A&A...478..659S}; 
(125)~\citet{2015AcA....65..297S}; 
(126)~\citet{2020AcA....70..101S}; 
(127)~\citet{2020AJ....159..198S}; 
(128)~\citet{2014AcA....64..177S,2016AcA....66..131S}; 
(129)~\citet{2019AcA....69..321S}; 
(130)~\citet{2017NatAs...1E.166P}; 
(131)~\citet{2014ApJ...788...48S,2017PASP..129j4502K}; 
(132)~Distefano \citep[catalogue GAIA\_BY\_DISTEFANO\_2019 in][]{DR3-DPACP-177};
(133)~\citet{2000A&AS..146..407B};
(134)~\citet{2019A&ARv..27....7C}.}
\end{table}  
\end{landscape}

Galaxies are the most common sources with few observations because they are the faintest classified objects and thus susceptible to reduced detectability at the faint end of the \gaia detection limit. Long-period variables rank second, as expected from their relatively high occurrence and some of their features (such as high intrinsic luminosity, red colours, and often high amplitudes), which ease their identification even at large distances. Active galactic nuclei appear among the top classes, given their magnitude distribution at the faint end. About 83\,\% of supernovae have less than 11~measurements in the \g~band, because the originating stars become detectable only after explosion and the subsequent brightness decay further reduces the time span of detectability \citep[for the distribution of the time intervals of \gaia observations, see appendix~A.4.3 in][]{2017arXiv170203295E}.

The following sub-sections present more details for each class, among which the number of selected classifiers (as an upper limit as not all sources were necessarily classified as a given class by all classifiers), the verification cuts, special considerations on completeness or contamination (when applicable), and the references to relevant articles. A selection of diagrams illustrating the general properties of classified sources for each class is presented in Appendix~\ref{app:plots}, followed by a visualisation of completeness versus contamination rates as a function of minimum classification score, $F_1$ score versus minimum classification score, and samples of light curves in the \g band versus time or phase (the latter is contingent upon the presence of a single dominant period).
An overview of all variability results (including classifications), with figures and tables combining metrics of several classes, is presented in \citet{DR3-DPACP-162}. 

\subsection{ACV|CP|MCP|ROAM|ROAP|SXARI stars\label{ssec:acv}}

This class group accounts for 10\,779 variable sources from a set of different types, which are particularly challenging and share similar features (according to the \gdr3 data): $\alpha^2$\,Canum Venaticorum, (Magnetic) Chemically Peculiar star, Rapidly Oscillating Am or Ap star, and SX\,Arietis variable.  Some types are physically unrelated to others, nevertheless they were grouped because classifiers often confused them. For example, the rapid and low amplitude oscillations of ROAM and ROAP could not be captured with \gaia's per-FoV photometry, so these types were modelled only with global properties, such as the location in the observational Hertzsprung--Russell diagram, thus causing significant contamination by constant stars, in addition to contaminants from other variability types such as DSCT and eclipsing binaries (see Table~\ref{tab:results_details}). Perhaps, their classification will improve with per-CCD photometry in the next \gaia data release. 
This group of variability types includes also some SPB stars, which are not mentioned in this class group denomination because of the attempt to split them into their own class (Sect.~\ref{ssec:spb}).  
This class was considered by the upper main-sequence oscillator SOS module, but no source satisfied its requirements \citep[see sect.~10.14 of the \gdr3 documentation;][]{2022gdr3.reptE..10R}. 

These candidates were selected from three multi-class and binary meta-classifiers (Sect.~\ref{ssec:classifier}) with some minimum probability level. The following additional conditions were required (employing field names in the \texttt{vari\_summary} table).
\begin{enumerate} 
    \item The values of \texttt{std\_dev\_mag\_g\_fov} were set above the third quartile of the standard deviations in \g, in 0.05\,mag intervals, of 1.6~million reference sources (see Appendix~\ref{app:plots}), as the ROAP training objects below this threshold (as mentioned in Appendix~\ref{app:training}) were most likely amplifying the contamination by constant stars.
    \item To further reduce contaminants, a higher level of variability probability than the one used for the general variability detection (Sect.~\ref{ssec:classifier}) was required.
    \item As predictions still tended to be less variable than those exhibited in the literature, a minimum threshold was set for the single-band Stetson index: \texttt{stetson\_mag\_g\_fov}\,$>$\,1.5.
    \item The reddened colour \texttt{median\_mag\_bp\,$-$\,median\_mag\_rp} was set to be bluer than 1\,mag, in order to remove sparse outliers,   whose presence among the training sources (see Fig.~\ref{fig:app:ACV_trn}d) was subsequently deemed questionable. In fact, the bulk of results was bluer than 0.5\,mag and redder objects were associated with low scores (Fig.~\ref{fig:app:ACV}b).
    \item While training objects reduced steeply beyond a median \g-band magnitude of about~10, most classifications extended towards fainter magnitudes; such dubious predictions were limited by the conservative condition \texttt{median\_mag\_g\_fov}\,$<$\,10\,mag.
\end{enumerate}

According to Table~\ref{tab:results_details}, additional candidates of this class group may be found among the false positives of the SPB~(221) and ACYG~(7) classes, where the numbers of known sources indicated in parentheses represent lower limits.


\subsection{$\alpha$\,Cygni stars (ACYG)\label{ssec:acyg}}

The classification of the $\alpha$\,Cygni-type variables comprised 329 candidates, selected from two binary meta-classifiers (Sect.~\ref{ssec:classifier}) with some minimum probability threshold, with no further condition. It is one of the rare types for which the training set sources were included in the assessment of completeness and contamination rates (in Table~\ref{tab:results}) because of the lack of known stars of this type. Among the contaminating classes, GCAS stars are the most common, as expected from the partial overlap in the bright and blue part of the colour--magnitude diagram (see Figs.~\ref{fig:app:ACYG}b and~\ref{fig:app:BE}b).

From the diagrams that include \texttt{median\_mag\_g\_fov} from the \texttt{vari\_summary} table (as in Figs.~\ref{fig:app:ACYG}b,c,g), two clumps of candidates separated at  \texttt{median\_mag\_g\_fov}\,$\approx$\,9.5\,mag are clearly visible; the fainter one is also associated with lower scores and, as apparent from the sky map in Fig.~\ref{fig:app:ACYG}a, 82\,\% of them (73/89) are projected in the direction of the Magellanic Clouds, with an average \texttt{best\_class\_score} of 0.3 (in the \texttt{vari\_classifier\_result} table). The bright and faint clumps were not so obvious from the training sources, as the relative occurrence in the Magellanic Clouds was strongly underrepresented and consequently related to lower scores than Galactic sources. 

The upper main-sequence oscillator SOS module did not consider ACYG classifications as input, but it implicitly did so as its candidates were selected before the implementation of the rules that reduced multi-class sources to a single class per source (Sect.~\ref{ssec:overlaps}) and eventually five ACYG candidates became upper main-sequence oscillator candidates too.


\subsection{Active galactic nuclei (AGN)\label{ssec:agn}}

The 1\,035\,207 AGN candidates were selected from 11~multi-class and binary meta-classifiers (Sect.~\ref{ssec:classifier}) following some minimum probability level and were filtered according to the following conditions (employing field names in the \texttt{vari\_summary} and \texttt{gaia\_source} tables), guided mostly by the \gaia celestial reference frame objects \citep[\gaia-CRF3;][]{EDR3-DPACP-133} among the classified AGN sources.
\begin{enumerate} 
    \item A higher level of variability probability than the one used for the general variability detection (Sect.~\ref{ssec:classifier}) was required in order to focus on truly variable AGNs, considering the possible contribution of artefacts such as the one described by \citet{DR3-DPACP-164}, when the host galaxy is detectable; the consequent offset with respect to the general threshold is visible in Fig.~\ref{fig:app:AGN}e.
    \item The renormalised unit weight error \texttt{ruwe} was set to be lower than~1.2, as the astrometric measurements of the vast majority of AGNs fit well the single-source model of the astrometric solution (97\,\% of \gaia-CRF3 objects satisfy \texttt{ruwe}\,$<$\,1.2). 
    \item The \texttt{abbe\_mag\_g\_fov} parameter was set to be lower than~0.9, as long timescale variations observed with \gaia's average sampling typically lead to low values for this parameter (see Fig.~\ref{fig:app:AGN}f, where the classification score mirrors the density of training objects in Fig.~\ref{fig:app:AGN_trn}f).
    \item The number of sources within 100~arcsec from each AGN candidate (computed by excluding the contribution of the AGN source at the centre) was set to be less than~126, in order to avoid crowded stellar fields in the foreground, as the steep increase in the number of stars per solid angle can lead to regions with excessive false positive rates, sometimes even where very few AGNs are detectable in the optical wavelengths following an enhanced interstellar extinction. The number density threshold corresponds to a suspicious increase of the fraction of unknown-to-known AGNs towards higher crowding. The most evident sky regions affected by this conditions are the Magellanic Clouds and the Galactic plane, as shown in Fig.~\ref{fig:app:AGN}a. Occasional AGN candidates appear at low Galactic latitudes, especially towards the Galactic anti-centre, most of which are associated with low classification scores.
    \item For sources with existing parallax, the latter was assumed to be insignificant, as for any extra-galactic object, after correcting it for a global systematic offset of $-$0.017\,mas \citep{2021A&A...649A...4L}, consequently its ratio with an assumed Gaussian uncertainty was expected to follow a standard normal distribution (with zero average and unit variance). We required parallax measurements not to deviate beyond five sigma: |\,\texttt{parallax}+0.017\,mas\,|\,$/$\,\texttt{parallax\_error}\,$<$\,5. Only 282 of the published AGN candidates had no parallax (nor proper motion) available, while the remaining 1\,034\,925 sources had (at least) five-parameter astrometric solutions.
    \item The same rationale of item~(5) was applied to AGN classifications with existing proper motion \citep[with negligible  offset; see][]{2021A&A...649A...4L,EDR3-DPACP-133}, leading to the following condition, which accounts also for the correlation between the measured proper motion along the right ascension and declination directions: \\ \texttt{pmra}$^2$/\,\texttt{pmra\_error}$^2+$\,\texttt{pmdec}$^2/$\,\texttt{pmdec\_error}$^2$+\newline $-$\,2\,(\texttt{pmra}\,/\,\texttt{pmra\_error})\,$\times$\,(\texttt{pmdec}\,$/$\,\texttt{pmdec\_error})\,$\times$\newline
    $\times$\,\texttt{pmra\_pmdec\_corr}\,$<$\,5$^2$\,(1\,$-$\,\texttt{pmra\_pmdec\_corr}$^2$).
    \item Three conditions were set in the \gaia reddened colour--colour diagrams as follows: 
    \begin{enumerate} 
        \item \texttt{median\_mag\_g\_fov}\,$-$\,\texttt{median\_mag\_rp}\,$>$\break $-$\,3.7\,(\texttt{median\_mag\_bp}\,$-$\,\texttt{median\_mag\_g\_fov})\,$-$\,0.7,
        \item \texttt{median\_mag\_g\_fov}\,$-$\,\texttt{median\_mag\_rp}\,$<$\break $-$\,0.75\,(\texttt{median\_mag\_bp}\,$-$\,\texttt{median\_mag\_g\_fov})\,$+$\,1.55,
        \item \texttt{median\_mag\_g\_fov}\,$-$\,\texttt{median\_mag\_rp}\,$>$\break 1.6\,(\texttt{median\_mag\_bp}\,$-$\,\texttt{median\_mag\_g\_fov})\,$-$\,0.6,
    \end{enumerate}
    whose impact is clearly visible in Fig.~\ref{fig:app:AGN}c.
\end{enumerate}

The set of conditions listed above proved efficient in the selection of the bulk of AGN candidates while keeping low contamination, however, peculiar objects may not conform to these general rules. In order to recuperate known AGNs with different behaviour from average and those of particular interest for which \gaia time series might provide useful complementary information, some criteria were relaxed and the following objects were included (provided they were classified as AGN): blazars \citep{2015Ap&SS.357...75M,2018MNRAS.478.1512B,2017A&A...598A..17C}, known and possible strong gravitationally lensed AGNs\footnote{\url{https://research.ast.cam.ac.uk/lensedquasars/index.html}}, the top-thousand brightest AGNs (among which 3C~273; \gaia~DR3 \texttt{source\_id} 3700386905605055360), and the supermassive binary black hole candidate OJ~287 (\gaia~DR3 \texttt{source\_id} 660820614442429056). 
Similarly, about 50~thousand AGN candidates were recovered following the feedback from the AGN SOS module.
According to Table~\ref{tab:results_details}, additional candidates may be found among the false positives of the GALAXY~(1884) and CV~(37) classes, where the numbers of known sources indicated in parentheses represent lower limits.

In comparison with the AGN SOS module requirements described in \citet{DR3-DPACP-167}, the conditions applied to classification results were generally more permissive, in particular:
\begin{enumerate}
    \item at least five (instead of 20)~FoV transits in the \g~band;
    \item no cut in the \texttt{ruwe} versus \texttt{abbe\_mag\_g\_fov} plane, but a slightly more restrictive limit on the \texttt{ruwe} parameter (less than~1.2 instead of~1.3);
    \item no constraint on the amount of scan-angle dependent signal \citep{DR3-DPACP-164};
    \item no requirements on the following parameters, which are defined and published in the \texttt{vari\_agn} table:
        \texttt{fractional\_variability\_g} \citep{2003MNRAS.345.1271V}, 
        \texttt{structure\_function\_index} \citep{1985ApJ...296...46S},
        \texttt{qso\_variability} and \texttt{non\_qso\_variability} \citep{2011AJ....141...93B}.
\end{enumerate}

\begin{figure}
\centering
\begin{overpic}[width=\hsize]{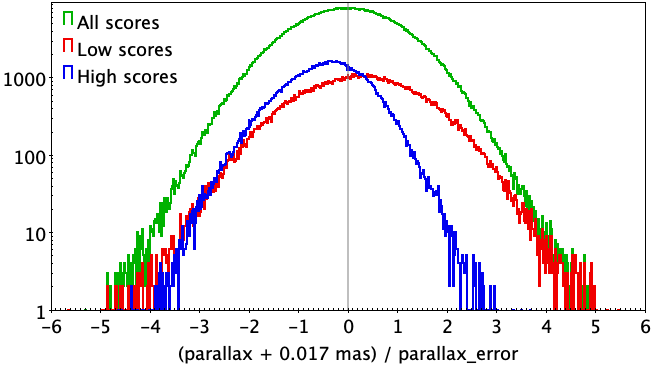}
 \put (0,1) {(a)}
\end{overpic} \\
\vspace{2mm}
\begin{overpic}[width=\hsize]{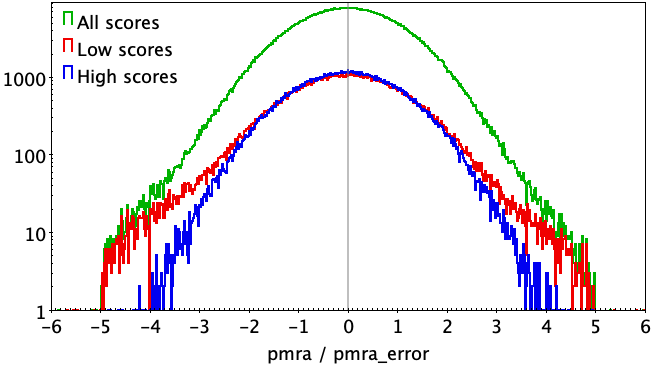}
 \put (0,1) {(b)}
\end{overpic} \\
\vspace{2mm}
\begin{overpic}[width=\hsize]{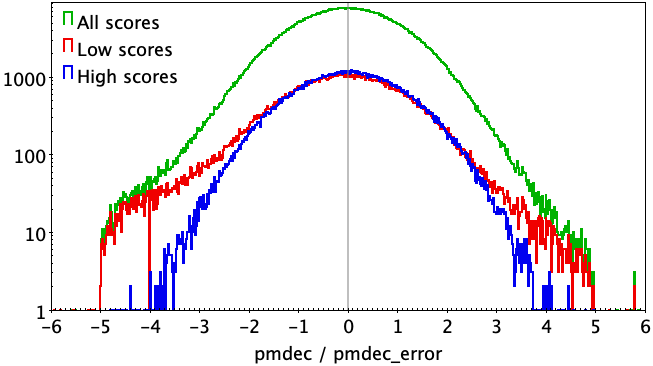}
 \put (0,1) {(c)}
\end{overpic}
\caption{Distribution of \texttt{parallax} (a) and proper motion components, \texttt{pmra} along the right ascension (b) and \texttt{pmdec} along the declination (c), normalised by the corresponding uncertainties and binned in intervals of 0.02, for the 1\,034\,925 AGN candidates with at least five-parameter astrometric solutions, for the top-150\,000 candidates (denoted as high scores; \texttt{best\_class\_score}\,$>$\,0.8995154), and for the bottom-150\,000 candidates (denoted as low scores; \texttt{best\_class\_score}\,$<$\,0.1615997), colour-coded as indicated in the legend. 
The bias from the inclusion of parallax among classification attributes is evident in panel (a) (see Sect.~\ref{ssec:agn} for details).} \label{fig:AGN_plx_pmra_pmdec}
\end{figure}

Parallax was used as classification attribute  (among others listed in Appendix~\ref{app:attributes}) and consequently it introduced a bias in the distribution of the parallax significance of AGN candidates with respect to \texttt{best\_class\_score} (in the \texttt{vari\_classifier\_result} table). Figure~\ref{fig:AGN_plx_pmra_pmdec}a illustrates a comparison between the top-150\,000 (high-score) AGN classifications and the bottom-150\,000 (low-score) candidates: there is a distinct deficit of sources with positive parallax among the most reliable AGN candidates, clearly yielding to stellar objects in the Galaxy (often associated with positive parallax), while the low-score sample is biased in the opposite direction, in addition to being the main contributor to thicker than Gaussian tails at both ends.
Eventually, the interplay among high and low score AGNs leads to an overall distribution that approximates a standard normal distribution.
As expected, no training AGN source and only 13~low-score AGN candidates fulfil the condition of \texttt{parallax\_over\_error}\,$>$\,5 required for  observational Hertzsprung--Russell diagrams (Figs.~\ref{fig:app:AGN_trn}d and~\ref{fig:app:AGN}d).
A different distribution of high versus low score candidates is expected for every classification attribute that can discriminate AGN from other classes, such as the \g-band magnitude and colours (Figs.~\ref{fig:app:AGN}b,c), the Abbe parameter (Fig.~\ref{fig:app:AGN}f), and the \citet{2011AJ....141...93B} metrics \citep{DR3-DPACP-167}.

Proper motion was not among the classification attributes and Figs.~\ref{fig:AGN_plx_pmra_pmdec}b,c do not show evident biases in the distribution of proper motion components, except for the expected thicker tails and more extreme values from low-score candidates with respect to those from sources in the high-score sample.
The asymmetric proper motion distribution of low-score AGNs, in particular in the declination direction (Fig.~\ref{fig:AGN_plx_pmra_pmdec}c), is characteristic of the motion of stars in the Galaxy \cite[see fig.~D.3 in][]{EDR3-DPACP-133} and thus it suggests stellar contamination.
Table~\ref{tab:agn} summarises the averages and standard deviations of the distributions illustrated in Fig.~\ref{fig:AGN_plx_pmra_pmdec}, in order to help assess biases and similarity with respect to a standard normal distribution.

\begin{table} 
\caption{Summary of the distributions of normalised astrometric parameters for the AGN candidates illustrated in Fig.~\ref{fig:AGN_plx_pmra_pmdec}. The definition of subsets with high and low \texttt{best\_class\_score} is the same as in Fig.~\ref{fig:AGN_plx_pmra_pmdec}. \label{tab:agn}} 
\centering    
\setlength{\tabcolsep}{4pt}
\begin{tabular}{lrc}   
\hline\hline & \\[-2.0ex]                 
Astrometric parameter normalised & Average & Standard\, \\
by its uncertainty (\texttt{gaia\_source})  &         & deviation \\
\hline & \\[-1.5ex]
(\texttt{parallax}+0.017)\,/\,\texttt{parallax\_error} & $-$0.007 & 1.067 \\
~~~~~Subset with high scores & $-$0.434 & 0.803\\
~~~~~Subset with low scores & 0.248 & 1.253\\
 & \\[-1.5ex]   
\texttt{pmra}\,/\,\texttt{pmra\_error} & $-$0.004 & 1.187 \\
~~~~~Subset with high scores & $-$0.007 & 1.048\\
~~~~~Subset with low scores & $-$0.022 & 1.752\\
 & \\[-1.5ex]  
\texttt{pmdec}\,/\,\texttt{pmdec\_error} & $-$0.014 & 1.280 \\
~~~~~Subset with high scores & 0.031 & 1.056\\
~~~~~Subset with low scores & $-$0.096 & 2.116\\
\hline    
\end{tabular}
\end{table}  

While the completeness rate of AGN candidates in Table~\ref{tab:results} is consistent with the assessment of \citet{DR3-DPACP-167}, the contamination rate is deemed underestimated, given the large set of 200\,081 `unknown' positives stated in Table~\ref{tab:results_details}. 
Comparing the 1\,034\,925 AGN candidates with at least five-parameter astrometric solutions with 17~catalogues from the literature as described in \citet{EDR3-DPACP-133}, only about 45~thousand sources were unknown. While many of them may still be genuine AGNs, literature catalogues may also have false positives, thus an approximate estimate for the contamination is $\sim$\,5\,\%.

A comparison of the sources from all of the classes identified by variability that overlap with those from QSO modules published by other \gaia coordination units (gathered in the \texttt{qso\_candidates} table) is presented in tables~12.9 and~12.12 of the \gdr3 documentation \citep{2022gdr3.reptE..12T}. It is noted that the unfiltered \texttt{qso\_candidates} table has significant stellar contamination \citep[estimated at 76\,\% in][where a query to select a purer  sub-sample is indicated]{DR3-DPACP-101}. 


\subsection{$\beta$\,Cephei stars (BCEP)\label{ssec:bcep}}

The classified $\beta$\,Cephei type candidates include 1475 sources, 174 of which are shared with the upper main-sequence oscillator SOS module \citep[see sect.~10.14 of the \gdr3 documentation;][]{2022gdr3.reptE..10R}, although multi-periodicity may not be easily detectable with the current average number of observations per source. More information on the upper main-sequence oscillators is provided in \citet{DR3-DPACP-79}.

Training sources for the $\beta$\,Cephei class included a tail of objects at the blue end with lower intrinsic luminosity than possible for this class (see Fig.~\ref{fig:app:BCEP}d), explaining the contamination by GCAS mentioned in Table~\ref{tab:results_details}.
Figures~\ref{fig:app:BCEP}b,e,g, show that candidates with fainter apparent magnitudes tend to have lower scores, mirroring the decreasing occurrence of training sources towards faint magnitudes (only three training objects had \g median magnitude fainter than 12).

The BCEP candidates were selected from 11 binary, multi-class, multi-stage, and meta-classifiers (Sect.~\ref{ssec:classifier}) according to some minimum probability threshold and two conditions were set in the \gaia reddened colour--colour diagrams, in order to remove sparse outliers with respect to the general colour--colour relation followed by all other sources of this class (with a model of \grp given \bprp, based on known objects; see Fig.~\ref{fig:app:BCEP}c): 
\begin{enumerate}
  \item \texttt{median\_mag\_g\_fov}\,$-$\,\texttt{median\_mag\_rp}\,$<$\break model(\grp~|~\bprp)\,$+$\,0.03\,mag,
  \item \texttt{median\_mag\_g\_fov}\,$-$\,\texttt{median\_mag\_rp}\,$>$\break model(\grp~|~\bprp)\,$-$\,0.03\,mag,
\end{enumerate}
employing field names in the \texttt{vari\_summary} table.

Although not listed in the top-6 false positive types of Table~\ref{tab:results_details}, additional BCEP candidates may be found among the false positives of the following classes: ACV|CP|MCP|ROAM|ROAP|SXARI~(3), BE|GCAS|SDOR|WR~(1), CV~(1), and RS~(1), where the numbers of known sources indicated in parentheses represent lower limits.


\subsection{BE|GCAS|SDOR|WR stars\label{ssec:be}}

This class group includes a set of 8560 eruptive variables of the following types: B-type emission line star, $\gamma$\,Cassiopeiae, S Doradus, and Wolf-Rayet star.
Stars of types BE|GCAS, SDOR, and WR can significantly overlap in the blue and bright end of the Hertzsprung--Russell diagram and share similarities in the irregular light changes. Therefore, although candidates of these types were selected independently (including the verification filters mentioned below), they were subsequently merged, preserving their original classification scores.
The bimodal distribution in apparent magnitude, clearly visible in Figs.~\ref{fig:app:BE_trn}b,e,g and~\ref{fig:app:BE}b,e,g, is not related to different class components, but to sources (mostly BE|GCAS) located in the Galaxy (bright clump) versus the Magellanic Clouds (faint clump). The latter is also associated with the clump with scattered colours in Figs.~\ref{fig:app:BE_trn}c and~\ref{fig:app:BE}c.

The BE|GCAS candidates were obtained from two binary and multi-class classifiers (Sect.~\ref{ssec:classifier}) with some minimum probability level. They were further constrained by the following criteria (employing field names in the \texttt{vari\_summary} and \texttt{gaia\_source} tables).
\begin{enumerate}
    \item A higher level of variability probability than the one used for the general variability detection (Sect.~\ref{ssec:classifier}) was required, in order to prioritise high-amplitude candidates. 
    \item The renormalised unit weight error \texttt{ruwe} was restricted to values lower than~1.4, which was consistent with the vast majority of known objects of this class.
    \item The values of the Abbe parameter were constrained for the three \gaia bands, to increase the contribution of correlated time-dependent variations, given the sampling of \gaia: 
    \begin{enumerate}
        \item \texttt{abbe\_mag\_g\_fov}\,$<$\,0.8,
        \item \texttt{abbe\_mag\_bp}\,$<$\,1,
        \item \texttt{abbe\_mag\_rp}\,$<$\,1.
    \end{enumerate}
\end{enumerate}
Exceptionally, probabilities from the general variability detection classifier were included in the computation of the classification score, to further increase the weight of high-amplitude candidates.
Given the similarities between Be pulsators and SPB stars and related confusion from classifiers, the upper main-sequence oscillator SOS module \citep[see sect.~10.14 of the \gdr3 documentation;][]{2022gdr3.reptE..10R} considered, among other input sources, classifications of BE|GCAS stars (223 of which were published as upper main-sequence oscillators). 

The SDOR classifications were derived from ten binary, multi-class, and meta-classifiers with minimum probability thresholds, some of which were combined with the maximum brightness limits of  \texttt{median\_mag\_g\_fov}\,$<$\,13 or 17\,mag, to exclude a suspicious component characterised by very faint and low-probability candidates, as some of them extended above the chosen minimum probability. The condition of \texttt{ruwe}\,$<$\,2 was applied to all candidates, as no known SDOR star existed above such a limit. Eight further questionable candidates were removed following visual inspection.

The WR candidates were selected from four binary and meta-classifiers according to minimum probability thresholds.
The number of sources within 100~arcsec from each WR candidate (computed by excluding the contribution of the WR source at the centre) was set to be greater than~157, as these relatively young stars were expected to be within the Galactic disc and no WR star was known to exist in lower star density fields.

According to Table~\ref{tab:results_details}, additional BE|GCAS|SDOR|WR candidates may be found among the false positives of the following classes: BCEP~(55), SPB~(42), ACV|CP|MCP|ROAM|ROAP|SXARI~(34), and ACYG~(20), where the numbers of known sources indicated in parentheses represent lower limits.


\subsection{Cepheids (CEP)\label{ssec:cep}}

The classification of Cepheid variables included 16\,141 stars of types $\delta$~Cephei, anomalous Cepheid, and type-II Cepheid. 
For this class, the selection of Cepheids from the relevant SOS module \citep{DR3-DPACP-169} was followed, with the addition of 1154 candidates from three binary and multi-class classifiers (Sect.~\ref{ssec:classifier}) with strict probability conditions, including about 400 sources that were rejected from the SOS module because reliable periods (and thus models) could not be achieved with the \gaia data, although such sources were believed to be genuine Cepheids. 
The pulsating nature of these objects is clearly visible in Figs.~\ref{fig:app:CEP_trn}g and~\ref{fig:app:CEP}g (until noise prevails at the faint end), indicating greater amplitudes in the \bp than in the \rp band. 
Low-amplitude candidates (\texttt{std\_dev\_mag\_g\_fov}\,$<$\,0.06\,mag) and those at the faint end (\texttt{median\_mag\_g\_fov}\,$<$\,20.3\,mag) are associated with low classification scores (Fig.~\ref{fig:app:CEP}e). 
The main classes responsible of the few percent of contamination are represented by RR\,Lyrae, eclipsing binaries, and long-period variables (Table~\ref{tab:results_details}). 

Although not listed in the top-6 false positive types of Table~\ref{tab:results_details}, additional CEP candidates may be found among the false positives of the following classes: RS~(494), RR~(274), LPV~(199), ECL~(114), DSCT|GDOR|SXPHE~(33), YSO~(13), BE|GCAS|SDOR|WR~(11), CV~(7), RCB~(4), S~(4), AGN~(3), ELL~(2), SYST~(2), ACV|CP|MCP|ROAM|ROAP|SXARI~(1), and BCEP~(1), where the numbers of known sources indicated in parentheses represent lower limits.


\subsection{Cataclysmic variables (CV)\label{ssec:cv}}

The 7306 candidates of cataclysmic variables (excluding supernovae and symbiotic stars, described in Sects.~\ref{ssec:sn} and~\ref{ssec:syst}, respectively) were selected from five binary, multi-class, and meta-classifiers (Sect.~\ref{ssec:classifier}), after the application of minimum probability thresholds. The verification of CV classifications lead to the following conditions (employing field names in the \texttt{vari\_summary} table). 
\begin{enumerate}
    \item A higher level of variability probability than the one used for the general variability detection (Sect.~\ref{ssec:classifier}) was required to omit doubtful low signal-to-noise candidates. 
    \item A suspicious clump of extremely blue candidates was filtered out by \texttt{median\_mag\_bp}\,$-$\,\texttt{median\_mag\_g\_fov}\,$>$\,$-$\,4\,mag.
    \item A thin tail of blue outliers was excluded by the condition  \texttt{median\_mag\_bp}\,$-$\,\texttt{median\_mag\_rp}\,$>$\,$-$\,0.5\,mag. 
    \item The number of sources within 100~arcsec from each CV candidate (computed by excluding the contribution of the CV source at the centre) was set to be less than~408, to prevent an increase of candidates towards the end of the tail of the distribution of CV from the literature. 
    \item The following two conditions served to exclude two clumps of candidates in the \texttt{outlier\_median\_g\_fov} versus \texttt{skewness\_mag\_g\_fov} space that were only marginally populated by literature instances:
    \begin{enumerate}
        \item \texttt{outlier\_median\_g\_fov}\,$>$\,20,
         \item \texttt{skewness\_mag\_g\_fov}\,$>$\,$-$\,4.
    \end{enumerate}
\end{enumerate}
The two features with skewness\,<\,$-$\,3 that appear in the plot of \texttt{skewness\_mag\_g\_fov} versus \texttt{abbe\_mag\_g\_fov} (Fig.~\ref{fig:app:CV}f) are related to faint sources, with median \g\,$\approx$\,20.7\,mag, that exhibit a single bright outlier; these CV candidates should be dismissed. 

According to Table~\ref{tab:results_details}, additional CV candidates may be found among the false positives of the SN~(22) and WD~(12) classes, where the numbers of known sources indicated in parentheses represent lower limits.


\subsection{DSCT|GDOR|SXPHE stars\label{ssec:dsct}}

This class group includes 748\,058 candidates of types $\delta$\,Scuti, $\gamma$\,Doradus, and SX\,Phoenicis, in particular as DSCT and SXPHE types could not be distinguished without metallicity or other indicators of type~I versus type~II star. These classifications were also the main contributors to the upper main-sequence oscillator SOS module \citep[see sect.~10.14 of the \gdr3 documentation;][]{2022gdr3.reptE..10R}, which were further analysed in \citet{DR3-DPACP-79}.

This sample is dominated by low amplitude candidates (\texttt{std\_dev\_mag\_g\_fov}\,$<$\,0.01\,mag for 537\,769 sources), which are expected to be significantly contaminated, as suggested from their distribution of \texttt{std\_dev\_mag\_bp}\,/\,\texttt{std\_dev\_mag\_rp}\,$<$\,1 between \texttt{median\_mag\_g\_fov} of 13 and 17\,mag, as shown in Fig.~\ref{fig:app:DSCT}g.
About seven thousand candidates with \bprp < 0.25\,mag, associated with low scores (0.01 on average and clearly visible in Fig.~\ref{fig:app:DSCT}b,d), should be disregarded. 

The DSCT|GDOR|SXPHE candidates were extracted from ten binary, multi-class, multi-stage, and meta-classifiers (Sect.~\ref{ssec:classifier}) with minimum probability thresholds. The following additional criteria were applied (employing field names in the \texttt{vari\_summary} table).
\begin{enumerate}
    \item Removal of outliers in the \gaia colours: 
    \begin{enumerate}
    \item \texttt{median\_mag\_bp}\,$-$\,\texttt{median\_mag\_rp}\,$>$\,$-$0.5\,mag,
    \item \texttt{median\_mag\_bp}\,$-$\,\texttt{median\_mag\_g\_fov}\,$>$\,$-$1\,mag,
    \item \texttt{median\_mag\_g\_fov}\,$-$\,\texttt{median\_mag\_rp}\,$<$\,1\,mag.
    \end{enumerate}
    \item Constraints on the colour--colour scatter around the mean relation (with a model of \grp given \bprp, based on known objects; see Fig.~\ref{fig:app:DSCT}c): 
    \begin{enumerate}
    \item \texttt{median\_mag\_g\_fov}\,$-$\,\texttt{median\_mag\_rp}\,$<$ \break model(\grp~|~\bprp)\,$+$\,0.025\,mag,
    \item \texttt{median\_mag\_g\_fov}\,$-$\,\texttt{median\_mag\_rp}\,$>$ \break model(\grp~|~\bprp)\,$-$\,0.025\,mag.
    \end{enumerate}
\end{enumerate}
The conditions described above were overridden by candidates that existed also in the upper main-sequence oscillator SOS module.

According to Table~\ref{tab:results_details}, additional DSCT|GDOR|SXPHE candidates may be found among the false positives of the following classes: RS~(792), AGN~(275),  ACV|CP|MCP|ROAM|ROAP|SXARI~(49), BCEP~(11), WD~(4), and SDB~(1), where the numbers of known sources indicated in parentheses represent lower limits.


\subsection{Eclipsing binaries (ECL)\label{ssec:ecl}}

The classification of eclipsing binaries included 2\,184\,356 systems of types $\beta$~Persei (Algol), $\beta$~Lyrae, and W\,Ursae Majoris. They were obtained from four meta-classifiers (Sect.~\ref{ssec:classifier}) and followed the selection of the corresponding SOS module described by \citet{DR3-DPACP-170}, including the following  conditions (employing field names in the \texttt{vari\_summary} table).
\begin{enumerate}
\item A minimum apparent brightness in the \g band as defined by  \texttt{median\_mag\_g\_fov}\,$<$\,20\,mag (see Figs.~\ref{fig:app:ECL}b,e,g).
\item A minimum value of the time series skewness, computed from \g-band magnitudes: \texttt{skewness\_mag\_g\_fov}\,$>$\,$-$\,0.2 (see Fig.~\ref{fig:app:ECL}f).
\item A minimum number of clean FoV transits in the \g~band: \texttt{num\_selected\_g\_fov}\,$>$\,15.
\item Further constraints on period and model properties listed in the \texttt{vari\_eclipsing\_binary} table.
\end{enumerate}

The ratio between \bp and \rp amplitudes is close to one (Fig.~\ref{fig:app:ECL}g), as expected from non-pulsating objects (apart from the increasing scatter towards faint magnitudes due to shot noise).
Low-amplitude and low-skewness candidates tend to be associated with low classification scores (Fig.~\ref{fig:app:ECL}e,f). 

According to Table~\ref{tab:results_details}, additional ECL candidates may be found among the false positives of the following classes: RS~(10\,908), S~(5713), RR~(4318), DSCT|GDOR|SXPHE~(3402), LPV~(2507), ELL~(2135), SOLAR\_LIKE~(286),  YSO~(219), BE|GCAS|SDOR|WR~(152), CEP~(104),  ACV|CP|MCP|ROAM|ROAP|SXARI~(61), CV~(28), ACYG~(6), EP~(3), SDB~(3), WD~(3), and MICROLENSING~(2), where the numbers of known sources indicated in parentheses represent lower limits.


\subsection{Ellipsoidal variables (ELL)\label{ssec:ell}}

The classification of 65\,300 ellipsoidal variables targeted input sources for the compact companion SOS module \citep{DR3-DPACP-174}. 
The training objects were not sufficiently representative of the whole sky (Fig.~\ref{fig:app:ELL_trn}a) and thus low scores were assigned to candidates beyond the Galactic bulge region (Fig.~\ref{fig:app:ELL}a). Most of them are likely contaminants, considering that low score classifications contribute negligibly to completeness while they double the contamination rate (see Fig.~\ref{fig:app:ELL_cc}a). As expected, most of the contaminants are represented by W\,Ursae Majoris eclipsing binaries (Table~\ref{tab:results_details}).

The ELL candidates were selected from ten binary, multi-class, and meta-classifiers (Sect.~\ref{ssec:classifier}) with minimum probability thresholds. They fulfil the following conditions (employing field names in the \texttt{vari\_summary} and \texttt{gaia\_source} tables).
\begin{enumerate}
    \item The renormalised unit weight error \texttt{ruwe} was set to values lower than~1.2, to restrict an excess of candidates with high \texttt{ruwe} values (and thus unreliable astrometric solutions).
    \item In order to better fit the distribution of known ELL stars in standard deviation versus median \g magnitude, the minimum level of variability was raised (see Fig.~\ref{fig:app:ELL}e):
    \begin{enumerate}
        \item \texttt{std\_dev\_mag\_g\_fov}\,$>$\,1.5 times the third quartile of the standard deviations in \g, in 0.05\,mag intervals, of 1.6~million reference sources (see Appendix~\ref{app:plots}),
        \item \texttt{std\_dev\_over\_rms\_err\_mag\_g\_fov}\,$>$\,4.
    \end{enumerate}
    \item Additional constraints removed candidates from colour and magnitude ranges that were poorly represented in the literature:
    \begin{enumerate}
        \item \texttt{median\_mag\_bp}\,$-$\,\texttt{median\_mag\_g\_fov}\,$>$\,0.3\,mag,
        \item \texttt{median\_mag\_bp}\,$-$\,\texttt{median\_mag\_g\_fov}\,$>$ \break 0.022\,(\texttt{median\_mag\_bp}\,$-$\,\texttt{median\_mag\_rp})$^2$\,$+$ \break 0.7\,(\texttt{median\_mag\_bp}\,$-$\,\texttt{median\_mag\_rp})\,$-$\,0.65\,mag,
        \item \texttt{median\_mag\_g\_fov}\,$<$\,19\,mag.
    \end{enumerate}
    \item The number of sources within 100~arcsec from each ELL candidate (computed by excluding the contribution of the ELL source at the centre) was set to be greater than~314, to avoid a suspicious concentration of candidates around the Galactic anti-centre.
\end{enumerate}
These criteria were overridden by candidates that existed also in the compact companion SOS module (including the intention of the last item above).

According to Table~\ref{tab:results_details}, additional ELL candidates may be found among the false positives of the ECL~(1056), LPV~(731), SPB~(9), BCEP~(3), and ACYG~(2) classes, where the numbers of known sources indicated in parentheses represent lower limits.


\subsection{Exoplanets (EP)\label{ssec:ep}}

The 214 stars classified with exoplanet transits have percent-level variations in the \g~band, above and below the general variability threshold, as shown in Fig.~\ref{fig:app:EP}e. These stars are relatively bright to allow for high per-FoV photometric precision, thus they are also nearby and distributed rather homogeneously across the sky (Fig.~\ref{fig:app:EP}a).

The weakness of the signal and the typically small number of observations in transit implied the necessity of a dedicated period search algorithm such as that used in the planetary transit SOS module \citep{DR3-DPACP-181}, rather than the generic computationally efficient generalised
Lomb-Scargle method used for all the classes \citep{1985A&AS...59...63H,2009A&A...496..577Z}.
Consequently, the EP classifications were limited to the selection of the corresponding SOS module and their scores were set to one.

A binary XGBoost classifier (Sect.~\ref{ssec:classifier}) was used to extract a sample of 18\,383 potential candidates, without the application of a minimum probability threshold.  All of these sources were processed by the planetary transit SOS module, which returned 214 objects, including known stars with  planetary transits as well as new candidates.

According to Table~\ref{tab:results_details}, at least 93 additional EP candidates may be found among the false positives of the SOLAR\_LIKE class.


\subsection{Long-period variables (LPV)\label{ssec:lpv}}

The classification of long-period variables included 2\,325\,775 stars, among which long secondary period variables, $o$\,Ceti (Mira) stars, (OGLE) small amplitude red giants, and semi-regular types.
They were obtained from four meta-classifiers (of binary and multi-class classifiers, see Sect.~\ref{ssec:classifier}) with minimum probability thresholds and selected according to the criteria defined in the LPV SOS module \citep{DR3-DPACP-171}, some of which were more permissive for classification candidates, as indicated in the following (employing field names in the \texttt{vari\_summary} and \texttt{gaia\_source} tables).
\begin{enumerate}
    \item A minimum number of clean FoV transits in the \g~band: \texttt{num\_selected\_g\_fov}\,$>$\,9 (instead of 12 in the SOS module).
    \item The \gaia reddened \bprp colour redder than 1~mag, estimated as  \texttt{median\_mag\_bp}\,$-$\,\texttt{median\_mag\_rp}\,$>$\,1\,mag.
    \item A minimum fraction of clean FoV transits in the \rp versus \g~band:   \texttt{num\_selected\_rp}\,$/$\,\texttt{num\_selected\_g\_fov}\,$>$\,0.5 (instead of 0.8 in the SOS module).
    \item A minimum amplitude assessed from the 5th to the 95th percentile of the magnitude distribution of \g-band time series: \texttt{trimmed\_range\_mag\_g\_fov}\,$>$\,0.1\,mag (Fig.~\ref{fig:app:LPV}c).
    \item Given the conditions above, LPV candidates satisfied in addition at least one of the following conditions: 
    \begin{enumerate}
       \item \texttt{std\_dev\_over\_rms\_err\_mag\_g\_fov}\,$>$\,4, for a significant signal-to-noise level,
       \item \texttt{median\_mag\_g\_fov}\,$<$\,14\,mag, for a substantial apparent brightness,
       \item \texttt{trimmed\_range\_mag\_g\_fov}\,$>$\,0.5\,mag, for a high amplitude,
       \item a high probability for a source to be classified as an LPV by any of two of the meta-classifiers.
    \end{enumerate}
    \item The number of groups of observations in the \g band (\texttt{visibility\_periods\_used}), separated from other groups by the absence of measurements for at least four days, was not constrained, while it was greater than ten in the SOS module. 
\end{enumerate}
For further details, including an analysis of the difference between the SOS module and the classification LPV candidates, see \citet{DR3-DPACP-171}.

Faint and/or blue candidates in Figs.~\ref{fig:app:LPV}b,d tend to be associated with low scores and might include spurious candidates, as their distribution in the sky suggests scanning law features (Fig.~\ref{fig:app:LPV}a) and the step around \g\,$\approx$\,20\,mag indicates marginal candidates bordering with other classes.

According to Table~\ref{tab:results_details}, additional LPV candidates may be found among the false positives of the following classes: RS~(5761), ELL~(443), YSO~(234),  SYST~(215),  CEP~(73),  AGN~(50), RCB~(42), BE|GCAS|SDOR|WR~(20), and MICROLENSING~(3),  where the numbers of known sources indicated in parentheses represent lower limits.


\subsection{Microlensing events\label{ssec:microlensing}}

The classification of 254 candidate microlensing events was performed independently from the microlensing SOS module \citep{DR3-DPACP-166} and the latter did not consider these classification results as input of potential candidates, as it already extracted all possible candidates with methods specific to this class. 
Nevertheless, the classification candidates had the advantage of fewer assumptions and requirements than the SOS counterparts, in particular regarding the time series modelling.
Among the microlensing candidates, 187 are in common with the SOS module and 67 are unique to classification.

Microlensing sources were selected from ten binary, multi-class, and meta-classifiers (Sect.~\ref{ssec:classifier}) with minimum probability thresholds. The following additional criteria were applied to such candidates (employing field names in the \texttt{vari\_summary} and \texttt{gaia\_source} tables).
\begin{enumerate}
    \item In order to focus on high amplitude events, the \texttt{std\_dev\_mag\_g\_fov} parameter was set to be greater than 0.06\,mag, as only a handful of the known sources (among the classified ones) were below this threshold.
    \item To further support the rationale of the previous item, at the cost of completeness, the most significant candidates were selected by \texttt{outlier\_median\_g\_fov}\,$>$\,100.
    \item The ratio of the mean magnitude, weighted by squared uncertainties, to the unweighted mean magnitude was required to be less than one, as expected from the bias introduced by weighting towards brighter measurements (this condition was fulfilled by almost all known events).
    \item The renormalised unit weight error \texttt{ruwe} was set to be lower than~1.4, considering the distribution of \texttt{ruwe} for known microlensing events peaked at close to one.
    \item The colour \texttt{median\_mag\_g\_fov}\,$-$\,\texttt{median\_mag\_rp} was set bluer than 1.7\,mag, following the distribution of known events. 
\end{enumerate}
Some of the criteria listed above were overridden by visual inspection of a selection of candidates.

As the probability of source-lens alignment along the line-of-sight to the observer increases in high-density regions, microlensing candidates are found, as expected,  prevalently towards the Galactic bulge and a minority of cases in the Galactic disc (see Fig.~\ref{fig:app:MICROLENSING}a). They populate the region of negative \texttt{skewness\_mag\_g\_fov} and low \texttt{abbe\_mag\_g\_fov} values, which  are typical of bright and time-dependent outburst-like events (Fig.~\ref{fig:app:MICROLENSING}f).

Although not listed in the top-6 false positive types of Table~\ref{tab:results_details}, at least three additional microlensing candidates may be found among the false positives of the LPV class.


\subsection{R\,Coronae Borealis stars (RCB)\label{ssec:rcb}}

The 153~stars classified as R\,Coronae Borealis variables were obtained from two meta-classifiers (of binary and multi-class classifiers, see Sect.~\ref{ssec:classifier}), after filtering out low probability candidates and  confirmation by visual inspection.
These high-amplitude variables are characterised by sudden fading by up to several magnitudes (Fig.~\ref{fig:app:RCB}e), followed by an irregular recovery (sample light curves are shown in Figs.~\ref{fig:app:RCB_cc}c,d,e,f). They are simultaneously eruptive and pulsating, although the amplitude of the latter is an order of magnitude lower. As a consequence of the long timescale variations, these objects are also characterised by low \texttt{abbe\_mag\_g\_fov} values (Fig.~\ref{fig:app:RCB}f) and long-period variables are the main source of contamination (Table~\ref{tab:results_details}).

Although not listed in the top-6 false positive types of Table~\ref{tab:results_details}, additional RCB candidates may be found among the false positives of the LPV~(5) and RS~(4) classes, where the numbers of known sources indicated in parentheses represent lower limits.


\subsection{RR\,Lyrae stars (RR)\label{ssec:rr}}

The classification of RR\,Lyrae stars includes 297\,778 variables of fundamental-mode, first-overtone, double mode (and anomalous double mode) types. The selection of  candidates followed the one of the relevant SOS module \citep{DR3-DPACP-168}. 
Additional candidates were obtained from four binary and multi-class classifiers (Sect.~\ref{ssec:classifier}) after the application of strict minimum probability thresholds.

Among the 26\,202 extra RR candidates in classification with respect to the RR\,Lyrae SOS module, 9239 are known RR\,Lyrae stars in the literature compilation of \citet{DR3-DPACP-177} (or 8276 if counting only those flagged by their \texttt{selection} field), 
for which the correct period could not be recovered with the \gdr3 data \citep{DR3-DPACP-168}. 
Other candidates were found to be eclipsing binaries (as already noted in Sect.~\ref{ssec:ecl}), which constitute the main source of contamination for this class (Table~\ref{tab:results_details}). 
About 800 AGN and galaxy contaminants, as well as dubious RR candidates, are listed in tables~5 and~6 of \citet{DR3-DPACP-168}.

Faint sources, affected by extinction in the Galactic disc and bulge, and low-amplitude candidates tend to be associated with low classification scores (Figs.~\ref{fig:app:RR}a,b,e).
Other low-score candidates are found between the main sequence and the white dwarf sequence (Fig.~\ref{fig:app:RR}d) and are located mainly in the Galactic bulge and disc (some are in the Magellanic Clouds too). They are characterised by higher values of the renormalised unit weight error (\texttt{ruwe}) than the other candidates: among all RR classifications with \texttt{parallax\_over\_error}\,$>$\,5, about 80\,\% have \texttt{ruwe}\,$<$\,1.22, while only half of the ones between the main and white-dwarf sequences fulfil this condition.
The pulsating nature of the RR candidates is confirmed in Fig.~\ref{fig:app:RR}g, until the contribution of noise overcomes the expected \bp to \rp~band amplitude ratio.

According to Table~\ref{tab:results_details}, additional RR candidates may be found among the false positives of the following classes: ECL~(5833), DSCT|GDOR|SXPHE~(525),  AGN~(264), GALAXY~(244), S~(168), CEP~(158), CV~(41), and BCEP~(3), where the numbers of known sources indicated in parentheses represent lower limits.


\subsection{RS\,Canum Venaticorum stars (RS)\label{ssec:rs}}

The classification of RS Canum Venaticorum type binary systems includes 742\,263 candidates. Although they are presented separately from SOLAR\_LIKE stars, any confusion between these two classes does not constitute real contamination, as at least one of the components of an RS binary system is a solar-like star.  
Thus, as expected, 20\,050 RS sources are included also in the  \texttt{vari\_rotation\_modulation} table of the solar-like SOS module \citep{DR3-DPACP-173} and the top RS false positive classes listed in Table~\ref{tab:results_details} are ROT and BY, which should not be considered as contaminants (as they are typical solar-like types).

The RS candidates were selected from ten binary, multi-class, and meta-classifiers (Sect.~\ref{ssec:classifier}) with minimum probability thresholds. The following additional criteria were applied (employing field names in the \texttt{vari\_summary} and \texttt{gaia\_source} tables).
\begin{enumerate}
    \item The renormalised unit weight error \texttt{ruwe} was set to be lower than~1.1, to exclude many candidates in a range sparsely populated by literature RS instances.
    \item To better fit the bulk of known RS stars in standard deviation versus median \g magnitude and exclude suspicious low-amplitude candidates, the minimum level of variability was raised (Fig.~\ref{fig:app:RS}e) by \texttt{std\_dev\_mag\_g\_fov}\,$>$\,10$^{0.2}$ times the third quartile of the standard deviations in \g, in 0.05\,mag intervals, of 1.6~million reference sources (see Appendix~\ref{app:plots}).
    \item To exclude a clump of negatively skewed candidates (by artefact) and for symmetry reasons, the following condition was applied:  |\,\texttt{skewness\_mag\_g\_fov}\,|\,$<$\,3 (as visible in Fig.~\ref{fig:app:RS}f).
    \item Two conditions in the \gaia reddened colour--colour diagrams excluded a significant excess of candidates where only few known RS were present and followed  the general colour--colour relation of all other known RS sources (with a model of \bpg given \grp, as shown in Fig.~\ref{fig:app:RS}c): 
    \begin{enumerate}
        \item \texttt{median\_mag\_bp}\,$-$\,\texttt{median\_mag\_g\_fov}\,$<$\break model(\bpg~|~\grp)\,$+$\,0.07\,mag,
        \item \texttt{median\_mag\_bp}\,$-$\,\texttt{median\_mag\_g\_fov}\,$>$\break model(\bpg~|~\grp)\,$-$\,0.06\,mag. 
    \end{enumerate}\end{enumerate}

According to Table~\ref{tab:results_details}, additional RS candidates may be found among the false positives of the following classes: ECL~(10\,791),  DSCT|GDOR|SXPHE~(1598), YSO~(1542), ELL~(210), RR~(165), SOLAR\_LIKE~(148), and CEP~(36),  where the numbers of known sources indicated in parentheses represent lower limits.


\subsection{Short-timescale objects (S)\label{ssec:s}}

This class was trained with stars exhibiting rapid light variations that however were not well studied in the literature. It resulted in 512\,005 short-timescale candidates, originally targeting possible input for the short-timescale SOS module \citep[described in sect.~10.12 of the \gdr3 documentation;][]{2022gdr3.reptE..10R}. Eventually, a similar goal was pursued independently (with different  data types, as the SOS module employed per-CCD photometry), leading to two rather complementary sets. 
This class is meant to be exploratory and considered as a sample where not well-defined short-timescale sources (including those from other classes published in \gdr3) can be found, such as, for instance, nine BLAP sources from \citet{2017NatAs...1E.166P}. 

The assessment of completeness and contamination in Table~\ref{tab:results} is not truly meaningful for this class, as completeness is relative to objects that might be better resolved in \gaia and thus classified accordingly, while the main source of contamination derives from EW-type eclipsing binaries (Table~\ref{tab:results_details}), which do vary on short timescales. 

The S candidates were obtained from seven binary, multi-class, and meta-classifiers (Sect.~\ref{ssec:classifier}) with minimum probability thresholds. The lowest probability candidates excluded by such thresholds corresponded also to the least sampled sources, causing the appearance of regions (intersected by the Ecliptic) almost devoid of candidates,  as a consequence of the \gaia scanning law (see Fig.~\ref{fig:app:S}a).
Additional selection criteria are described in the following (employing field names in the \texttt{vari\_summary} table).
\begin{enumerate}
    \item To remove a suspicious peak of low-amplitude candidates without counterpart in the literature, a minimum standard deviation threshold was set: \texttt{std\_dev\_mag\_g\_fov}\,$>$\,0.05\,mag, as noticeable in Fig.~\ref{fig:app:S}e.
    \item The tails of colour distributions of S candidates that were overly represented in proportion to the literature were reduced by the following conditions: 
    \begin{enumerate}
     \item \texttt{median\_mag\_bp}\,$-$\,\texttt{median\_mag\_rp}\,$<$\,2.6\,mag,
     \item \texttt{median\_mag\_bp}\,$-$\,\texttt{median\_mag\_g\_fov}\,$<$\,1.3\,mag,
     \item \texttt{median\_mag\_g\_fov}\,$-$\,\texttt{median\_mag\_rp}\,$<$\,1.6\,mag (see Fig.~\ref{fig:app:S}c).
    \end{enumerate} 
    \item To exclude a population of candidates with magnitude time series associated with extremely negative skewness and a few positively skewed outliers, only \texttt{skewness\_mag\_g\_fov} values between $-$1.4 and 4 were accepted (Fig.~\ref{fig:app:S}f). 
    \item The number of sources within a radius of 3~arcsec of each S candidate (excluding the contribution of the candidate at the centre) was set to be less than~5, to exclude candidates in the most crowded environments.
    \item The minimum ratio of two (unpublished) spectral shape components in the \bp band (SSC ids~0 and~2)\footnote{\url{https://gea.esac.esa.int/archive/documentation/GDR3/Data_processing/chap_cu5pho/sec_cu5pho_intro/cu5_pho_intro_ssc.html}\label{foot:ssc}} was set to match approximately the one from S objects in the literature. 
    \item A constraint on the amount of scan-angle dependent signal \citep{DR3-DPACP-164}, quantified by the Spearman correlation $r_{{\rm ipd},G}$ between the \g-band epoch photometry and the model of the Image Parameter Determination \citep[IPD; see sect.~3.3.6 of the \gdr3 documentation in][]{2022gdr3.reptE...3C}, was applied as $r_{{\rm ipd},G}$\,$<$\,0.45, where the upper limit corresponded to the minimum between two apparent distributions.  This condition halved the number of S sources in the literature, but it was necessary to avoid a sample otherwise dominated by candidates with spurious signals.
\end{enumerate}

Although not listed in the top-6 false positive types of Table~\ref{tab:results_details}, additional S sources from the literature are found among the following classes: ECL~(24), RR~(12), DSCT|GDOR|SXPHE~(9), CV~(3), RS~(2), WD~(2), and AGN~(1), where the numbers of known sources indicated in parentheses represent lower limits. 


\subsection{Subdwarf B-type stars (SDB)\label{ssec:sdb}}

The classification of subdwarf~B variables returned 893 candidates meant to represent stars of types V1093\,Herculis and V361\,Hydrae. While they occupy the expected location in the observational Hertzsprung--Russell diagram (Fig.~\ref{fig:app:SDB}d), their variability might not be due to pulsation only and the presence of the latter cannot be assured when concurrent high-amplitude phenomena exist, such as:
\begin{enumerate}
\item the reflection effect due to irradiation of a cool companion by a hot subdwarf primary and consequent re-radiation from the illuminated side of the cool companion;
\item the tidal distortion of the cool companion, which generates ellipsoidal-like flux variations (and possible mass transfer);
\item the possible presence of spots on a rotating SDB star.
\end{enumerate}
The first two items are related to binary systems with a hot subdwarf and a cool companion. They cause larger variations in the red band than in the blue band. The difference between the \bp and \rp amplitudes of the reflection and tidal effects arises from the fact that the blue part of the flux variations of the cool companion is diluted in the blue-band flux of the hot subdwarf, leading to a smaller percentage of flux variation in \bp than in \rp \citep[for example, see][]{2014A&A...570A..70S}.
A multi-periodic analysis of the SDB candidates should be performed to account for the effects of binarity and/or spots, in addition to the stellar oscillations.

These candidates were selected from ten binary, multi-class, and meta-classifiers (Sect.~\ref{ssec:classifier}) with minimum probability thresholds. 
Additional conditions are described as follows (employing field names in the \texttt{vari\_summary} and \texttt{gaia\_source} tables).
\begin{enumerate}
    \item To highlight candidates with clear signals, the minimum level of variability was raised by requiring \texttt{std\_dev\_mag\_g\_fov} to be greater than the third quartile of the standard deviations in \g, in 0.05\,mag intervals, of 1.6~million reference sources (see Appendix~\ref{app:plots}). Additionally, a higher level of variability probability than the one used for the general variability detection (Sect.~\ref{ssec:classifier}) was required (Fig.~\ref{fig:app:SDB}e).
    \item Sparse outliers with respect to the general colour--colour relation followed by all other candidates were removed by the condition \texttt{median\_mag\_bp}\,$-$\,\texttt{median\_mag\_g\_fov}\,$>$ 0.33\,(\texttt{median\_mag\_bp}\,$-$\,\texttt{median\_mag\_rp})\,$-$\,0.03\,mag.
    \item The renormalised unit weight error \texttt{ruwe} was restricted to values lower than~1.15, to remove a tail of SDB candidates at high \texttt{ruwe} values, which was marginally represented in the literature.
    \item  The possible association of SDB candidates with crowded regions (mostly around the Galactic bulge) was limited by setting the number of sources within 100~arcsec from each SDB candidate (computed by excluding the contribution of the SDB source at the centre) to be less than~471.  
\end{enumerate}

Although not listed in the top-6 false positive types of Table~\ref{tab:results_details}, at least one additional SDB candidate may be found among the false positives of the WD class.


\subsection{Supernovae (SN)\label{ssec:sn}}

The 3029 classified supernovae represent some of the most extreme types of cataclysmic variables.
As anticipated in Sect.~\ref{sec:results}, SN candidates are the least sampled sources, given their transient detectability, with an average number of clean observations in the \g band \texttt{num\_selected\_g\_fov} of 8 (versus 45) within an average time interval \texttt{time\_duration\_g\_fov} of 100 (versus 941)~days, where the comparison in parentheses refers to all sources in the \texttt{vari\_classifier\_result} table.

The \grp versus~\bpg diagram in Fig.~\ref{fig:app:SN}c suggests that the galaxies hosting the SN candidates are detected, because the extra flux that is collected for extended sources in the \bp and \rp bands with respect to \g (see Sect.~\ref{ssec:galaxy}) causes the overall negative slope of the colour--colour distribution.
Classification scores of the candidates tend to be high for sources with high signal-to-noise (Fig.~\ref{fig:app:SN}e).
The linear features observed in the \texttt{skewness\_mag\_g\_fov} versus \texttt{abbe\_mag\_g\_fov} diagram (Fig.~\ref{fig:app:SN}f) are due to sources with the least amount of observations (typically five).

The SN candidates were obtained from ten binary, multi-class, and meta-classifiers (Sect.~\ref{ssec:classifier}) with minimum probability thresholds. Additional verification filters are described as follows (employing field names in the \texttt{vari\_summary} and \texttt{gaia\_source} tables).
\begin{enumerate}
    \item Given the slow SN luminosity decay with respect to \gaia's average sampling, most candidates have an Abbe value lower than about~0.5. A tail of candidates with Abbe  greater than~1 (with no counterpart in the literature) was excluded by the condition \texttt{abbe\_mag\_g\_fov}\,$<$\,1 (Fig.~\ref{fig:app:SN}f).
    \item Obvious contamination-dominated candidates at low Galactic latitudes ($b$) were removed by requiring |\,\texttt{b}\,|\,$>$\,7\,degrees (Fig.~\ref{fig:app:SN}a). 
    \item In order to favour candidates with a clear signal, the single-band Stetson index \texttt{stetson\_mag\_g\_fov} was set to be greater than~8. 
    \item To remove a small number of outliers in the \bprp distribution, \texttt{median\_mag\_bp}\,$-$\,\texttt{median\_mag\_rp} was restricted between $-$1 and~2\,mag.
    \item The minimum number of clean measurements in the \g band was raised to five by \texttt{num\_selected\_g\_fov}\,$>$\,4 (it was 3). 
    \item The number of sources within 100~arcsec from each SN candidate (computed by excluding the contribution of the SN at the centre) was set to be less than~314, to remove a high-density tail of candidates prone to  contamination.
\end{enumerate}
    
According to Table~\ref{tab:results_details}, at least 106 additional SN candidates may be found among the false positives of the GALAXY class.


\subsection{Solar-like stars (SOLAR\_LIKE)\label{ssec:solar_like}}

The classification of 1\,934\,844 stars with solar-like variability, such as flaring and rotating spotted stars, was obtained independently of the rotational modulation SOS module \citep{DR3-DPACP-173}.
Further insights on stellar chromospheric activity using \gaia's radial velocity spectrometer are described in \citet{DR3-DPACP-175}.

The scanning law features in the sky map of candidates (Fig.~\ref{fig:app:SOLAR}a) reflect the dependence of the solar-like identifications on the number of observations, which was learnt from the distribution of training sources that included \gaia~DR2 results, as shown in Fig.~\ref{fig:app:SOLAR_trn}a. The average number of \g-band clean observations (\texttt{num\_selected\_g\_fov}) is~58 for solar-like candidates, with respect to~45 for all classification results.

The contamination by constant stars (Table~\ref{tab:results_details}) is not unexpected, given the low-amplitude signal (for example, see Figs.~\ref{fig:app:SOLAR_LIKE_cc}c--f), and the true nature of such objects depends on the photometric precision of those sources in \gdr3 with respect to that of other surveys.

The solar-like candidates were obtained from 11 binary, multi-class, and meta-classifiers (Sect.~\ref{ssec:classifier}) with minimum probability thresholds and two additional conditions in the \gaia reddened colour--colour diagrams, to enforce the general colour--colour relation followed by most solar-like sources in the literature (with a model of \grp given \bprp, based on known objects; see Fig.~\ref{fig:app:SOLAR}c): 
\begin{enumerate}
  \item \texttt{median\_mag\_g\_fov}\,$-$\,\texttt{median\_mag\_rp}\,$<$\break model(\grp~|~\bprp)\,$+$\,0.01\,mag,
  \item \texttt{median\_mag\_g\_fov}\,$-$\,\texttt{median\_mag\_rp}\,$>$\break model(\grp~|~\bprp)\,$-$\,0.02\,mag,
\end{enumerate}
employing field names in the \texttt{vari\_summary} table.
The impact of these conditions are also apparent in other colour--colour diagrams, such as the one in Fig.~\ref{fig:app:SOLAR}c.

According to Table~\ref{tab:results_details}, additional SOLAR\_LIKE candidates may be found among the false positives of the following classes: RS~(41\,913), ECL~(5804),  YSO~(1922), DSCT|GDOR|SXPHE~(1052), RR~(314), WD~(3), and SDB~(1), where the numbers of known sources indicated in parentheses represent lower limits.


\subsection{Slowly pulsating B-type stars (SPB)\label{ssec:spb}}

The classified slowly pulsating B~stars include 1228 sources, among which 434 contributed to the upper main-sequence oscillator SOS module \citep[see sect.~10.14 of the \gdr3 documentation;][]{2022gdr3.reptE..10R}. Detailed analyses on this type of objects are presented in \citet{DR3-DPACP-79}.

The classification SPB candidates were obtained from eight binary, multi-class, multi-stage, and meta-classifiers (Sect.~\ref{ssec:classifier}), after the application of minimum probability thresholds and of an additional condition on the \bprp colour, namely \texttt{median\_mag\_bp}\,$-$\,\texttt{median\_mag\_rp}\,$<$\,0.15\,mag (employing field names in the \texttt{vari\_summary} table). The main reason of the colour cut was to remove suspicious candidates, employing a stricter colour range than the one used for training (see Figs.~\ref{fig:app:SPB_trn}b versus~\ref{fig:app:SPB}b), according to the SPB candidates that were also part of the results of the upper main-sequence oscillator SOS module. 
Most SPB contaminants originated from classes in the ACV|CP|MCP|ROAM|ROAP|SXARI group (Table~\ref{tab:results_details}), as expected from the presence of known SPB stars contaminating this group of types.

Although not listed in the top-6 false positive types of Table~\ref{tab:results_details}, additional SPB candidates may be found among the false positives of the following classes: DSCT|GDOR|SXPHE~(17), ACV|CP|MCP|ROAM|ROAP|SXARI~(13), and RS~(1), where the numbers of known sources indicated in parentheses represent lower limits.


\subsection{Symbiotic stars (SYST)\label{ssec:syst}}

The classification of 649 symbiotic variable stars was obtained from a single binary meta-classifier (Sect.~\ref{ssec:classifier}), after the application of a minimum probability threshold that selected the high probability component (to which almost all known symbiotic stars belonged) of a bimodal distribution.

The SYST candidates are prevalently distributed in high stellar density regions, such as the Galactic bulge, disc, and Magellanic Clouds (Fig.~\ref{fig:app:SYST}a).
The stars in the red clump in the colour--magnitude diagram in Fig.~\ref{fig:app:SYST}b are mostly in the Galactic disc and bulge. Given the long timescale variations typical of this class (see Figs.~\ref{fig:app:SYST_cc}c--f for a sample of light curves), the SYST candidates tend to have small values of the \texttt{abbe\_mag\_g\_fov} parameter (Fig.~\ref{fig:app:SYST}f) and long-period variables constitute their main source of contamination   (Table~\ref{tab:results_details}).

Although not listed in the top-6 false positive types of Table~\ref{tab:results_details}, additional SYST candidates may be found among the false positives of the LPV~(6) and RS~(1) classes, where the numbers of known sources indicated in parentheses represent lower limits.


\subsection{Variable white dwarfs (WD)\label{ssec:wd}}

The classification of 910 white dwarf variables intended to target pulsating stars of types ZZ\,Ceti, V777\,Herculis, and GW\,Virginis. However, training sources, in particular the ones based simply on variability and location in the observational Hertzsprung--Russell diagram \citep[such as][]{2020svos.conf...11E}, most likely included photometric variations due to binarity and spots (see Sect.~\ref{ssec:sdb}), 
and not necessarily pulsation. Because of this reason, there are fewer WD candidates than the corresponding training sources (which account for less than a quarter of the selected candidates). 
The reflection and tidal effects in binary sytems with a WD and a cooler companion cause larger variations in \rp than in \bp, as observed in many training and classified sources (Figs.~\ref{fig:app:WD_trn}g and~\ref{fig:app:WD}g). 
The selection of the most variable WD candidates implicitly favoured effects of binarity or spots with respect to those from stellar oscillations.

Given the low intrinsic brightness of WDs, the ones detectable by \gaia are nearby \citep[for example, see fig.~2 of][]{2019A&A...623A.110G} and thus they are distributed rather homogeneously in the sky, as shown in Figs.~\ref{fig:app:WD_trn}a and \ref{fig:app:WD}a.
The computation of the classification score included probabilities from one multi-class meta-classifier and from the general variability detection classifier, to increase the relevance of high-amplitude candidates.
The main contaminating classes are represented by those of cataclysmic variables and post-common envelope binaries (Table~\ref{tab:results_details}), both of which consist of systems that include a WD.

The variable WD candidates were obtained from 11 binary, multi-class, and meta-classifiers (Sect.~\ref{ssec:classifier}) with minimum probability thresholds. Additional verification filters are described in the following (employing field names in the \texttt{vari\_summary} and \texttt{gaia\_source} tables).
\begin{enumerate}
    \item Since WD candidates tended to be on average less variable than instances from the literature, a higher level of variability probability than the one used for the general variability detection (Sect.~\ref{ssec:classifier}) was required (Fig.~\ref{fig:app:WD}e).
    \item A significant clump of candidates at \bprp$\approx$\,1\,mag, with no WD counterpart in the literature and directed towards the Galactic bulge, was excluded by the condition \texttt{median\_mag\_bp}$-$\texttt{median\_mag\_rp}$<$0.5\,mag\,(Figs.\,\ref{fig:app:WD}b,d). 
    \item The renormalised unit weight error \texttt{ruwe} was set to be lower than~1.15, to remove contaminants towards the Magellanic Clouds (only one known WD lost among the classified ones). 
    \item The corrected \bp and \rp flux excess factor \citep{2021A&A...649A...3R} divided by its scatter was set to be less than~3.6, to exclude outliers with significant flux excess in \bp and \rp typical of extended objects (for example, see Sect.~\ref{ssec:galaxy}). The removed objects corresponded to the blue-end of the \bpg distribution, which was devoid of literature WD counterparts.
\end{enumerate}

Although not listed in the top-6 false positive types of Table~\ref{tab:results_details}, additional variable WD candidates may be found among the false positives of the DSCT|GDOR|SXPHE~(4) and ECL~(2) classes, where the numbers of known sources indicated in parentheses represent lower limits.


\subsection{Young stellar objects (YSO)\label{ssec:yso}}

The classification of young stellar objects included 79\,375 sources of several types (listed in item~\ref{itm:yso} of Sect.~\ref{sssec:classes}). These candidates were validated in detail by \citet{DR3-DPACP-172}. 
The low completeness and high contamination (mainly by RS and SOLAR\_LIKE stars; see Table~\ref{tab:results_details}) are expected for YSOs identified in the optical wavelengths. While complementary observations in the infrared bands would help reduce the confusion between YSO and solar-like stars, the contamination rate and false positives listed in Tables~\ref{tab:results} and~\ref{tab:results_details}, respectively, are significantly overestimated, considering subsequent studies on the RS and BY classifications of \citet{2020ApJS..249...18C}, as reported in \citet{DR3-DPACP-172}.

The YSO candidates were selected from three binary and multi-class classifiers (Sect.~\ref{ssec:classifier}) with minimum probability thresholds. Additional verification filters are described as follows (employing field names in the \texttt{vari\_summary} and \texttt{gaia\_source} tables), often with a common criterion that trimmed one or both ends of a distribution, when a high fraction of unknown YSO candidates was removed at the cost of a low fraction of known YSOs (among the classified sources).
\begin{enumerate}
    \item The minimum variability probability of the general variability detection (Sect.~\ref{ssec:classifier}) was set to be greater than the 5th percentile of known YSOs (Fig.~\ref{fig:app:YSO}e).
    \item The \g-band $\chi^2$ value was required to be greater than the 5th percentile of known YSOs.
    \item The \texttt{std\_dev\_over\_rms\_err\_mag\_g\_fov} parameter was set to be greater than the 5th percentile of known YSOs.
    \item The single-band \texttt{stetson\_mag\_g\_fov} index was set to be greater than the 5th percentile of known YSOs.
    \item The proper motion components in the right ascension and declination directions (\texttt{pmra} and \texttt{pmdec}, respectively), were restricted between the 1st and the 99th percentiles of known YSOs.
    \item The \texttt{parallax\_over\_error} ratio was set to be greater than~3, for a minimal significance of parallax and also for a selection of sources that were not too distant for YSOs to be observable.
    \item Two conditions in \bpg versus \bprp selected the YSO candidates close to the general colour--colour relation followed by most YSOs in the literature (with a model of \bpg given \bprp, based on known objects; see Fig.~\ref{fig:app:YSO}c): 
    \begin{enumerate}
        \item \texttt{median\_mag\_bp}\,$-$\,\texttt{median\_mag\_g\_fov}\,$<$\break model(\bpg~|~\bprp)\,$+$\,0.1\,mag,
        \item \texttt{median\_mag\_bp}\,$-$\,\texttt{median\_mag\_g\_fov}\,$>$\break model(\bpg~|~\bprp)\,$-$\,0.04\,mag.
    \end{enumerate}
\end{enumerate}

According to Table~\ref{tab:results_details}, additional YSO candidates may be found among the false positives of the following classes: AGN~(328), CV~(20),  BE|GCAS|SDOR|WR~(9), and RCB~(5), where the numbers of known sources indicated in parentheses represent lower limits.


\subsection{Galaxies (GALAXY, in \texttt{galaxy\_candidates})\label{ssec:galaxy}}

The classification of 2\,451\,364 galaxies was made possible by their apparent variability in the \gaia photometry.
As mentioned in Sect.~\ref{sec:data}, galaxies can be affected by spurious signals peculiar to \gaia's detection and measurement strategy \citep{DR3-DPACP-164}. 
The main role of galaxies for the classification of variable objects was to reduce the impact of artificial variations on the identification of candidates of genuine variability types, as already noticed in \gaia~DR2 \citep{2019A&A...622A..60C}.

Unlike stars, galaxies occupy the red end in \grp and the blue end in \bpg (see Fig.~\ref{fig:app:GALAXY_ALL}c), because the \bp and \rp window size is much larger than the \g~one, thus more flux from extended objects is included in the sum of \bp and \rp bands than in the \g~band. This discrepancy is estimated by the field \texttt{gaia\_source.phot\_bp\_rp\_excess\_factor} \citep{2021A&A...649A...3R}.
As expected, only a few training sources or galaxy candidates fulfil the condition of \texttt{parallax\_over\_error}\,$>$\,5 required for  observational Hertzsprung--Russell diagrams (Figs.~\ref{fig:app:GALAXY_trn}d and~\ref{fig:app:GALAXY_ALL}d).

Galaxy candidates were obtained from three binary, multi-class, and meta-classifiers (Sect.~\ref{ssec:classifier}) with minimum probability thresholds. They were further filtered by the following conditions (employing field names in the \texttt{vari\_summary} and \texttt{gaia\_source} tables).
\begin{enumerate}
    \item Removal of extremely blue and red outliers:
    \begin{enumerate}
        \item \texttt{median\_mag\_bp}\,$-$\,\texttt{median\_mag\_rp} in the range from 0 to 3\,mag,
         \item \texttt{median\_mag\_bp}\,$-$\,\texttt{median\_mag\_g\_fov}\,$>$\,$-$\,4\,mag,
         \item \texttt{median\_mag\_g\_fov}\,$-$\,\texttt{median\_mag\_rp} in the range from 0.5 to 5.5\,mag.
    \end{enumerate}
   \item The distribution of the \bp and \rp flux excess factor \texttt{phot\_bp\_rp\_excess\_factor}  showed a bimodal distribution and known galaxies corresponded to the mode at high values, so \texttt{phot\_bp\_rp\_excess\_factor}  was set to be greater than five (excluding the first peak of the distribution). 
    \item With respect to point sources, extended objects are often associated with higher positional uncertainty (further amplified by the causes of the spurious photometric variations) and the condition \texttt{astrometric\_excess\_noise}\,$>$\,7\,mas matched the distribution of known galaxies.
    \item The number of sources within 100~arcsec from each GALAXY candidate (computed by excluding the contribution of the galaxy at the centre) was set to be less than~314, affecting mostly candidates around the Galactic plane and behind the Magellanic Clouds (Fig.~\ref{fig:app:GALAXY_ALL}a).
    \item It was further required that \texttt{num\_selected\_rp}\,$>$\,5.
    \item Two (unpublished) spectral shape components in the \bp band (SSC ids~2 and~3)$^{\rm \ref{foot:ssc}}$ were set to be greater than minimum thresholds, according to the galaxy distributions in the literature. 
\end{enumerate}

According to Table~\ref{tab:results_details}, additional GALAXY candidates may be found among the false positives of the following classes: LPV~(1364), S~(244), AGN~(48), and SN~(2), where the numbers of known sources indicated in parentheses represent lower limits.
In addition to those of galaxy contaminants in other classes, as mentioned in the beginning of Sect.~\ref{sec:results}, only the light curves of the sources included in the \gaia Andromeda Photometric Survey \citep{DR3-DPACP-142} are published (Fig.~\ref{fig:app:GALAXY}).

A comparison of the sources from all of the classes identified by variability that overlap with those from galaxy modules published by other \gaia coordination units (gathered in the \texttt{galaxy\_candidates} table) is presented in tables~12.15 and~12.18 of the \gdr3 documentation \citep{2022gdr3.reptE..12T}. It is noted that the unfiltered \texttt{galaxy\_candidates} table has significant stellar contamination \citep[see][where a query to select a purer sub-sample is indicated]{DR3-DPACP-101}. 
An example of a query to extract our GALAXY candidates from the \texttt{galaxy\_candidates} table is presented in Appendix~\ref{app:queries}.


\section{Conclusions\label{sec:conclusions}}

The \gdr3 photometric time series provided sufficient information to classify ten million variable objects into two dozen variability class groups across the whole sky. 
This combination of number of sources and classes made it one of the largest and most uniformly constructed variable source catalogues in the literature.
The cross-match of \gaia sources with an extensive compilation of known variability types \citep{DR3-DPACP-177} enabled a detailed exploitation of the knowledge in the literature for supervised machine learning and for the assessment of the results.
A multi-classifier approach made it possible to obtain suitable models for a large variety of variability classes, involving several types of pulsating stars, eclipsing binaries, ellipsoidal variables, spotted stars, eruptive and cataclysmic phenomena, stochastic variations of AGNs,  microlensing events, and planetary transits. Almost half of the genuine variable sources (4.7~million) and several classes are available uniquely as classification results (in the \texttt{vari\_classifier\_result} table), while the other variable sources are (also) included among the SOS module results. Galaxies were detected by an artificial signal of \gaia, which might have led to a biased yet, nevertheless, numerous addition to the extra-galactic content of this data release.

In \gaia~DR4, the number of photometric epochs will double and additional input data types, such as the \bp and \rp spectra (as time series, as well as averaged in time) and radial velocities, will be available. Together with ongoing developments in our attribute extraction and classification techniques, the discernibility of variability types is expected to further improve and allow for a significant increase in the number of classified sources and related classes. 
 

\begin{acknowledgements}
This work presents results from the European Space Agency (ESA) space mission \gaia. \gaia\ data are being processed by the \gaia\ Data Processing and Analysis Consortium (DPAC). Funding for the DPAC is provided by national institutions, in particular the institutions participating in the \gaia\ MultiLateral Agreement (MLA). The \gaia\ mission website is \url{https://www.cosmos.esa.int/gaia}. The \gaia\ archive website is \url{https://archives.esac.esa.int/gaia}.

The \gaia\ mission and data processing have financially been supported by, in alphabetical order by country:

the Algerian Centre de Recherche en Astronomie, Astrophysique et G\'{e}ophysique of Bouzareah Observatory;

the Austrian Fonds zur F\"{o}rderung der wissenschaftlichen Forschung (FWF) Hertha Firnberg Programme through grants T359, P20046, and P23737;

the BELgian federal Science Policy Office (BELSPO) through various PROgramme de D\'{e}veloppement d'Exp\'{e}riences scientifiques (PRODEX) grants, the Research Foundation Flanders (Fonds Wetenschappelijk Onderzoek) through grant VS.091.16N, the Fonds de la Recherche Scientifique (FNRS), and the Research Council of Katholieke Universiteit (KU) Leuven through grant C16/18/005 (Pushing AsteRoseismology to the next level with TESS, GaiA, and the Sloan DIgital Sky SurvEy -- PARADISE);

the Brazil-France exchange programmes Funda\c{c}\~{a}o de Amparo \`{a} Pesquisa do Estado de S\~{a}o Paulo (FAPESP) and Coordena\c{c}\~{a}o de Aperfeicoamento de Pessoal de N\'{\i}vel Superior (CAPES) - Comit\'{e} Fran\c{c}ais d'Evaluation de la Coop\'{e}ration Universitaire et Scientifique avec le Br\'{e}sil (COFECUB);

the Chilean Agencia Nacional de Investigaci\'{o}n y Desarrollo (ANID) through Fondo Nacional de Desarrollo Cient\'{\i}fico y Tecnol\'{o}gico (FONDECYT) Regular Project 1210992 (L.~Chemin);

the National Natural Science Foundation of China (NSFC) through grants 11573054, 11703065, and 12173069, the China Scholarship Council through grant 201806040200, and the Natural Science Foundation of Shanghai through grant 21ZR1474100;  

the Tenure Track Pilot Programme of the Croatian Science Foundation and the \'{E}cole Polytechnique F\'{e}d\'{e}rale de Lausanne and the project TTP-2018-07-1171 `Mining the Variable Sky', with the funds of the Croatian-Swiss Research Programme;

the Czech-Republic Ministry of Education, Youth, and Sports through grant LG 15010 and INTER-EXCELLENCE grant LTAUSA18093, and the Czech Space Office through ESA PECS contract 98058;

the Danish Ministry of Science;

the Estonian Ministry of Education and Research through grant IUT40-1;

the European Commission's Sixth Framework Programme through the European Leadership in Space Astrometry (\href{https://www.cosmos.esa.int/web/gaia/elsa-rtn-programme}{ELSA}) Marie Curie Research Training Network (MRTN-CT-2006-033481), through Marie Curie project PIOF-GA-2009-255267 (Space AsteroSeismology \& RR Lyrae stars, SAS-RRL), and through a Marie Curie Transfer-of-Knowledge (ToK) fellowship (MTKD-CT-2004-014188); the European Commission's Seventh Framework Programme through grant FP7-606740 (FP7-SPACE-2013-1) for the \gaia\ European Network for Improved data User Services (\href{https://gaia.ub.edu/twiki/do/view/GENIUS/}{GENIUS}) and through grant 264895 for the \gaia\ Research for European Astronomy Training (\href{https://www.cosmos.esa.int/web/gaia/great-programme}{GREAT-ITN}) network;

the European Cooperation in Science and Technology (COST) through COST Action CA18104 `Revealing the Milky Way with \gaia\ (MW-\gaia)';

the European Research Council (ERC) through grants 320360, 647208, and 834148 and through the European Union's Horizon 2020 research and innovation and excellent science programmes through Marie Sk{\l}odowska-Curie grant 745617 (Our Galaxy at full HD -- Gal-HD) and 895174 (The build-up and fate of self-gravitating systems in the Universe) as well as grants 687378 (Small Bodies: Near and Far), 682115 (Using the Magellanic Clouds to Understand the Interaction of Galaxies), 695099 (A sub-percent distance scale from binaries and Cepheids -- CepBin), 716155 (Structured ACCREtion Disks -- SACCRED), 951549 (Sub-percent calibration of the extragalactic distance scale in the era of big surveys -- UniverScale), and 101004214 (Innovative Scientific Data Exploration and Exploitation Applications for Space Sciences -- EXPLORE);

the European Science Foundation (ESF), in the framework of the \gaia\ Research for European Astronomy Training Research Network Programme (\href{https://www.cosmos.esa.int/web/gaia/great-programme}{GREAT-ESF});

the European Space Agency (ESA) in the framework of the \gaia\ project, through the Plan for European Cooperating States (PECS) programme through contracts C98090 and 4000106398/12/NL/KML for Hungary, through contract 4000115263/15/NL/IB for Germany, and through PROgramme de D\'{e}veloppement d'Exp\'{e}riences scientifiques (PRODEX) grant 4000127986 for Slovenia;  

the Academy of Finland through grants 299543, 307157, 325805, 328654, 336546, and 345115 and the Magnus Ehrnrooth Foundation;

the French Centre National d'\'{E}tudes Spatiales (CNES), the Agence Nationale de la Recherche (ANR) through grant ANR-10-IDEX-0001-02 for the `Investissements d'avenir' programme, through grant ANR-15-CE31-0007 for project `Modelling the Milky Way in the \gaia\ era' (MOD4\gaia), through grant ANR-14-CE33-0014-01 for project `The Milky Way disc formation in the \gaia\ era' (ARCHEOGAL), through grant ANR-15-CE31-0012-01 for project `Unlocking the potential of Cepheids as primary distance calibrators' (UnlockCepheids), through grant ANR-19-CE31-0017 for project `Secular evolution of galaxies' (SEGAL), and through grant ANR-18-CE31-0006 for project `Galactic Dark Matter' (GaDaMa), the Centre National de la Recherche Scientifique (CNRS) and its SNO \gaia\ of the Institut des Sciences de l'Univers (INSU), its Programmes Nationaux: Cosmologie et Galaxies (PNCG), Gravitation R\'{e}f\'{e}rences Astronomie M\'{e}trologie (PNGRAM), Plan\'{e}tologie (PNP), Physique et Chimie du Milieu Interstellaire (PCMI), and Physique Stellaire (PNPS), the `Action F\'{e}d\'{e}ratrice \gaia' of the Observatoire de Paris, the R\'{e}gion de Franche-Comt\'{e}, the Institut National Polytechnique (INP) and the Institut National de Physique nucl\'{e}aire et de Physique des Particules (IN2P3) co-funded by CNES;

the German Aerospace Agency (Deutsches Zentrum f\"{u}r Luft- und Raumfahrt e.V., DLR) through grants 50QG0501, 50QG0601, 50QG0602, 50QG0701, 50QG0901, 50QG1001, 50QG1101, 50\-QG1401, 50QG1402, 50QG1403, 50QG1404, 50QG1904, 50QG2101, 50QG2102, and 50QG2202, and the Centre for Information Services and High Performance Computing (ZIH) at the Technische Universit\"{a}t Dresden for generous allocations of computer time;

the Hungarian Academy of Sciences through the Lend\"{u}let Programme grants LP2014-17 and LP2018-7 and the Hungarian National Research, Development, and Innovation Office (NKFIH) through grant KKP-137523 (`SeismoLab');

the Science Foundation Ireland (SFI) through a Royal Society - SFI University Research Fellowship (M.~Fraser);

the Israel Ministry of Science and Technology through grant 3-18143 and the Tel Aviv University Center for Artificial Intelligence and Data Science (TAD) through a grant;

the Agenzia Spaziale Italiana (ASI) through contracts I/037/08/0, I/058/10/0, 2014-025-R.0, 2014-025-R.1.2015, and 2018-24-HH.0 to the Italian Istituto Nazionale di Astrofisica (INAF), contract 2014-049-R.0/1/2 to INAF for the Space Science Data Centre (SSDC, formerly known as the ASI Science Data Center, ASDC), contracts I/008/10/0, 2013/030/I.0, 2013-030-I.0.1-2015, and 2016-17-I.0 to the Aerospace Logistics Technology Engineering Company (ALTEC S.p.A.), INAF, and the Italian Ministry of Education, University, and Research (Ministero dell'Istruzione, dell'Universit\`{a} e della Ricerca) through the Premiale project `MIning The Cosmos Big Data and Innovative Italian Technology for Frontier Astrophysics and Cosmology' (MITiC);

the Netherlands Organisation for Scientific Research (NWO) through grant NWO-M-614.061.414, through a VICI grant (A.~Helmi), and through a Spinoza prize (A.~Helmi), and the Netherlands Research School for Astronomy (NOVA);

the Polish National Science Centre through HARMONIA grant 2018/30/M/ST9/00311 and DAINA grant 2017/27/L/ST9/03221 and the Ministry of Science and Higher Education (MNiSW) through grant DIR/WK/2018/12;

the Portuguese Funda\c{c}\~{a}o para a Ci\^{e}ncia e a Tecnologia (FCT) through national funds, grants SFRH/\-BD/128840/2017 and PTDC/FIS-AST/30389/2017, and work contract DL 57/2016/CP1364/CT0006, the Fundo Europeu de Desenvolvimento Regional (FEDER) through grant POCI-01-0145-FEDER-030389 and its Programa Operacional Competitividade e Internacionaliza\c{c}\~{a}o (COMPETE2020) through grants UIDB/04434/2020 and UIDP/04434/2020, and the Strategic Programme UIDB/\-00099/2020 for the Centro de Astrof\'{\i}sica e Gravita\c{c}\~{a}o (CENTRA);  

the Slovenian Research Agency through grant P1-0188;

the Spanish Ministry of Economy (MINECO/FEDER, UE), the Spanish Ministry of Science and Innovation (MICIN), the Spanish Ministry of Education, Culture, and Sports, and the Spanish Government through grants BES-2016-078499, BES-2017-083126, BES-C-2017-0085, ESP2016-80079-C2-1-R, ESP2016-80079-C2-2-R, FPU16/03827, PDC2021-121059-C22, RTI2018-095076-B-C22, and TIN2015-65316-P (`Computaci\'{o}n de Altas Prestaciones VII'), the Juan de la Cierva Incorporaci\'{o}n Programme (FJCI-2015-2671 and IJC2019-04862-I for F.~Anders), the Severo Ochoa Centre of Excellence Programme (SEV2015-0493), and MICIN/AEI/10.13039/501100011033 (and the European Union through European Regional Development Fund `A way of making Europe') through grant RTI2018-095076-B-C21, the Institute of Cosmos Sciences University of Barcelona (ICCUB, Unidad de Excelencia `Mar\'{\i}a de Maeztu') through grant CEX2019-000918-M, the University of Barcelona's official doctoral programme for the development of an R+D+i project through an Ajuts de Personal Investigador en Formaci\'{o} (APIF) grant, the Spanish Virtual Observatory through project AyA2017-84089, the Galician Regional Government, Xunta de Galicia, through grants ED431B-2021/36, ED481A-2019/155, and ED481A-2021/296, the Centro de Investigaci\'{o}n en Tecnolog\'{\i}as de la Informaci\'{o}n y las Comunicaciones (CITIC), funded by the Xunta de Galicia and the European Union (European Regional Development Fund -- Galicia 2014-2020 Programme), through grant ED431G-2019/01, the Red Espa\~{n}ola de Supercomputaci\'{o}n (RES) computer resources at MareNostrum, the Barcelona Supercomputing Centre - Centro Nacional de Supercomputaci\'{o}n (BSC-CNS) through activities AECT-2017-2-0002, AECT-2017-3-0006, AECT-2018-1-0017, AECT-2018-2-0013, AECT-2018-3-0011, AECT-2019-1-0010, AECT-2019-2-0014, AECT-2019-3-0003, AECT-2020-1-0004, and DATA-2020-1-0010, the Departament d'Innovaci\'{o}, Universitats i Empresa de la Generalitat de Catalunya through grant 2014-SGR-1051 for project `Models de Programaci\'{o} i Entorns d'Execuci\'{o} Parallels' (MPEXPAR), and Ramon y Cajal Fellowship RYC2018-025968-I funded by MICIN/AEI/10.13039/501100011033 and the European Science Foundation (`Investing in your future');

the Swedish National Space Agency (SNSA/Rymdstyrelsen);

the Swiss State Secretariat for Education, Research, and Innovation through the Swiss Activit\'{e}s Nationales Compl\'{e}mentaires and the Swiss National Science Foundation through an Eccellenza Professorial Fellowship (award PCEFP2\_194638 for R.~Anderson);

the United Kingdom Particle Physics and Astronomy Research Council (PPARC), the United Kingdom Science and Technology Facilities Council (STFC), and the United Kingdom Space Agency (UKSA) through the following grants to the University of Bristol, the University of Cambridge, the University of Edinburgh, the University of Leicester, the Mullard Space Sciences Laboratory of University College London, and the United Kingdom Rutherford Appleton Laboratory (RAL): PP/D006511/1, PP/D006546/1, PP/D006570/1, ST/I000852/1, ST/J005045/1, ST/K00056X/1, ST/\-K000209/1, ST/K000756/1, ST/L006561/1, ST/N000595/1, ST/N000641/1, ST/N000978/1, ST/\-N001117/1, ST/S000089/1, ST/S000976/1, ST/S000984/1, ST/S001123/1, ST/S001948/1, ST/\-S001980/1, ST/S002103/1, ST/V000969/1, ST/W002469/1, ST/W002493/1, ST/W002671/1, ST/W002809/1, and EP/V520342/1.

The Ground Based Optical Tracking (GBOT) programme uses observations collected at (i) the European Organisation for Astronomical Research in the Southern Hemisphere (ESO) with the VLT Survey Telescope (VST), under ESO programmes
092.B-0165,
093.B-0236,
094.B-0181,
095.B-0046,
096.B-0162,
097.B-0304,
098.B-0030,
099.B-0034,
0100.B-0131,
0101.B-0156,
0102.B-0174, and
0103.B-0165;
and (ii) the Liverpool Telescope, which is operated on the island of La Palma by Liverpool John Moores University in the Spanish Observatorio del Roque de los Muchachos of the Instituto de Astrof\'{\i}sica de Canarias with financial support from the United Kingdom Science and Technology Facilities Council, and (iii) telescopes of the Las Cumbres Observatory Global Telescope Network.

This work made use of software from H2O \citep{h2o_platform}, Postgres-XL (\url{https://www.postgres-xl.org}), Java (\url{https://www.oracle.com/java/}), R~\citep{R-citation}, TBase database management system (\url{https://github.com/Tencent/TBase}), and TOPCAT/STILTS \citep{2005ASPC..347...29T}. 

\end{acknowledgements}

\bibliographystyle{aa}
\raggedbottom 
\bibliography{Gaia_DR3_Classification_of_variable_sources}


\begin{appendix}

\section{Special training selections\label{app:training}}
In addition to the general selections applied to all training-set sources (Sect.~\ref{sssec:selection}), (sub)class-specific filtering conditions are listed in the following, often as a function of literature catalogue, employing fields defined in the \gdr3 archive tables \texttt{gaia\_source} and \texttt{vari\_summary}. For brevity, the version of the International Variable Star Index (VSX) is 2019-11-12 and it is not repeated at each mention of \citet{2006SASS...25...47W}.
\begin{enumerate}
  \item BE: conditions reflecting their bright and blue nature.
    \begin{enumerate}
         \item \texttt{parallax\_over\_error}\,$>$\,2, \texttt{median\_mag\_bp}\,$-$\,\texttt{median\_mag\_rp}\,$<$\,0.5\,mag, and \texttt{median\_mag\_g\_fov}\,$-$\,19\,$<$\,0\,mag, for sources in \citet{2002A&A...393..887M} (Small Magellanic Cloud);
         \item \texttt{parallax\_over\_error}\,$>$\,5, \texttt{median\_mag\_bp}\,$-$\,\texttt{median\_mag\_rp}\,$<$\,0.5\,mag, and absolute \g~magnitude~$<$\,0\,mag, for sources in \citet{2012ApJS..203...32R};
         \item \texttt{parallax\_over\_error}\,$>$\,2, for sources in \citet{2005MNRAS.361.1055S} (Large Magellanic Cloud);
         \item \texttt{parallax\_over\_error}\,$>$\,5, \texttt{median\_mag\_bp}\,$-$\,\texttt{median\_mag\_rp}\,$<$\,1\,mag, and absolute \g~magnitude~$<$\,0\,mag, for sources in \citet{2006SASS...25...47W}.
   \end{enumerate}
   \item Cepheids:
    \begin{enumerate}
        \item ACEP: period\,$>$1\,d, for sources in \citet{2014ApJS..213....9D};
        \item CEP: \texttt{std\_dev\_mag\_g\_fov} above the median standard deviations in \g, in 0.05\,mag intervals, of 1.6~million reference sources (see Appendix~\ref{app:plots});
        \item DCEP: \texttt{std\_dev\_mag\_g\_fov}\,$>$\,0.01\,mag, for sources in \citet{2012AcA....62..219S,2015AcA....65..297S,2017AcA....67..297S,2020AcA....70..101S} and \citet{2018AcA....68..315U};
        \item RV: \texttt{median\_mag\_bp}\,$-$\,\texttt{median\_mag\_rp}\,$>$\,0.5\,mag and \texttt{std\_dev\_mag\_g\_fov}\,$>$\,0.1\,mag;
        \item T2CEP:  \texttt{std\_dev\_mag\_g\_fov}\,$>$\,0.03\,mag.
    \end{enumerate}
    \item CV: \texttt{std\_dev\_mag\_g\_fov}\,$>$\,0.1\,mag.
    \item DSCT|SXPHE: \texttt{std\_dev\_mag\_g\_fov}\,$>$\,0.02\,mag and a lower limit at the median standard deviations in \g, in 0.05\,mag intervals, of 1.6~million reference sources (see Appendix~\ref{app:plots}).
    \item Eclipsing binaries and ellipsoidals: \texttt{std\_dev\_mag\_g\_fov} above custom percentiles (as listed below) of the  standard deviations in \g, in 0.05\,mag intervals, of 1.6~million reference sources (see Appendix~\ref{app:plots}), according to the comparison between the literature period and that recovered by the relevant SOS modules.
    \begin{enumerate}
    \item EA: 
     \begin{enumerate}
 \item the 95th percentile, for sources in \citet{2014ApJS..213....9D,2017MNRAS.469.3688D}, \citet{2013AJ....146..101P}, and Rybizki \citep[catalogue GAIA\_ECL\_RYBIZKI\_2018 in][]{DR3-DPACP-177};
 \item the 90th percentile, for sources in \citet{2020ApJS..249...18C}, \citet{1997ESASP1200.....E}, \citet{2013AcA....63..323P,2016AcA....66..421P}, \citet{2009AcA....59...33P}, and \citet{2016AcA....66..405S};
 \item the 85th percentile, if the period from \gaia data matches that of the literature, for sources in \citet{2016AJ....151...68K}, \citet{2016AcA....66..421P}, and \citet{2016AcA....66..405S};
 \item the 80th percentile, if the period from \gaia data matches that of the literature, otherwise the 95th percentile, for sources in \citet{2018MNRAS.477.3145J,2019MNRAS.486.1907J,2019MNRAS.485..961J} and \citet{2006SASS...25...47W};
 \item \texttt{skewness\_mag\_g\_fov}\,$>$\,0.9, for sources in \citet{2020ApJS..249...18C}, \citet{1997ESASP1200.....E}, \citet{2016AJ....151...68K}, \citet{2018MNRAS.477.3145J,2019MNRAS.486.1907J,2019MNRAS.485..961J},  \citet{2014ApJS..213....9D,2017MNRAS.469.3688D}, \citet{2013AJ....146..101P}, \citet{2009AcA....59...33P}, Rybizki \citep[catalogue GAIA\_ECL\_RYBIZKI\_2018 in][]{DR3-DPACP-177},  and \citet{2006SASS...25...47W}.
 \end{enumerate}
    \item EB: 
 \begin{enumerate}
  \item the 95th percentile, for sources in \citet{2014ApJS..213....9D} and Rybizki \citep[catalogue GAIA\_ECL\_RYBIZKI\_2018 in][]{DR3-DPACP-177};
  \item the 90th percentile, for sources in \citet{2013AcA....63..323P};
  \item the 85th percentile, if the period from \gaia data matches that of the literature, for sources in \citet{2016AJ....151...68K};
  \item the 85th percentile, if the period from \gaia data matches that of the literature, otherwise the 95th percentile, for sources in \citet{2018MNRAS.477.3145J,2019MNRAS.486.1907J,2019MNRAS.485..961J}, \citet{2002AcA....52..397P}, and \citet{2006SASS...25...47W};
  \item the 80th percentile, if the period from \gaia data matches that of the literature, otherwise the 90th percentile, for sources in \citet{1997ESASP1200.....E}.
  \end{enumerate}
    \item ECL: 
    \begin{enumerate}
      \item the 75th percentile (same as the general level), but only if \texttt{std\_dev\_mag\_g\_fov}\,$>$\,0.01\,mag and the period from \gaia data matches that of the literature, for sources in \citet{2006SASS...25...47W};
     \item the 75th percentile (same as the general level), but only if \texttt{std\_dev\_mag\_g\_fov}\,$>$\,0.01\,mag and (the period from \gaia data matches that of the literature or \texttt{std\_dev\_mag\_g\_fov}\,$>$\,0.025\,mag), for sources in \citet{2012AcA....62..219S}.
    \end{enumerate}
    \item EW:
    \begin{enumerate}
      \item the 95th percentile, for sources in \citet{2014ApJS..213....9D} and  \citet{2009AcA....59...33P};
  \item the 90th percentile, for sources in \citet{1997ESASP1200.....E}, \citet{2013AcA....63..323P}, and Rybizki \citep[catalogue GAIA\_ECL\_RYBIZKI\_2018 in][]{DR3-DPACP-177};
  \item the 90th percentile, if the period from \gaia data matches that of the literature, for sources in \citet{2016AcA....66..421P}, \citet{2016AcA....66..405S}, and \citet{2006SASS...25...47W};
  \item the 80th percentile, if the period from \gaia data matches that of the literature, otherwise the 95th percentile, for sources in \citet{2016AJ....151...68K}, \citet{2018MNRAS.477.3145J,2019MNRAS.486.1907J,2019MNRAS.485..961J}, and  \citet{2002AcA....52..397P};
  \item the 80th percentile, if the period from \gaia data matches that of the literature, otherwise the 90th percentile, for sources in \citet{2016AcA....66..421P}, \citet{2020ApJS..249...18C}.
  \end{enumerate}
    \item ELL: 
    \begin{enumerate}
  \item the 75th percentile (same as the general level), but only if \texttt{std\_dev\_mag\_g\_fov}\,$>$\,0.03\,mag, for sources in \citet{2018MNRAS.477.3145J,2019MNRAS.486.1907J,2019MNRAS.485..961J};
  \item the 75th percentile (same as the general level), but only if \texttt{std\_dev\_mag\_g\_fov}\,$>$\,0.01\,mag and (the period from \gaia data matches that of the literature or \texttt{std\_dev\_mag\_g\_fov}\,$>$\,0.025\,mag), for all other literature catalogues.
    \end{enumerate}
  \end{enumerate}
  \item EP: \texttt{skewness\_mag\_g\_fov}\,$>0$ and signal in the \gaia data confirmed by the planetary transit SOS module.
  \item GALAXY: objects at the brightest and reddest ends, overlapping with known stellar distributions, were filtered out by the condition  \texttt{median\_mag\_g\_fov}\,$>$\linebreak 3\,[(\,\texttt{median\_mag\_bp}\,$-$\,\texttt{median\_mag\_rp}\,)\,$-$\,1.9]\,$+$\,20\,mag. Also, a few extremely blue outliers were excluded by \texttt{median\_mag\_bp}\,$-$\,\texttt{median\_mag\_rp}\,$>$\,0\,mag.
   \item GCAS: conditions that reflect their bright and blue nature include \texttt{parallax\_over\_error}\,$>$\,5, \texttt{median\_mag\_bp}\,$-$\,\texttt{median\_mag\_rp}\,$<$\,1\,mag, and absolute \g~magnitude less than 1.5 and 1.0\,mag, for sources in \citet{2018MNRAS.477.3145J,2019MNRAS.486.1907J,2019MNRAS.485..961J} and \citet{2006SASS...25...47W}, respectively.
  \item Long-period variables: empirical relations were used to identify bright red giant stars in the observational Hertzsprung--Russell diagram, namely \texttt{median\_mag\_bp}\,$-$\,\texttt{median\_mag\_rp}\,$>$1\,mag and a threshold for the absolute \g~magnitude expressed in terms of parallax, colour, and apparent magnitude as follows. Stars in the (reddened) red clump have an absolute \g~magnitude of about $1.8\,(\texttt{median\_mag\_bp}\,$-$\,\texttt{median\_mag\_rp})\,-$\,1.8\,mag, which can be expressed as a \texttt{parallax} (in mas) of approximately 10\string^[1.7$+$0.36\,(\texttt{median\_mag\_bp}\,$-$\,\texttt{median\_mag\_rp})\,$-$\,0.2\,$\times$\linebreak \texttt{median\_mag\_g\_fov}]. It was found that, generally, stars brighter than the ones in the red clump could be identified by \texttt{parallax}\,(star)\,$-$\,0.12\,mas\,$<$\,\texttt{parallax}\,(red clump), where the offset of $-$\,0.12\,mas served to include the scatter from parallax noise of stars in the Magellanic Clouds. This expression was adapted to include red giant branch stars fainter than the red clump, as follows: \texttt{parallax}\,$-$\,0.12\,mas\,$<$\,10\string^[1.8\,$+$\,0.6\,(\texttt{median\_mag\_bp}\,$-$\,\linebreak\texttt{median\_mag\_rp})\,$-$\,0.2\,\texttt{median\_mag\_g\_fov}]. 
  \begin{enumerate}
      \item M: \texttt{std\_dev\_mag\_g\_fov}\,$>$\,0.1\,mag;
      \item SRA, SRB, SRC: the cross-match with  \citet{2012A&A...548A..79A} and \citet{2006SASS...25...47W} was limited to angular distances less than 1.5~and 2.5\,arcsec, respectively, unless \texttt{parallax}\,$>$\,2\,mas;
      \item SRS: the cross-match with \citet{2006SASS...25...47W} was limited to angular distances less than 1.5\,arcsec, unless \texttt{parallax}\,$>$\,2\,mas.
  \end{enumerate}
  \item MICROLENSING: \texttt{skewness\_mag\_g\_fov}\,$<$\,0 and confirmation of the signal presence in the \gaia data by the microlensing SOS module.
  \item RCB: visual inspection and selection were applied to this rare class.
  \item ROAP: \texttt{std\_dev\_mag\_g\_fov} above the median standard deviations in \g, in 0.05\,mag intervals, of 1.6~million reference sources (see Appendix~\ref{app:plots}).
  \item RR\,Lyrae stars:
  \begin{enumerate}
  \item RRAB, RRD: \texttt{std\_dev\_mag\_g\_fov}\,$>$\,0.056\,mag;
  \item RRC: \texttt{std\_dev\_mag\_g\_fov}\,$>$\,0.056\,mag, first overtone period\,$>$\,0.2\,d, and, for \texttt{median\_mag\_g\_fov}\,$<$\,16.5\,mag, |\,\texttt{std\_dev\_mag\_bp}$/$\texttt{std\_dev\_mag\_rp}\,$-$\,1\,|\,$>$\,0.1, where the latter was meant to remove a typical feature of eclipsing binaries, when sources had sufficient signal-to-noise ratio in the \bp and \rp bands;
  \item a lower minimum threshold of 0.03\,mag was applied to the \texttt{std\_dev\_mag\_g\_fov} for hundreds of RR\,Lyrae stars that were wished not to be missed.
  \end{enumerate}
  \item RS: 
    \begin{enumerate}
    \item cross-match angular distance $<$\,0.1\,\texttt{parallax}\,+\,0.2\,mas, for sources in \citet{2020ApJS..249...18C}, \citet{2008MNRAS.389.1722E}, and  \citet{2006SASS...25...47W};
    \item removal of outliers with respect to the \bpg versus \grp relation followed by all other sources of this class.
    \end{enumerate}
  \item SN: \texttt{std\_dev\_mag\_g\_fov}\,$>$\,0.1\,mag.
  \item Solar-like stars:
    \begin{enumerate}
    \item FLARES: 
      \begin{enumerate}
        \item cross-match angular distance less than 0.1\,\texttt{parallax}\,+\,0.2\,mas, for sources in \citet{2013ApJS..209....5S}, \citet{2011AJ....141...50W}, and \citet{2015ApJ...798...92W};
        \item removal of outliers with respect to the \bpg versus \grp relation followed by all other sources of this class.
      \end{enumerate}
    \item ROT:
      \begin{enumerate}
        \item cross-match angular distance less than 0.1\,\texttt{parallax}+0.4\,mas and 0.1\,\texttt{parallax}+0.6\,mas, for sources in Distefano \citep[catalogue GAIA\_ROT\_GAIA\_2017 in][]{DR3-DPACP-177} and \citet{2006SASS...25...47W}, respectively;
        \item removal of outliers with respect to the \bpg versus \grp relation followed by all other sources of this class.
      \end{enumerate}    
    \item SOLAR\_LIKE:
      \begin{enumerate}
        \item cross-match angular distance less than 0.05\,\texttt{parallax}\,+\,0.25 and 0.1\,\texttt{parallax}\,+\,0.7\,mas, for sources in \citet{2007A&A...469..713M} and \citet{2017ApJ...835...61Z}, respectively;
        \item removal of outliers with respect to the \bpg versus \grp relation followed by all other sources of this class.
      \end{enumerate}        
    \end{enumerate}
    \item White dwarfs:
      \begin{enumerate}
        \item GWVIR: \texttt{std\_dev\_mag\_g\_fov} above the 85th percentile of the  standard deviations in \g, in 0.05\,mag intervals, of 1.6~million reference sources (see Appendix~\ref{app:plots}), for sources in \citet{2020svos.conf...11E};
         \item ELM\_ZZA: absolute \g~magnitude greater than 5\,mag.
      \end{enumerate}        
    \item Young stellar objects:
      \begin{enumerate}
      \item DIP, WTTS:
        \begin{enumerate}
            \item \texttt{parallax}\,$>$\,1\,mas;
            \item removal of outliers with respect to the \bpg versus \grp relation followed by all other sources of this class.
        \end{enumerate}
        \item HAEBE:
        \begin{enumerate}
            \item \texttt{median\_mag\_g\_fov}\,$<$\,16\,mag;
            \item removal of outliers with respect to the \bpg versus \grp relation followed by all other sources of this class.
        \end{enumerate}
        \item UXOR: removal of outliers with respect to the \bpg versus \grp relation followed by all other sources of this class.
        \item TTS: cross-match angular distance less than 0.1\,\texttt{parallax}\,+\,0.7\,mas, for sources in \citet{2020IAUS..345..378V} and \citet{2006SASS...25...47W}.
        \item YSO:
          \begin{enumerate}
          \item \texttt{parallax}\,$>$\,0.9\,mas;
          \item |\,Galactic latitude\,|\,$<$\,30$^{\circ}$;
          \item cross-match angular distance less than 0.05\,\texttt{parallax}\,+\,0.65\,mas, for sources in \citet{2020IAUS..345..378V} and \citet{2006SASS...25...47W};
          \item removal of outliers with respect to the \bpg versus \grp relation followed by all other sources of this class.
           \end{enumerate}        
    \end{enumerate}        
\end{enumerate}

\newpage

\section{Classification attributes\label{app:attributes}}
The classification attributes selected to characterise training set sources for classifier models are listed in terms of parameters in the \texttt{vari\_summary} table, unless a different table is mentioned or prepended to field names:
\begin{enumerate}
\item the Abbe value ({\texttt{abbe\_mag\_g\_fov}}) of FoV transit magnitudes in the \g band; 
\item the astrometry-based luminosity \citep{1999ASPC..167...13A} as {\texttt{gaia\_source.parallax}}\,10\string^(0.2\,{\texttt{median\_mag\_g\_fov}}\,$-$\,2); 
\item the possibly reddened colour index \bprp, estimated by  {\texttt{median\_mag\_bp}}\,$-$\,{\texttt{median\_mag\_rp}}; 
\item the possibly reddened colour index \grp, estimated by {\texttt{median\_mag\_g\_fov}}\,$-$\,{\texttt{median\_mag\_rp}}; 
\item the sample-size unbiased unweighted variance and kurtosis (central moments) of FoV-transit magnitudes in the \g~band, denoised assuming Gaussian uncertainties \citep{2014A&C.....5....1R}; 
\item the duration of the time series ({\texttt{time\_duration\_g\_fov}}), from the first to the last FoV~transit in the \g~band; 
\item the unweighted 95th percentile of magnitude changes per time interval between successive FoV~transits in the \g~band; 
\item the {\texttt{qso\_variability}} and {\texttt{non\_qso\_variability}} parameters from \citet{2011AJ....141...93B}, computed from FoV-transit magnitudes in the \g~band, after adaptations to the \gaia data \citep[these values are published only in the \texttt{vari\_agn} table, see][]{DR3-DPACP-167};
\item the ratio between the sample-size biased unweighted standard deviation of FoV-transit magnitudes in the \g~band and the root-mean-square of the corresponding uncertainties ({\texttt{std\_dev\_over\_rms\_err\_mag\_g\_fov}});
\item the square root of the sample-size unbiased unweighted variance ({\texttt{std\_dev\_mag\_g\_fov}}) of FoV-transit magnitudes in the \g~band; 
\item the source parallax ({\texttt{gaia\_source.parallax}}); 
\item the Pearson correlation coefficient for the magnitudes of FoV transits in the \bp and \rp~bands;
\item the sample-size unbiased unweighted skewness moment of FoV~transit magnitudes in the \g~band, standardised by the variance of such measurements ({\texttt{skewness\_mag\_g\_fov}});
\item the ratio between the third spectral shape coefficients$^{\rm \ref{foot:ssc}}$ in the \bp and \rp bands;  
\item the ratio between the standard deviations in magnitude of FoV~transit observations in the \bp and \rp~bands, that is {\texttt{std\_dev\_mag\_bp}}\,$/$\,{\texttt{std\_dev\_mag\_rp}}; 
\item the single-band Stetson variability index \citep{1996PASP..108..851S} computed from FoV~transit magnitudes in the \g~band, pairing observations within 0.1~days (\texttt{stetson\_mag\_g\_fov});
\item a Wesenheit-like magnitude of FoV~transits in the \g~band as {\texttt{median\_mag\_g\_fov}}$-$2({\texttt{median\_mag\_bp}}$-${\texttt{median\_mag\_rp}});
\item parameters derived from the Least Square periodogram \citep{1985A&AS...59...63H,2009A&A...496..577Z} configured as described in sect.~10.2.3 of the \gdr3 documentation \citep{2022gdr3.reptE..10R}: 
\begin{enumerate}
    \item the top frequencies (corresponding to the highest periodogram amplitudes) in the frequency ranges 0.1--1 and 1--25\,d$^{-1}$; 
    \item the signal detection efficiencies (the difference between the maximum and mean periodogram amplitudes, divided by the standard deviation of such amplitudes) in the frequency ranges 0.1--1 and 1--25\,d$^{-1}$;  
    \item the false alarm probabilities \citep{2009MNRAS.395.1541B} of the top frequencies in the ranges 0.0007--0.1, 0.1--1, and 1--25\,d$^{-1}$;
    \item the highest periodogram amplitudes in the frequency ranges 0.0007--0.1, 0.1--1, and 1--25\,d$^{-1}$.
\end{enumerate}
\end{enumerate}

\section{Additional class labels\label{app:FP_labels}}
Class labels that were not targeted for publication in \gdr3 but that appear among the false positives in Table~\ref{tab:results_details} are defined as follows \citep[consistently with][]{DR3-DPACP-177}:
\begin{enumerate}
    \item CST: non-variable object such as a constant star, including former suspect variable with undetected variability in subsequent observations; 
    \item HMXB: high-mass X-ray binary system with a massive star and a compact companion;
    \item L: slow irregular variable or insufficiently studied object that could belong to other classes (such as SR);
    \item PCEB: post-common envelope binary (or pre-cataclysmic variable);
    \item RAD\_VEL\_VAR: object with variable radial velocity; 
    \item SB: spectroscopic binary.
\end{enumerate}

\section{Sample ADQL queries\label{app:queries}}

Source identifiers and classification scores of AGN and SN candidates can be queried as follows.

\begin{verbatim}
SELECT source_id, best_class_name, 
   best_class_score
FROM gaiadr3.vari_classifier_result
WHERE best_class_name in ('AGN', 'SN') 
\end{verbatim}

Source identifiers and classification scores of GALAXY candidates can be queried as follows.

\begin{verbatim}
SELECT source_id, vari_best_class_name, 
   vari_best_class_score
FROM gaiadr3.galaxy_candidates
WHERE vari_best_class_name = 'GALAXY'
\end{verbatim}

All possible candidates of RR\,Lyrae and Cepheid classes can be retrieved from the union of classification and of the corresponding SOS modules as illustrated in the following query.

\begin{verbatim}
SELECT s.source_id, ra, dec,
   best_classification AS RR_class, 
   type_best_classification AS CEP_class, 
   best_class_name AS CLS_class
FROM gaiadr3.gaia_source AS s
LEFT OUTER JOIN gaiadr3.vari_rrlyrae AS rr 
   ON s.source_id=rr.source_id
LEFT OUTER JOIN gaiadr3.vari_cepheid AS cep 
   ON s.source_id=cep.source_id
LEFT OUTER JOIN gaiadr3.vari_classifier_result 
   AS cls ON s.source_id=cls.source_id
WHERE best_class_name in ('RR', 'CEP')
\end{verbatim}

Classified source identifiers of objects that are in `common', `extra', `other', or `missed' with respect to SOS modules (see Table~\ref{tab:results}) can be retrieved as shown in the following examples (assuming the AGN class). 
\begin{enumerate}
\item Classified sources in `common' with the SOS module:
\begin{verbatim}
SELECT c.source_id
FROM gaiadr3.vari_classifier_result AS c 
INNER JOIN gaiadr3.vari_agn AS x 
   ON c.source_id = x.source_id
WHERE best_class_name = 'AGN'
\end{verbatim}
\item `Extra' classified sources with respect to the SOS module:
\begin{verbatim}
SELECT c.source_id
FROM gaiadr3.vari_classifier_result AS c 
LEFT JOIN gaiadr3.vari_agn AS x 
   ON c.source_id = x.source_id
WHERE best_class_name = 'AGN' 
   AND x.source_id IS NULL
\end{verbatim}
\item Sources in the SOS module but classified as `other' classes:
\begin{verbatim}
SELECT c.source_id, best_class_name
FROM gaiadr3.vari_classifier_result AS c 
RIGHT JOIN gaiadr3.vari_agn AS x 
   ON c.source_id = x.source_id
WHERE best_class_name != 'AGN'
\end{verbatim}
\item Sources in the SOS module but `missed' by classification:
\begin{verbatim}
SELECT c.source_id 
FROM gaiadr3.vari_classifier_result AS c
RIGHT JOIN gaiadr3.vari_agn AS x 
   ON c.source_id = x.source_id
WHERE c.source_id IS NULL
\end{verbatim}
\end{enumerate}


\section{Common diagrams for all classes\label{app:plots}}
Sources from training and classification results are shown for each class in different diagrams, on top of a set of background sources for reference purposes (depicted in grey). Such diagrams are described in terms of \texttt{vari\_summary} parameters (unless stated otherwise) and labelled according to the following items: 
\begin{itemize}
\item[(a)] sky maps in Aitoff projection, in Galactic coordinates (with Galactic longitude of zero at the centre and increasing towards the left); 
\item[(b)] \g versus \mbox{\g\,$-$\,\rp} colour--magnitude diagrams as \texttt{median\_mag\_g\_fov}\,vs\,\texttt{median\_mag\_bp}\,$-$\,\texttt{median\_mag\_rp};
\item[(c)] \mbox{\bp\,$-$\,\g} versus \mbox{\g\,$-$\,\rp} colour--colour diagrams as \texttt{median\_mag\_bp}\,$-$\,\texttt{median\_mag\_g\_fov} versus \texttt{median\_mag\_g\_fov}\,$-$\,\texttt{median\_mag\_rp};
\item[(d)] absolute \g magnitude versus the reddened \mbox{\bp\,$-$\,\rp} colour for observational Hertzsprung--Russell diagrams  as {\small \texttt{median\_mag\_g\_fov}$+$5[$1+\log_{10}$({\small\texttt{gaia\_source.parallax}}/1000)]} versus \texttt{median\_mag\_bp}\,$-$\,\texttt{median\_mag\_rp} (for sources with \texttt{gaia\_source.parallax\_over\_error}\,$>$\,5);
\item[(e)] time series standard deviation as a function of magnitude as \texttt{std\_dev\_mag\_g\_fov} versus \texttt{median\_mag\_g\_fov} (with a white curve illustrating the third quartile of the standard deviations in \g, in 0.05\,mag intervals, of 1.6~million reference sources, defined in subsequent paragraphs of this Appendix);
\item[(f)] metrics targeting non-periodic variations, such as \texttt{skewness\_mag\_g\_fov}
 versus \texttt{abbe\_mag\_g\_fov};
\item[(g)] metrics targeting periodic variations of pulsating stars, such as $\log_{10}$\,(\texttt{std\_dev\_mag\_bp}$/$\texttt{std\_dev\_mag\_rp}) versus \texttt{median\_mag\_g\_fov}. 
\end{itemize}
Training sources are illustrated with red points, with darker shades corresponding to higher number density, while classification results are colour-coded by \texttt{best\_class\_score} (in the \texttt{vari\_classifier\_result} table). 

Classification results (in the \texttt{vari\_classifier\_result} table) include additional plots on:
 \begin{itemize}
\item[(a)] completeness versus contamination, colour-coded by the minimum \texttt{best\_class\_score};
\item[(b)] $F_1$~score versus the minimum \texttt{best\_class\_score};
\item[(c)] sample light curves in the \g band as a function of time or phase (folded by the most significant period that was published in the corresponding SOS module, in absence of which the literature period was used), after the application of the operators described in sect.~10.2.3 of the \gdr3 documentation \citep{2022gdr3.reptE..10R} and sect.~3.1 of \citet{DR3-DPACP-162}.
\end{itemize}

For all but the observational Hertzsprung--Russell diagrams, about 1.6~million reference background sources (depicted in grey) were selected by randomly sampling sources from the full range of magnitude, with an upper limit of 6000~objects per 0.05\,mag interval, and then by filtering out sources with less than five~FoV transits in the \g band and those without any measurement in both \bp and \rp.

For the observational Hertzsprung--Russell diagrams, about 4.25~million background sources were derived from the following concurrent conditions on parameters available in the  \texttt{gaiadr3.gaia\_source} table:
\begin{verbatim}
parallax_over_error > 10
ruwe < 1.2
visibility_periods_used > 11
phot_g_mean_flux > 0
phot_bp_mean_flux_over_error > 10
phot_rp_mean_flux_over_error > 10
\end{verbatim}
and in table \texttt{gaiadr3.vari\_summary} (including sources with unpublished values for the following fields):
\begin{verbatim}
num_selected_bp > 10
num_selected_rp > 10;
\end{verbatim}
in order to limit the amount of sources, their distribution in log(\texttt{parallax}) was binned in 100~intervals and distributed evenly by random source sampling \citep[for more details, see][]{DR3-DPACP-177}.

The diagrams presented in this Appendix represent only a selection of the verification metrics employed during training and classification assessment. Nevertheless, these figures often capture salient characteristics that are peculiar to each class.
Figures related to the training set might include sources with unpublished photometric time series, as classification results did not necessarily include all training objects and not all classified sources were included in the \gaia archive \citep{DR3-CU9}.

\begin{figure*}
\centering
\stackinset{c}{-0.7cm}{c}{2.7cm}{(a)}{} \includegraphics[width=0.6\hsize]{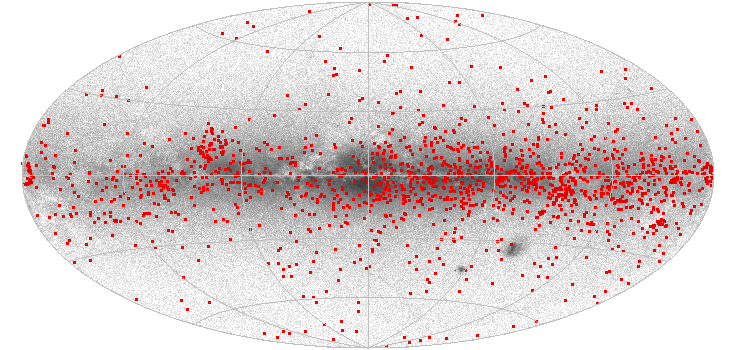} \\ 
\vspace{4mm}
\stackinset{c}{-0.3cm}{c}{3cm}{(b)}{} \includegraphics[width=0.45\hsize]{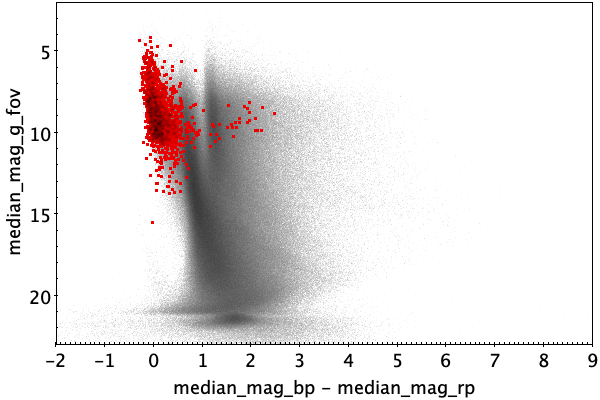}  
\hspace{2mm}
\stackinset{c}{8.8cm}{c}{3cm}{(c)}{} \includegraphics[width=0.45\hsize]{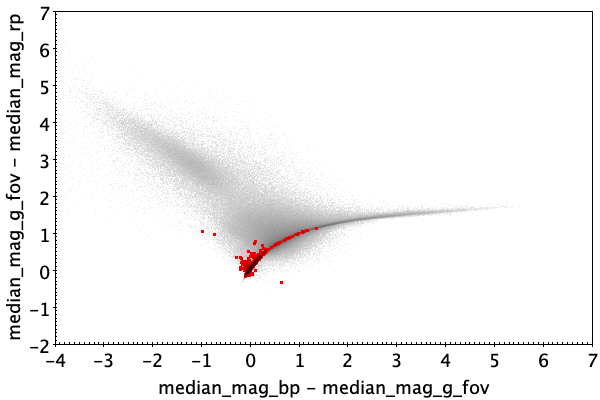} \\ 
\vspace{4mm}
\stackinset{c}{-0.3cm}{c}{3cm}{(d)}{} \includegraphics[width=0.45\hsize]{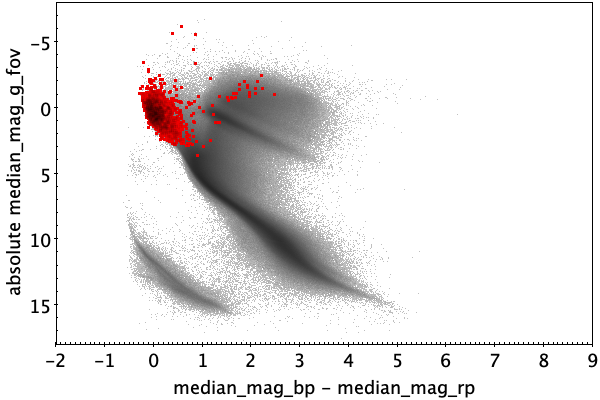}  
\hspace{2mm}
\stackinset{c}{8.8cm}{c}{3cm}{(e)}{} \includegraphics[width=0.45\hsize]{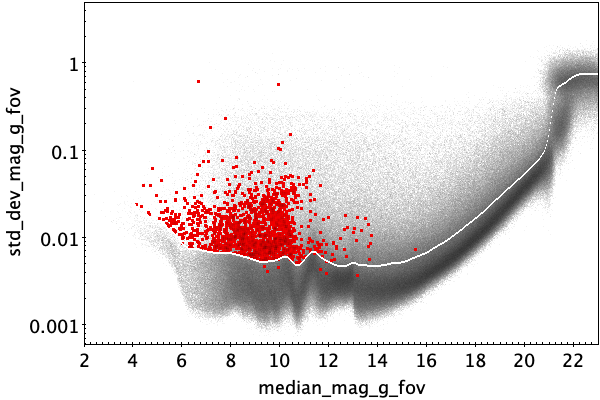} \\ 
\vspace{4mm}
\stackinset{c}{-0.3cm}{c}{3cm}{(f)}{} \includegraphics[width=0.45\hsize]{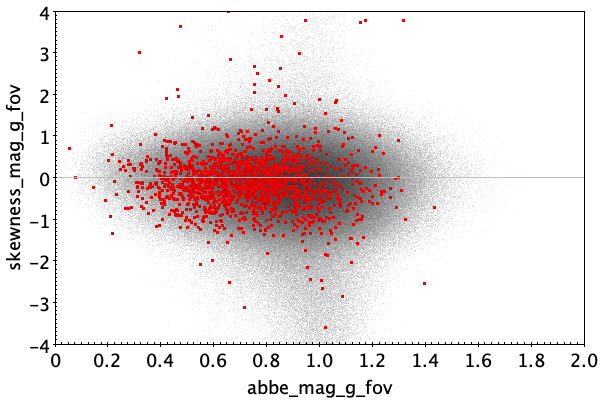}  
\hspace{2mm}
\stackinset{c}{8.8cm}{c}{3cm}{(g)}{} \includegraphics[width=0.45\hsize]{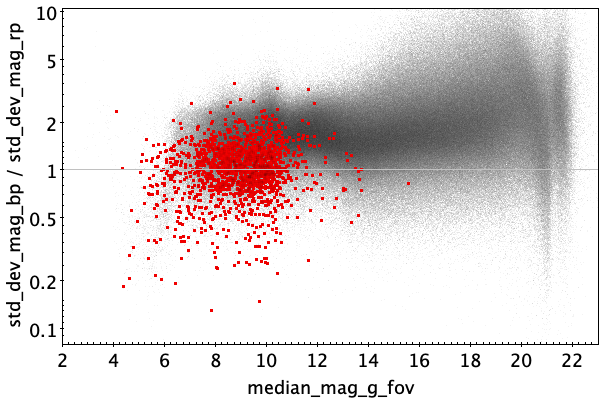}  \\ 
\vspace{4mm}
 \caption{ACV|CP|MCP|ROAM|ROAP|SXARI: 1572 training sources.}  
 \label{fig:app:ACV_trn}
\end{figure*}

\begin{figure*}
\centering
\stackinset{c}{-0.7cm}{c}{2.7cm}{(a)}{}
\includegraphics[width=0.6\hsize]{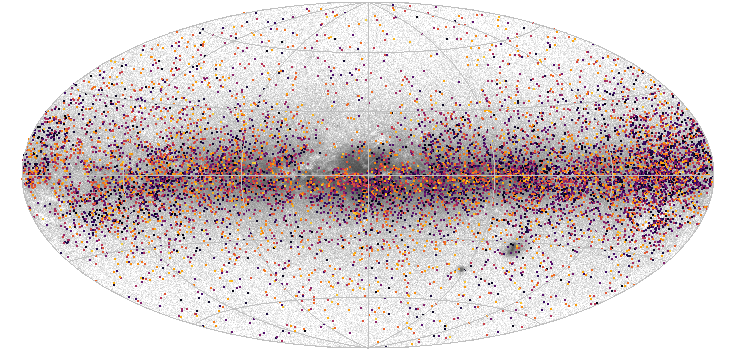} 
\stackinset{c}{2.2cm}{c}{2.7cm}{\includegraphics[height=5.5cm]{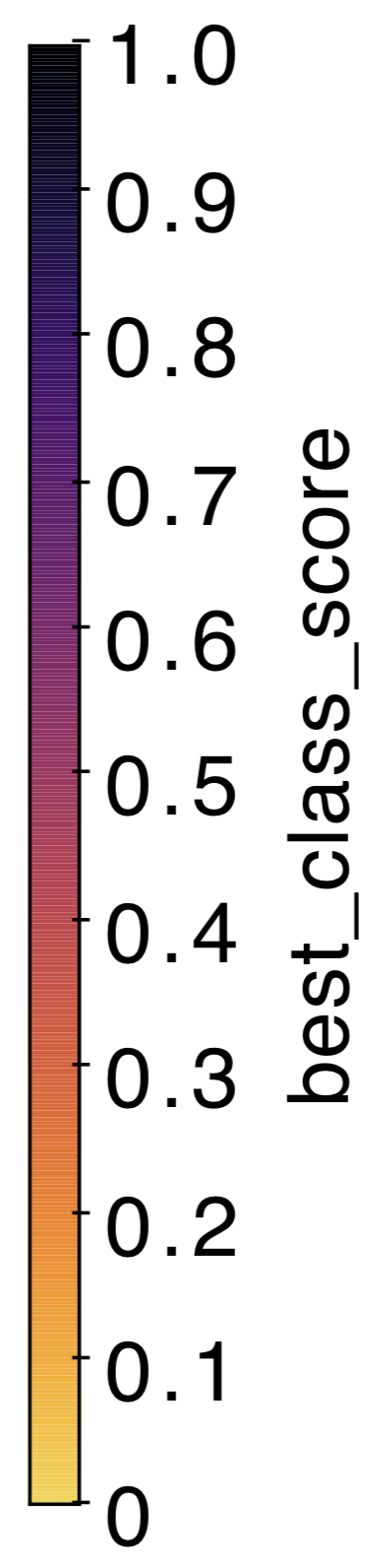}}{} \\ 
\vspace{4mm}
\stackinset{c}{-0.3cm}{c}{3cm}{(b)}{} \includegraphics[width=0.45\hsize]{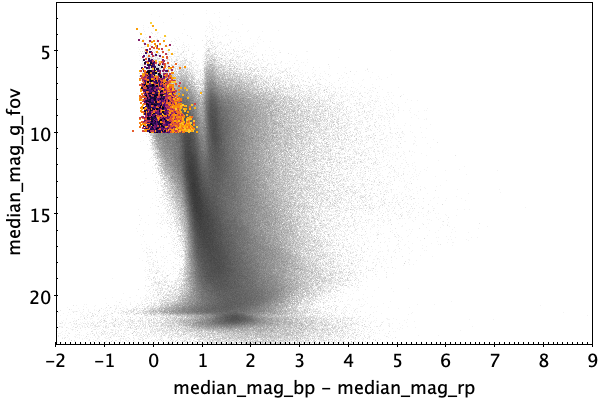}  
\hspace{2mm}
\stackinset{c}{8.8cm}{c}{3cm}{(c)}{} \includegraphics[width=0.45\hsize]{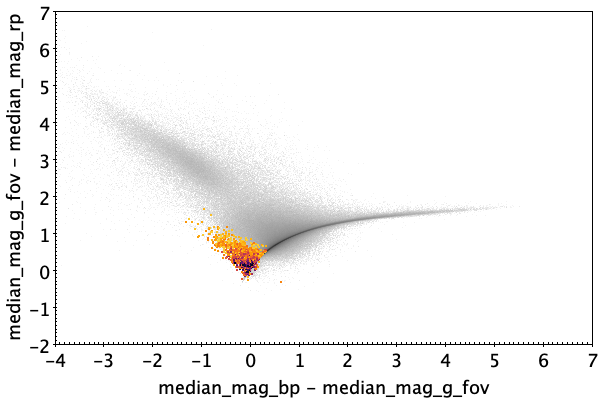} \\ 
\vspace{4mm}
\stackinset{c}{-0.3cm}{c}{3cm}{(d)}{} \includegraphics[width=0.45\hsize]{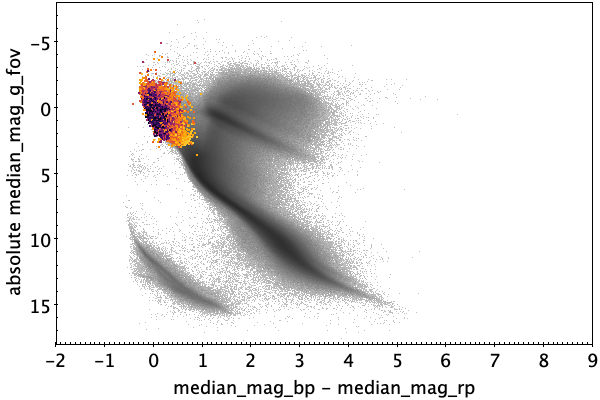}  
\hspace{2mm}
\stackinset{c}{8.8cm}{c}{3cm}{(e)}{} \includegraphics[width=0.45\hsize]{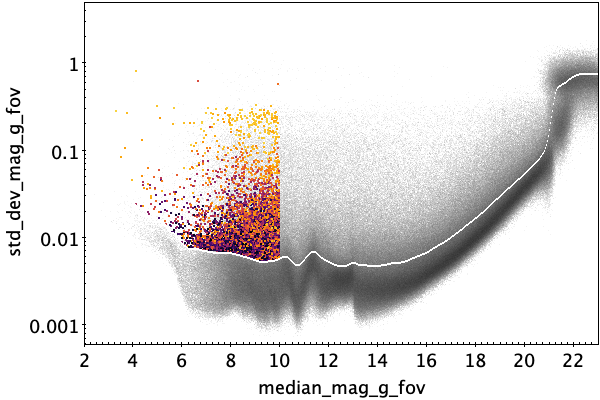} \\ 
\vspace{4mm}
\stackinset{c}{-0.3cm}{c}{3cm}{(f)}{} \includegraphics[width=0.45\hsize]{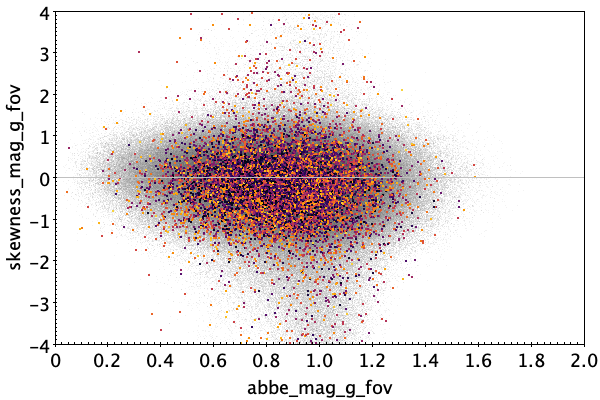}  
\hspace{2mm}
\stackinset{c}{8.8cm}{c}{3cm}{(g)}{} \includegraphics[width=0.45\hsize]{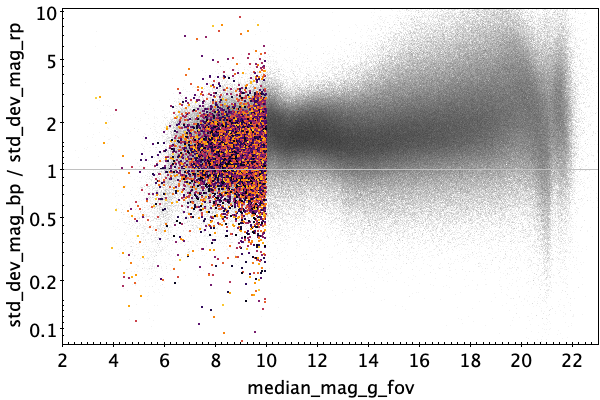}  \\ 
\vspace{4mm}
 \caption{ACV|CP|MCP|ROAM|ROAP|SXARI: 10\,779 classified sources.}  
 \label{fig:app:ACV}
\end{figure*}

\begin{figure*}
\centering
\stackinset{c}{-0.3cm}{c}{3cm}{(a)}{} \includegraphics[width=0.45\hsize]{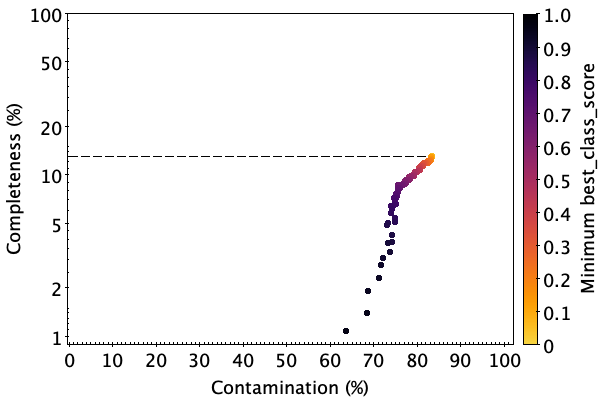}  
\hspace{2mm}
\stackinset{c}{8.8cm}{c}{3cm}{(b)}{} \includegraphics[width=0.45\hsize]{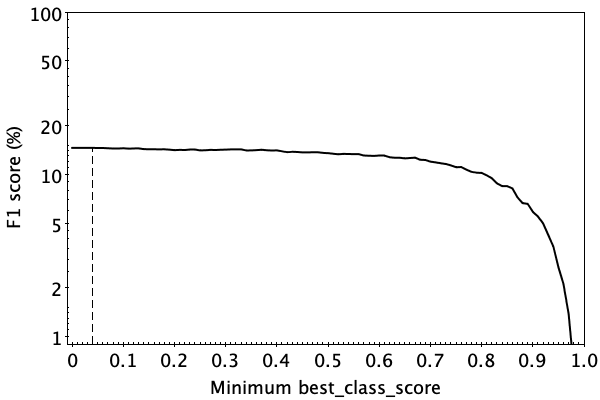} \\
\vspace{4mm}
\stackinset{c}{-0.3cm}{c}{3cm}{(c)}{} \includegraphics[width=0.45\hsize]{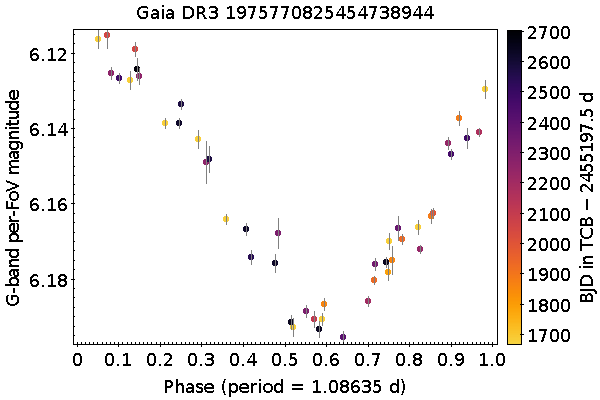}  
\hspace{2mm}
\stackinset{c}{8.8cm}{c}{3cm}{(d)}{} \includegraphics[width=0.45\hsize]{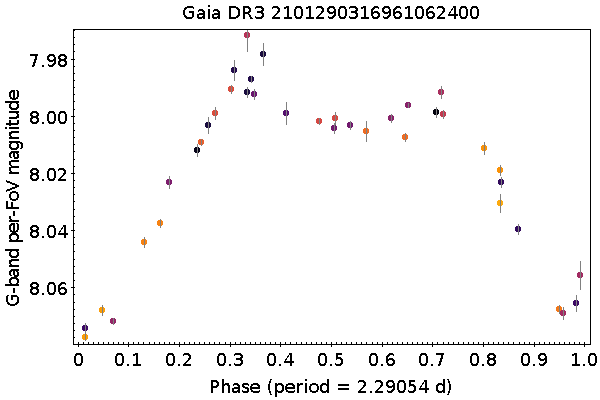} \\
\vspace{4mm}
\stackinset{c}{-0.3cm}{c}{3cm}{(e)}{} \includegraphics[width=0.45\hsize]{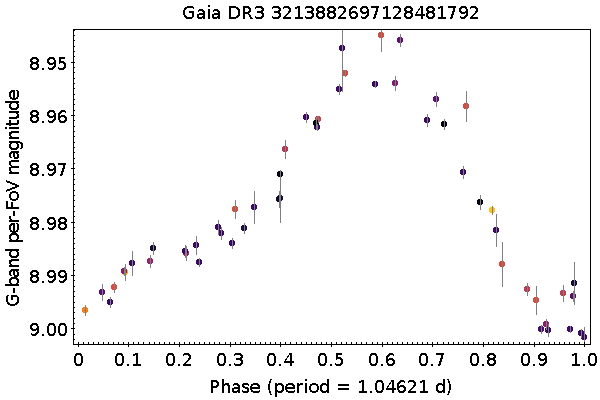}  
\hspace{2mm}
\stackinset{c}{8.8cm}{c}{3cm}{(f)}{} \includegraphics[width=0.45\hsize]{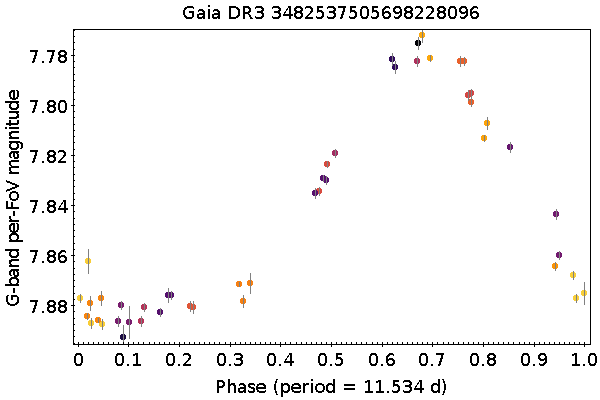} \\
\vspace{4mm}
 \caption{ACV|CP|MCP|ROAM|ROAP|SXARI: completeness, contamination, $F_1$-score, and sample light curves. The dashed lines indicate the maximum completeness (with minimum \texttt{best\_class\_score} of zero) in panel~(a) and the minimum \texttt{best\_class\_score} that maximises the $F_1$-score (for an optimal balance between completeness and contamination) in panel~(b). For all period-folded light curves, times are colour-coded according to the same legend as shown in panel~(c).}
 \label{fig:app:ACV_cc}
\end{figure*}

\begin{figure*}
\centering
\stackinset{c}{-0.7cm}{c}{2.7cm}{(a)}{} \includegraphics[width=0.6\hsize]{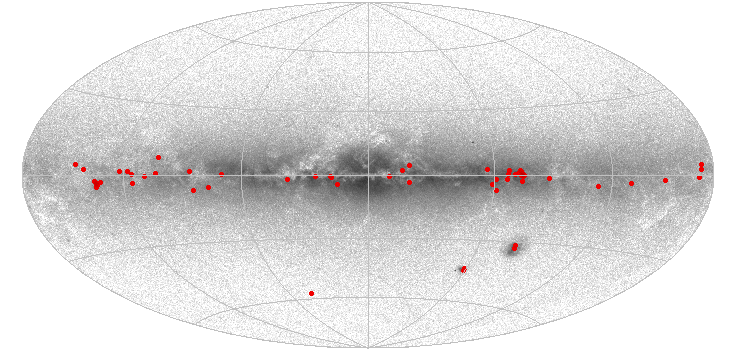} \\ 
\vspace{4mm}
\stackinset{c}{-0.3cm}{c}{3cm}{(b)}{} \includegraphics[width=0.45\hsize]{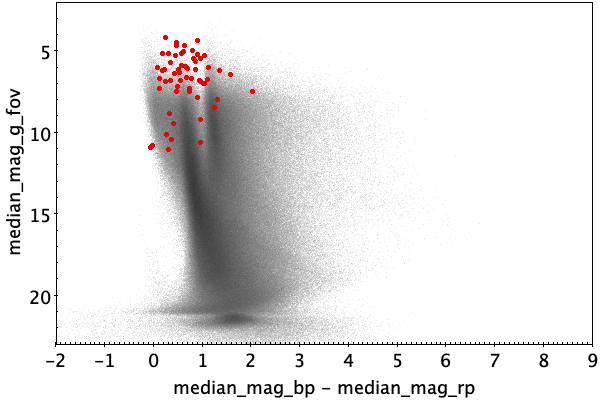}  
\hspace{2mm}
\stackinset{c}{8.8cm}{c}{3cm}{(c)}{} \includegraphics[width=0.45\hsize]{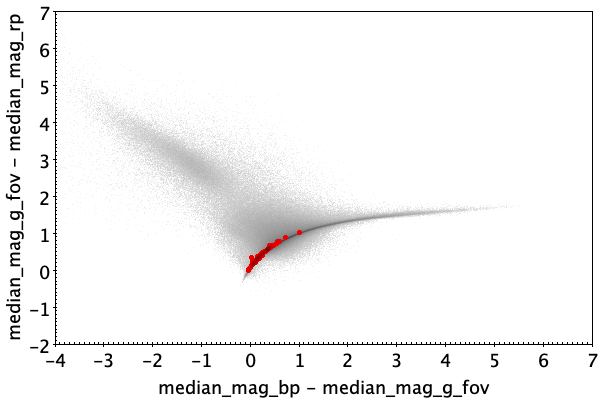} \\ 
\vspace{4mm}
\stackinset{c}{-0.3cm}{c}{3cm}{(d)}{} \includegraphics[width=0.45\hsize]{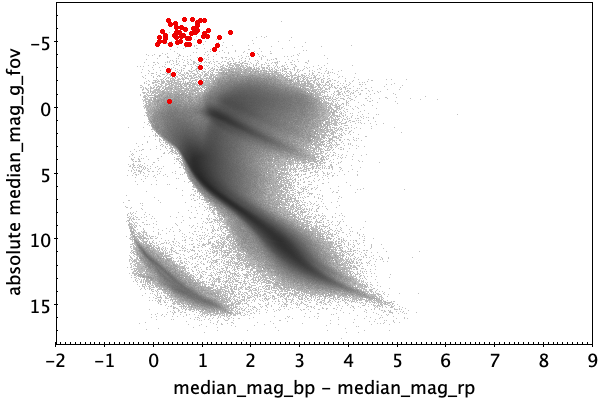}  
\hspace{2mm}
\stackinset{c}{8.8cm}{c}{3cm}{(e)}{} \includegraphics[width=0.45\hsize]{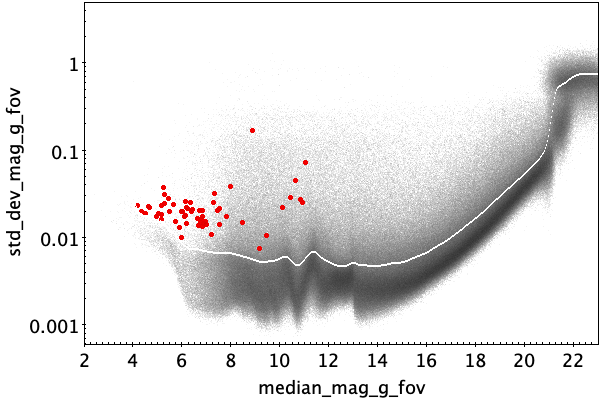} \\ 
\vspace{4mm}
\stackinset{c}{-0.3cm}{c}{3cm}{(f)}{} \includegraphics[width=0.45\hsize]{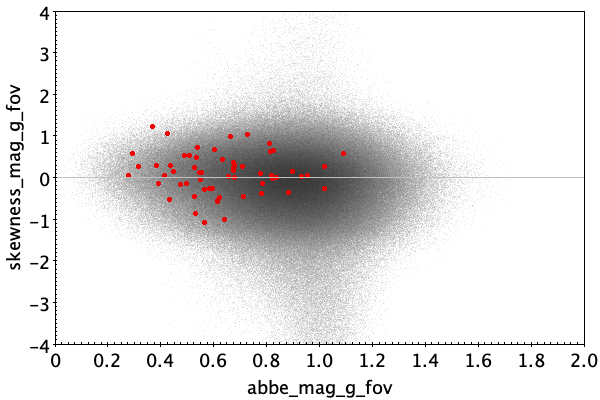}  
\hspace{2mm}
\stackinset{c}{8.8cm}{c}{3cm}{(g)}{} \includegraphics[width=0.45\hsize]{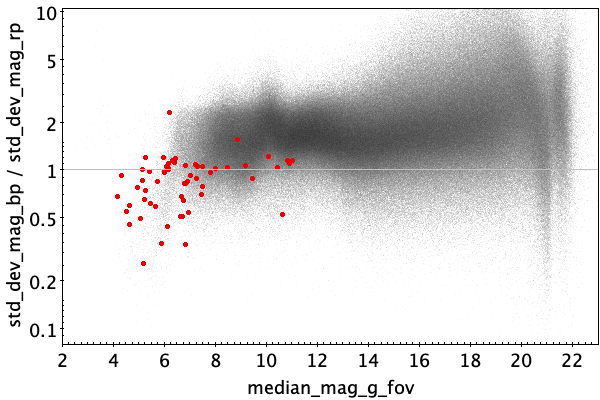}  \\ 
\vspace{4mm}
 \caption{ACYG: 59 training sources.}  
 \label{fig:app:ACYG_trn}
\end{figure*}

\begin{figure*}
\centering
\stackinset{c}{-0.7cm}{c}{2.7cm}{(a)}{}
\includegraphics[width=0.6\hsize]{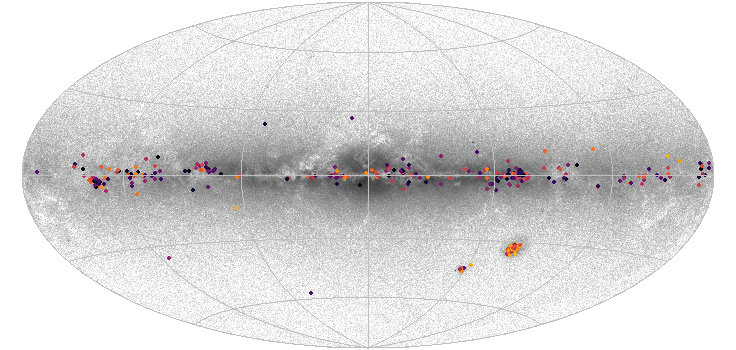} 
\stackinset{c}{2.2cm}{c}{2.7cm}{\includegraphics[height=5.5cm]{figures/appendix/vertical_best_class_score.png}}{} \\ 
\vspace{4mm}
\stackinset{c}{-0.3cm}{c}{3cm}{(b)}{} \includegraphics[width=0.45\hsize]{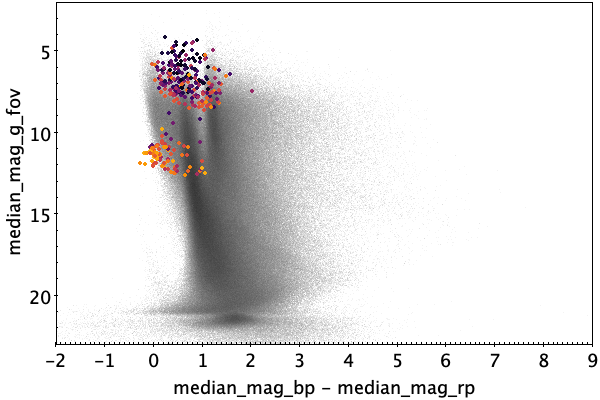}  
\hspace{2mm}
\stackinset{c}{8.8cm}{c}{3cm}{(c)}{} \includegraphics[width=0.45\hsize]{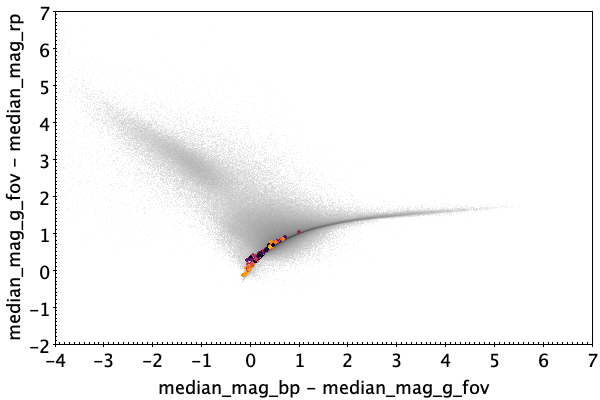} \\ 
\vspace{4mm}
\stackinset{c}{-0.3cm}{c}{3cm}{(d)}{} \includegraphics[width=0.45\hsize]{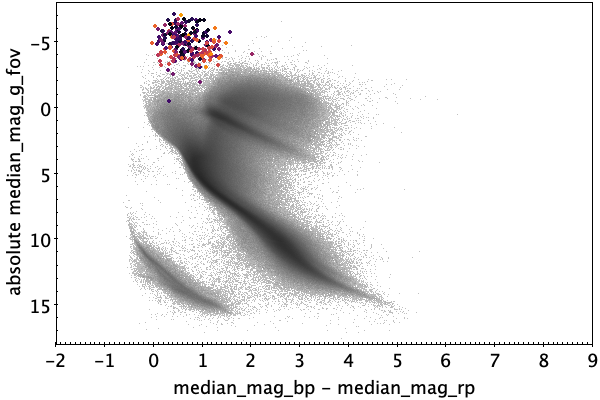}  
\hspace{2mm}
\stackinset{c}{8.8cm}{c}{3cm}{(e)}{} \includegraphics[width=0.45\hsize]{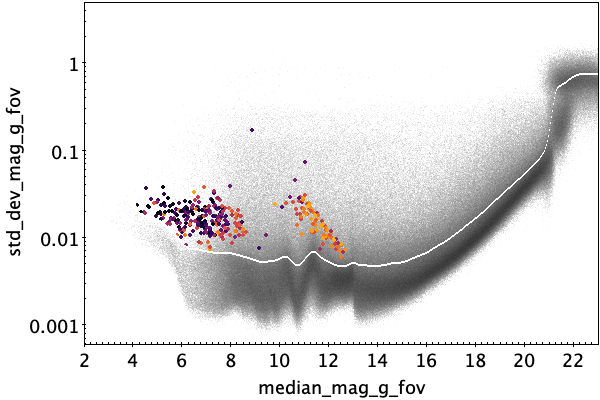} \\ 
\vspace{4mm}
\stackinset{c}{-0.3cm}{c}{3cm}{(f)}{} \includegraphics[width=0.45\hsize]{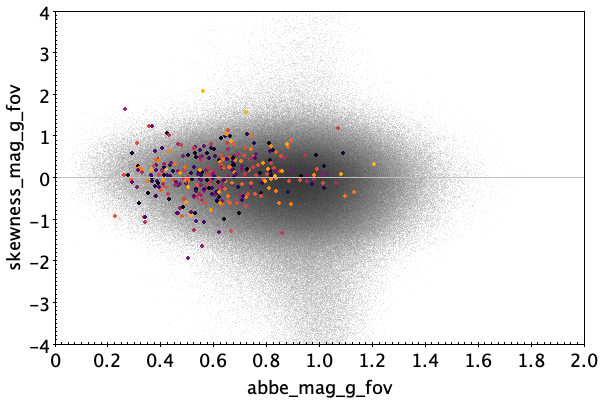}  
\hspace{2mm}
\stackinset{c}{8.8cm}{c}{3cm}{(g)}{} \includegraphics[width=0.45\hsize]{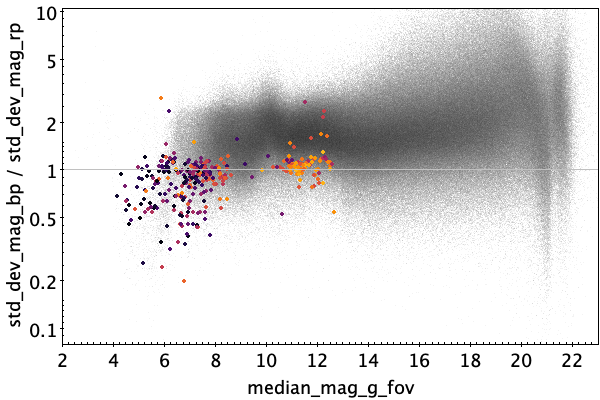}  \\ 
\vspace{4mm}
 \caption{ACYG: 329 classified sources.}  
 \label{fig:app:ACYG}
\end{figure*}

\begin{figure*}
\centering
\stackinset{c}{-0.3cm}{c}{3cm}{(a)}{} \includegraphics[width=0.45\hsize]{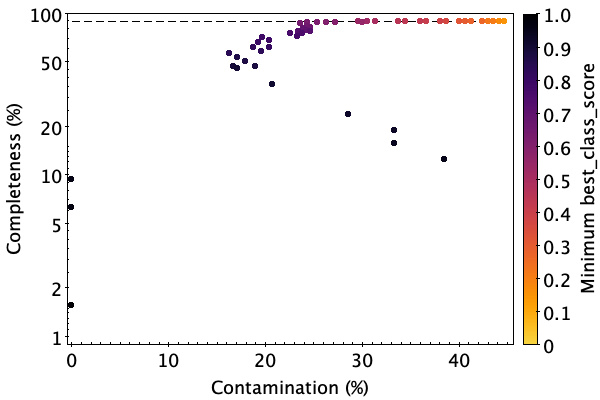}  
\hspace{2mm}
\stackinset{c}{8.8cm}{c}{3cm}{(b)}{} \includegraphics[width=0.45\hsize]{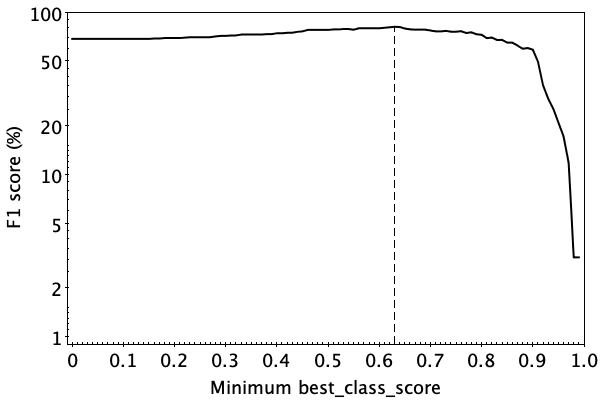} \\ 
\vspace{4mm}
\stackinset{c}{-0.3cm}{c}{3cm}{(c)}{} \includegraphics[width=0.45\hsize]{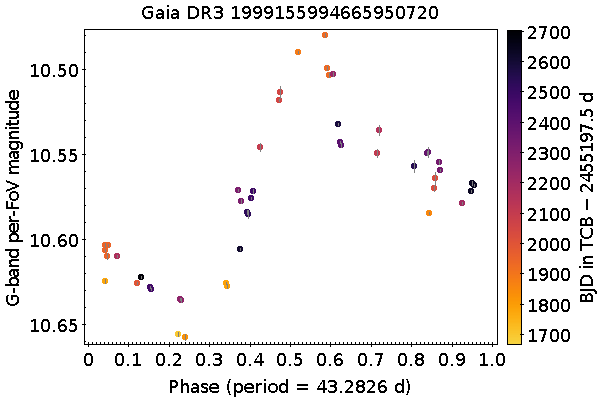}  
\hspace{2mm}
\stackinset{c}{8.8cm}{c}{3cm}{(d)}{} \includegraphics[width=0.45\hsize]{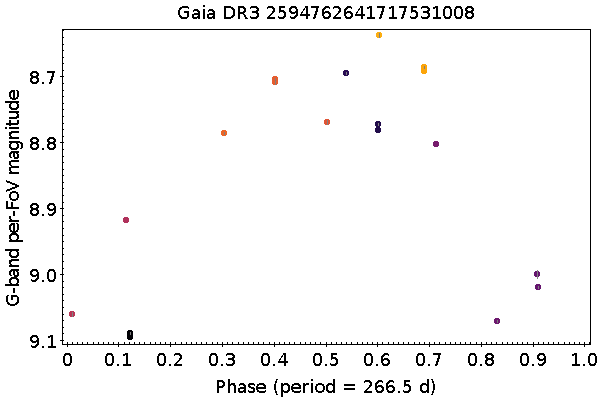} \\
\vspace{4mm}
\stackinset{c}{-0.3cm}{c}{3cm}{(e)}{} \includegraphics[width=0.45\hsize]{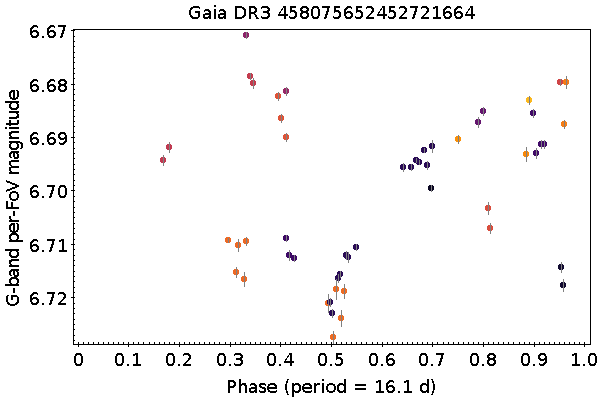}  
\hspace{2mm}
\stackinset{c}{8.8cm}{c}{3cm}{(f)}{} \includegraphics[width=0.45\hsize]{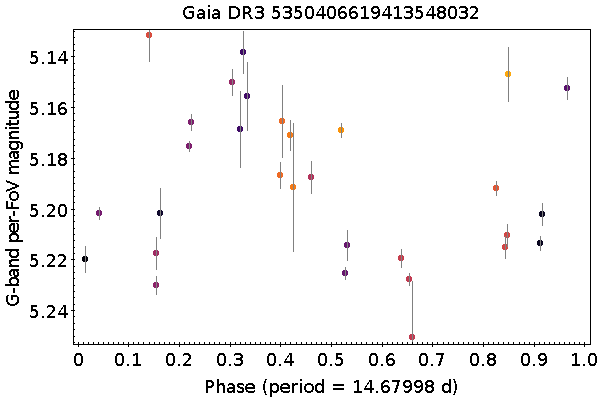} \\
\vspace{4mm}
 \caption{Same as Fig.~\ref{fig:app:ACV_cc}, but for ACYG.}
 \label{fig:app:ACYG_cc}
\end{figure*}

\begin{figure*}
\centering
\stackinset{c}{-0.7cm}{c}{2.7cm}{(a)}{} \includegraphics[width=0.6\hsize]{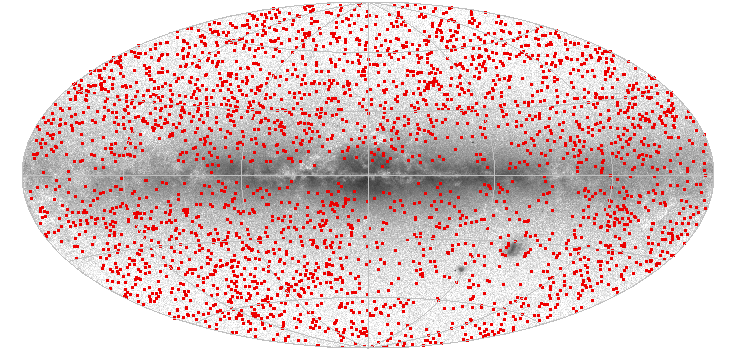} \\ 
\vspace{4mm}
\stackinset{c}{-0.3cm}{c}{3cm}{(b)}{} \includegraphics[width=0.45\hsize]{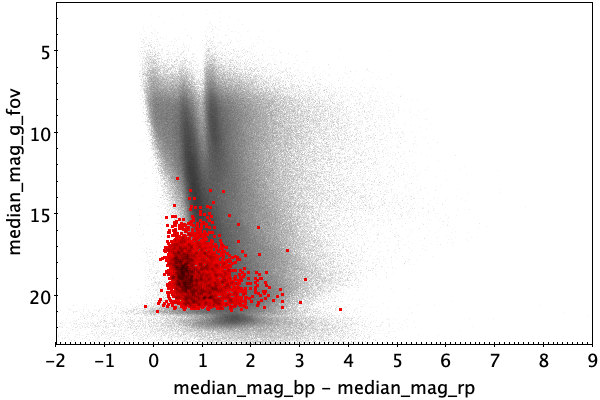}  
\hspace{2mm}
\stackinset{c}{8.8cm}{c}{3cm}{(c)}{} \includegraphics[width=0.45\hsize]{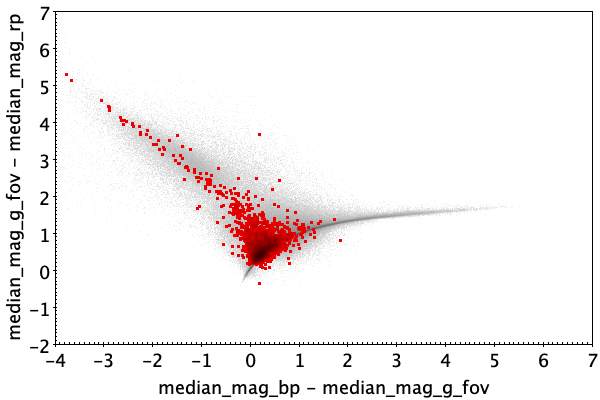} \\ 
\vspace{4mm}
\stackinset{c}{-0.3cm}{c}{3cm}{(d)}{} \includegraphics[width=0.45\hsize]{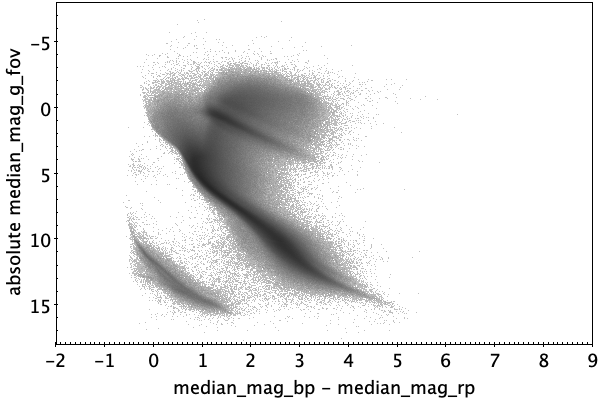}  
\hspace{2mm}
\stackinset{c}{8.8cm}{c}{3cm}{(e)}{} \includegraphics[width=0.45\hsize]{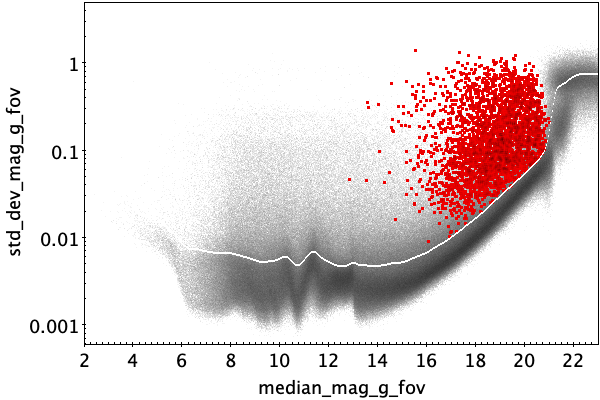} \\ 
\vspace{4mm}
\stackinset{c}{-0.3cm}{c}{3cm}{(f)}{} \includegraphics[width=0.45\hsize]{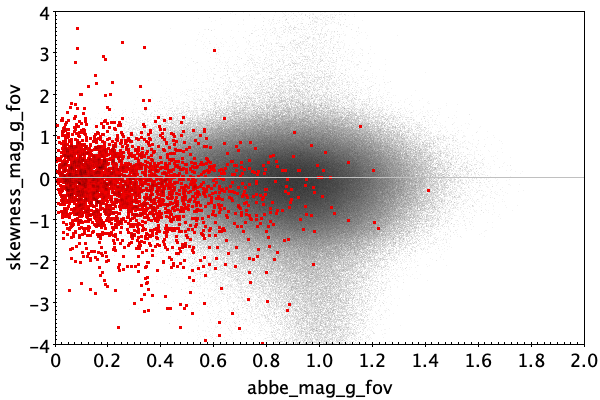}  
\hspace{2mm}
\stackinset{c}{8.8cm}{c}{3cm}{(g)}{} \includegraphics[width=0.45\hsize]{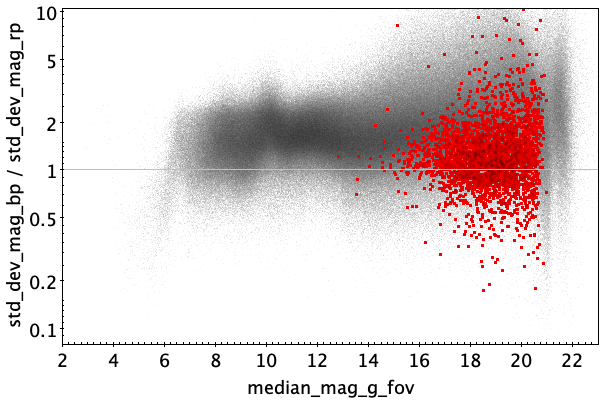}  \\ 
\vspace{4mm}
 \caption{AGN: 3089 training sources.}  
 \label{fig:app:AGN_trn}
\end{figure*}

\begin{figure*}
\centering
\stackinset{c}{-0.7cm}{c}{2.7cm}{(a)}{}
\includegraphics[width=0.6\hsize]{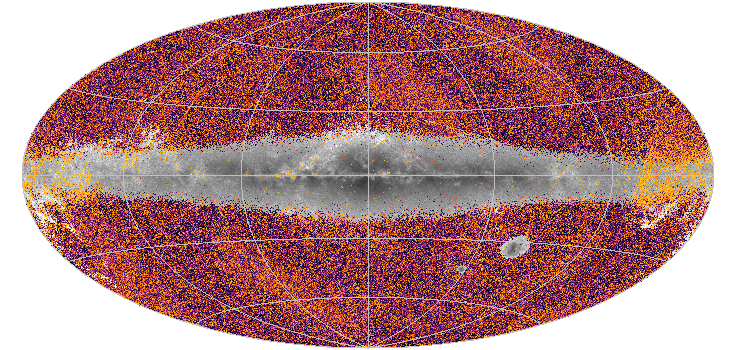} 
\stackinset{c}{2.2cm}{c}{2.7cm}{\includegraphics[height=5.5cm]{figures/appendix/vertical_best_class_score.png}}{} \\ 
\vspace{4mm}
\stackinset{c}{-0.3cm}{c}{3cm}{(b)}{} \includegraphics[width=0.45\hsize]{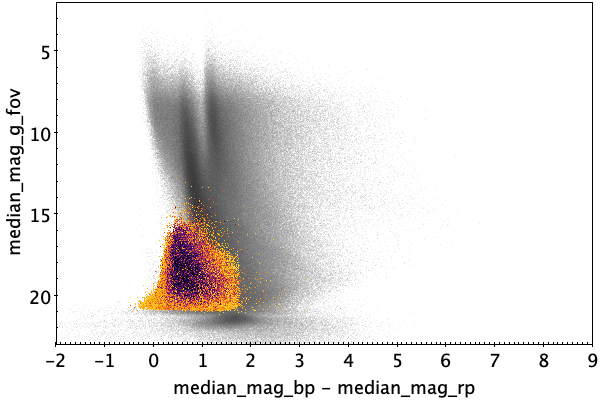}  
\hspace{2mm}
\stackinset{c}{8.8cm}{c}{3cm}{(c)}{} \includegraphics[width=0.45\hsize]{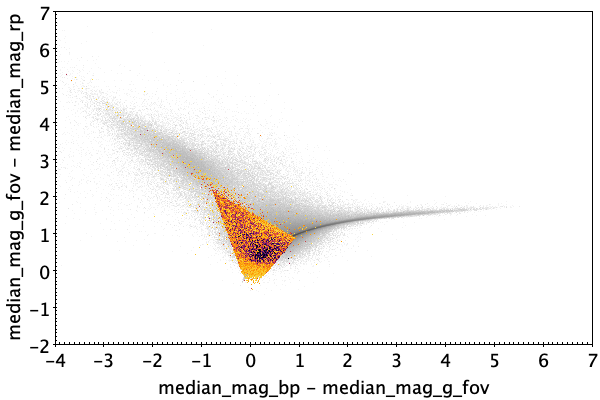} \\ 
\vspace{4mm}
\stackinset{c}{-0.3cm}{c}{3cm}{(d)}{} \includegraphics[width=0.45\hsize]{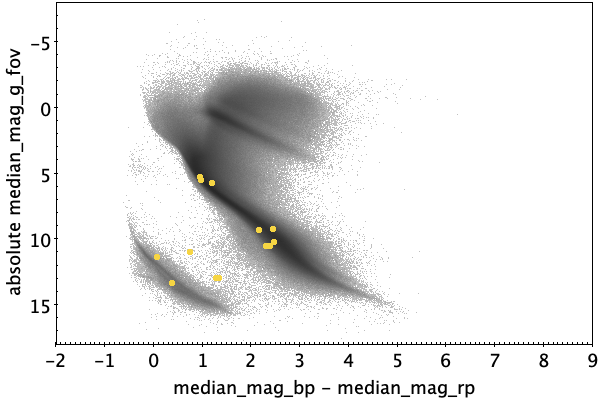}  
\hspace{2mm}
\stackinset{c}{8.8cm}{c}{3cm}{(e)}{} \includegraphics[width=0.45\hsize]{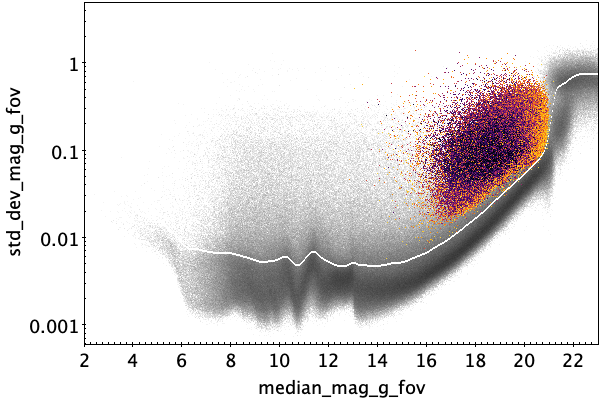} \\ 
\vspace{4mm}
\stackinset{c}{-0.3cm}{c}{3cm}{(f)}{} \includegraphics[width=0.45\hsize]{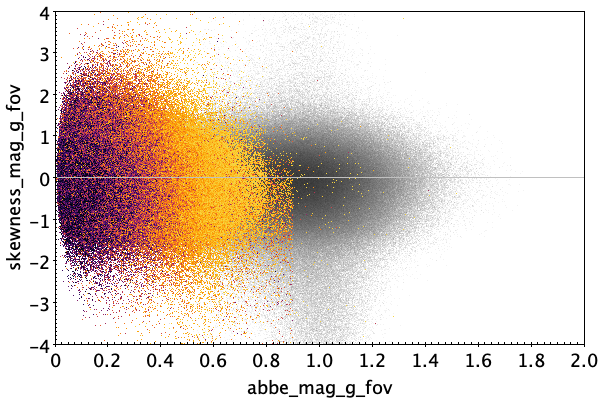}  
\hspace{2mm}
\stackinset{c}{8.8cm}{c}{3cm}{(g)}{} \includegraphics[width=0.45\hsize]{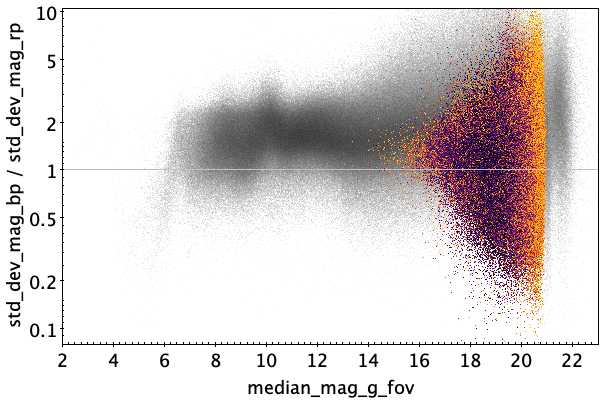}  \\ 
\vspace{4mm}
 \caption{AGN: 1\,035\,207 classified sources.}  
 \label{fig:app:AGN}
\end{figure*}

\begin{figure*}
\centering
\stackinset{c}{-0.3cm}{c}{3cm}{(a)}{} \includegraphics[width=0.45\hsize]{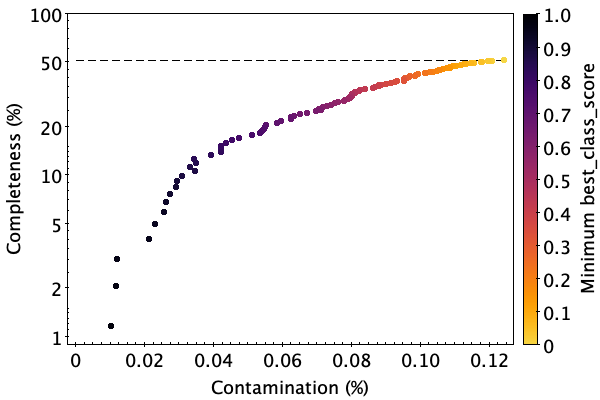}  
\hspace{2mm}
\stackinset{c}{8.8cm}{c}{3cm}{(b)}{} \includegraphics[width=0.45\hsize]{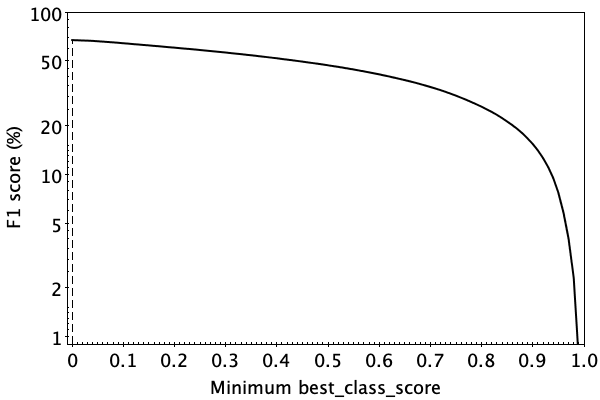} \\ 
\vspace{4mm}
\stackinset{c}{-0.3cm}{c}{3cm}{(c)}{} \includegraphics[width=0.45\hsize]{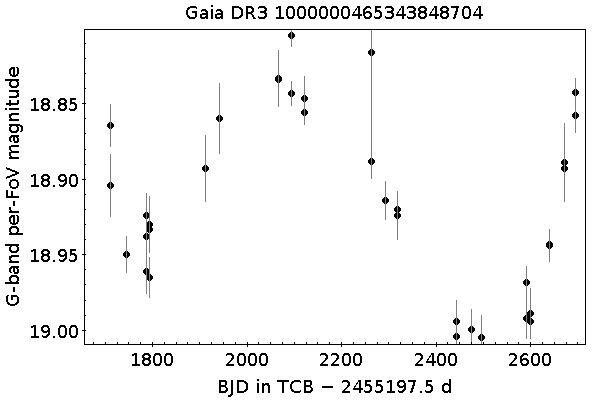}  
\hspace{2mm}
\stackinset{c}{8.8cm}{c}{3cm}{(d)}{} \includegraphics[width=0.45\hsize]{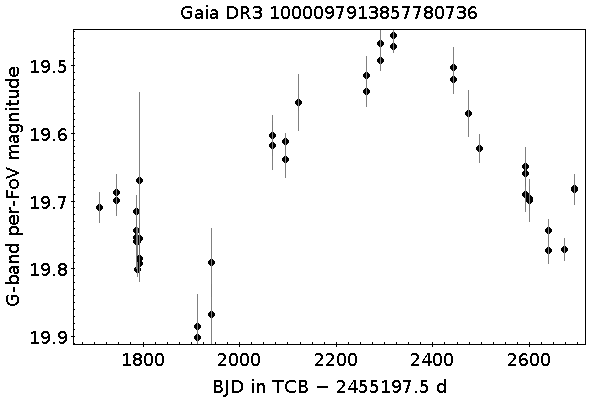} \\
\vspace{4mm}
\stackinset{c}{-0.3cm}{c}{3cm}{(e)}{} \includegraphics[width=0.45\hsize]{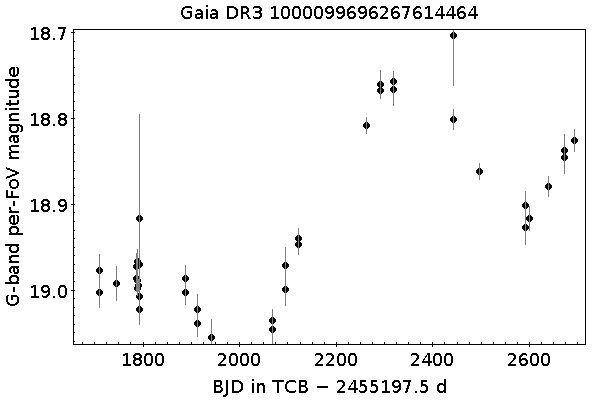}  
\hspace{2mm}
\stackinset{c}{8.8cm}{c}{3cm}{(f)}{} \includegraphics[width=0.45\hsize]{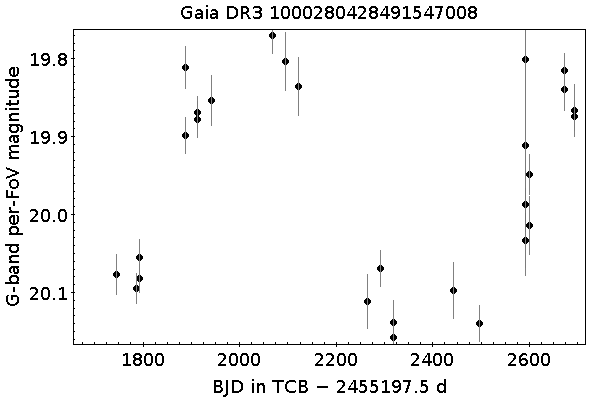} \\
\vspace{4mm}
 \caption{Same as Fig.~\ref{fig:app:ACV_cc}, but for AGN.}
 \label{fig:app:AGN_cc}
\end{figure*}

\begin{figure*}
\centering
\stackinset{c}{-0.7cm}{c}{2.7cm}{(a)}{} \includegraphics[width=0.6\hsize]{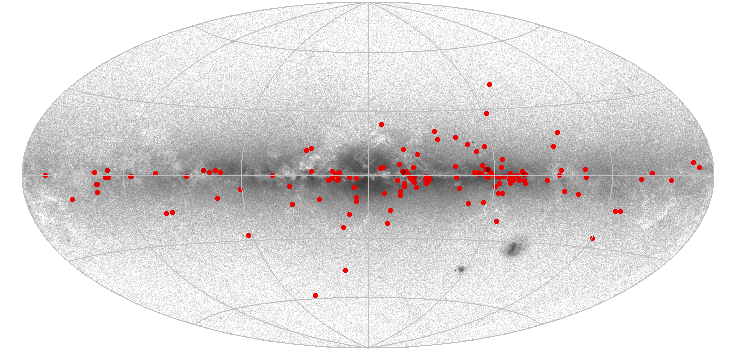} \\ 
\vspace{4mm}
\stackinset{c}{-0.3cm}{c}{3cm}{(b)}{} \includegraphics[width=0.45\hsize]{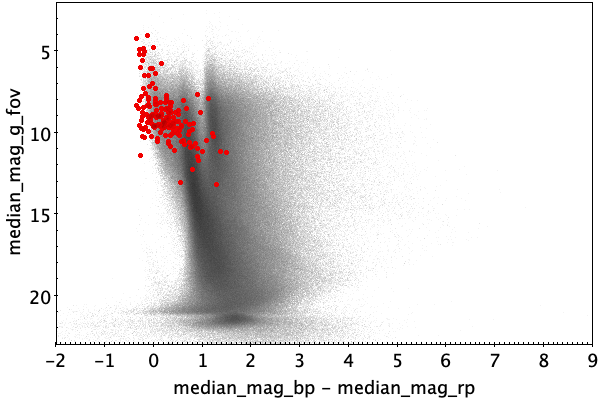}  
\hspace{2mm}
\stackinset{c}{8.8cm}{c}{3cm}{(c)}{} \includegraphics[width=0.45\hsize]{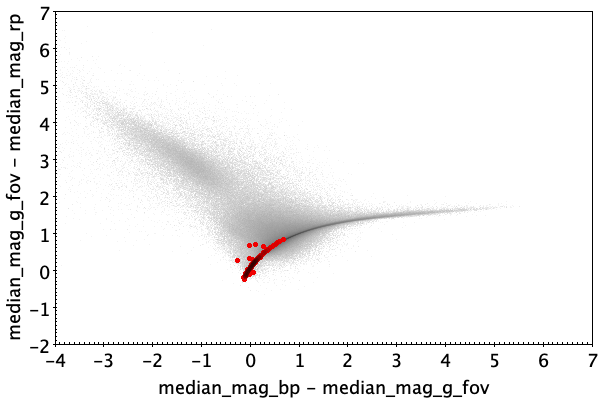} \\ 
\vspace{4mm}
\stackinset{c}{-0.3cm}{c}{3cm}{(d)}{} \includegraphics[width=0.45\hsize]{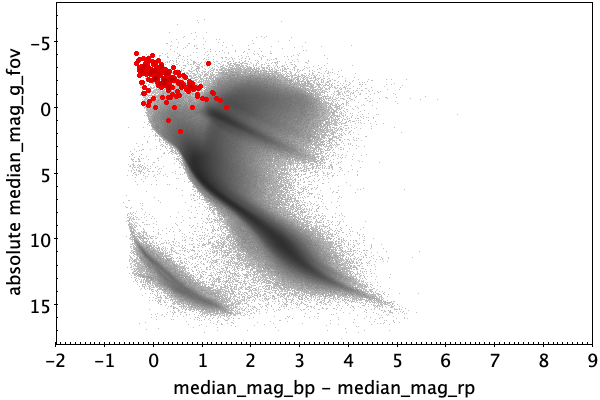}  
\hspace{2mm}
\stackinset{c}{8.8cm}{c}{3cm}{(e)}{} \includegraphics[width=0.45\hsize]{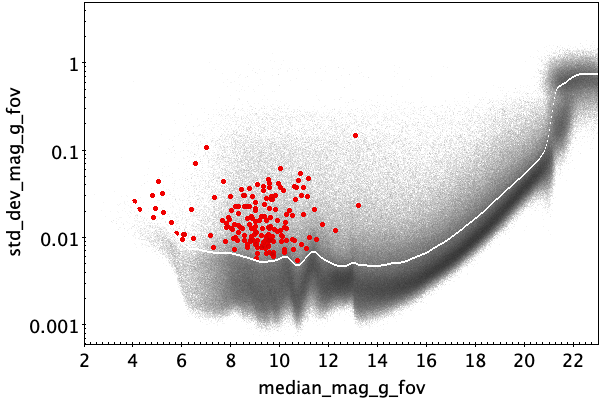} \\ 
\vspace{4mm}
\stackinset{c}{-0.3cm}{c}{3cm}{(f)}{} \includegraphics[width=0.45\hsize]{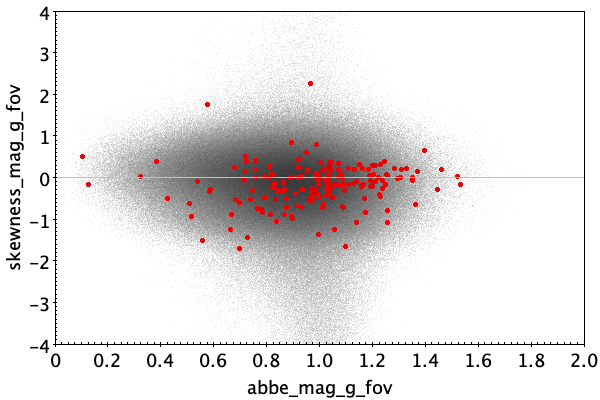}  
\hspace{2mm}
\stackinset{c}{8.8cm}{c}{3cm}{(g)}{} \includegraphics[width=0.45\hsize]{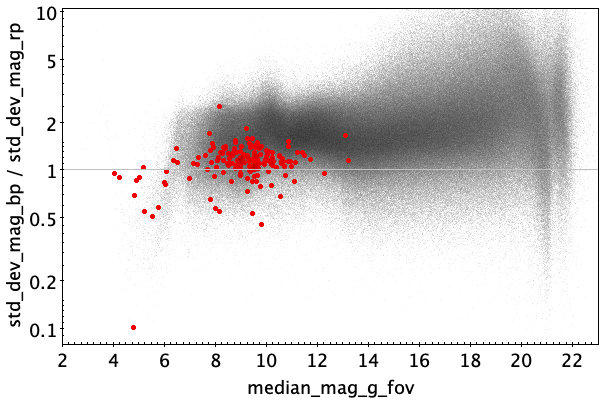}  \\ 
\vspace{4mm}
 \caption{BCEP: 173 training sources.}  
 \label{fig:app:BCEP_trn}
\end{figure*}

\begin{figure*}
\centering
\stackinset{c}{-0.7cm}{c}{2.7cm}{(a)}{}
\includegraphics[width=0.6\hsize]{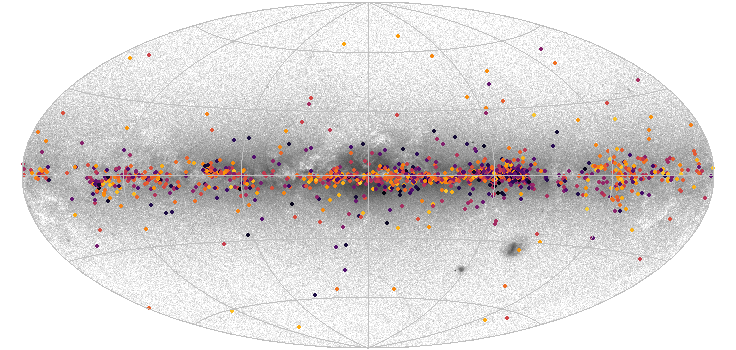} 
\stackinset{c}{2.2cm}{c}{2.7cm}{\includegraphics[height=5.5cm]{figures/appendix/vertical_best_class_score.png}}{} \\ 
\vspace{4mm}
\stackinset{c}{-0.3cm}{c}{3cm}{(b)}{} \includegraphics[width=0.45\hsize]{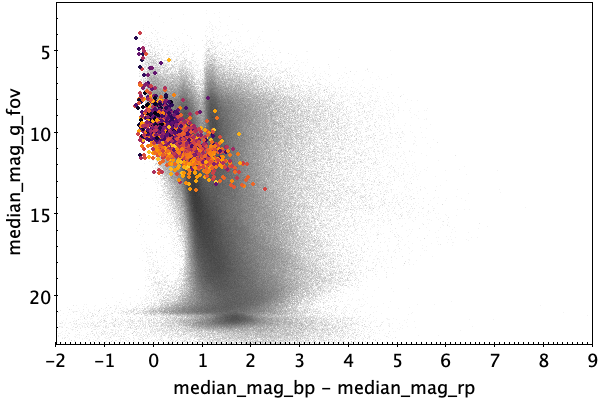}  
\hspace{2mm}
\stackinset{c}{8.8cm}{c}{3cm}{(c)}{} \includegraphics[width=0.45\hsize]{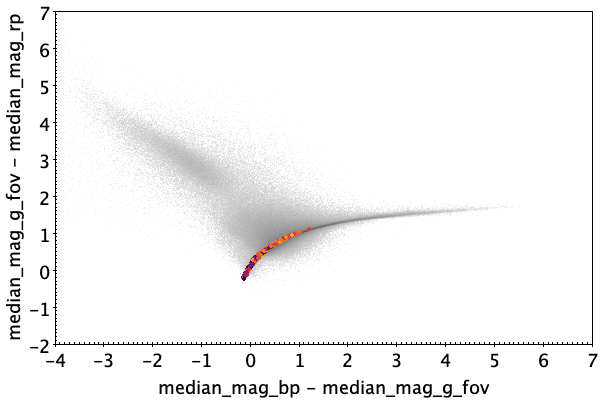} \\ 
\vspace{4mm}
\stackinset{c}{-0.3cm}{c}{3cm}{(d)}{} \includegraphics[width=0.45\hsize]{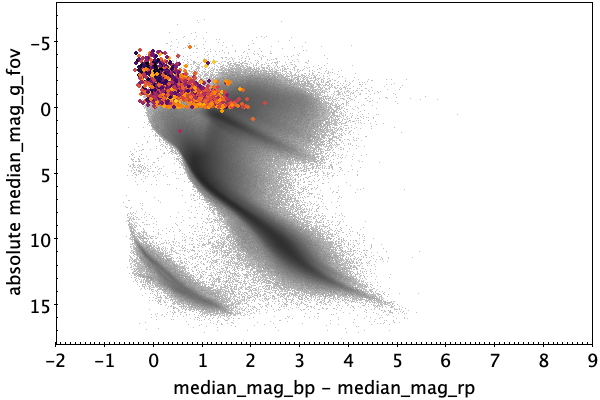}  
\hspace{2mm}
\stackinset{c}{8.8cm}{c}{3cm}{(e)}{} \includegraphics[width=0.45\hsize]{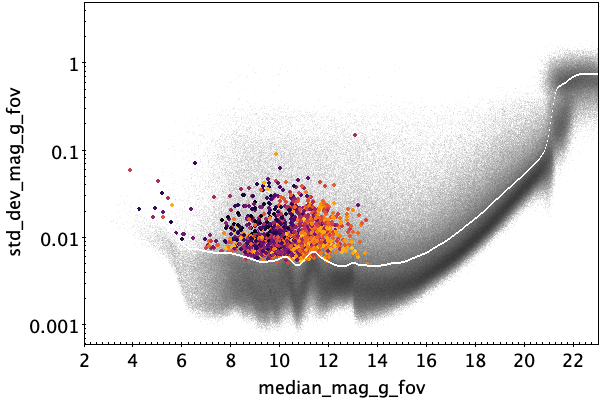} \\ 
\vspace{4mm}
\stackinset{c}{-0.3cm}{c}{3cm}{(f)}{} \includegraphics[width=0.45\hsize]{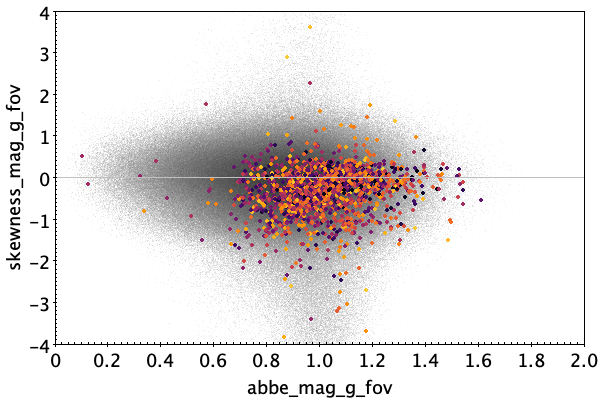}  
\hspace{2mm}
\stackinset{c}{8.8cm}{c}{3cm}{(g)}{} \includegraphics[width=0.45\hsize]{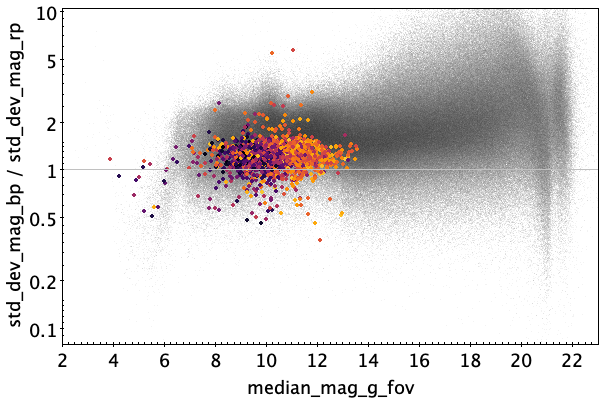}  \\ 
\vspace{4mm}
 \caption{BCEP: 1475 classified sources.}  
 \label{fig:app:BCEP}
\end{figure*}

\begin{figure*}
\centering
\stackinset{c}{-0.3cm}{c}{3cm}{(a)}{} \includegraphics[width=0.45\hsize]{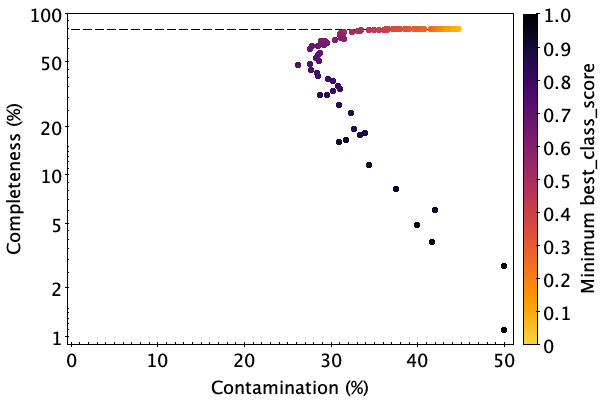}  
\hspace{2mm}
\stackinset{c}{8.8cm}{c}{3cm}{(b)}{} \includegraphics[width=0.45\hsize]{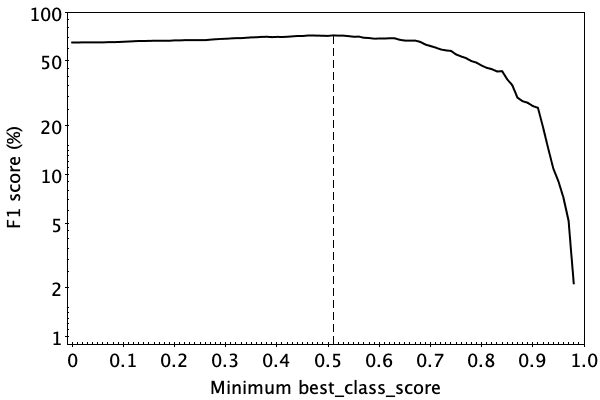} \\ 
\vspace{4mm}
\stackinset{c}{-0.3cm}{c}{3cm}{(c)}{} \includegraphics[width=0.45\hsize]{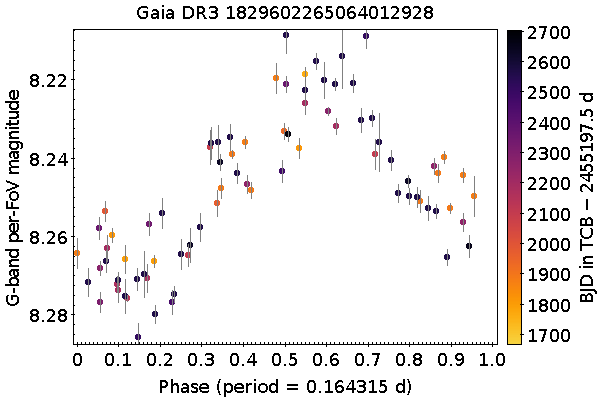}  
\hspace{2mm}
\stackinset{c}{8.8cm}{c}{3cm}{(d)}{} \includegraphics[width=0.45\hsize]{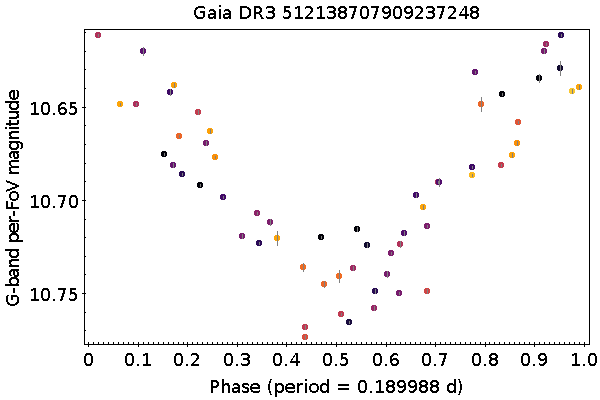} \\
\vspace{4mm}
\stackinset{c}{-0.3cm}{c}{3cm}{(e)}{} \includegraphics[width=0.45\hsize]{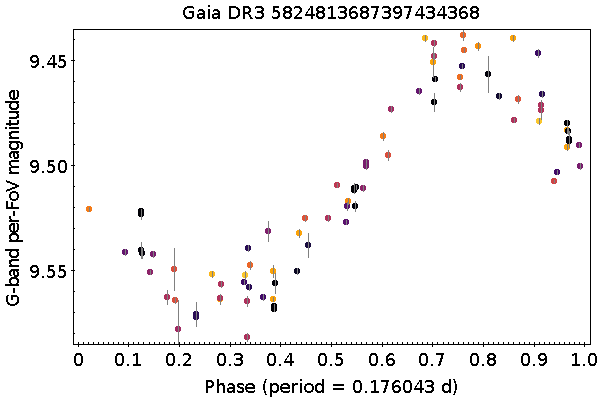}  
\hspace{2mm}
\stackinset{c}{8.8cm}{c}{3cm}{(f)}{} \includegraphics[width=0.45\hsize]{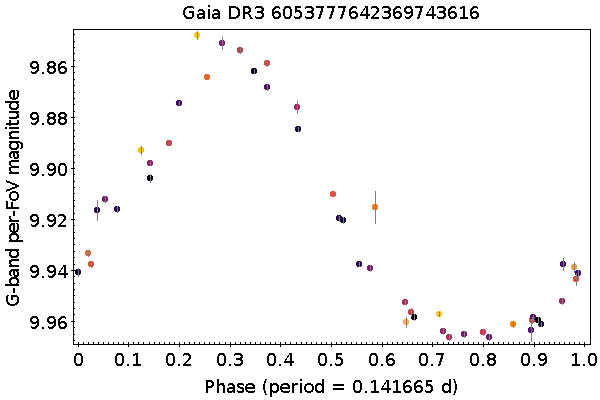} \\
\vspace{4mm}
 \caption{Same as Fig.~\ref{fig:app:ACV_cc}, but for BCEP.}
 \label{fig:app:BCEP_cc}
\end{figure*}

\begin{figure*}
\centering
\stackinset{c}{-0.7cm}{c}{2.7cm}{(a)}{} \includegraphics[width=0.6\hsize]{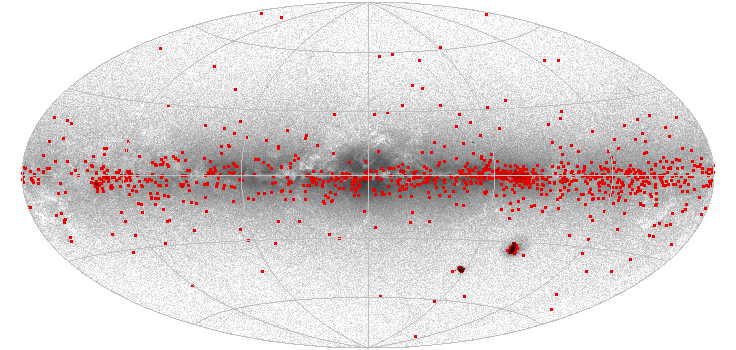} \\ 
\vspace{4mm}
\stackinset{c}{-0.3cm}{c}{3cm}{(b)}{} \includegraphics[width=0.45\hsize]{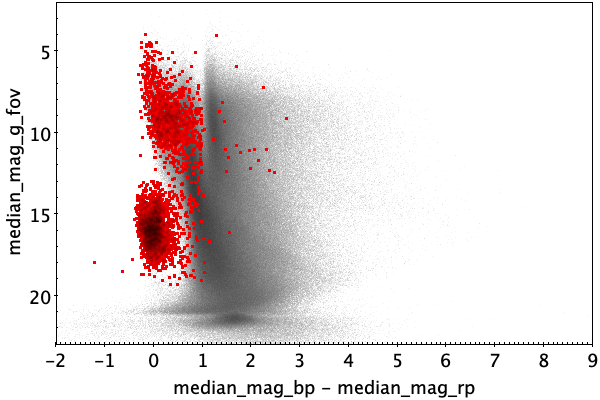}  
\hspace{2mm}
\stackinset{c}{8.8cm}{c}{3cm}{(c)}{} \includegraphics[width=0.45\hsize]{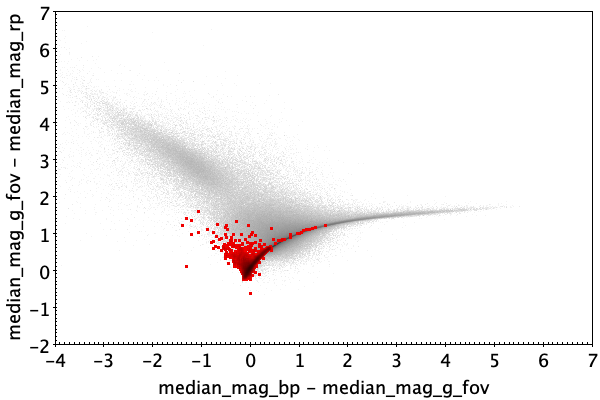} \\ 
\vspace{4mm}
\stackinset{c}{-0.3cm}{c}{3cm}{(d)}{} \includegraphics[width=0.45\hsize]{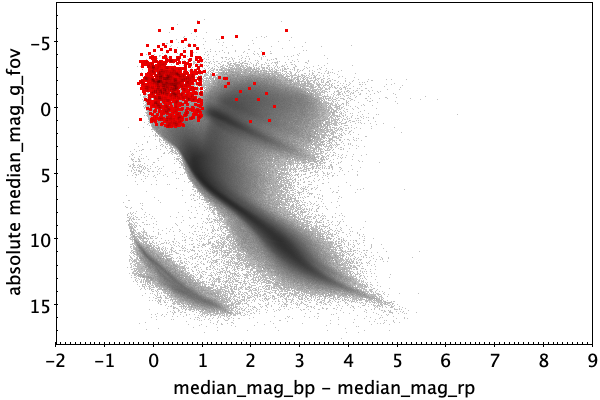}  
\hspace{2mm}
\stackinset{c}{8.8cm}{c}{3cm}{(e)}{} \includegraphics[width=0.45\hsize]{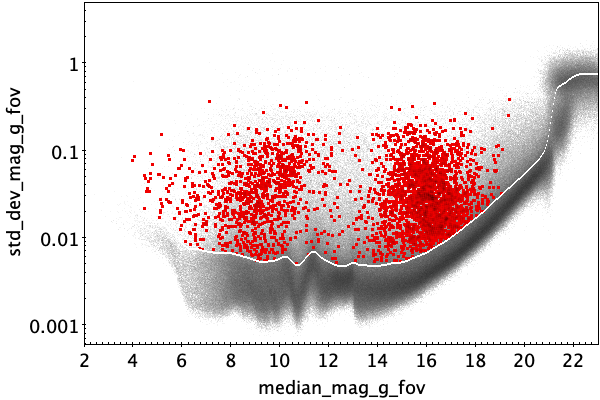} \\ 
\vspace{4mm}
\stackinset{c}{-0.3cm}{c}{3cm}{(f)}{} \includegraphics[width=0.45\hsize]{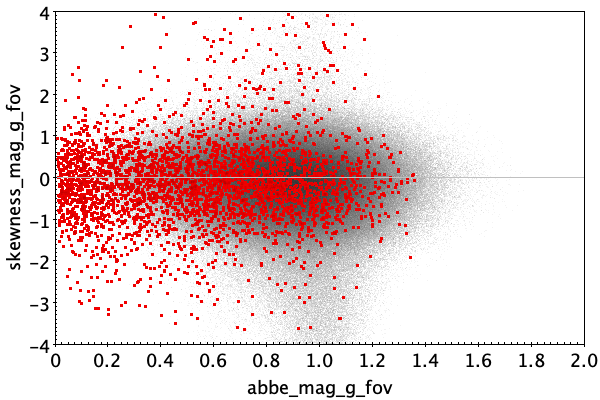}  
\hspace{2mm}
\stackinset{c}{8.8cm}{c}{3cm}{(g)}{} \includegraphics[width=0.45\hsize]{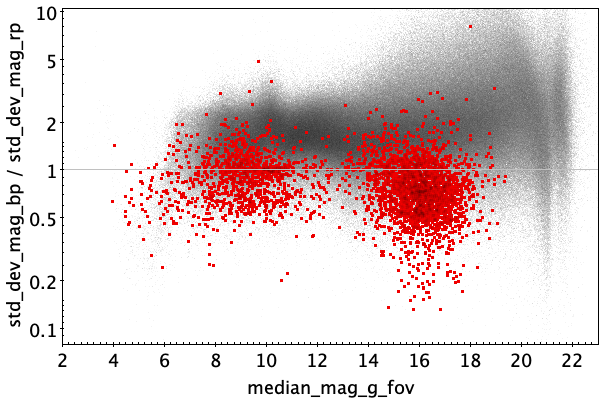}  \\ 
\vspace{4mm}
 \caption{BE|GCAS|SDOR|WR: 3546 training sources.}  
 \label{fig:app:BE_trn}
\end{figure*}

\begin{figure*}
\centering
\stackinset{c}{-0.7cm}{c}{2.7cm}{(a)}{}
\includegraphics[width=0.6\hsize]{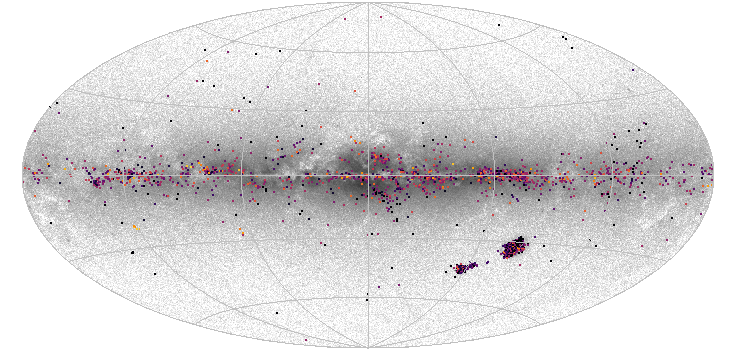} 
\stackinset{c}{2.2cm}{c}{2.7cm}{\includegraphics[height=5.5cm]{figures/appendix/vertical_best_class_score.png}}{} \\ 
\vspace{4mm}
\stackinset{c}{-0.3cm}{c}{3cm}{(b)}{} \includegraphics[width=0.45\hsize]{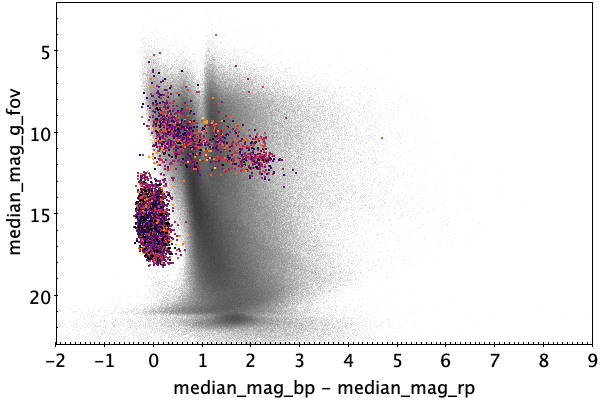}  
\hspace{2mm}
\stackinset{c}{8.8cm}{c}{3cm}{(c)}{} \includegraphics[width=0.45\hsize]{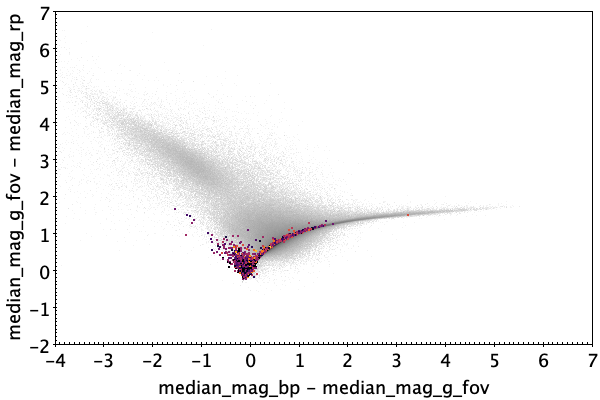} \\ 
\vspace{4mm}
\stackinset{c}{-0.3cm}{c}{3cm}{(d)}{} \includegraphics[width=0.45\hsize]{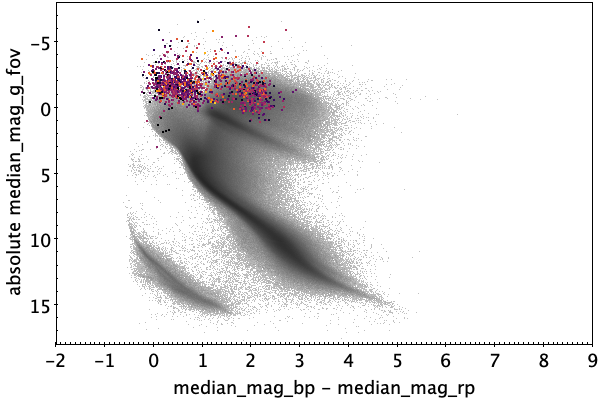}  
\hspace{2mm}
\stackinset{c}{8.8cm}{c}{3cm}{(e)}{} \includegraphics[width=0.45\hsize]{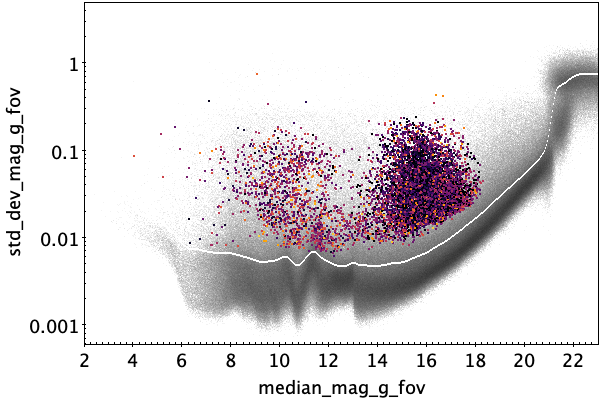} \\ 
\vspace{4mm}
\stackinset{c}{-0.3cm}{c}{3cm}{(f)}{} \includegraphics[width=0.45\hsize]{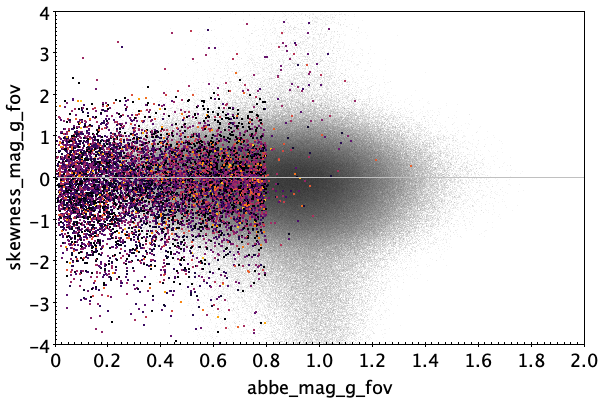}  
\hspace{2mm}
\stackinset{c}{8.8cm}{c}{3cm}{(g)}{} \includegraphics[width=0.45\hsize]{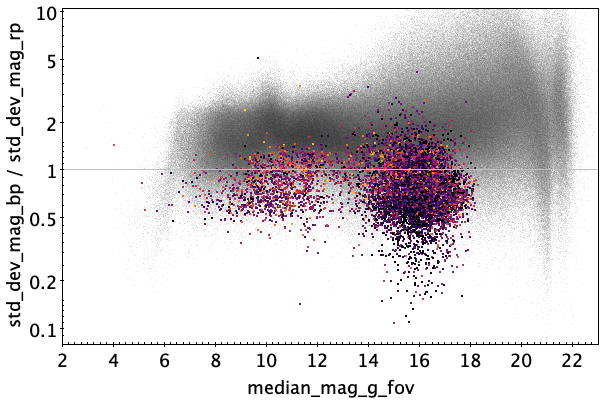}  \\ 
\vspace{4mm}
 \caption{BE|GCAS|SDOR|WR: 8560 classified sources.}  
 \label{fig:app:BE}
\end{figure*}

\begin{figure*}
\centering
\stackinset{c}{-0.3cm}{c}{3cm}{(a)}{} \includegraphics[width=0.45\hsize]{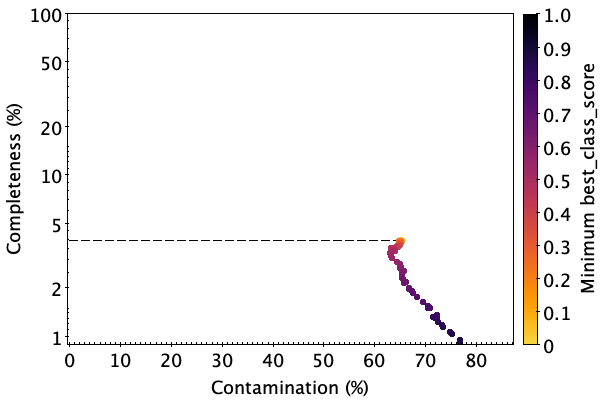}  
\hspace{2mm}
\stackinset{c}{8.8cm}{c}{3cm}{(b)}{} \includegraphics[width=0.45\hsize]{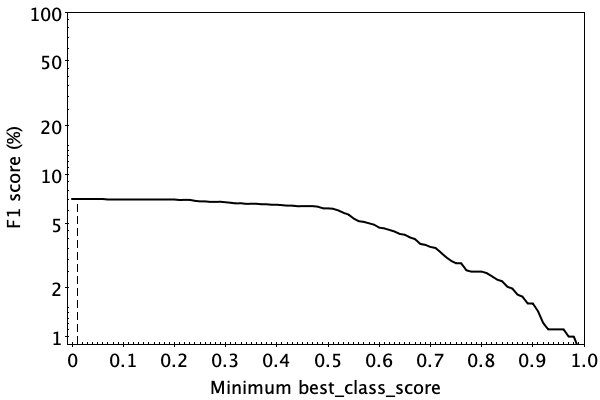} \\ 
\vspace{4mm}
\stackinset{c}{-0.3cm}{c}{3cm}{(c)}{} \includegraphics[width=0.45\hsize]{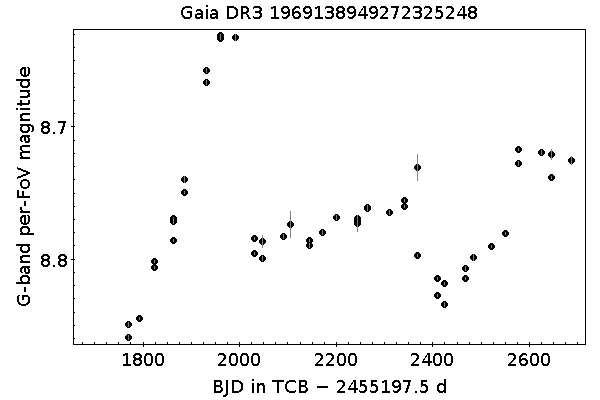}  
\hspace{2mm}
\stackinset{c}{8.8cm}{c}{3cm}{(d)}{} \includegraphics[width=0.45\hsize]{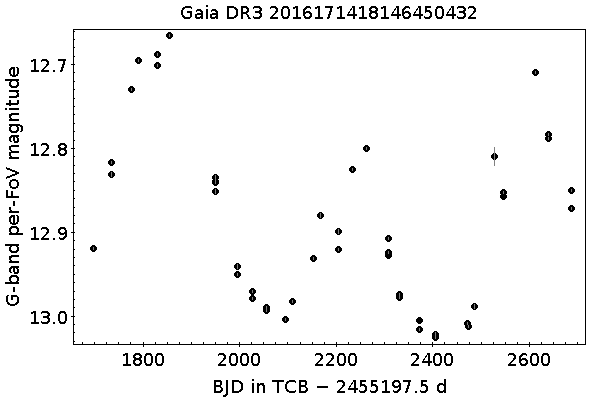} \\
\vspace{4mm}
\stackinset{c}{-0.3cm}{c}{3cm}{(e)}{} \includegraphics[width=0.45\hsize]{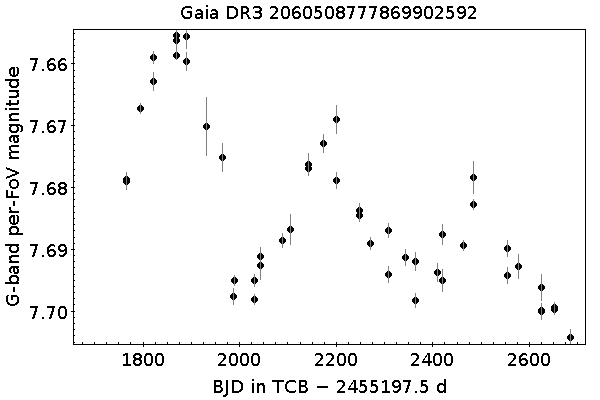}  
\hspace{2mm}
\stackinset{c}{8.8cm}{c}{3cm}{(f)}{} \includegraphics[width=0.45\hsize]{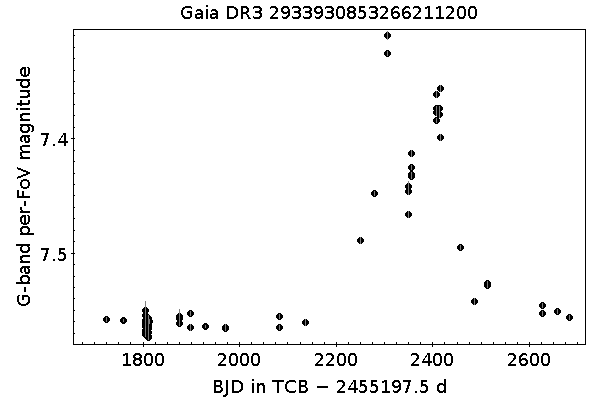} \\
\vspace{4mm}
 \caption{Same as Fig.~\ref{fig:app:ACV_cc}, but for BE|GCAS|SDOR|WR.}
 \label{fig:app:BE_cc}
\end{figure*}

\begin{figure*}
\centering
\stackinset{c}{-0.7cm}{c}{2.7cm}{(a)}{} \includegraphics[width=0.6\hsize]{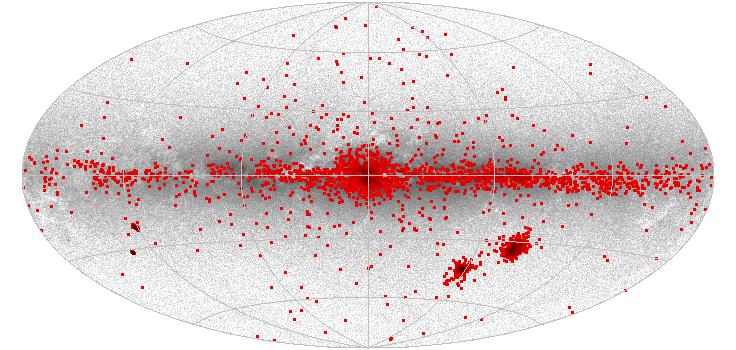} \\ 
\vspace{4mm}
\stackinset{c}{-0.3cm}{c}{3cm}{(b)}{} \includegraphics[width=0.45\hsize]{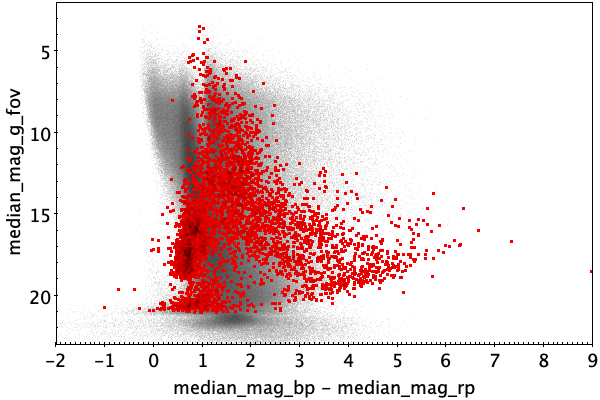}  
\hspace{2mm}
\stackinset{c}{8.8cm}{c}{3cm}{(c)}{} \includegraphics[width=0.45\hsize]{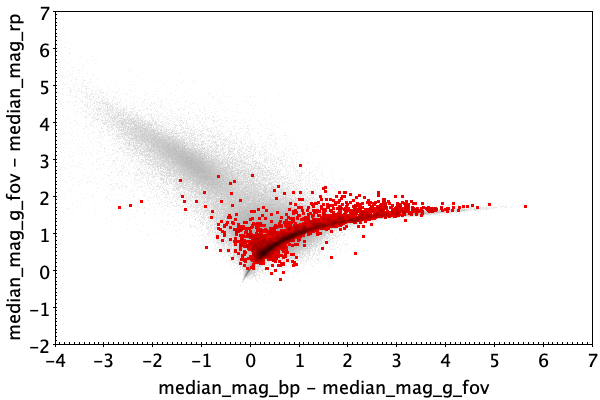} \\ 
\vspace{4mm}
\stackinset{c}{-0.3cm}{c}{3cm}{(d)}{} \includegraphics[width=0.45\hsize]{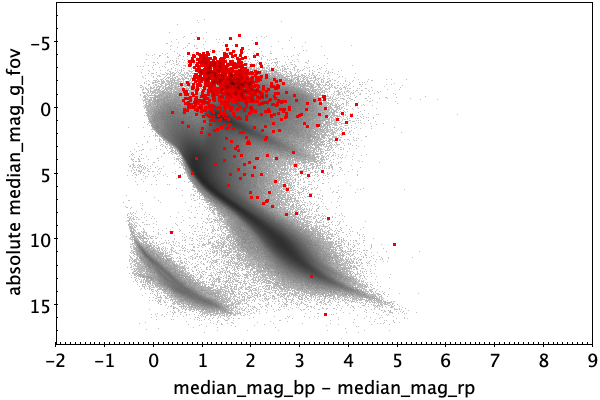}  
\hspace{2mm}
\stackinset{c}{8.8cm}{c}{3cm}{(e)}{} \includegraphics[width=0.45\hsize]{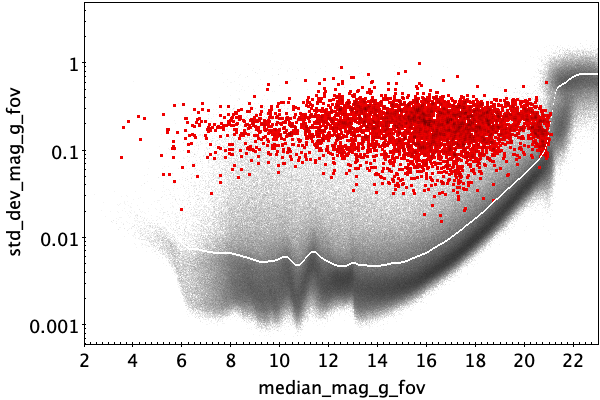} \\ 
\vspace{4mm}
\stackinset{c}{-0.3cm}{c}{3cm}{(f)}{} \includegraphics[width=0.45\hsize]{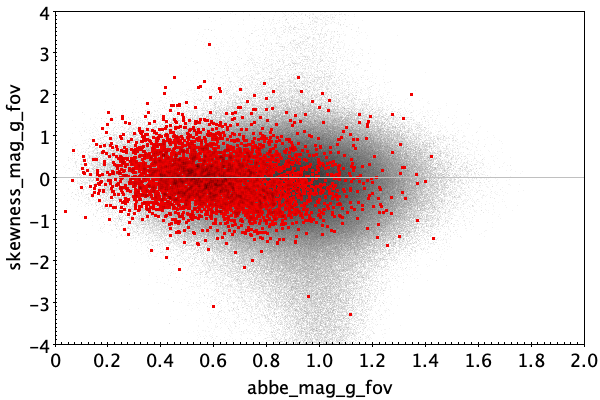}  
\hspace{2mm}
\stackinset{c}{8.8cm}{c}{3cm}{(g)}{} \includegraphics[width=0.45\hsize]{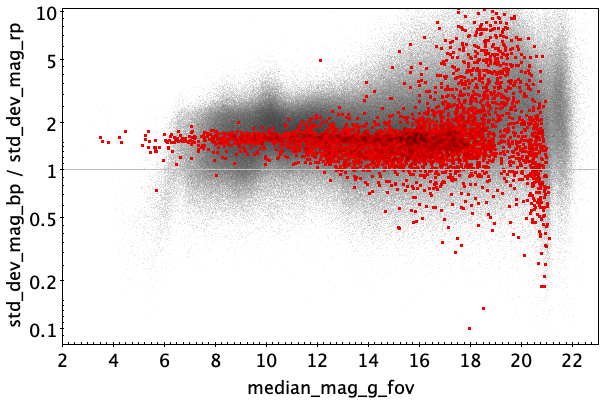}  \\ 
\vspace{4mm}
 \caption{CEP: 4448 training sources.}  
 \label{fig:app:CEP_trn}
\end{figure*}

\begin{figure*}
\centering
\stackinset{c}{-0.7cm}{c}{2.7cm}{(a)}{}
\includegraphics[width=0.6\hsize]{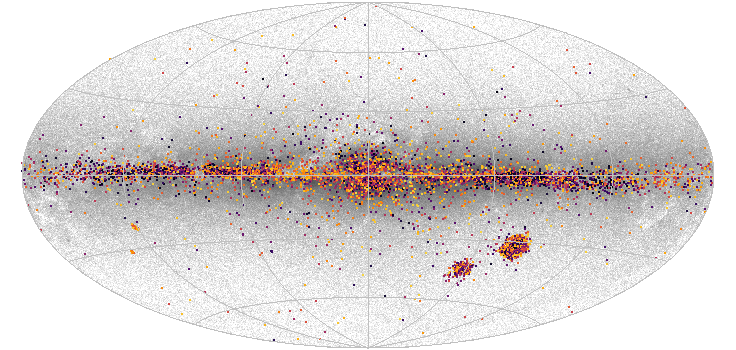} 
\stackinset{c}{2.2cm}{c}{2.7cm}{\includegraphics[height=5.5cm]{figures/appendix/vertical_best_class_score.png}}{} \\ 
\vspace{4mm}
\stackinset{c}{-0.3cm}{c}{3cm}{(b)}{} \includegraphics[width=0.45\hsize]{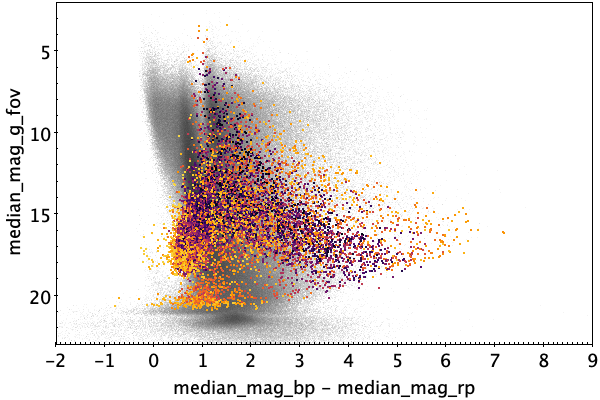}  
\hspace{2mm}
\stackinset{c}{8.8cm}{c}{3cm}{(c)}{} \includegraphics[width=0.45\hsize]{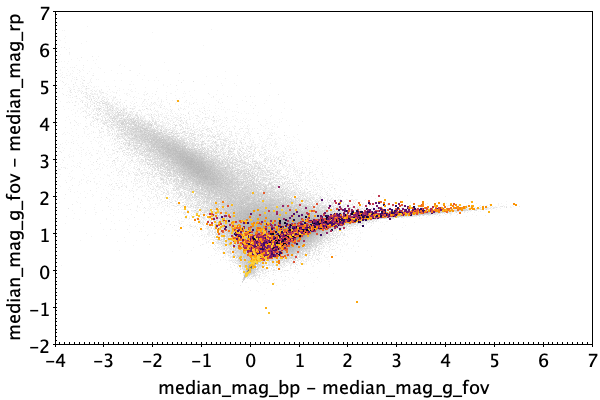} \\ 
\vspace{4mm}
\stackinset{c}{-0.3cm}{c}{3cm}{(d)}{} \includegraphics[width=0.45\hsize]{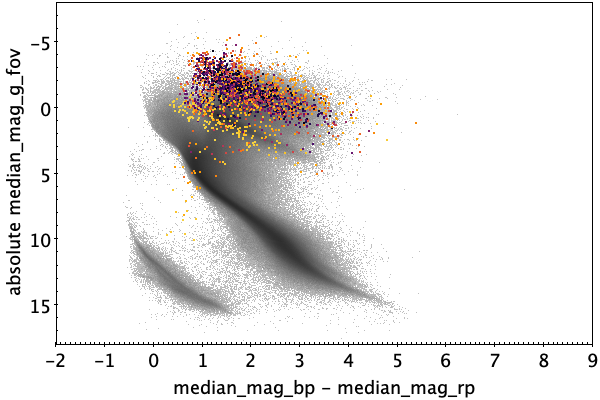}  
\hspace{2mm}
\stackinset{c}{8.8cm}{c}{3cm}{(e)}{} \includegraphics[width=0.45\hsize]{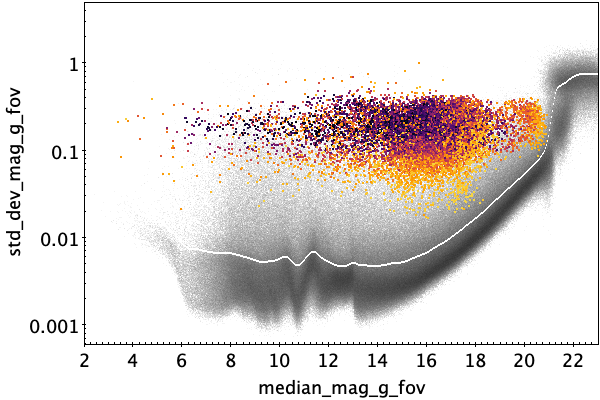} \\ 
\vspace{4mm}
\stackinset{c}{-0.3cm}{c}{3cm}{(f)}{} \includegraphics[width=0.45\hsize]{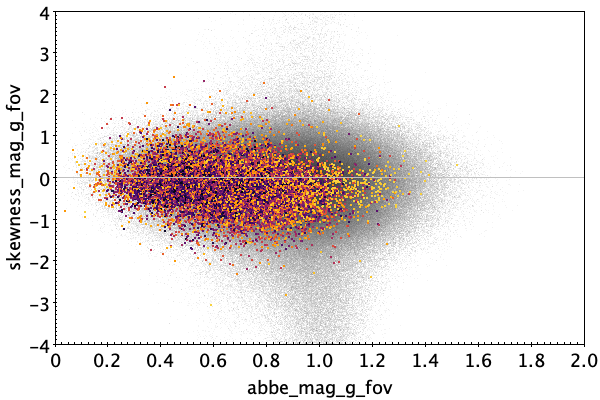}  
\hspace{2mm}
\stackinset{c}{8.8cm}{c}{3cm}{(g)}{} \includegraphics[width=0.45\hsize]{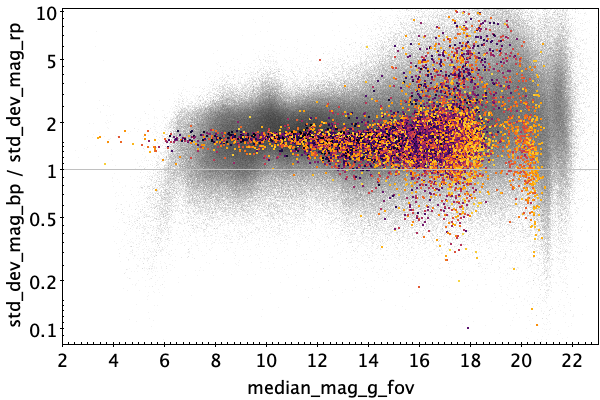}  \\ 
\vspace{4mm}
 \caption{CEP: 16\,141 classified sources.}  
 \label{fig:app:CEP}
\end{figure*}

\begin{figure*}
\centering
\stackinset{c}{-0.3cm}{c}{3cm}{(a)}{} \includegraphics[width=0.45\hsize]{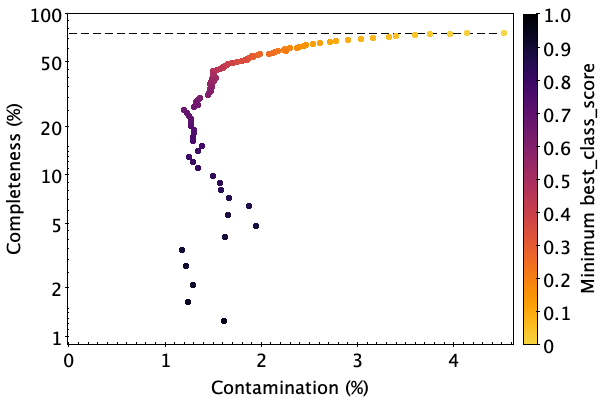}  
\hspace{2mm}
\stackinset{c}{8.8cm}{c}{3cm}{(b)}{} \includegraphics[width=0.45\hsize]{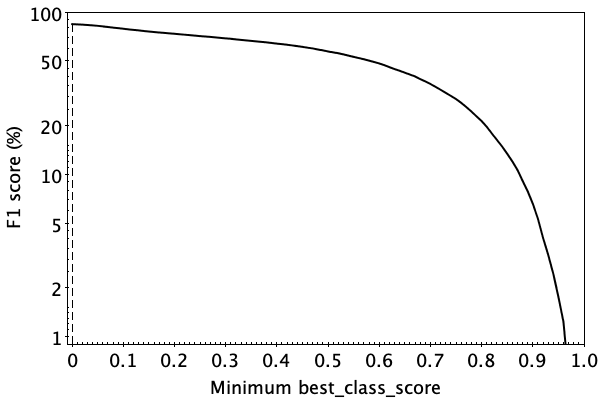} \\ 
\vspace{4mm}
\stackinset{c}{-0.3cm}{c}{3cm}{(c)}{} \includegraphics[width=0.45\hsize]{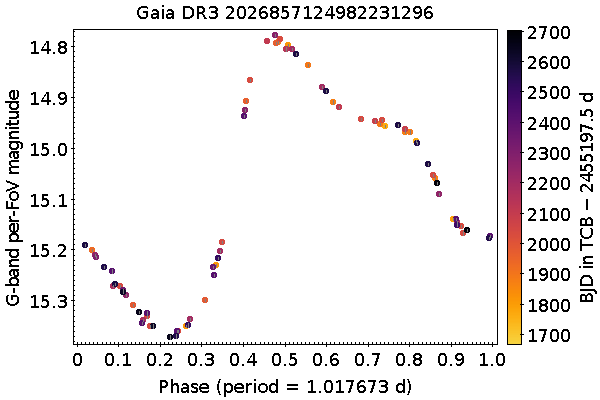}  
\hspace{2mm}
\stackinset{c}{8.8cm}{c}{3cm}{(d)}{} \includegraphics[width=0.45\hsize]{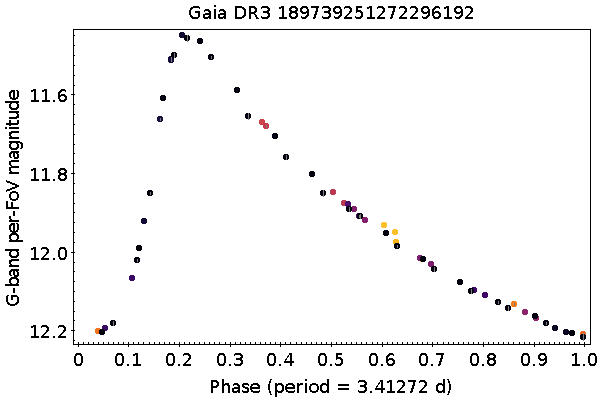} \\
\vspace{4mm}
\stackinset{c}{-0.3cm}{c}{3cm}{(e)}{} \includegraphics[width=0.45\hsize]{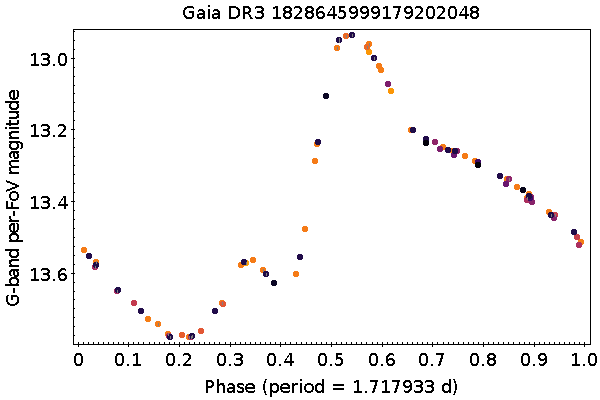}  
\hspace{2mm}
\stackinset{c}{8.8cm}{c}{3cm}{(f)}{} \includegraphics[width=0.45\hsize]{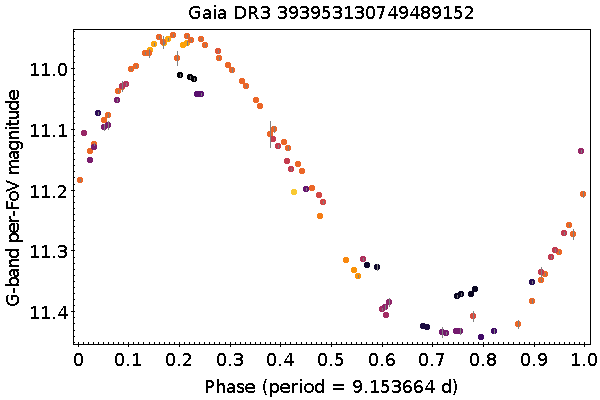} \\
\vspace{4mm}
 \caption{Same as Fig.~\ref{fig:app:ACV_cc}, but for CEP: (c) anomalous Cepheid, (d) $\delta$~Cephei, (e) BL\,Herculis, and (f) W\,Virginis stars.}
 \label{fig:app:CEP_cc}
\end{figure*}

\begin{figure*}
\centering
\stackinset{c}{-0.7cm}{c}{2.7cm}{(a)}{} \includegraphics[width=0.6\hsize]{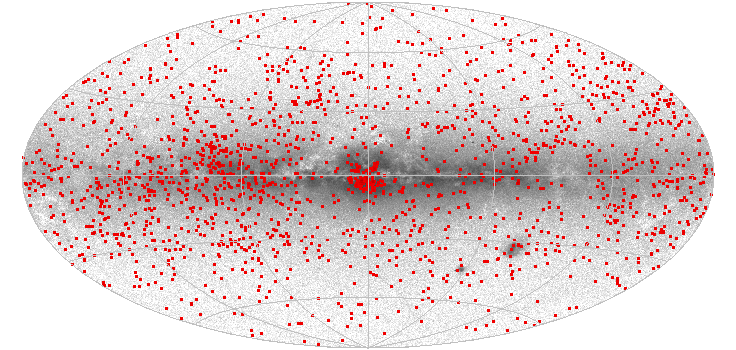} \\ 
\vspace{4mm}
\stackinset{c}{-0.3cm}{c}{3cm}{(b)}{} \includegraphics[width=0.45\hsize]{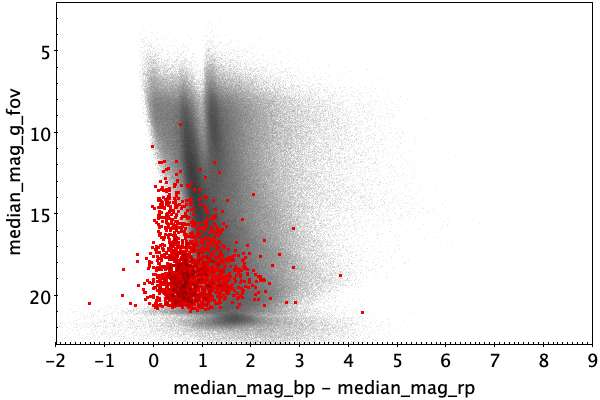}  
\hspace{2mm}
\stackinset{c}{8.8cm}{c}{3cm}{(c)}{} \includegraphics[width=0.45\hsize]{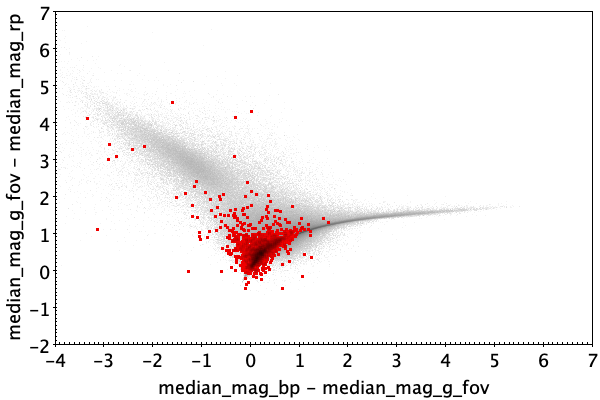} \\ 
\vspace{4mm}
\stackinset{c}{-0.3cm}{c}{3cm}{(d)}{} \includegraphics[width=0.45\hsize]{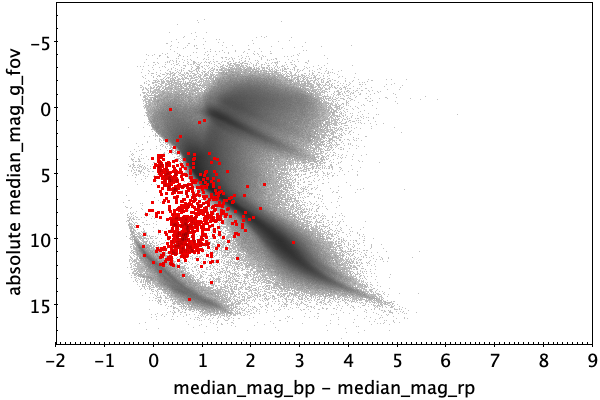}  
\hspace{2mm}
\stackinset{c}{8.8cm}{c}{3cm}{(e)}{} \includegraphics[width=0.45\hsize]{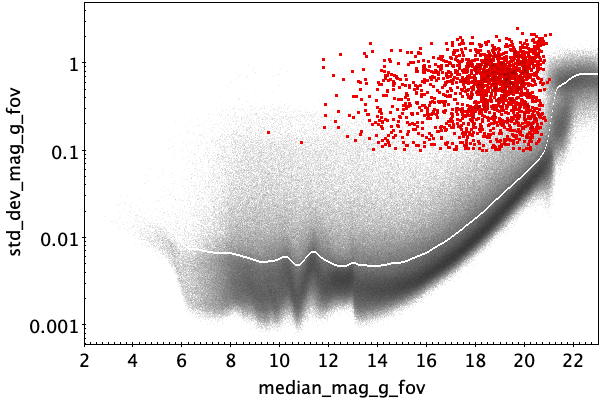} \\ 
\vspace{4mm}
\stackinset{c}{-0.3cm}{c}{3cm}{(f)}{} \includegraphics[width=0.45\hsize]{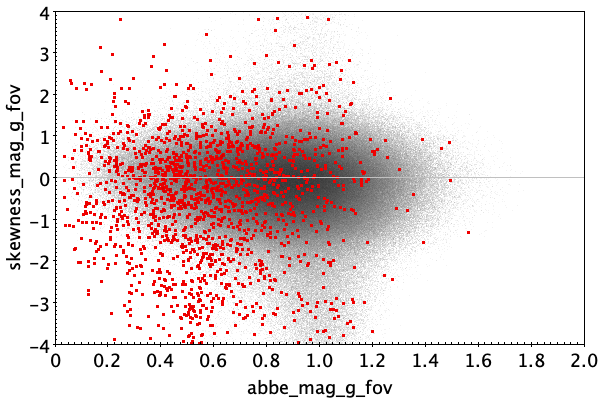}  
\hspace{2mm}
\stackinset{c}{8.8cm}{c}{3cm}{(g)}{} \includegraphics[width=0.45\hsize]{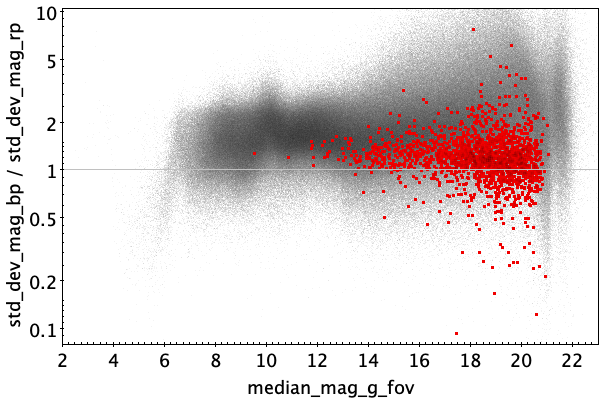}  \\ 
\vspace{4mm}
 \caption{CV: 1815 training sources.}  
 \label{fig:app:CV_trn}
\end{figure*}

\begin{figure*}
\centering
\stackinset{c}{-0.7cm}{c}{2.7cm}{(a)}{}
\includegraphics[width=0.6\hsize]{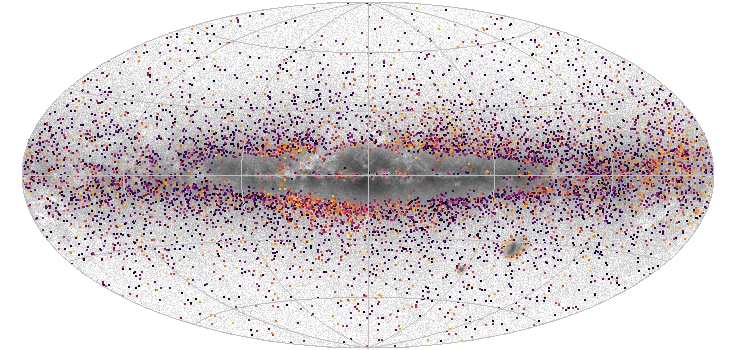} 
\stackinset{c}{2.2cm}{c}{2.7cm}{\includegraphics[height=5.5cm]{figures/appendix/vertical_best_class_score.png}}{} \\ 
\vspace{4mm}
\stackinset{c}{-0.3cm}{c}{3cm}{(b)}{} \includegraphics[width=0.45\hsize]{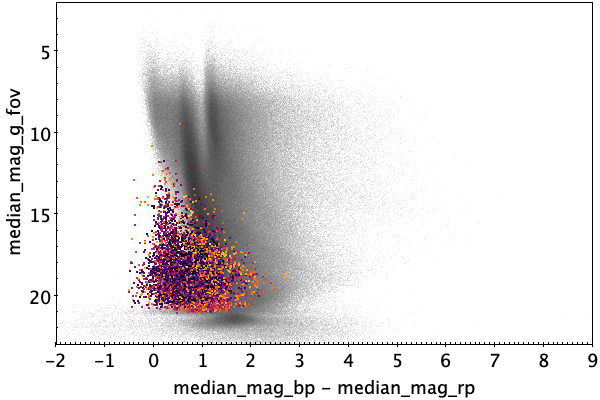}  
\hspace{2mm}
\stackinset{c}{8.8cm}{c}{3cm}{(c)}{} \includegraphics[width=0.45\hsize]{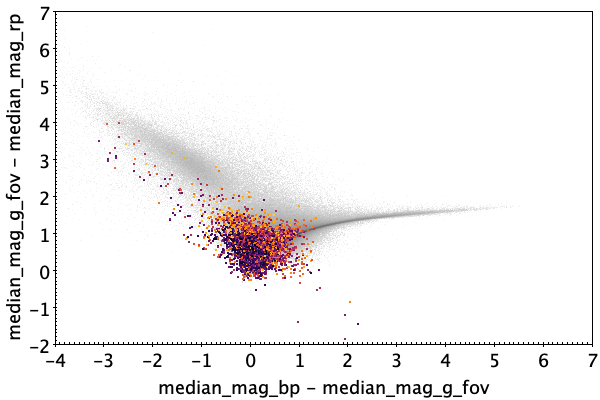} \\ 
\vspace{4mm}
\stackinset{c}{-0.3cm}{c}{3cm}{(d)}{} \includegraphics[width=0.45\hsize]{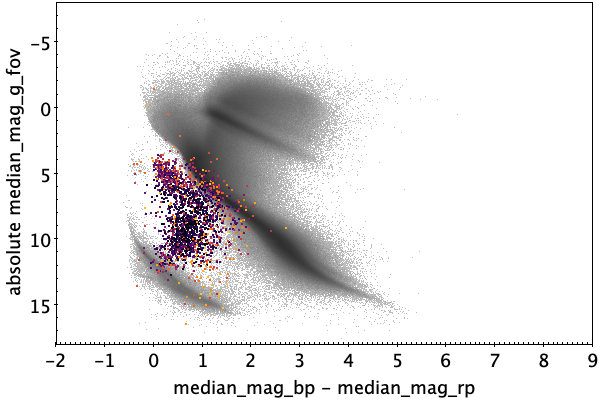}  
\hspace{2mm}
\stackinset{c}{8.8cm}{c}{3cm}{(e)}{} \includegraphics[width=0.45\hsize]{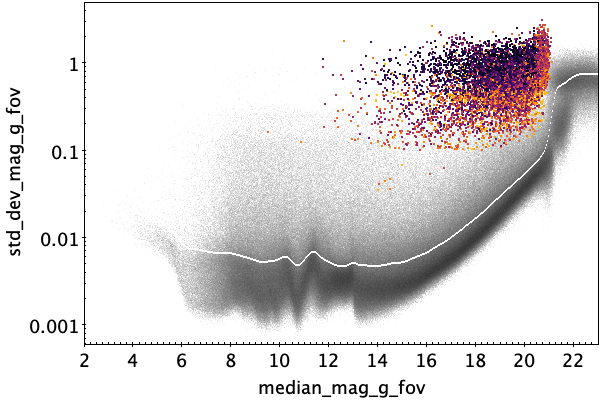} \\ 
\vspace{4mm}
\stackinset{c}{-0.3cm}{c}{3cm}{(f)}{} \includegraphics[width=0.45\hsize]{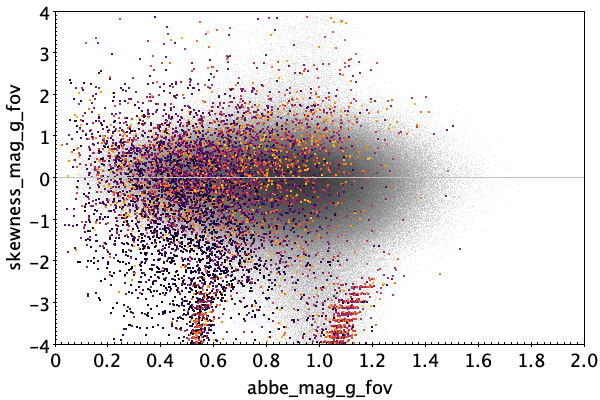}  
\hspace{2mm}
\stackinset{c}{8.8cm}{c}{3cm}{(g)}{} \includegraphics[width=0.45\hsize]{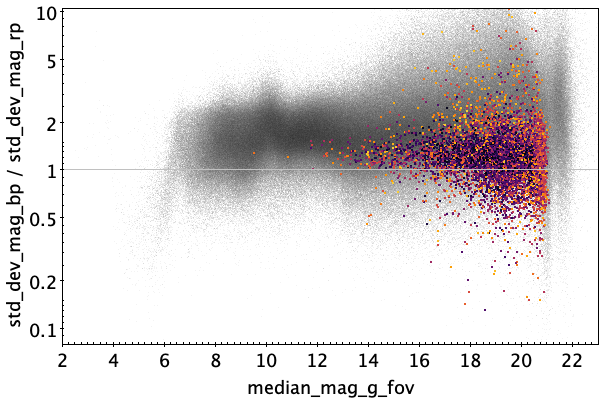}  \\ 
\vspace{4mm}
 \caption{CV: 7306 classified sources.}  
 \label{fig:app:CV}
\end{figure*}

\begin{figure*}
\centering
\stackinset{c}{-0.3cm}{c}{3cm}{(a)}{} \includegraphics[width=0.45\hsize]{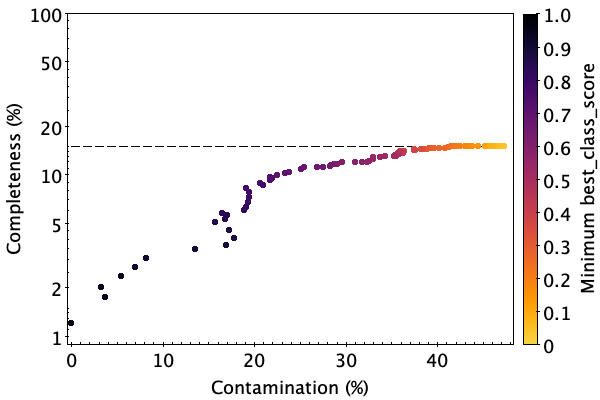}  
\hspace{2mm}
\stackinset{c}{8.8cm}{c}{3cm}{(b)}{} \includegraphics[width=0.45\hsize]{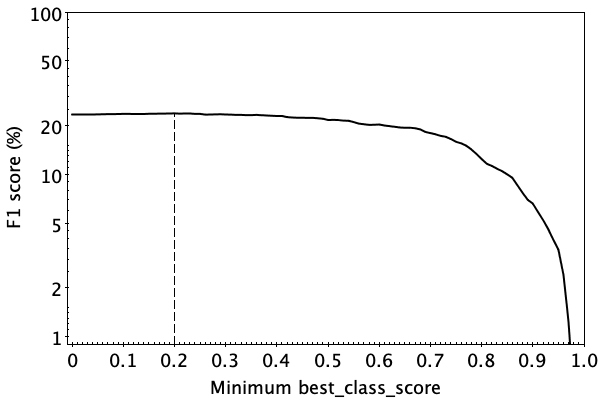} \\ 
\vspace{4mm}
\stackinset{c}{-0.3cm}{c}{3cm}{(c)}{} \includegraphics[width=0.45\hsize]{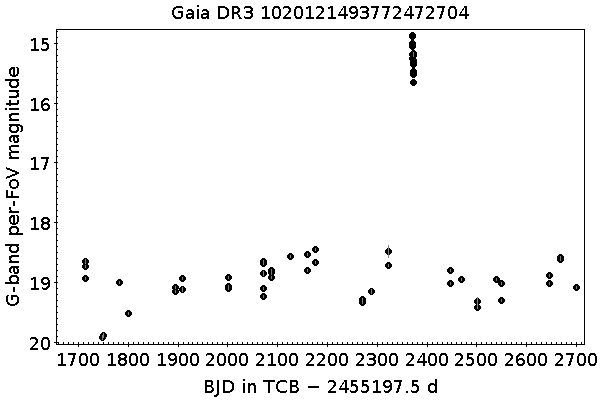}  
\hspace{2mm}
\stackinset{c}{8.8cm}{c}{3cm}{(d)}{} \includegraphics[width=0.45\hsize]{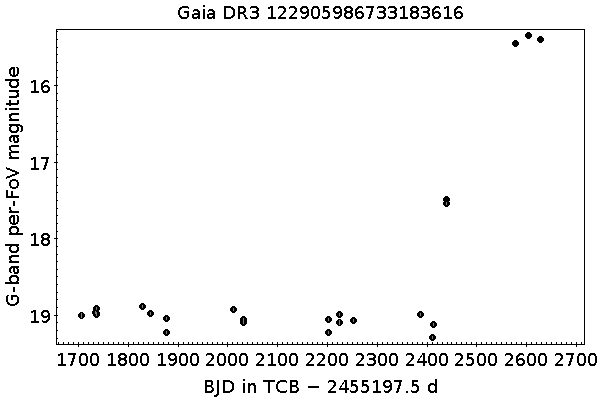} \\
\vspace{4mm}
\stackinset{c}{-0.3cm}{c}{3cm}{(e)}{} \includegraphics[width=0.45\hsize]{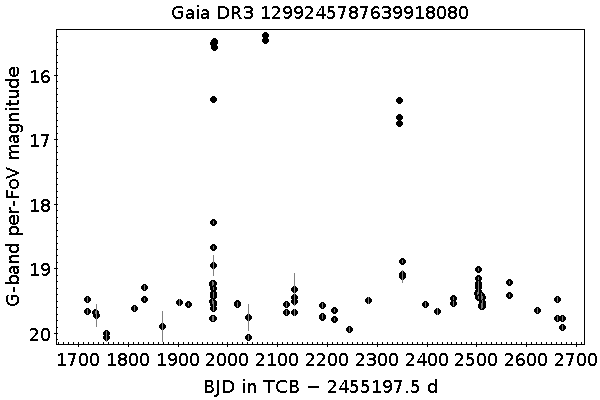}  
\hspace{2mm}
\stackinset{c}{8.8cm}{c}{3cm}{(f)}{} \includegraphics[width=0.45\hsize]{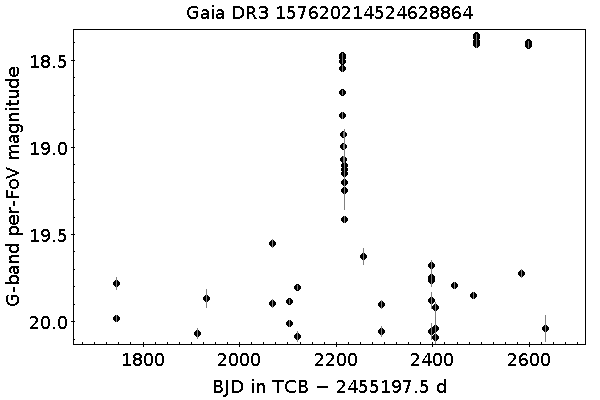} \\
\vspace{4mm}
 \caption{Same as Fig.~\ref{fig:app:ACV_cc}, but for CV.}
 \label{fig:app:CV_cc}
\end{figure*}

\begin{figure*}
\centering
\stackinset{c}{-0.7cm}{c}{2.7cm}{(a)}{} \includegraphics[width=0.6\hsize]{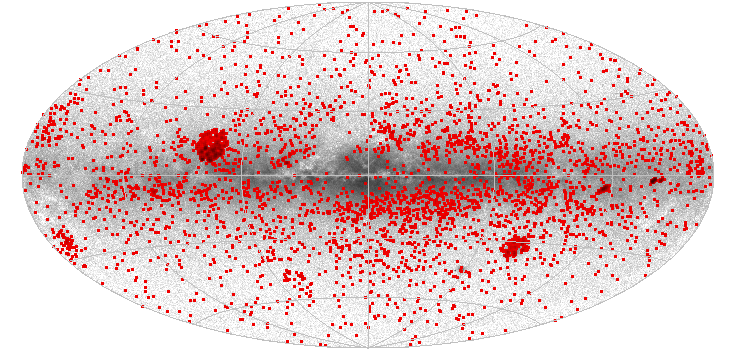} \\ 
\vspace{4mm}
\stackinset{c}{-0.3cm}{c}{3cm}{(b)}{} \includegraphics[width=0.45\hsize]{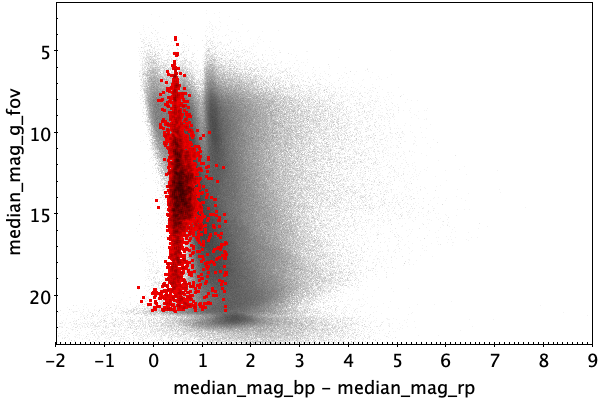}  
\hspace{2mm}
\stackinset{c}{8.8cm}{c}{3cm}{(c)}{} \includegraphics[width=0.45\hsize]{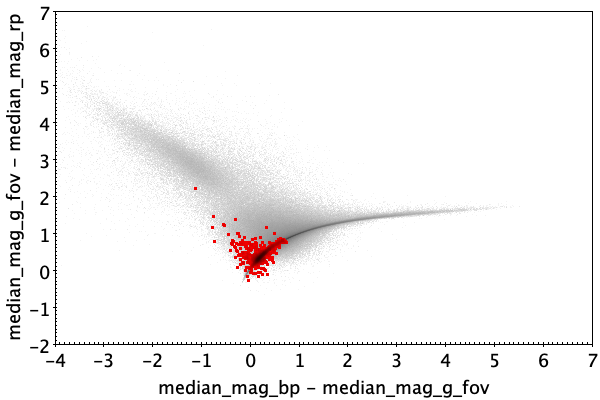} \\ 
\vspace{4mm}
\stackinset{c}{-0.3cm}{c}{3cm}{(d)}{} \includegraphics[width=0.45\hsize]{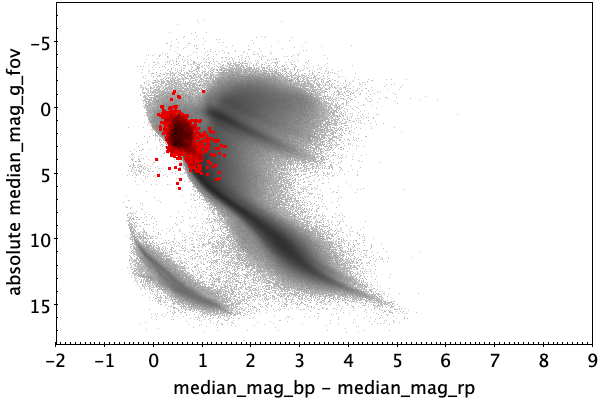}  
\hspace{2mm}
\stackinset{c}{8.8cm}{c}{3cm}{(e)}{} \includegraphics[width=0.45\hsize]{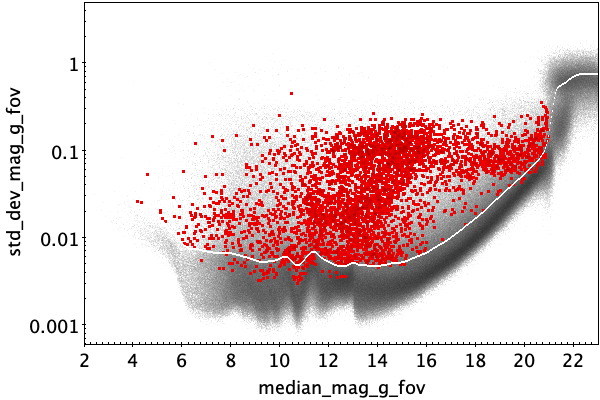} \\ 
\vspace{4mm}
\stackinset{c}{-0.3cm}{c}{3cm}{(f)}{} \includegraphics[width=0.45\hsize]{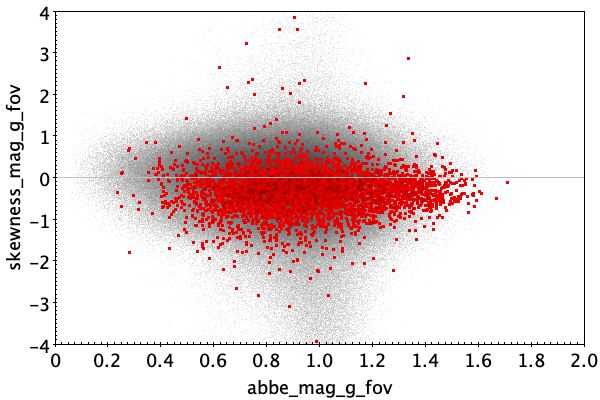}  
\hspace{2mm}
\stackinset{c}{8.8cm}{c}{3cm}{(g)}{} \includegraphics[width=0.45\hsize]{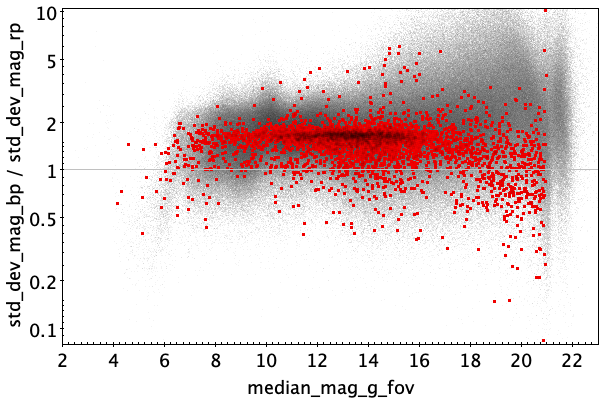}  \\ 
\vspace{4mm}
 \caption{DSCT|GDOR|SXPHE: 4259 training sources.}  
 \label{fig:app:DSCT_trn}
\end{figure*}

\begin{figure*}
\centering
\stackinset{c}{-0.7cm}{c}{2.7cm}{(a)}{}
\includegraphics[width=0.6\hsize]{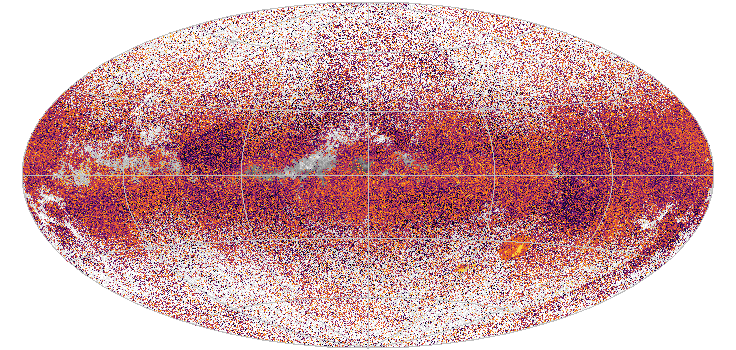} 
\stackinset{c}{2.2cm}{c}{2.7cm}{\includegraphics[height=5.5cm]{figures/appendix/vertical_best_class_score.png}}{} \\ 
\vspace{4mm}
\stackinset{c}{-0.3cm}{c}{3cm}{(b)}{} \includegraphics[width=0.45\hsize]{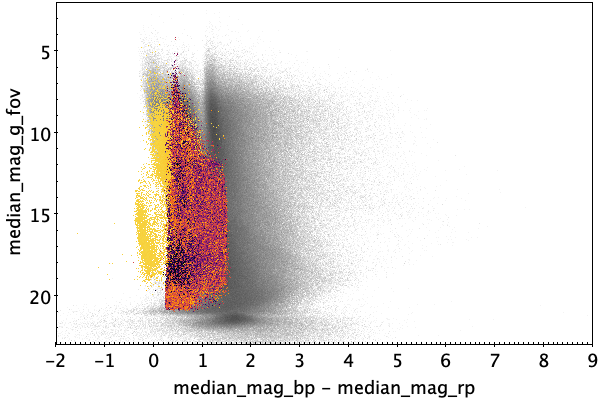}  
\hspace{2mm}
\stackinset{c}{8.8cm}{c}{3cm}{(c)}{} \includegraphics[width=0.45\hsize]{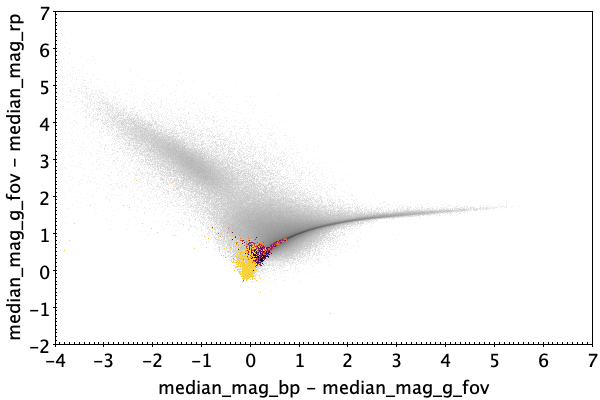} \\ 
\vspace{4mm}
\stackinset{c}{-0.3cm}{c}{3cm}{(d)}{} \includegraphics[width=0.45\hsize]{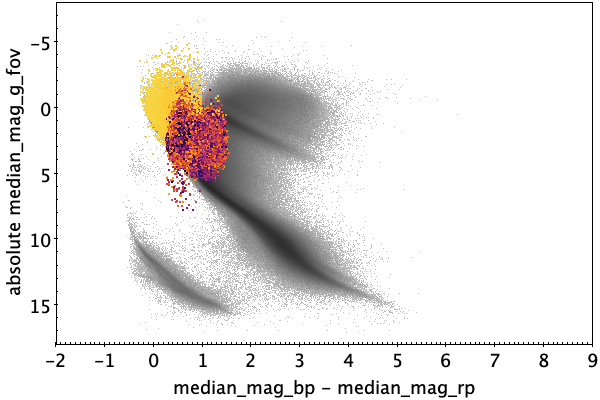}  
\hspace{2mm}
\stackinset{c}{8.8cm}{c}{3cm}{(e)}{} \includegraphics[width=0.45\hsize]{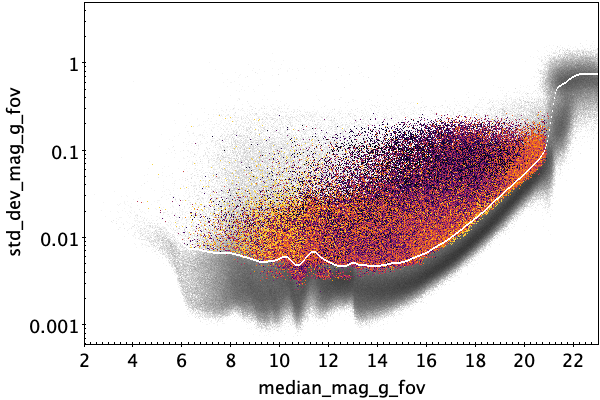} \\ 
\vspace{4mm}
\stackinset{c}{-0.3cm}{c}{3cm}{(f)}{} \includegraphics[width=0.45\hsize]{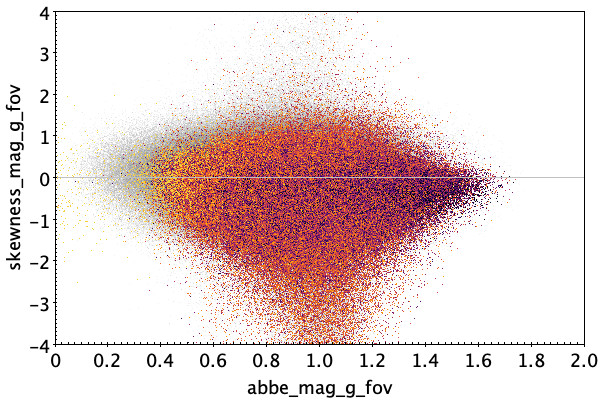}  
\hspace{2mm}
\stackinset{c}{8.8cm}{c}{3cm}{(g)}{} \includegraphics[width=0.45\hsize]{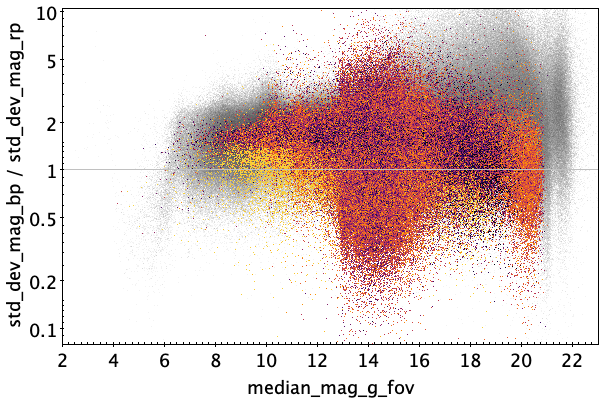}  \\ 
\vspace{4mm}
 \caption{DSCT|GDOR|SXPHE: 748\,058 classified sources.}  
 \label{fig:app:DSCT}
\end{figure*}

\begin{figure*}
\centering
\stackinset{c}{-0.3cm}{c}{3cm}{(a)}{} \includegraphics[width=0.45\hsize]{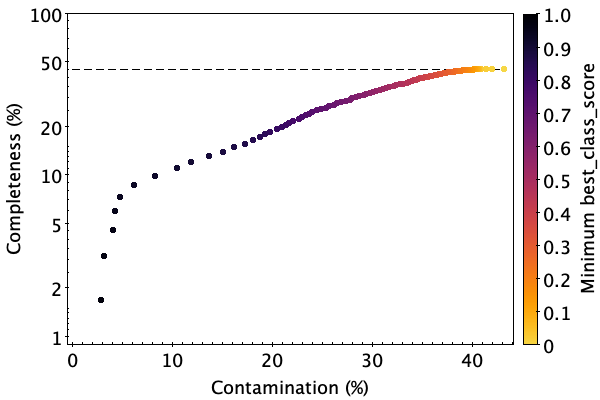}  
\hspace{2mm}
\stackinset{c}{8.8cm}{c}{3cm}{(b)}{} \includegraphics[width=0.45\hsize]{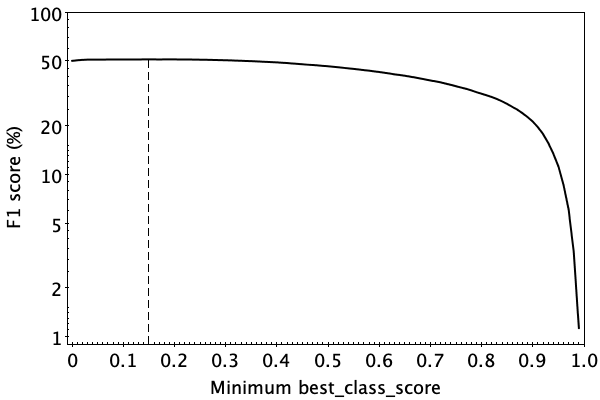} \\ 
\vspace{4mm}
\stackinset{c}{-0.3cm}{c}{3cm}{(c)}{} \includegraphics[width=0.45\hsize]{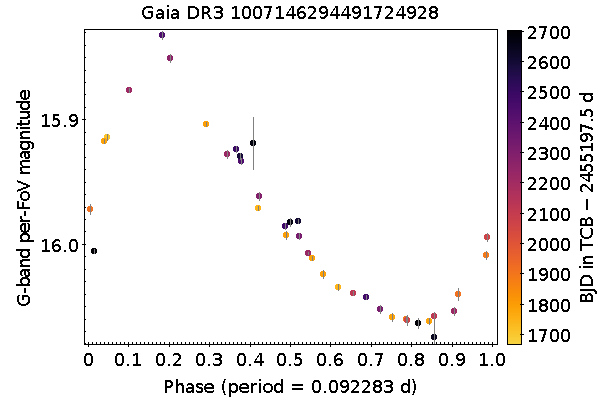}  
\hspace{2mm}
\stackinset{c}{8.8cm}{c}{3cm}{(d)}{} \includegraphics[width=0.45\hsize]{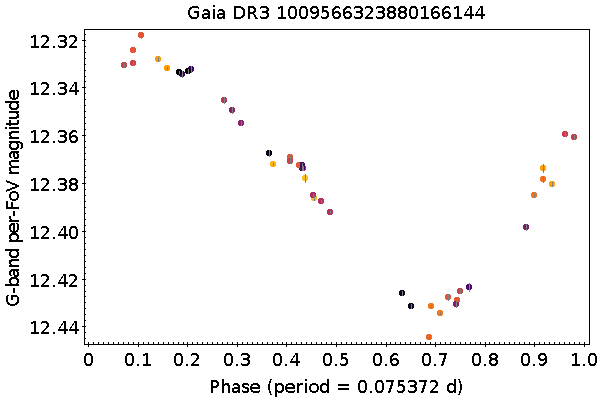} \\
\vspace{4mm}
\stackinset{c}{-0.3cm}{c}{3cm}{(e)}{} \includegraphics[width=0.45\hsize]{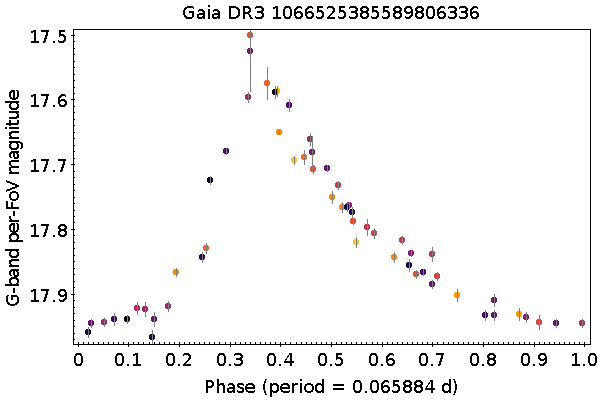}  
\hspace{2mm}
\stackinset{c}{8.8cm}{c}{3cm}{(f)}{} \includegraphics[width=0.45\hsize]{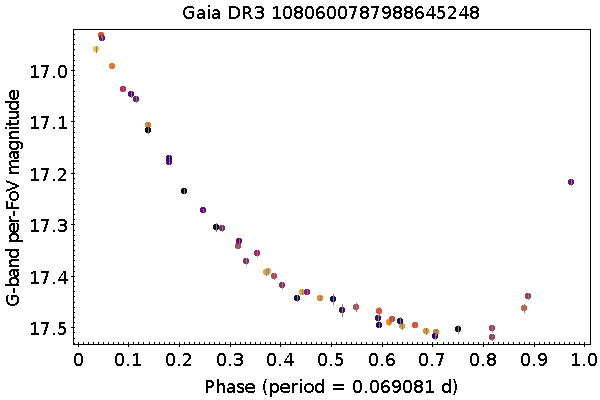} \\
\vspace{4mm}
 \caption{Same as Fig.~\ref{fig:app:ACV_cc}, but for DSCT|GDOR|SXPHE.}
 \label{fig:app:DSCT_cc}
\end{figure*}

\begin{figure*}
\centering
\stackinset{c}{-0.7cm}{c}{2.7cm}{(a)}{} \includegraphics[width=0.6\hsize]{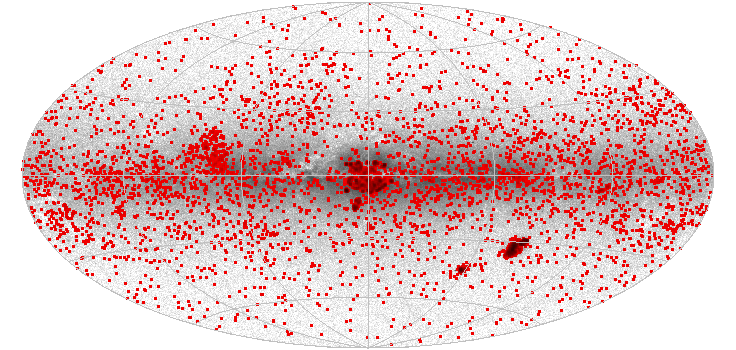} \\ 
\vspace{4mm}
\stackinset{c}{-0.3cm}{c}{3cm}{(b)}{} \includegraphics[width=0.45\hsize]{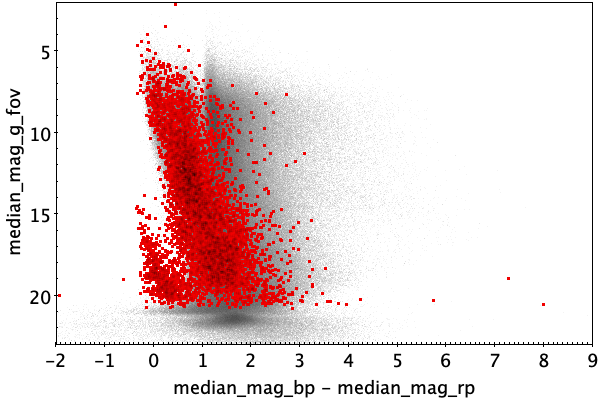}  
\hspace{2mm}
\stackinset{c}{8.8cm}{c}{3cm}{(c)}{} \includegraphics[width=0.45\hsize]{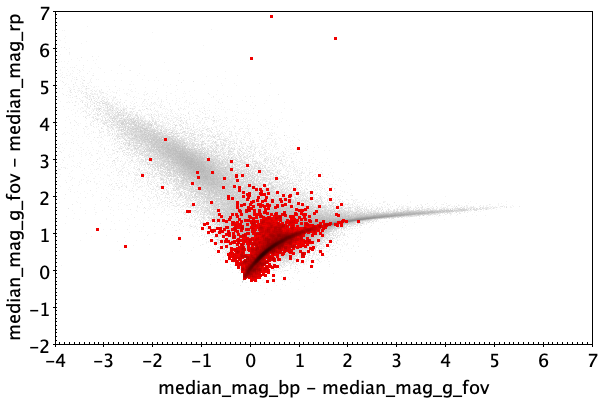} \\ 
\vspace{4mm}
\stackinset{c}{-0.3cm}{c}{3cm}{(d)}{} \includegraphics[width=0.45\hsize]{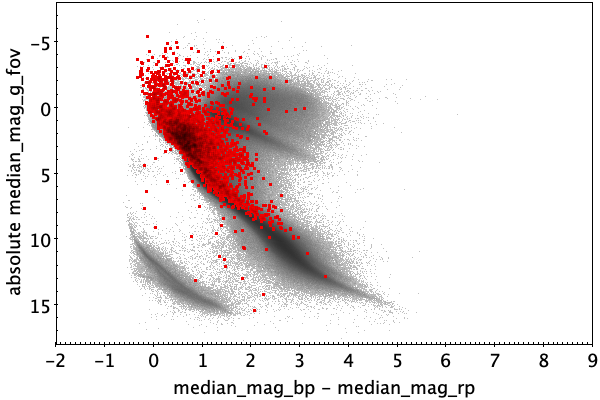}  
\hspace{2mm}
\stackinset{c}{8.8cm}{c}{3cm}{(e)}{} \includegraphics[width=0.45\hsize]{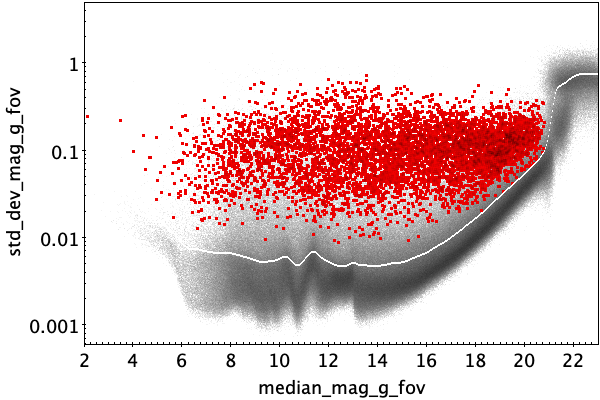} \\ 
\vspace{4mm}
\stackinset{c}{-0.3cm}{c}{3cm}{(f)}{} \includegraphics[width=0.45\hsize]{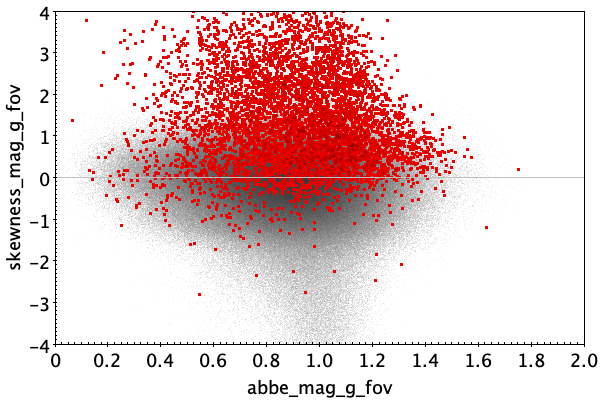}  
\hspace{2mm}
\stackinset{c}{8.8cm}{c}{3cm}{(g)}{} \includegraphics[width=0.45\hsize]{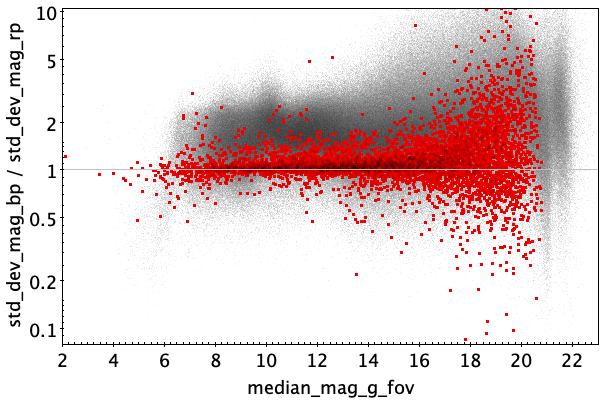}  \\ 
\vspace{4mm}
 \caption{ECL: 6360 training sources.}  
 \label{fig:app:ECL_trn}
\end{figure*}

\begin{figure*}
\centering
\stackinset{c}{-0.7cm}{c}{2.7cm}{(a)}{}
\includegraphics[width=0.6\hsize]{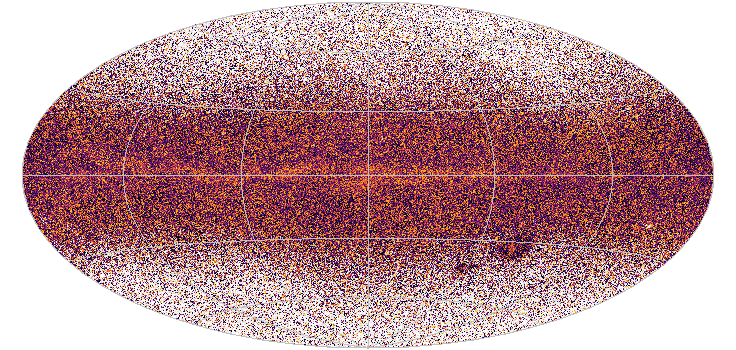} 
\stackinset{c}{2.2cm}{c}{2.7cm}{\includegraphics[height=5.5cm]{figures/appendix/vertical_best_class_score.png}}{} \\ 
\vspace{4mm}
\stackinset{c}{-0.3cm}{c}{3cm}{(b)}{} \includegraphics[width=0.45\hsize]{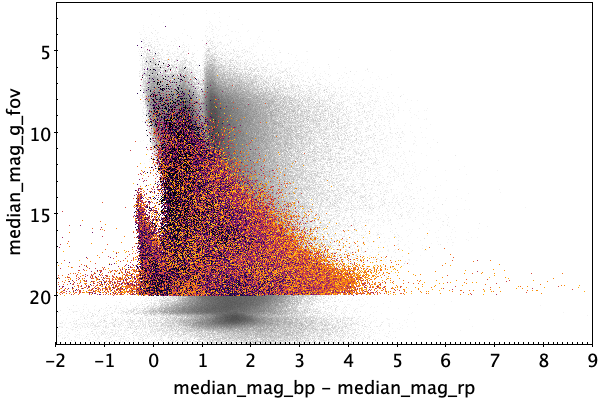}  
\hspace{2mm}
\stackinset{c}{8.8cm}{c}{3cm}{(c)}{} \includegraphics[width=0.45\hsize]{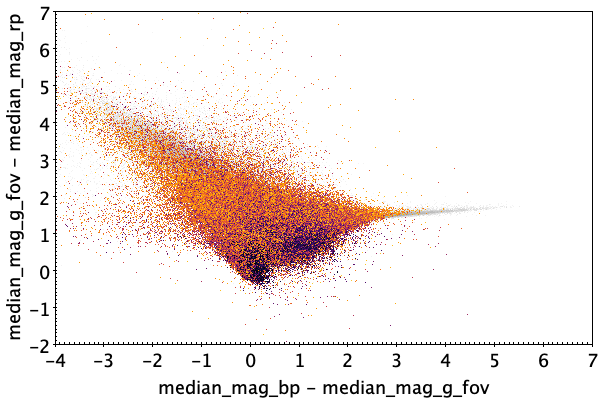} \\ 
\vspace{4mm}
\stackinset{c}{-0.3cm}{c}{3cm}{(d)}{} \includegraphics[width=0.45\hsize]{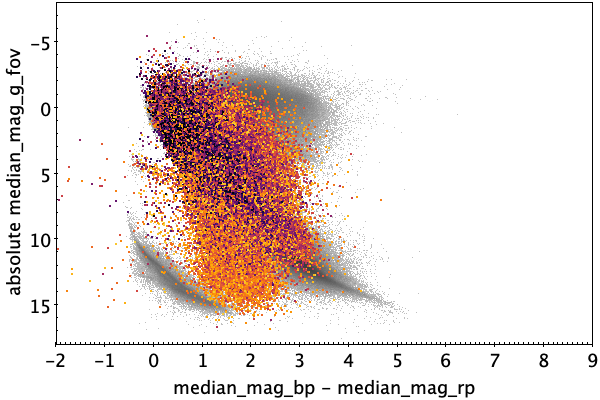}  
\hspace{2mm}
\stackinset{c}{8.8cm}{c}{3cm}{(e)}{} \includegraphics[width=0.45\hsize]{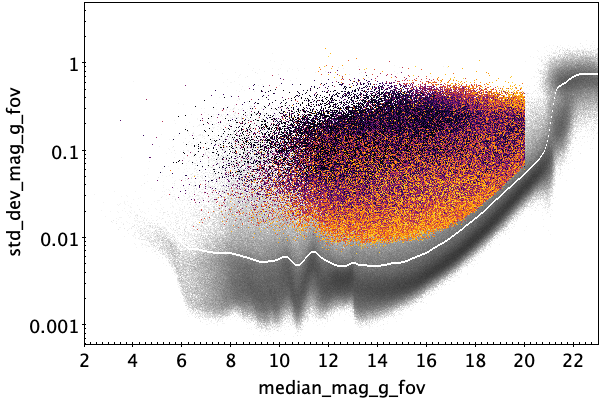} \\ 
\vspace{4mm}
\stackinset{c}{-0.3cm}{c}{3cm}{(f)}{} \includegraphics[width=0.45\hsize]{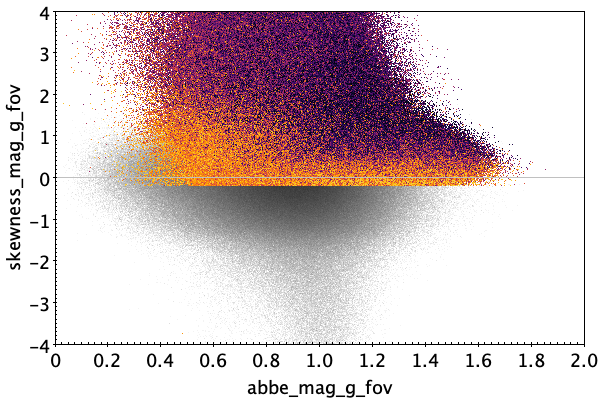}  
\hspace{2mm}
\stackinset{c}{8.8cm}{c}{3cm}{(g)}{} \includegraphics[width=0.45\hsize]{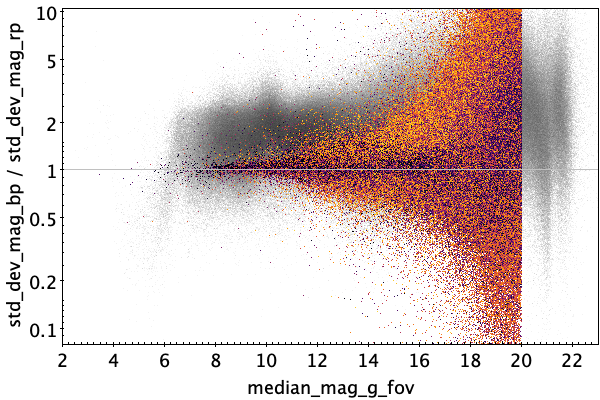}  \\ 
\vspace{4mm}
 \caption{ECL: 2\,184\,356 classified sources.}  
 \label{fig:app:ECL}
\end{figure*}

\begin{figure*}
\centering
\stackinset{c}{-0.3cm}{c}{3cm}{(a)}{} \includegraphics[width=0.45\hsize]{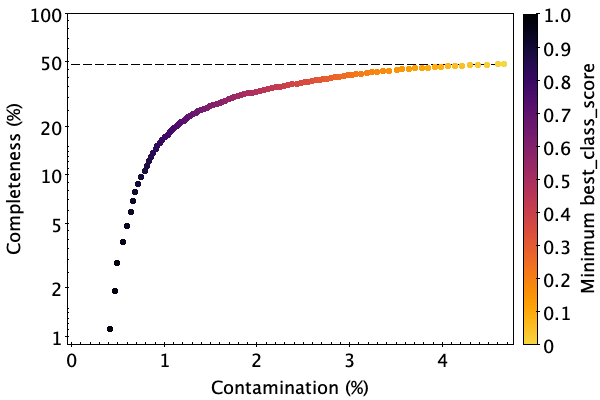}  
\hspace{2mm}
\stackinset{c}{8.8cm}{c}{3cm}{(b)}{} \includegraphics[width=0.45\hsize]{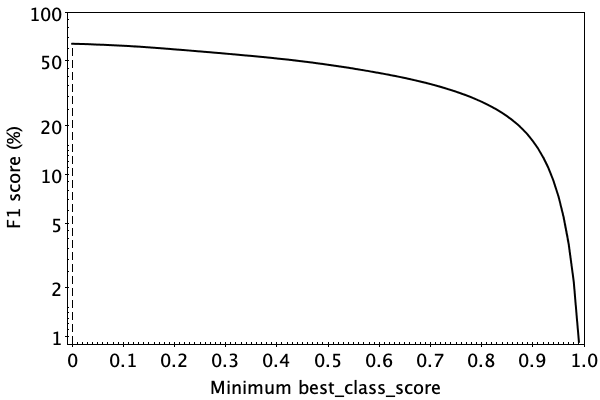} \\ 
\vspace{4mm}
\stackinset{c}{-0.3cm}{c}{3cm}{(c)}{} \includegraphics[width=0.45\hsize]{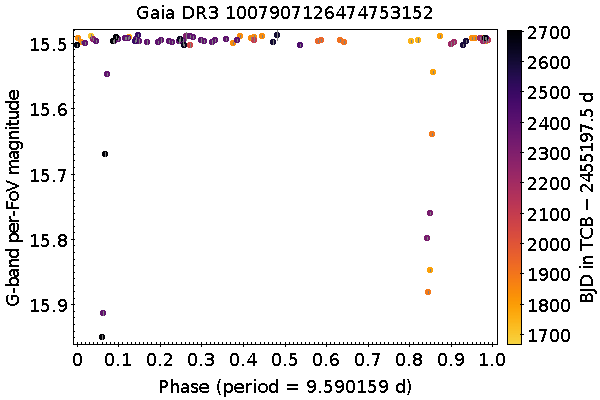}  
\hspace{2mm}
\stackinset{c}{8.8cm}{c}{3cm}{(d)}{} \includegraphics[width=0.45\hsize]{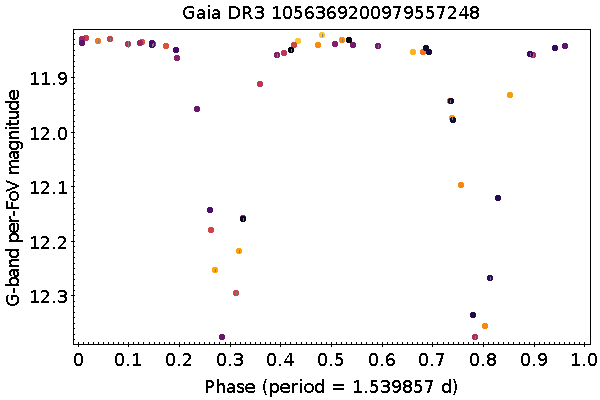} \\
\vspace{4mm}
\stackinset{c}{-0.3cm}{c}{3cm}{(e)}{} \includegraphics[width=0.45\hsize]{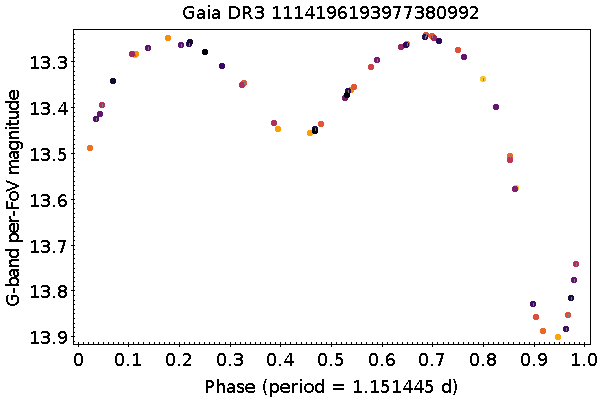}  
\hspace{2mm}
\stackinset{c}{8.8cm}{c}{3cm}{(f)}{} \includegraphics[width=0.45\hsize]{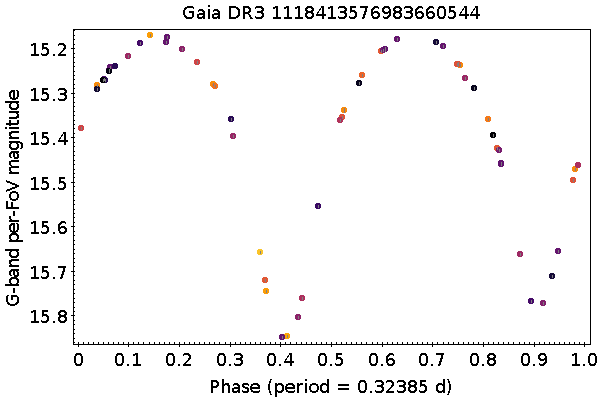} \\
\vspace{4mm}
 \caption{Same as Fig.~\ref{fig:app:ACV_cc}, but for ECL.}
 \label{fig:app:ECL_cc}
\end{figure*}

\begin{figure*}
\centering
\stackinset{c}{-0.7cm}{c}{2.7cm}{(a)}{} \includegraphics[width=0.6\hsize]{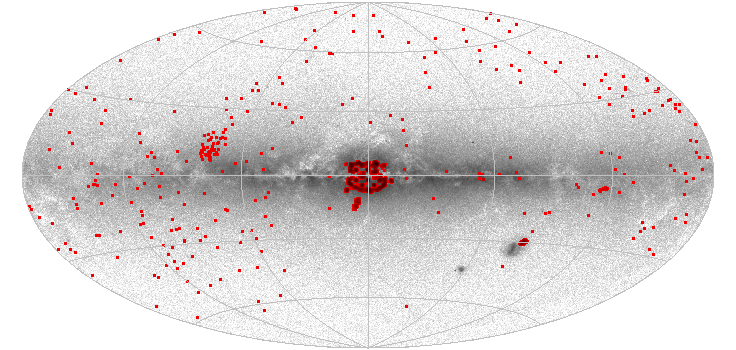} \\ 
\vspace{4mm}
\stackinset{c}{-0.3cm}{c}{3cm}{(b)}{} \includegraphics[width=0.45\hsize]{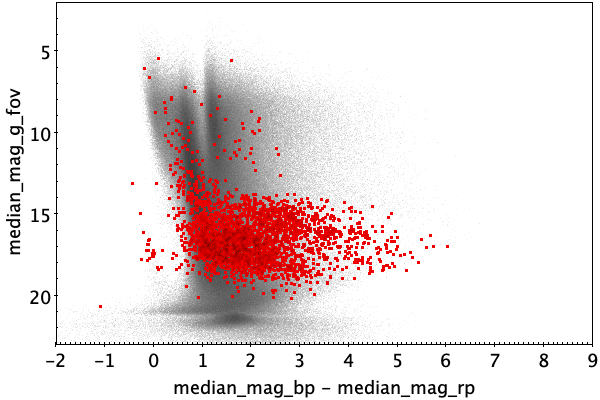}  
\hspace{2mm}
\stackinset{c}{8.8cm}{c}{3cm}{(c)}{} \includegraphics[width=0.45\hsize]{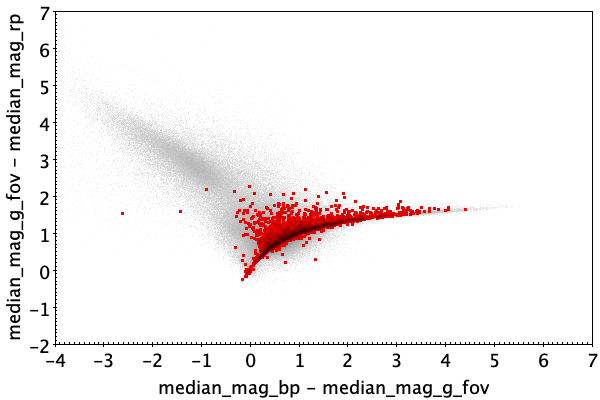} \\ 
\vspace{4mm}
\stackinset{c}{-0.3cm}{c}{3cm}{(d)}{} \includegraphics[width=0.45\hsize]{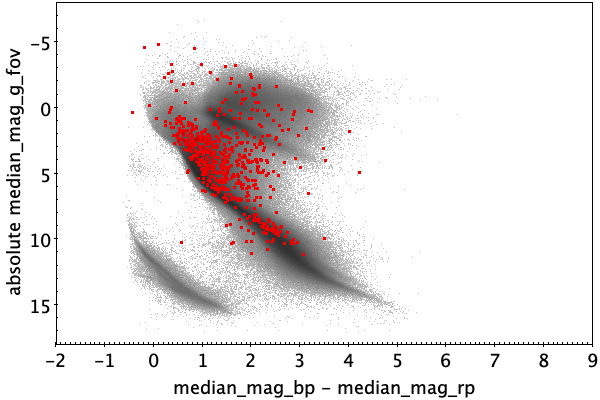}  
\hspace{2mm}
\stackinset{c}{8.8cm}{c}{3cm}{(e)}{} \includegraphics[width=0.45\hsize]{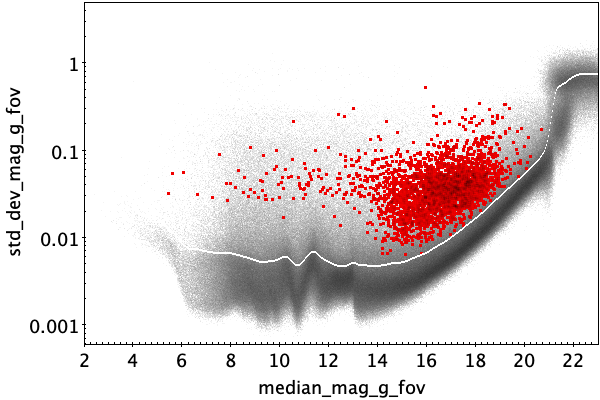} \\ 
\vspace{4mm}
\stackinset{c}{-0.3cm}{c}{3cm}{(f)}{} \includegraphics[width=0.45\hsize]{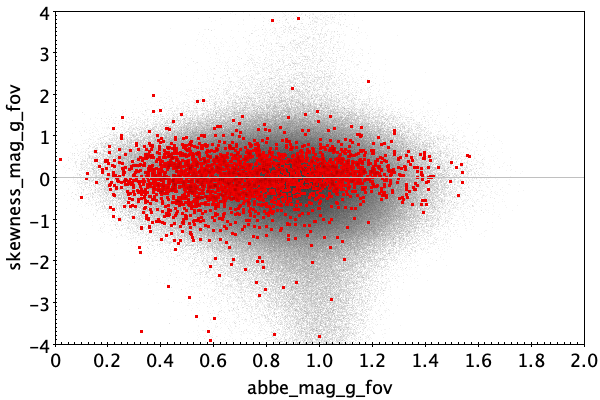}  
\hspace{2mm}
\stackinset{c}{8.8cm}{c}{3cm}{(g)}{} \includegraphics[width=0.45\hsize]{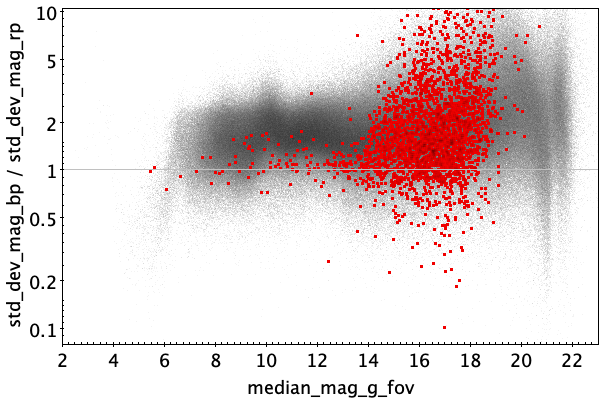}  \\ 
\vspace{4mm}
 \caption{ELL: 2864 training sources.}  
 \label{fig:app:ELL_trn}
\end{figure*}

\begin{figure*}
\centering
\stackinset{c}{-0.7cm}{c}{2.7cm}{(a)}{}
\includegraphics[width=0.6\hsize]{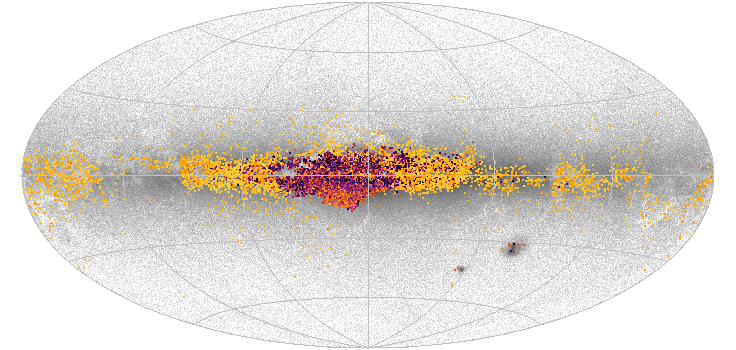} 
\stackinset{c}{2.2cm}{c}{2.7cm}{\includegraphics[height=5.5cm]{figures/appendix/vertical_best_class_score.png}}{} \\ 
\vspace{4mm}
\stackinset{c}{-0.3cm}{c}{3cm}{(b)}{} \includegraphics[width=0.45\hsize]{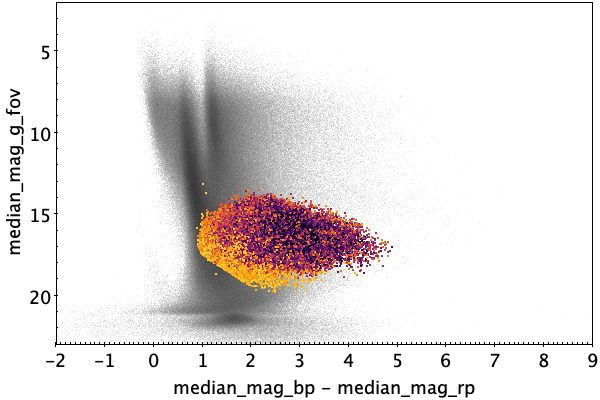}  
\hspace{2mm}
\stackinset{c}{8.8cm}{c}{3cm}{(c)}{} \includegraphics[width=0.45\hsize]{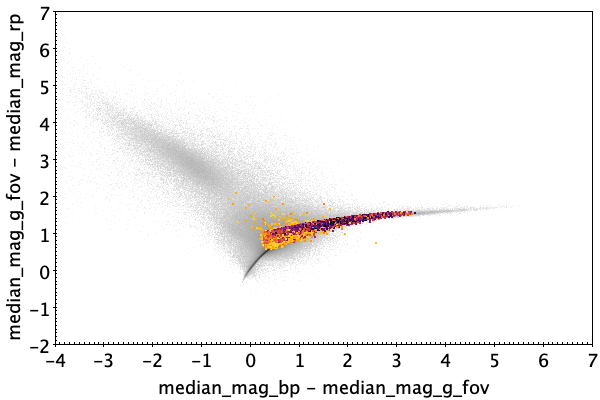} \\ 
\vspace{4mm}
\stackinset{c}{-0.3cm}{c}{3cm}{(d)}{} \includegraphics[width=0.45\hsize]{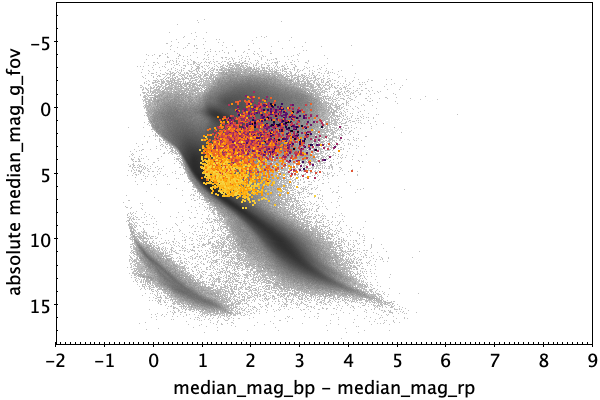}  
\hspace{2mm}
\stackinset{c}{8.8cm}{c}{3cm}{(e)}{} \includegraphics[width=0.45\hsize]{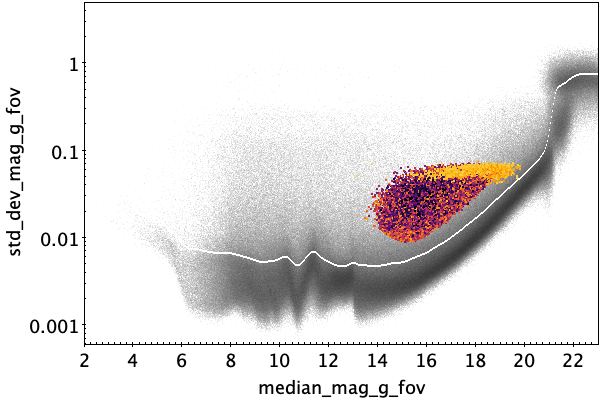} \\ 
\vspace{4mm}
\stackinset{c}{-0.3cm}{c}{3cm}{(f)}{} \includegraphics[width=0.45\hsize]{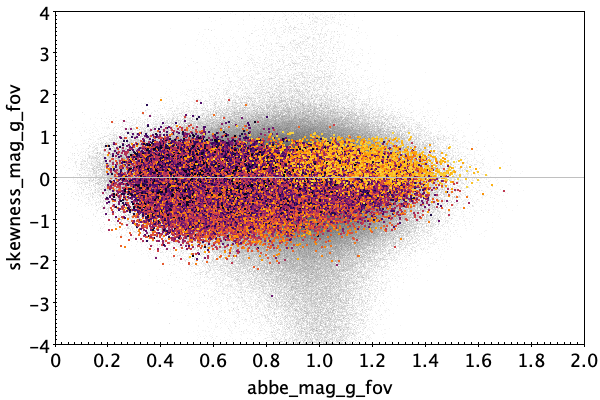}  
\hspace{2mm}
\stackinset{c}{8.8cm}{c}{3cm}{(g)}{} \includegraphics[width=0.45\hsize]{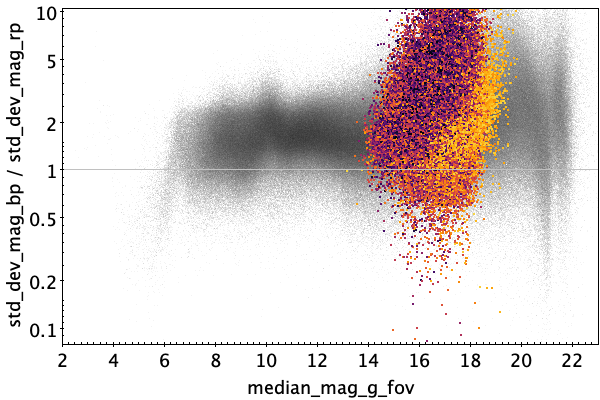}  \\ 
\vspace{4mm}
 \caption{ELL: 65\,300 classified sources.}  
 \label{fig:app:ELL}
\end{figure*}

\begin{figure*}
\centering
\stackinset{c}{-0.3cm}{c}{3cm}{(a)}{} \includegraphics[width=0.45\hsize]{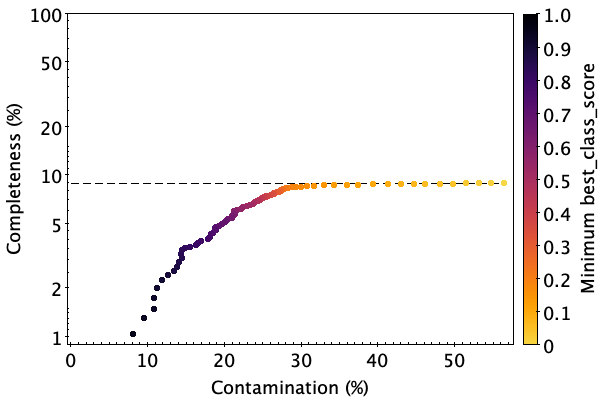}  
\hspace{2mm}
\stackinset{c}{8.8cm}{c}{3cm}{(b)}{} \includegraphics[width=0.45\hsize]{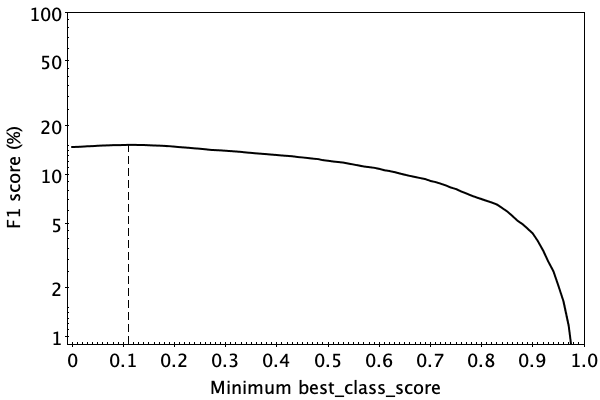} \\ 
\vspace{4mm}
\stackinset{c}{-0.3cm}{c}{3cm}{(c)}{} \includegraphics[width=0.45\hsize]{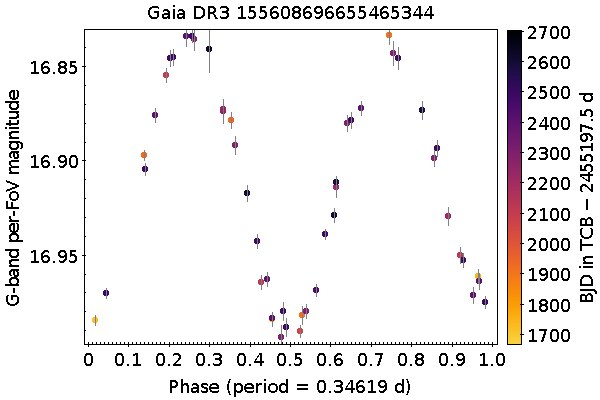}  
\hspace{2mm}
\stackinset{c}{8.8cm}{c}{3cm}{(d)}{} \includegraphics[width=0.45\hsize]{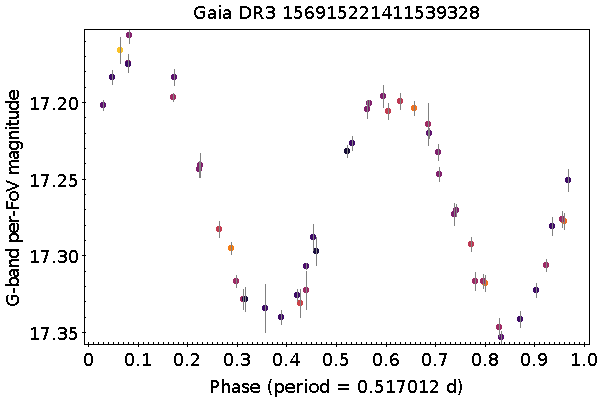} \\
\vspace{4mm}
\stackinset{c}{-0.3cm}{c}{3cm}{(e)}{} \includegraphics[width=0.45\hsize]{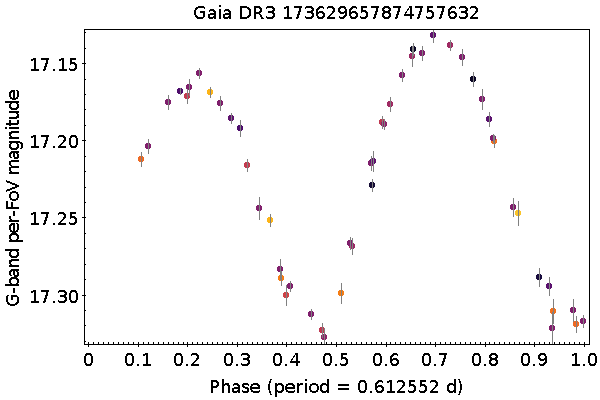}  
\hspace{2mm}
\stackinset{c}{8.8cm}{c}{3cm}{(f)}{} \includegraphics[width=0.45\hsize]{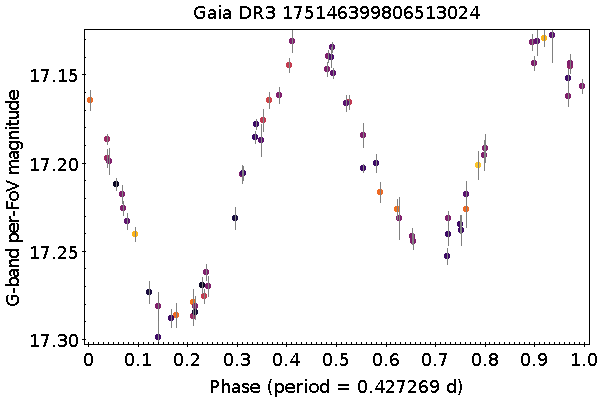} \\
\vspace{4mm}
 \caption{Same as Fig.~\ref{fig:app:ACV_cc}, but for ELL.}
 \label{fig:app:ELL_cc}
\end{figure*}

\begin{figure*}
\centering
\stackinset{c}{-0.7cm}{c}{2.7cm}{(a)}{} \includegraphics[width=0.6\hsize]{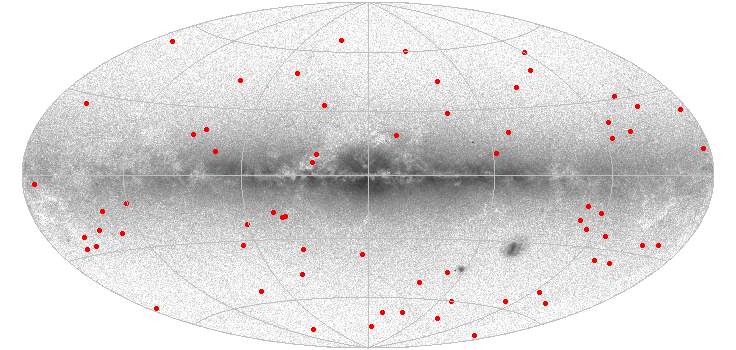} \\ 
\vspace{4mm}
\stackinset{c}{-0.3cm}{c}{3cm}{(b)}{} \includegraphics[width=0.45\hsize]{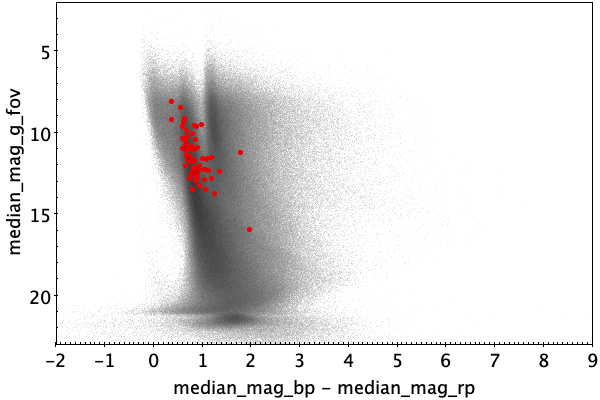}  
\hspace{2mm}
\stackinset{c}{8.8cm}{c}{3cm}{(c)}{} \includegraphics[width=0.45\hsize]{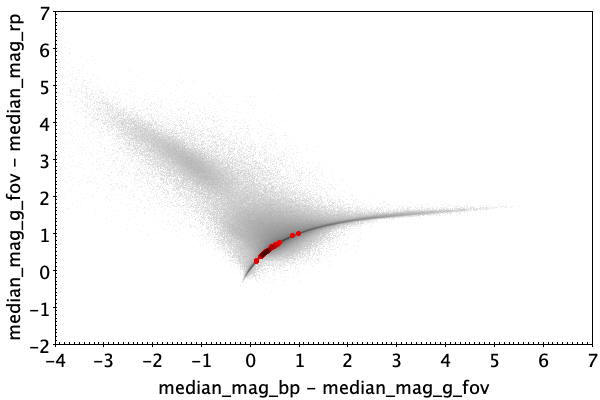} \\ 
\vspace{4mm}
\stackinset{c}{-0.3cm}{c}{3cm}{(d)}{} \includegraphics[width=0.45\hsize]{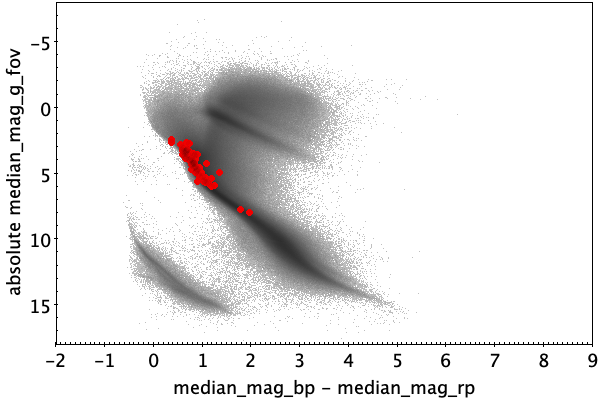}  
\hspace{2mm}
\stackinset{c}{8.8cm}{c}{3cm}{(e)}{} \includegraphics[width=0.45\hsize]{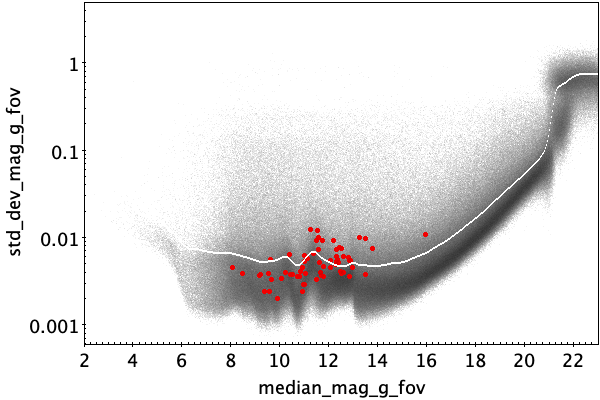} \\ 
\vspace{4mm}
\stackinset{c}{-0.3cm}{c}{3cm}{(f)}{} \includegraphics[width=0.45\hsize]{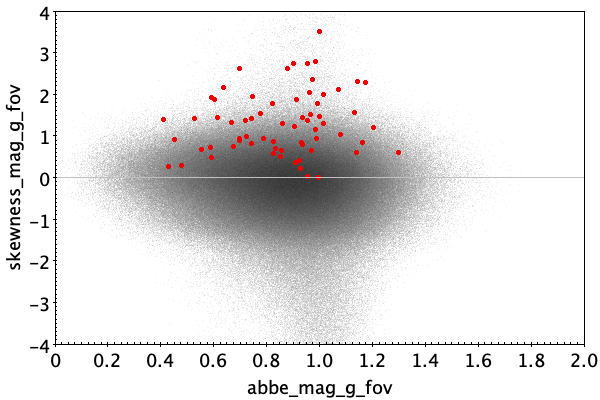}  
\hspace{2mm}
\stackinset{c}{8.8cm}{c}{3cm}{(g)}{} \includegraphics[width=0.45\hsize]{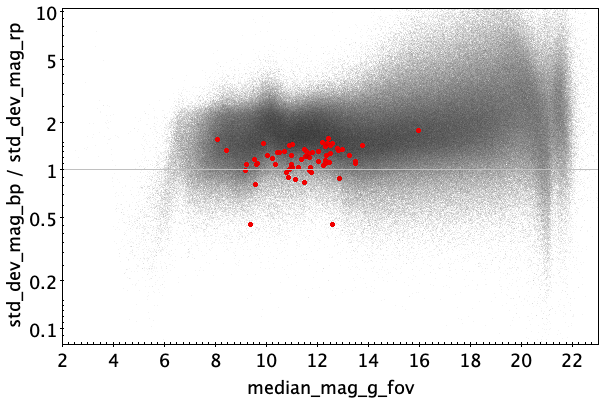}  \\ 
\vspace{4mm}
 \caption{EP: 66 training sources.}  
 \label{fig:app:EP_trn}
\end{figure*}

\begin{figure*}
\centering
\stackinset{c}{-0.7cm}{c}{2.7cm}{(a)}{}
\includegraphics[width=0.6\hsize]{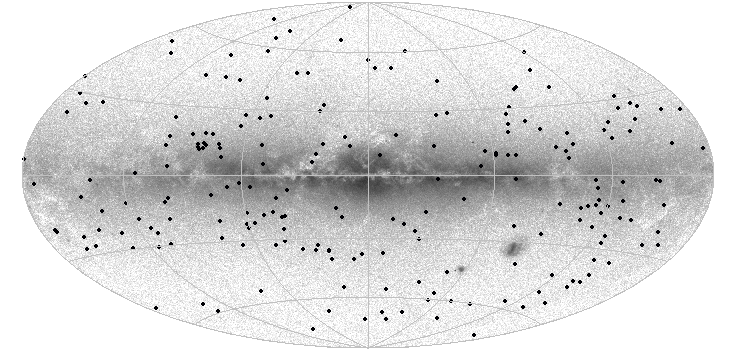} 
\stackinset{c}{2.2cm}{c}{2.7cm}{\includegraphics[height=5.5cm]{figures/appendix/vertical_best_class_score.png}}{} \\ 
\vspace{4mm}
\stackinset{c}{-0.3cm}{c}{3cm}{(b)}{} \includegraphics[width=0.45\hsize]{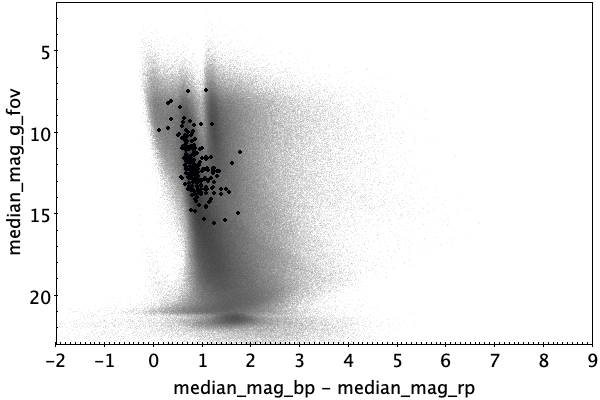}  
\hspace{2mm}
\stackinset{c}{8.8cm}{c}{3cm}{(c)}{} \includegraphics[width=0.45\hsize]{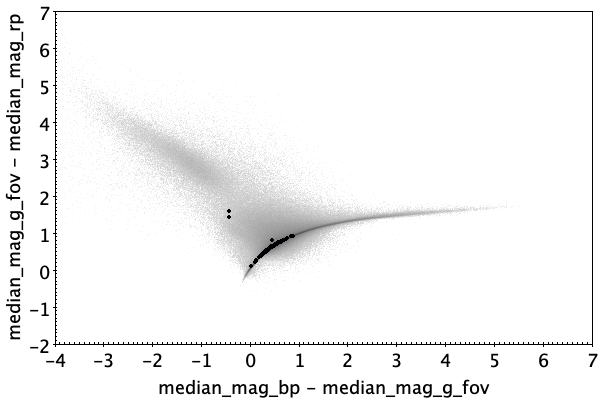} \\ 
\vspace{4mm}
\stackinset{c}{-0.3cm}{c}{3cm}{(d)}{} \includegraphics[width=0.45\hsize]{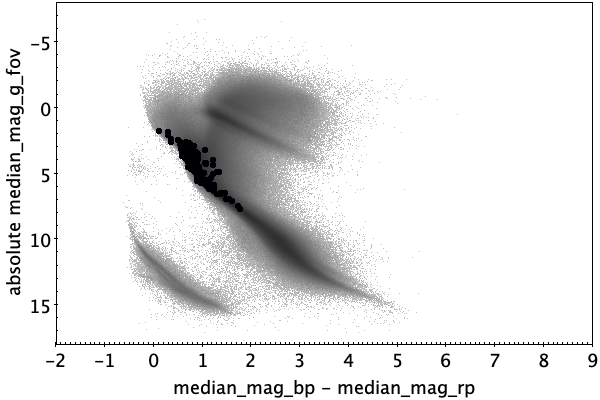}  
\hspace{2mm}
\stackinset{c}{8.8cm}{c}{3cm}{(e)}{} \includegraphics[width=0.45\hsize]{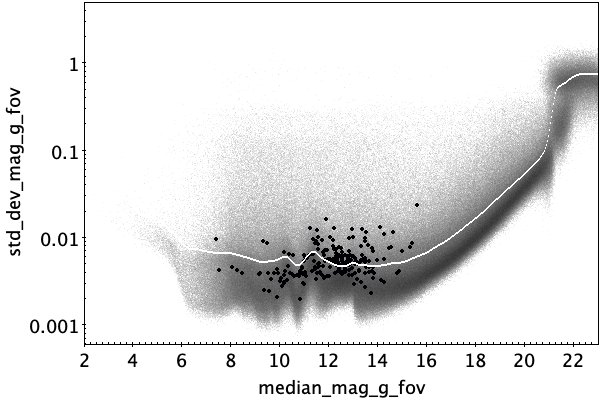} \\ 
\vspace{4mm}
\stackinset{c}{-0.3cm}{c}{3cm}{(f)}{} \includegraphics[width=0.45\hsize]{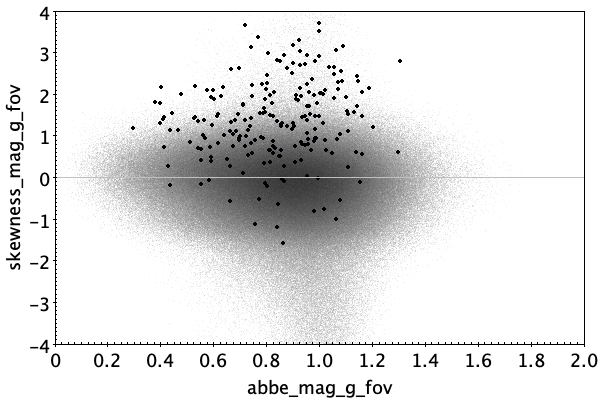}  
\hspace{2mm}
\stackinset{c}{8.8cm}{c}{3cm}{(g)}{} \includegraphics[width=0.45\hsize]{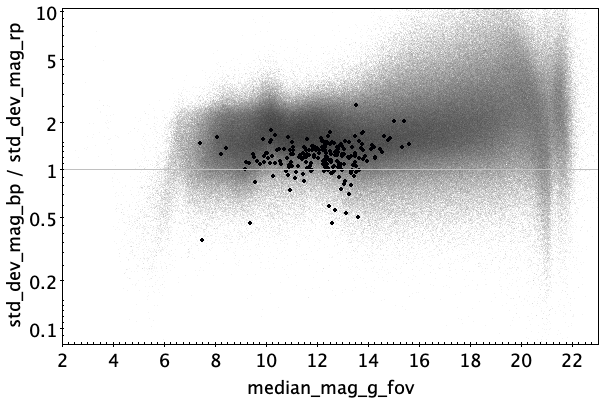}  \\ 
\vspace{4mm}
 \caption{EP: 214 classified sources.}  
 \label{fig:app:EP}
\end{figure*}

\begin{figure*}
\centering
\stackinset{c}{-0.3cm}{c}{3cm}{(a)}{} \includegraphics[width=0.45\hsize]{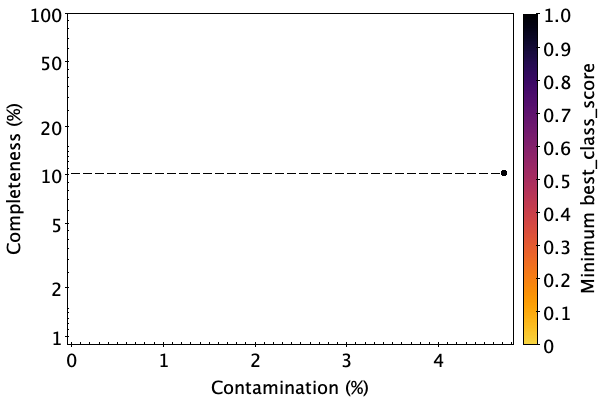}  
\hspace{2mm}
\stackinset{c}{8.8cm}{c}{3cm}{(b)}{} \includegraphics[width=0.45\hsize]{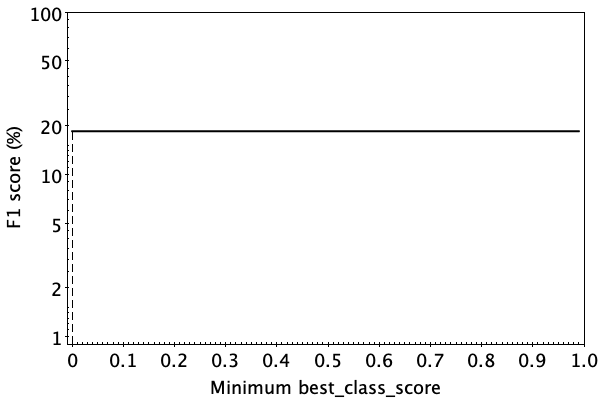} \\ 
\vspace{4mm}
\stackinset{c}{-0.3cm}{c}{3cm}{(c)}{} \includegraphics[width=0.45\hsize]{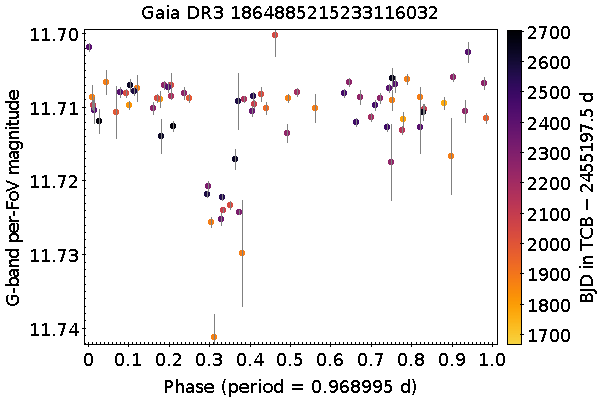}  
\hspace{2mm}
\stackinset{c}{8.8cm}{c}{3cm}{(d)}{} \includegraphics[width=0.45\hsize]{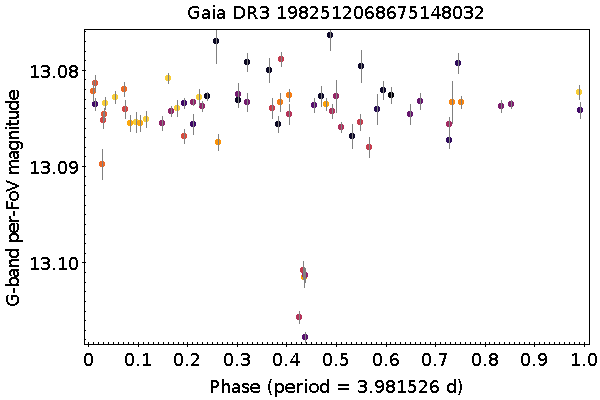} \\
\vspace{4mm}
\stackinset{c}{-0.3cm}{c}{3cm}{(e)}{} \includegraphics[width=0.45\hsize]{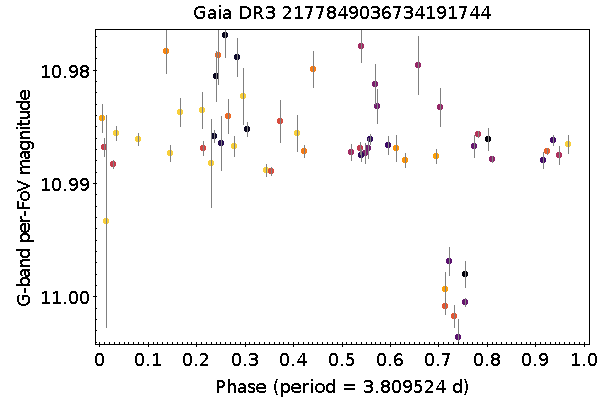}  
\hspace{2mm}
\stackinset{c}{8.8cm}{c}{3cm}{(f)}{} \includegraphics[width=0.45\hsize]{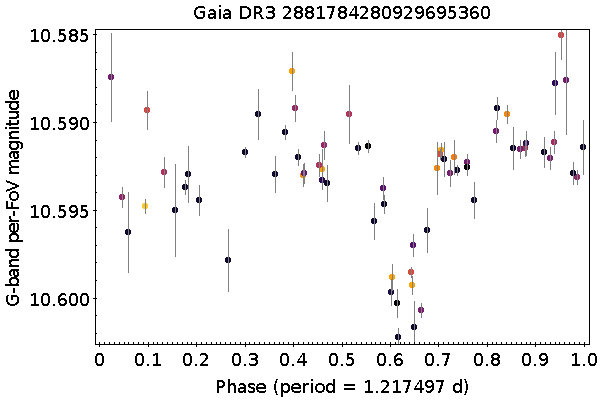} \\
\vspace{4mm}
 \caption{Same as Fig.~\ref{fig:app:ACV_cc}, but for EP.} 
 \label{fig:app:EP_cc}
\end{figure*}

\begin{figure*}
\centering
\stackinset{c}{-0.7cm}{c}{2.7cm}{(a)}{} \includegraphics[width=0.6\hsize]{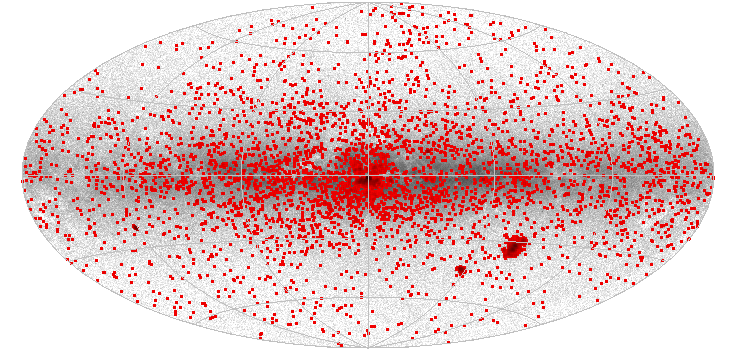} \\ 
\vspace{4mm}
\stackinset{c}{-0.3cm}{c}{3cm}{(b)}{} \includegraphics[width=0.45\hsize]{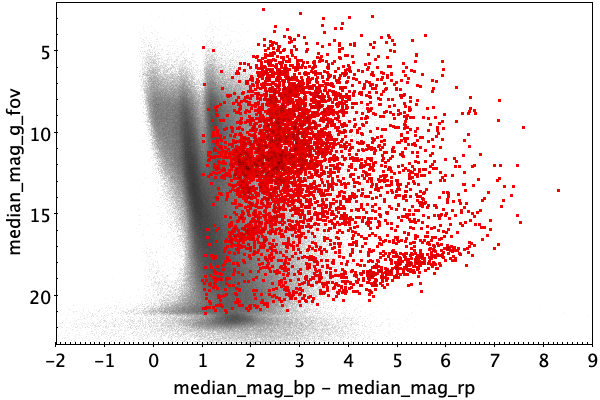}  
\hspace{2mm}
\stackinset{c}{8.8cm}{c}{3cm}{(c)}{} \includegraphics[width=0.45\hsize]{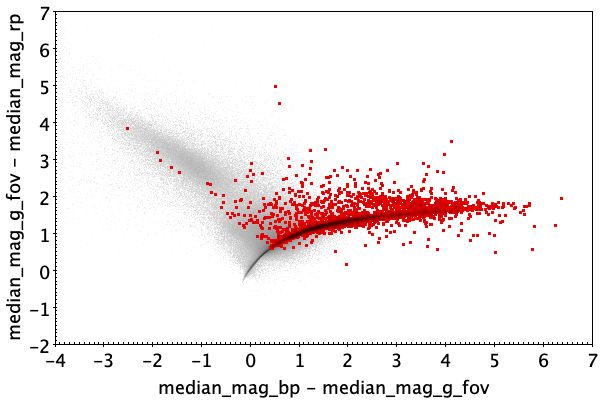} \\ 
\vspace{4mm}
\stackinset{c}{-0.3cm}{c}{3cm}{(d)}{} \includegraphics[width=0.45\hsize]{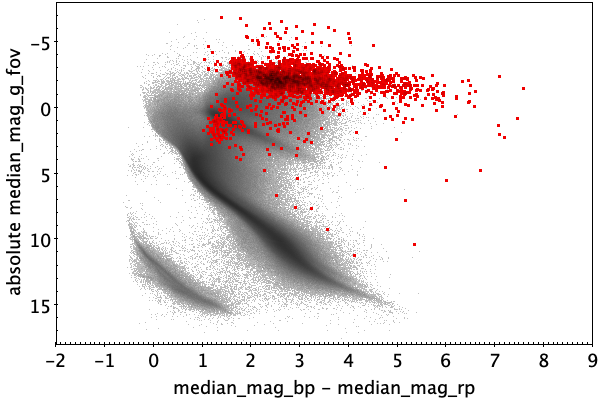}  
\hspace{2mm}
\stackinset{c}{8.8cm}{c}{3cm}{(e)}{} \includegraphics[width=0.45\hsize]{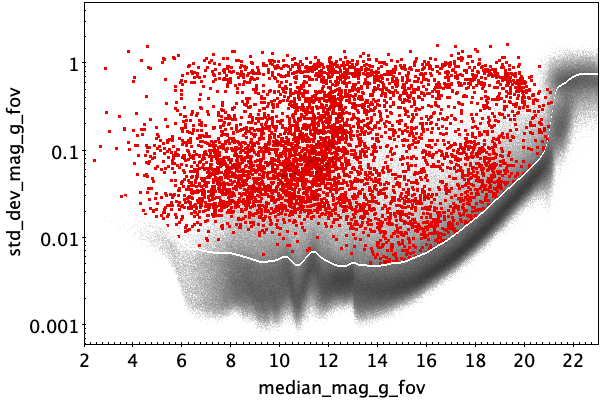} \\ 
\vspace{4mm}
\stackinset{c}{-0.3cm}{c}{3cm}{(f)}{} \includegraphics[width=0.45\hsize]{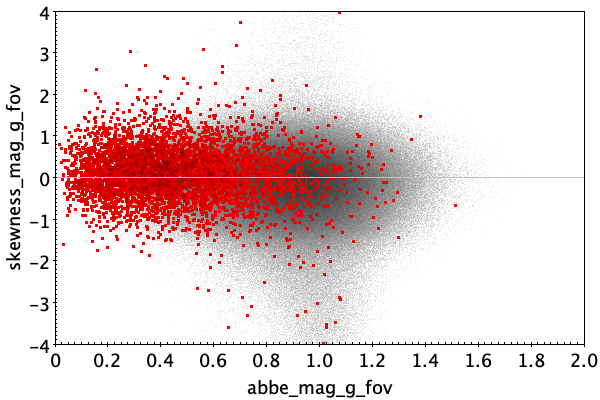}  
\hspace{2mm}
\stackinset{c}{8.8cm}{c}{3cm}{(g)}{} \includegraphics[width=0.45\hsize]{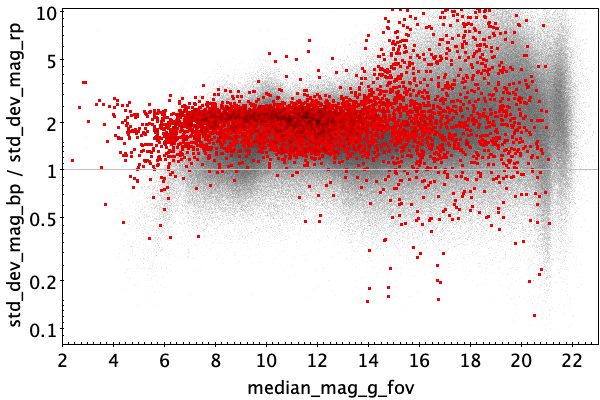}  \\ 
\vspace{4mm}
 \caption{LPV: 5353 training sources.}  
 \label{fig:app:LPV_trn}
\end{figure*}

\begin{figure*}
\centering
\stackinset{c}{-0.7cm}{c}{2.7cm}{(a)}{}
\includegraphics[width=0.6\hsize]{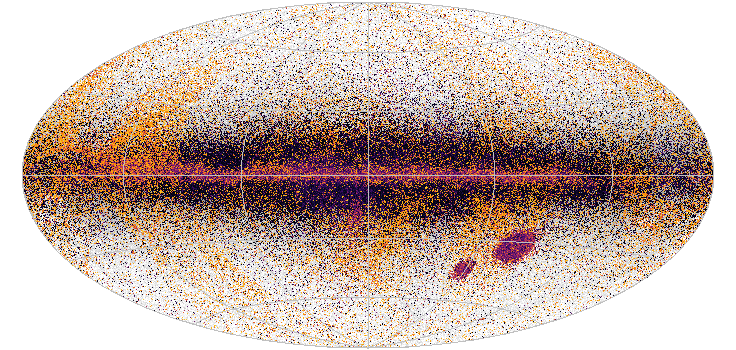} 
\stackinset{c}{2.2cm}{c}{2.7cm}{\includegraphics[height=5.5cm]{figures/appendix/vertical_best_class_score.png}}{} \\ 
\vspace{4mm}
\stackinset{c}{-0.3cm}{c}{3cm}{(b)}{} \includegraphics[width=0.45\hsize]{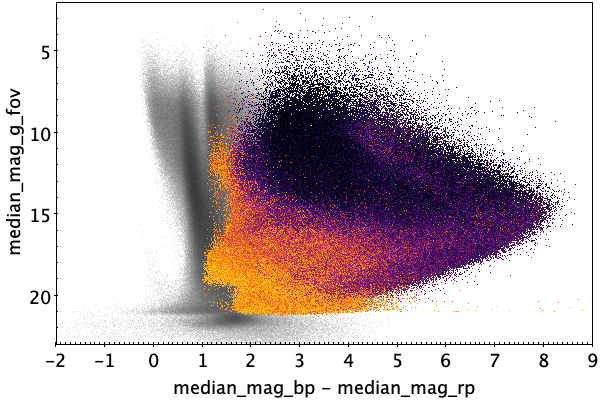}  
\hspace{2mm}
\stackinset{c}{8.8cm}{c}{3cm}{(c)}{} \includegraphics[width=0.45\hsize]{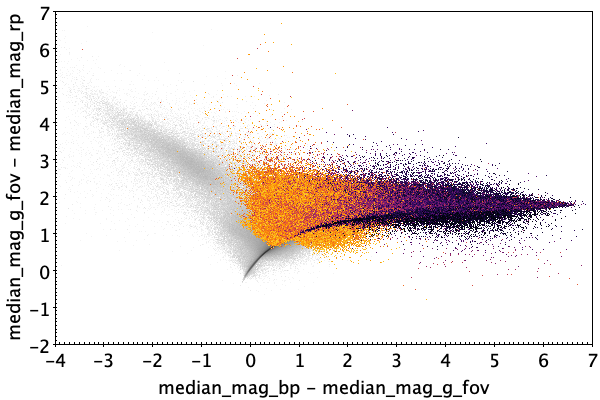} \\ 
\vspace{4mm}
\stackinset{c}{-0.3cm}{c}{3cm}{(d)}{} \includegraphics[width=0.45\hsize]{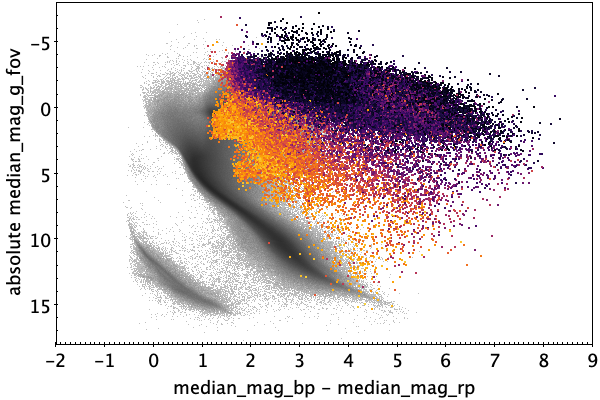}  
\hspace{2mm}
\stackinset{c}{8.8cm}{c}{3cm}{(e)}{} \includegraphics[width=0.45\hsize]{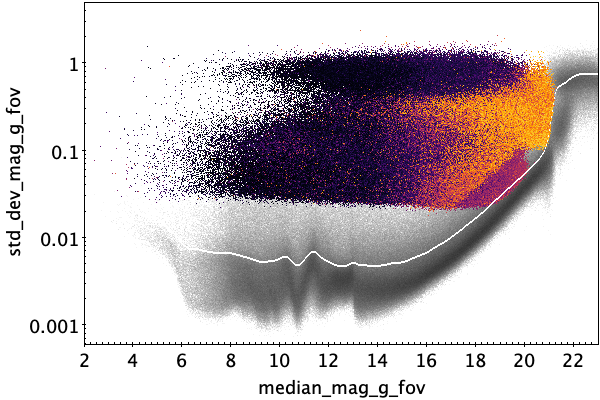} \\ 
\vspace{4mm}
\stackinset{c}{-0.3cm}{c}{3cm}{(f)}{} \includegraphics[width=0.45\hsize]{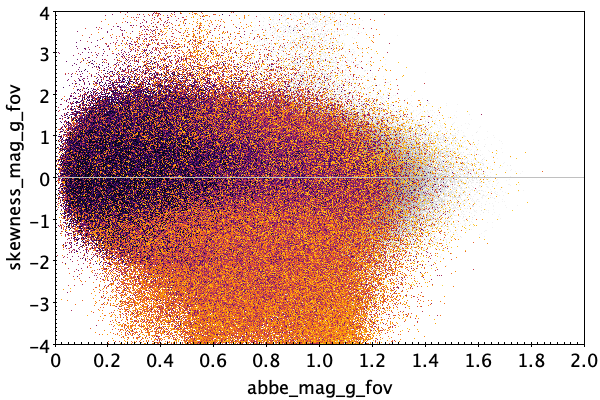}  
\hspace{2mm}
\stackinset{c}{8.8cm}{c}{3cm}{(g)}{} \includegraphics[width=0.45\hsize]{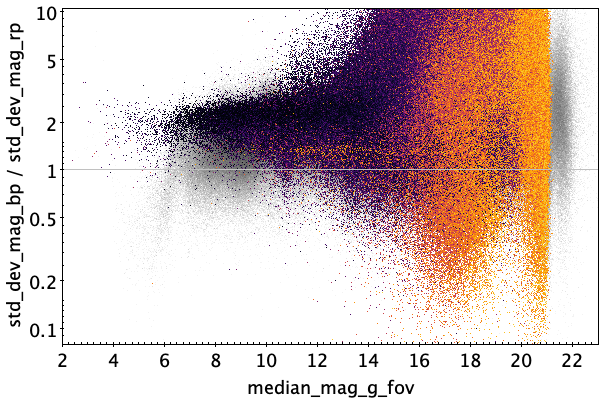}  \\ 
\vspace{4mm}
 \caption{LPV: 2\,325\,775 classified sources.}  
 \label{fig:app:LPV}
\end{figure*}

\begin{figure*}
\centering
\stackinset{c}{-0.3cm}{c}{3cm}{(a)}{} \includegraphics[width=0.45\hsize]{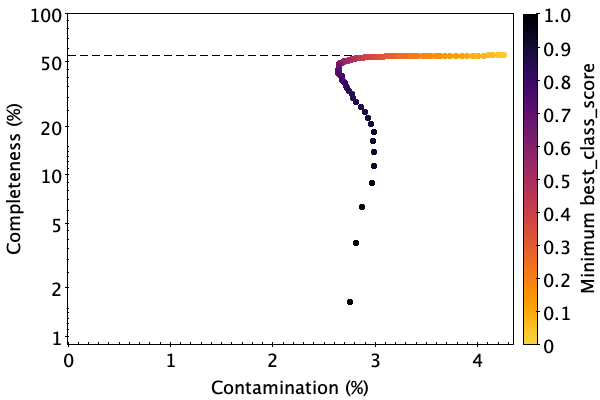}  
\hspace{2mm}
\stackinset{c}{8.8cm}{c}{3cm}{(b)}{} \includegraphics[width=0.45\hsize]{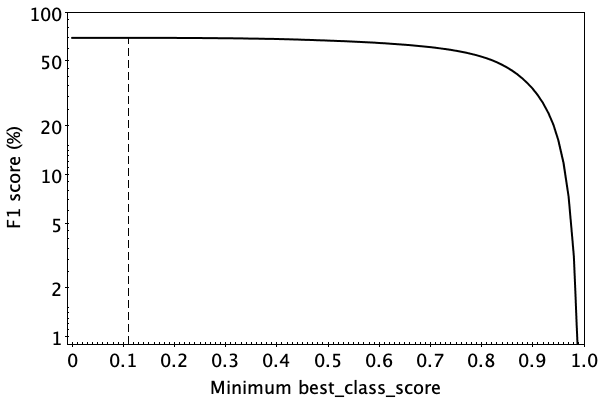} \\  
\vspace{4mm}
\stackinset{c}{-0.3cm}{c}{3cm}{(c)}{} \includegraphics[width=0.45\hsize]{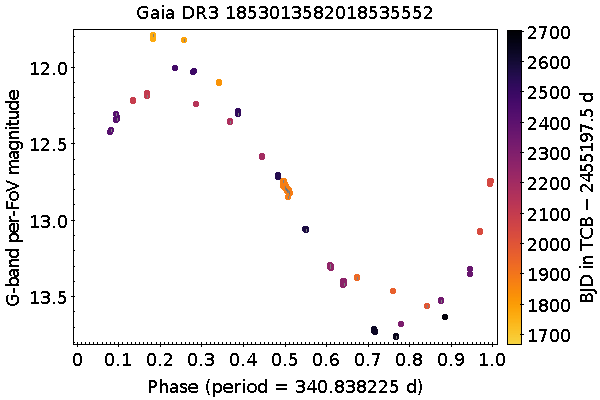}  
\hspace{2mm}
\stackinset{c}{8.8cm}{c}{3cm}{(d)}{} \includegraphics[width=0.45\hsize]{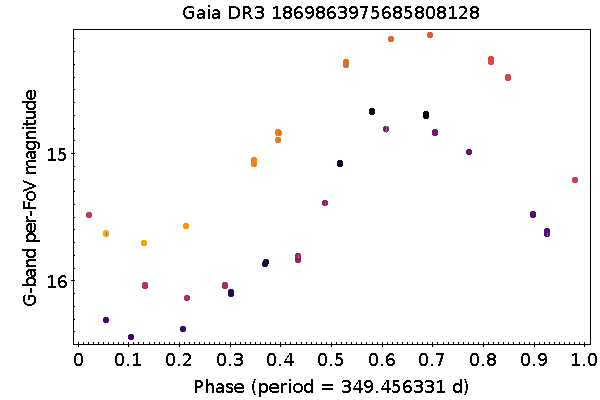} \\
\vspace{4mm}
\stackinset{c}{-0.3cm}{c}{3cm}{(e)}{} \includegraphics[width=0.45\hsize]{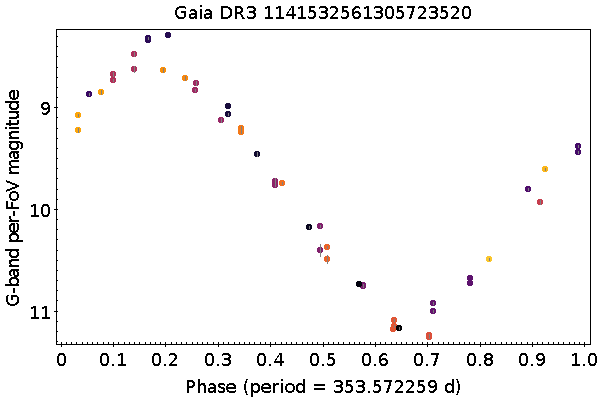}  
\hspace{2mm}
\stackinset{c}{8.8cm}{c}{3cm}{(f)}{} \includegraphics[width=0.45\hsize]{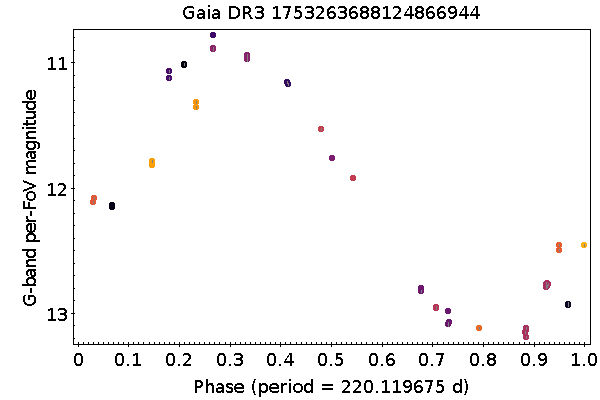} \\
\vspace{4mm}
 \caption{Same as Fig.~\ref{fig:app:ACV_cc}, but for LPV: (c,d) C-rich and (e,f) O-rich long-period variables.}
 \label{fig:app:LPV_cc}
\end{figure*}

\begin{figure*}
\centering
\stackinset{c}{-0.7cm}{c}{2.7cm}{(a)}{} \includegraphics[width=0.6\hsize]{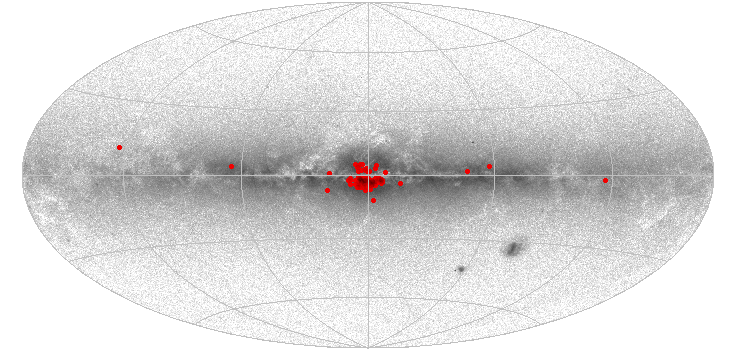} \\ 
\vspace{4mm}
\stackinset{c}{-0.3cm}{c}{3cm}{(b)}{} \includegraphics[width=0.45\hsize]{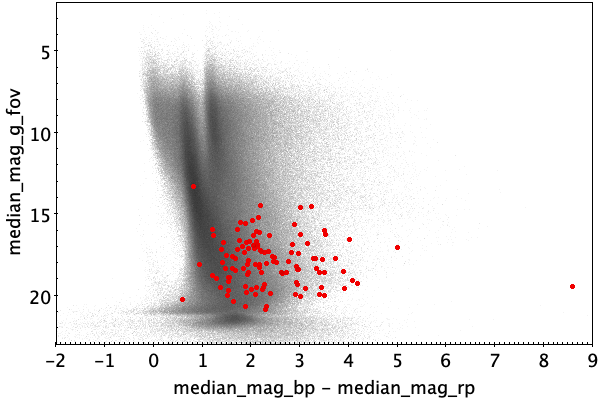}  
\hspace{2mm}
\stackinset{c}{8.8cm}{c}{3cm}{(c)}{} \includegraphics[width=0.45\hsize]{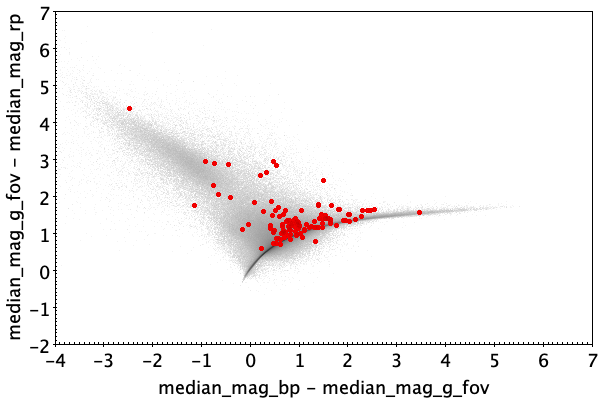} \\ 
\vspace{4mm}
\stackinset{c}{-0.3cm}{c}{3cm}{(d)}{} \includegraphics[width=0.45\hsize]{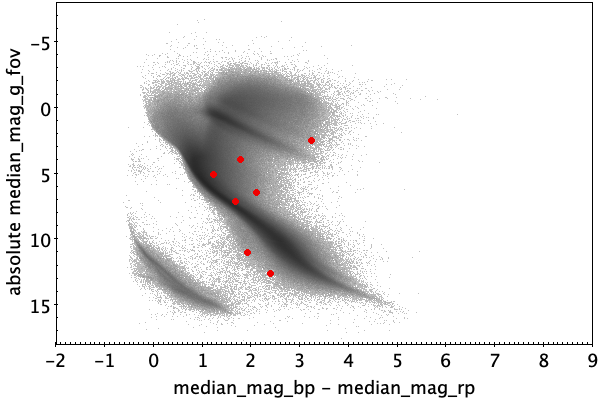}  
\hspace{2mm}
\stackinset{c}{8.8cm}{c}{3cm}{(e)}{} \includegraphics[width=0.45\hsize]{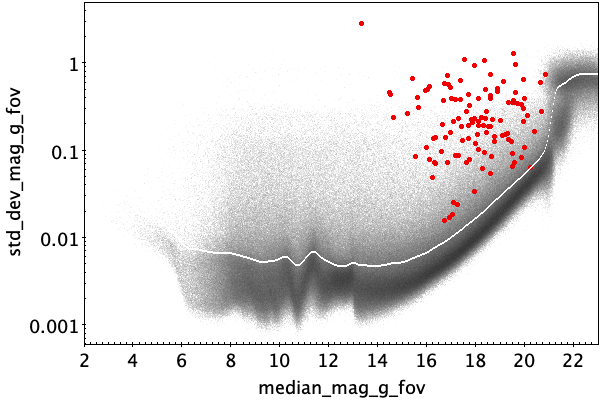} \\ 
\vspace{4mm}
\stackinset{c}{-0.3cm}{c}{3cm}{(f)}{} \includegraphics[width=0.45\hsize]{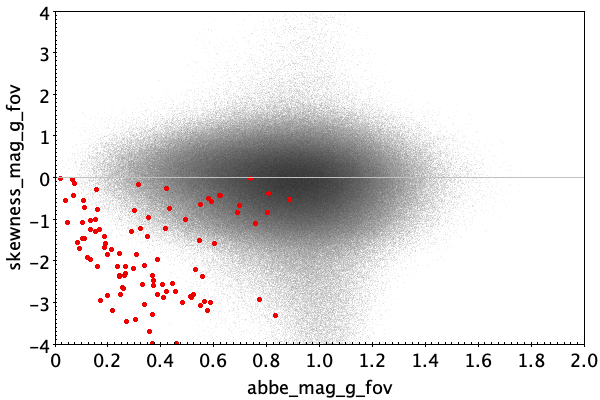}  
\hspace{2mm}
\stackinset{c}{8.8cm}{c}{3cm}{(g)}{} \includegraphics[width=0.45\hsize]{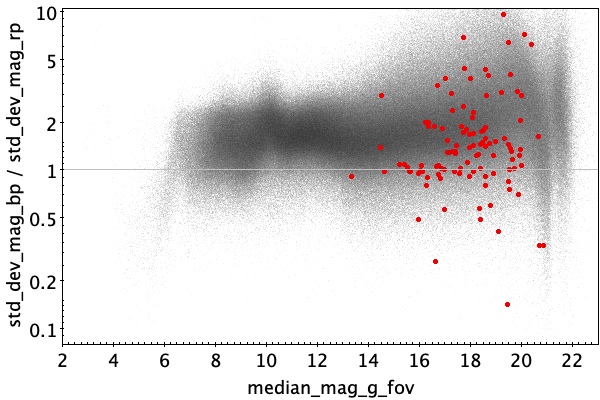}  \\ 
\vspace{4mm}
 \caption{MICROLENSING: 116 training sources.}  
 \label{fig:app:MICROLENSING_trn}
\end{figure*}

\begin{figure*}
\centering
\stackinset{c}{-0.7cm}{c}{2.7cm}{(a)}{}
\includegraphics[width=0.6\hsize]{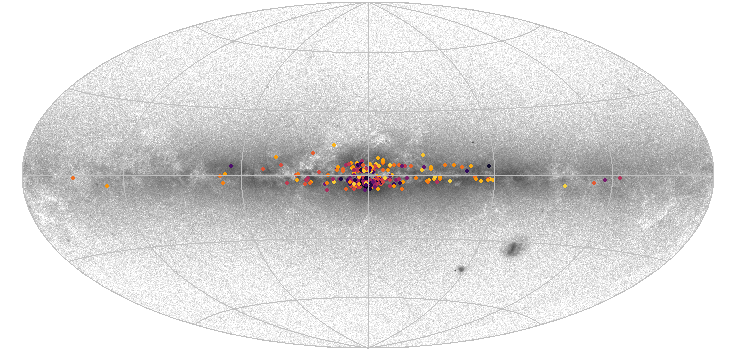} 
\stackinset{c}{2.2cm}{c}{2.7cm}{\includegraphics[height=5.5cm]{figures/appendix/vertical_best_class_score.png}}{} \\ 
\vspace{4mm}
\stackinset{c}{-0.3cm}{c}{3cm}{(b)}{} \includegraphics[width=0.45\hsize]{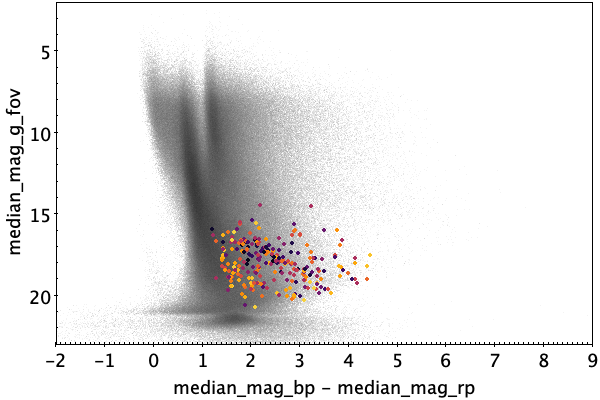}  
\hspace{2mm}
\stackinset{c}{8.8cm}{c}{3cm}{(c)}{} \includegraphics[width=0.45\hsize]{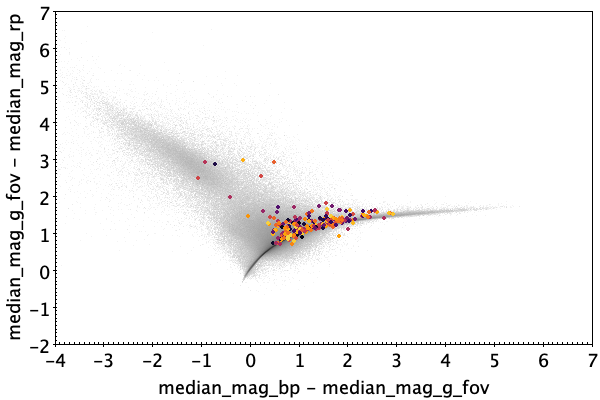} \\ 
\vspace{4mm}
\stackinset{c}{-0.3cm}{c}{3cm}{(d)}{} \includegraphics[width=0.45\hsize]{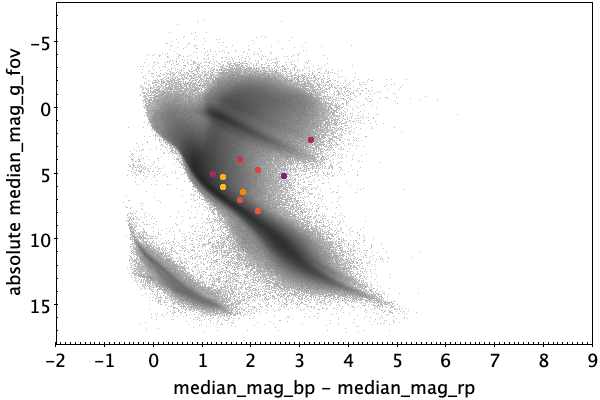}  
\hspace{2mm}
\stackinset{c}{8.8cm}{c}{3cm}{(e)}{} \includegraphics[width=0.45\hsize]{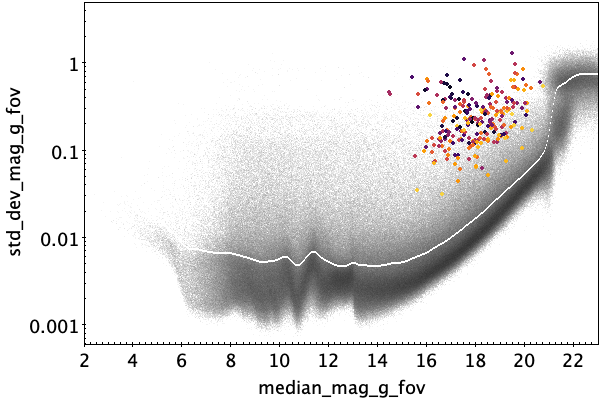} \\ 
\vspace{4mm}
\stackinset{c}{-0.3cm}{c}{3cm}{(f)}{} \includegraphics[width=0.45\hsize]{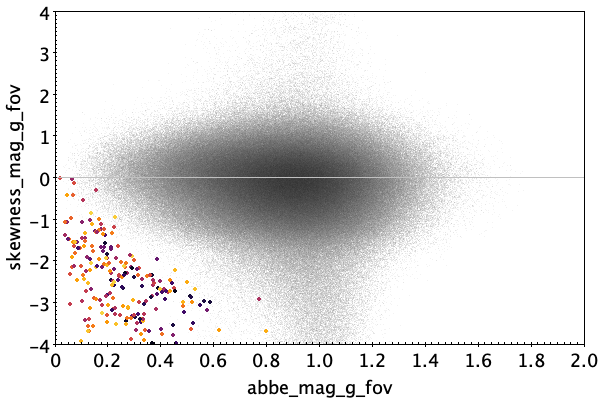}  
\hspace{2mm}
\stackinset{c}{8.8cm}{c}{3cm}{(g)}{} \includegraphics[width=0.45\hsize]{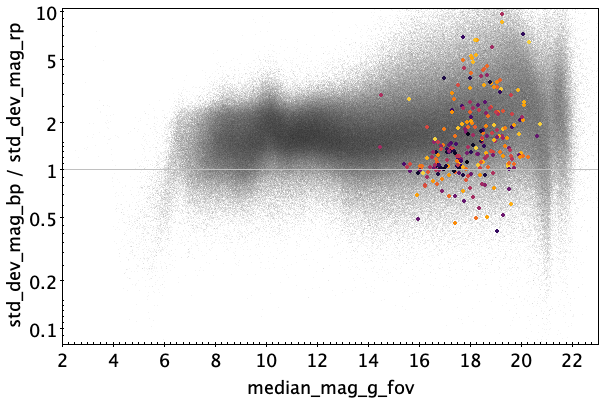}  \\ 
\vspace{4mm}
 \caption{MICROLENSING: 254 classified sources.}  
 \label{fig:app:MICROLENSING}
\end{figure*}

\begin{figure*}
\centering
\stackinset{c}{-0.3cm}{c}{3cm}{(a)}{} \includegraphics[width=0.45\hsize]{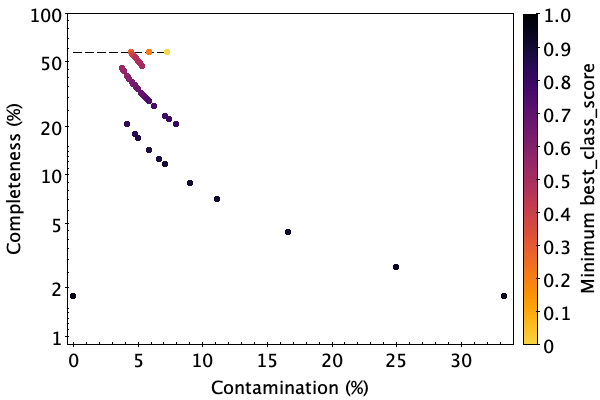}   
\hspace{2mm}
\stackinset{c}{8.8cm}{c}{3cm}{(b)}{} \includegraphics[width=0.45\hsize]{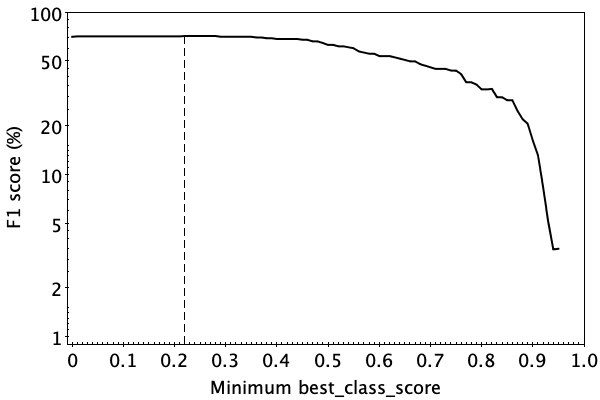} \\  
\vspace{4mm}
\stackinset{c}{-0.3cm}{c}{3cm}{(c)}{} \includegraphics[width=0.45\hsize]{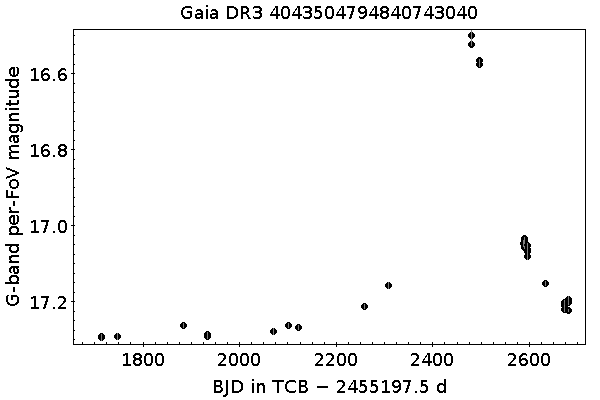}  
\hspace{2mm}
\stackinset{c}{8.8cm}{c}{3cm}{(d)}{} \includegraphics[width=0.45\hsize]{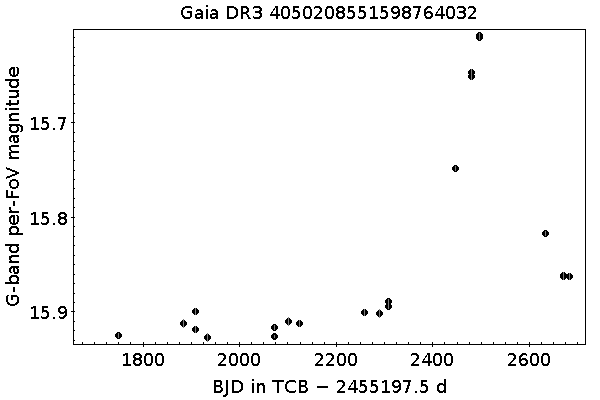} \\
\vspace{4mm}
\stackinset{c}{-0.3cm}{c}{3cm}{(e)}{} \includegraphics[width=0.45\hsize]{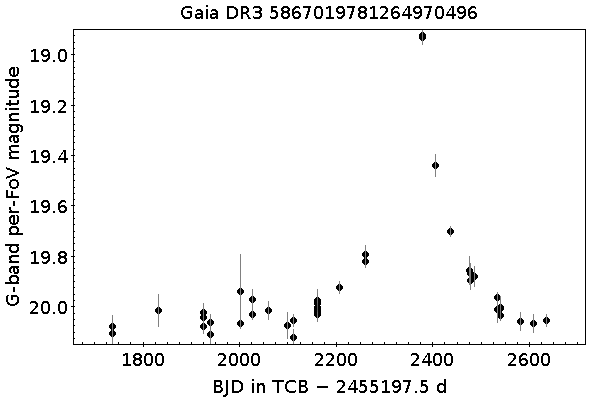}  
\hspace{2mm}
\stackinset{c}{8.8cm}{c}{3cm}{(f)}{} \includegraphics[width=0.45\hsize]{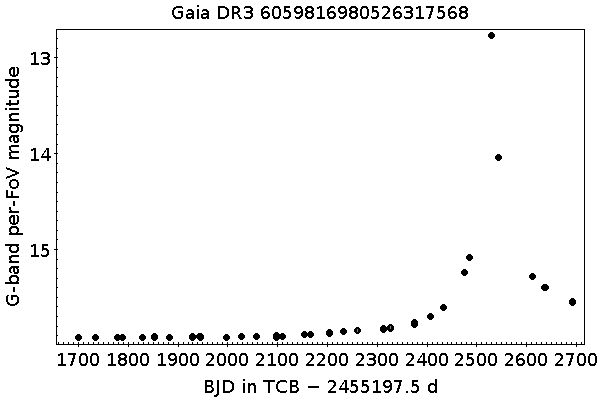} \\
\vspace{4mm}
 \caption{Same as Fig.~\ref{fig:app:ACV_cc}, but for MICROLENSING.}
 \label{fig:app:MICROLENSING_cc}
\end{figure*}

\begin{figure*}
\centering
\stackinset{c}{-0.7cm}{c}{2.7cm}{(a)}{} \includegraphics[width=0.6\hsize]{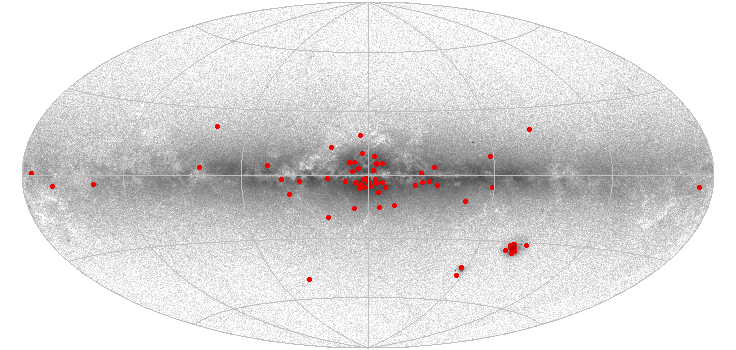} \\ 
\vspace{4mm}
\stackinset{c}{-0.3cm}{c}{3cm}{(b)}{} \includegraphics[width=0.45\hsize]{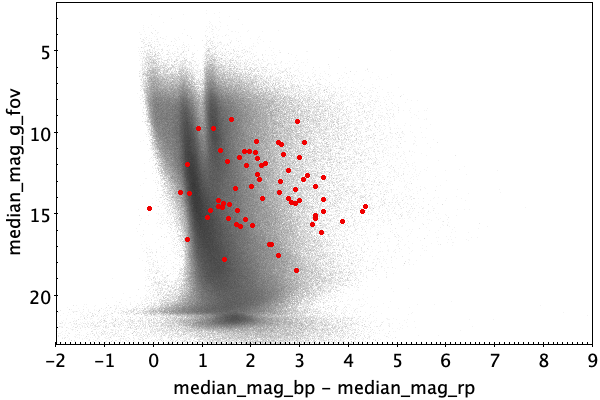}  
\hspace{2mm}
\stackinset{c}{8.8cm}{c}{3cm}{(c)}{} \includegraphics[width=0.45\hsize]{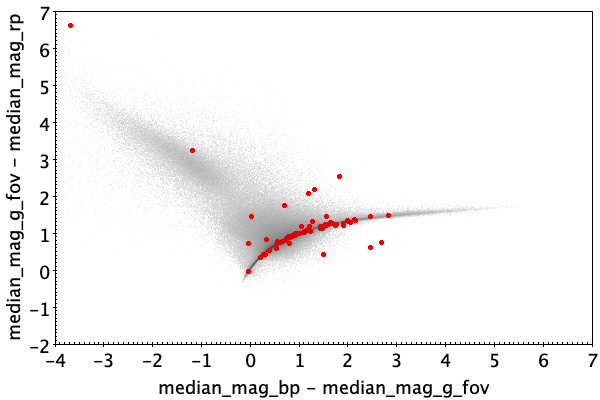} \\ 
\vspace{4mm}
\stackinset{c}{-0.3cm}{c}{3cm}{(d)}{} \includegraphics[width=0.45\hsize]{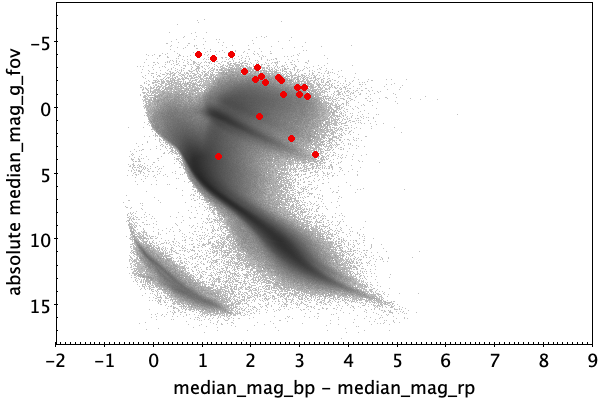}  
\hspace{2mm}
\stackinset{c}{8.8cm}{c}{3cm}{(e)}{} \includegraphics[width=0.45\hsize]{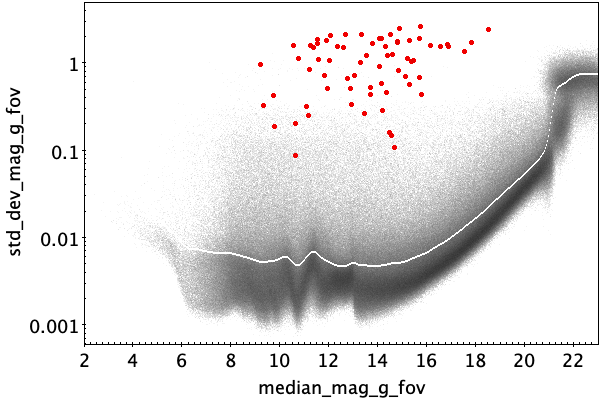} \\ 
\vspace{4mm}
\stackinset{c}{-0.3cm}{c}{3cm}{(f)}{} \includegraphics[width=0.45\hsize]{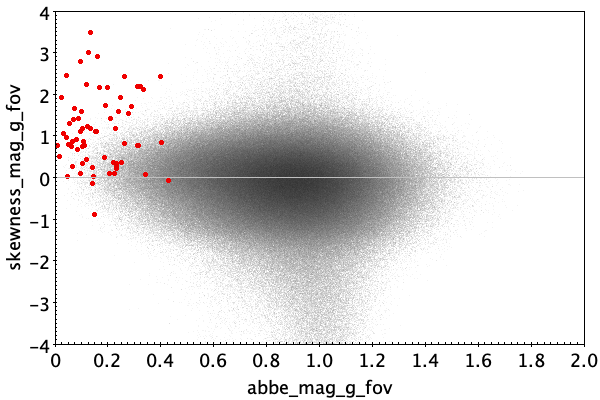}  
\hspace{2mm}
\stackinset{c}{8.8cm}{c}{3cm}{(g)}{} \includegraphics[width=0.45\hsize]{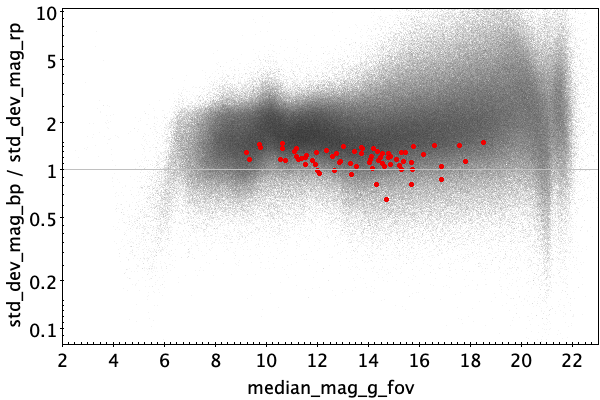}  \\ 
\vspace{4mm}
 \caption{RCB: 69 training sources.}  
 \label{fig:app:RCB_trn}
\end{figure*}

\begin{figure*}
\centering
\stackinset{c}{-0.7cm}{c}{2.7cm}{(a)}{}
\includegraphics[width=0.6\hsize]{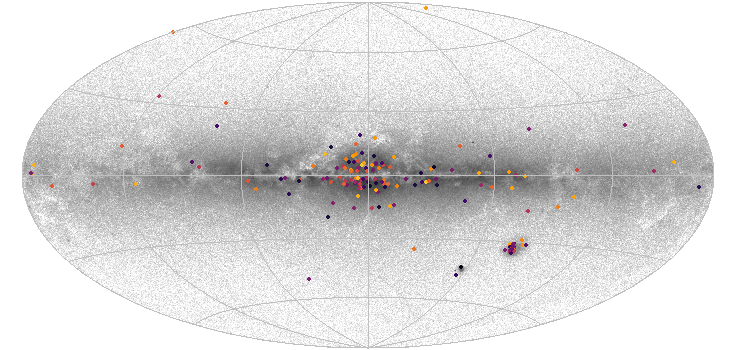} 
\stackinset{c}{2.2cm}{c}{2.7cm}{\includegraphics[height=5.5cm]{figures/appendix/vertical_best_class_score.png}}{} \\ 
\vspace{4mm}
\stackinset{c}{-0.3cm}{c}{3cm}{(b)}{} \includegraphics[width=0.45\hsize]{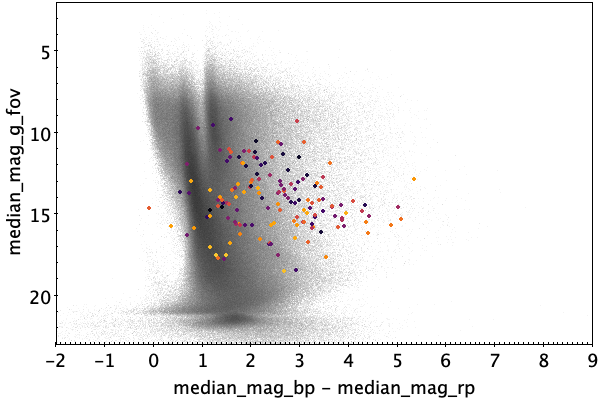}  
\hspace{2mm}
\stackinset{c}{8.8cm}{c}{3cm}{(c)}{} \includegraphics[width=0.45\hsize]{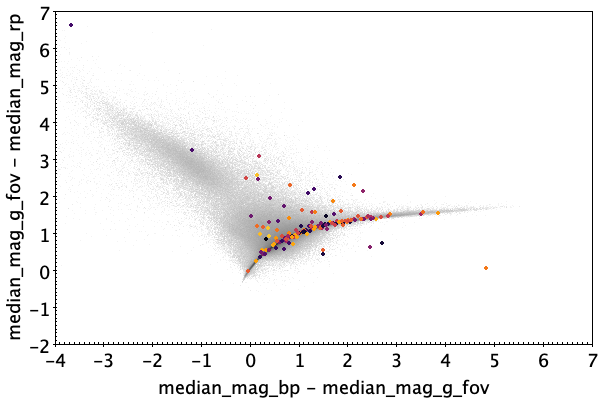} \\ 
\vspace{4mm}
\stackinset{c}{-0.3cm}{c}{3cm}{(d)}{} \includegraphics[width=0.45\hsize]{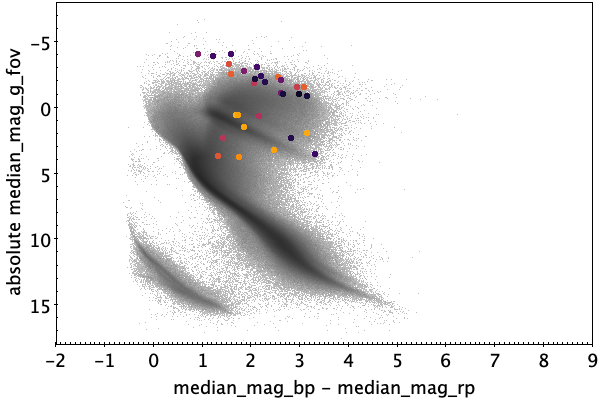}  
\hspace{2mm}
\stackinset{c}{8.8cm}{c}{3cm}{(e)}{} \includegraphics[width=0.45\hsize]{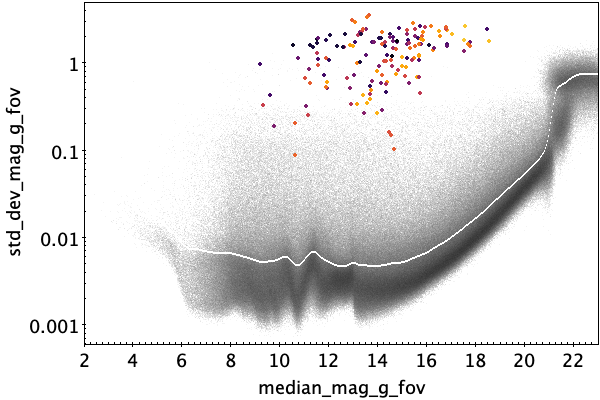} \\ 
\vspace{4mm}
\stackinset{c}{-0.3cm}{c}{3cm}{(f)}{} \includegraphics[width=0.45\hsize]{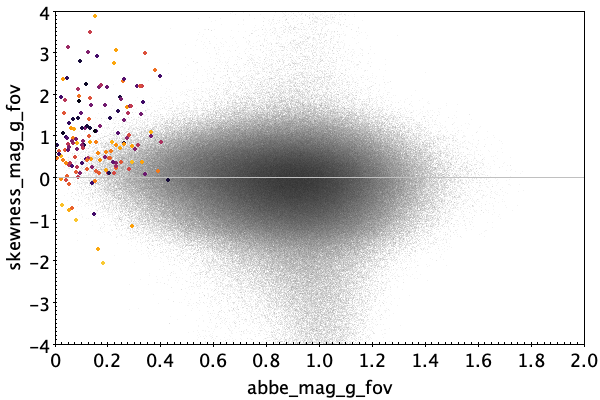}  
\hspace{2mm}
\stackinset{c}{8.8cm}{c}{3cm}{(g)}{} \includegraphics[width=0.45\hsize]{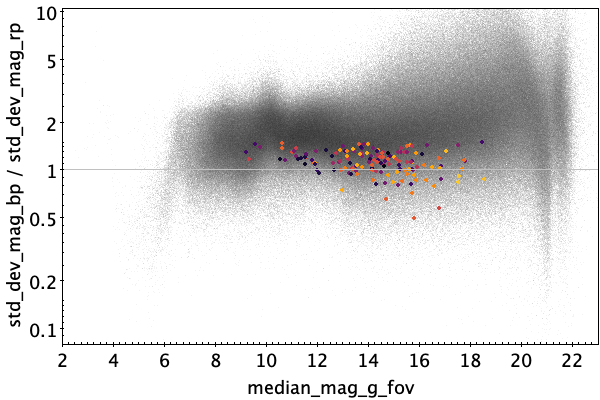}  \\ 
\vspace{4mm}
 \caption{RCB: 153 classified sources.}  
 \label{fig:app:RCB}
\end{figure*}

\begin{figure*}
\centering
\stackinset{c}{-0.3cm}{c}{3cm}{(a)}{} \includegraphics[width=0.45\hsize]{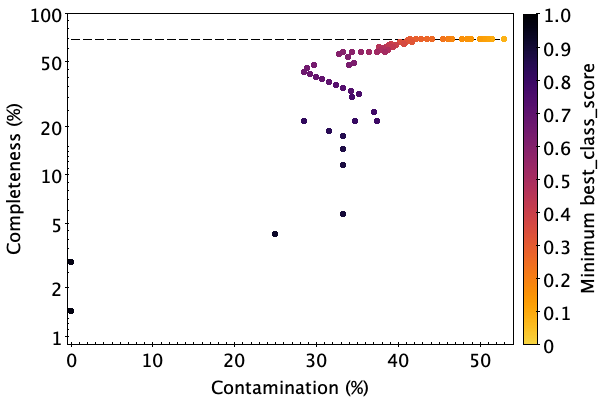}  
\hspace{2mm}
\stackinset{c}{8.8cm}{c}{3cm}{(b)}{} \includegraphics[width=0.45\hsize]{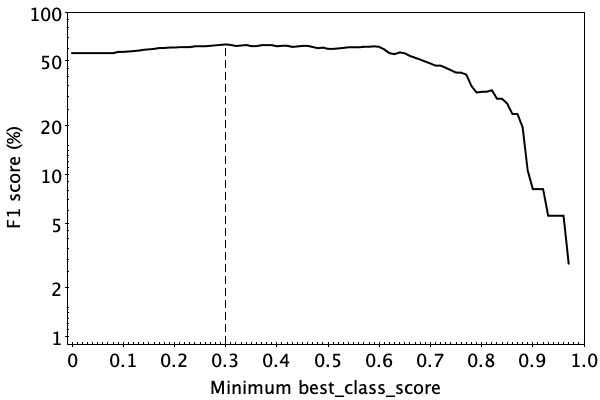} \\  
\vspace{4mm}
\stackinset{c}{-0.3cm}{c}{3cm}{(c)}{} \includegraphics[width=0.45\hsize]{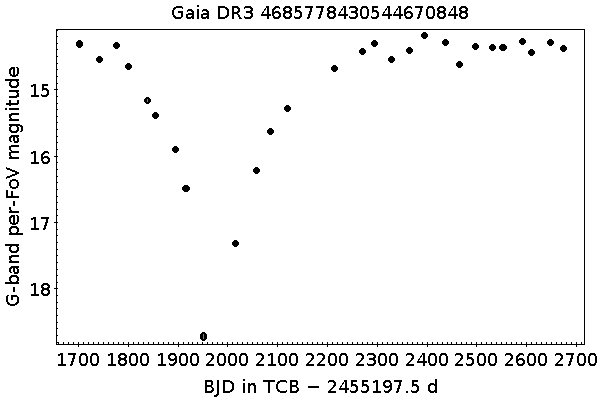}  
\hspace{2mm}
\stackinset{c}{8.8cm}{c}{3cm}{(d)}{} \includegraphics[width=0.45\hsize]{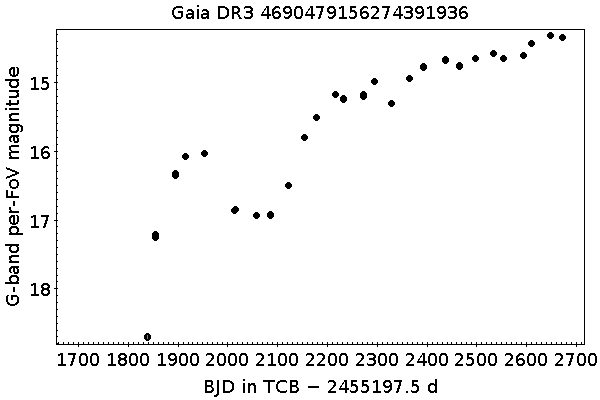} \\
\vspace{4mm}
\stackinset{c}{-0.3cm}{c}{3cm}{(e)}{} \includegraphics[width=0.45\hsize]{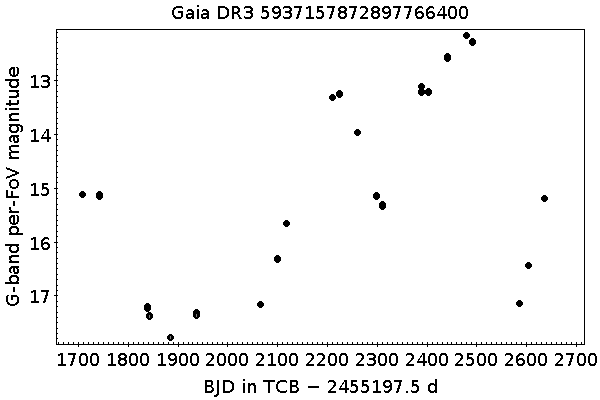}  
\hspace{2mm}
\stackinset{c}{8.8cm}{c}{3cm}{(f)}{} \includegraphics[width=0.45\hsize]{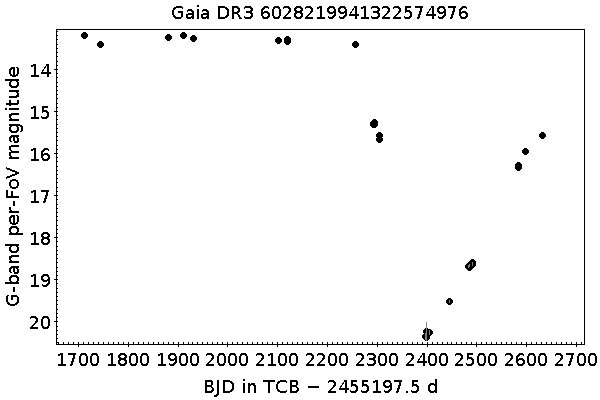} \\
\vspace{4mm}
 \caption{Same as Fig.~\ref{fig:app:ACV_cc}, but for RCB.}
 \label{fig:app:RCB_cc}
\end{figure*}

\begin{figure*}
\centering
\stackinset{c}{-0.7cm}{c}{2.7cm}{(a)}{} \includegraphics[width=0.6\hsize]{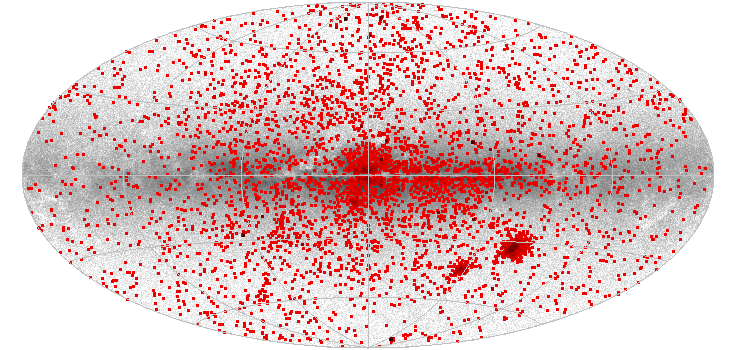} \\ 
\vspace{4mm}
\stackinset{c}{-0.3cm}{c}{3cm}{(b)}{} \includegraphics[width=0.45\hsize]{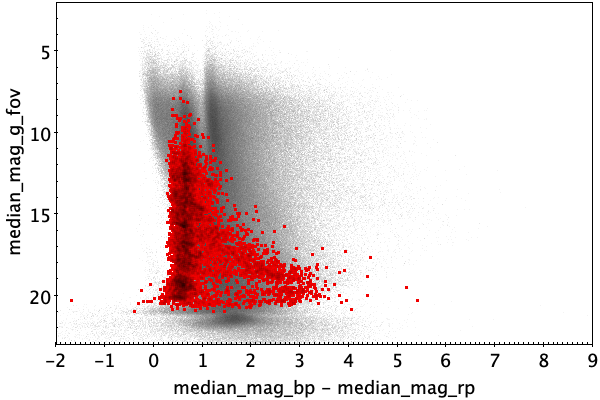}  
\hspace{2mm}
\stackinset{c}{8.8cm}{c}{3cm}{(c)}{} \includegraphics[width=0.45\hsize]{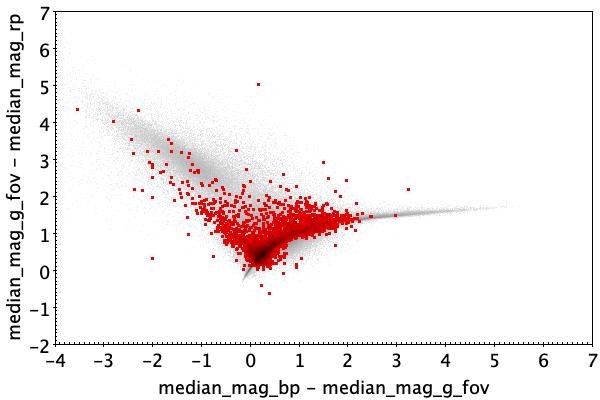} \\ 
\vspace{4mm}
\stackinset{c}{-0.3cm}{c}{3cm}{(d)}{} \includegraphics[width=0.45\hsize]{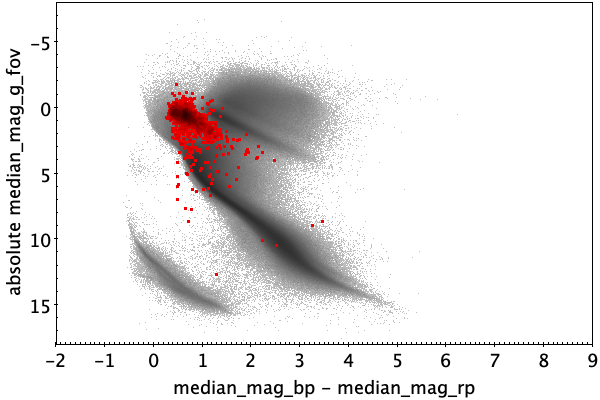}  
\hspace{2mm}
\stackinset{c}{8.8cm}{c}{3cm}{(e)}{} \includegraphics[width=0.45\hsize]{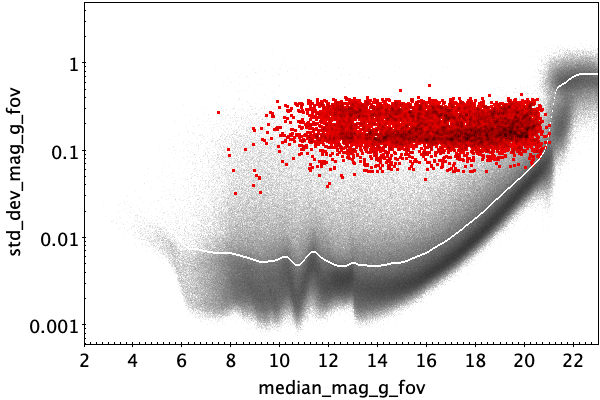} \\ 
\vspace{4mm}
\stackinset{c}{-0.3cm}{c}{3cm}{(f)}{} \includegraphics[width=0.45\hsize]{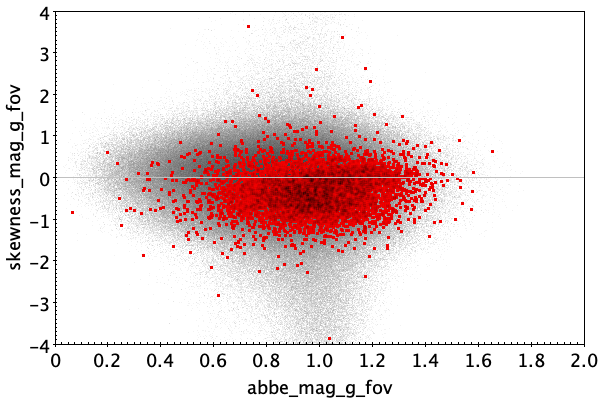}  
\hspace{2mm}
\stackinset{c}{8.8cm}{c}{3cm}{(g)}{} \includegraphics[width=0.45\hsize]{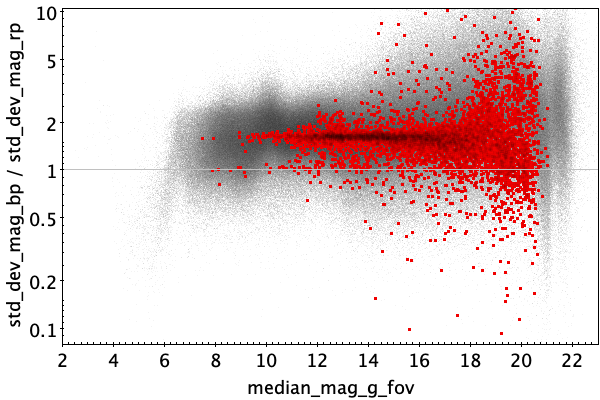}  \\ 
\vspace{4mm}
 \caption{RR: 6377 training sources.}  
 \label{fig:app:RR_trn}
\end{figure*}

\begin{figure*}
\centering
\stackinset{c}{-0.7cm}{c}{2.7cm}{(a)}{}
\includegraphics[width=0.6\hsize]{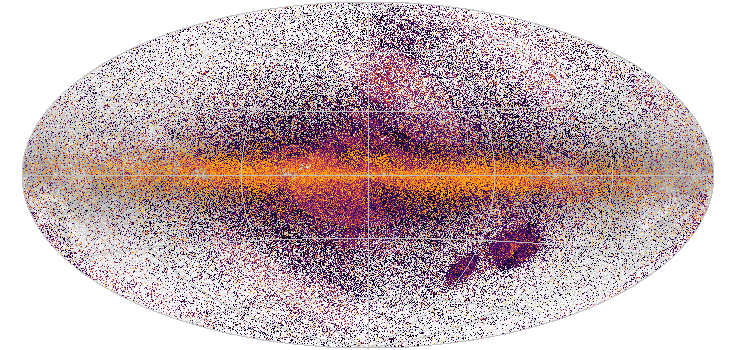} 
\stackinset{c}{2.2cm}{c}{2.7cm}{\includegraphics[height=5.5cm]{figures/appendix/vertical_best_class_score.png}}{} \\ 
\vspace{4mm}
\stackinset{c}{-0.3cm}{c}{3cm}{(b)}{} \includegraphics[width=0.45\hsize]{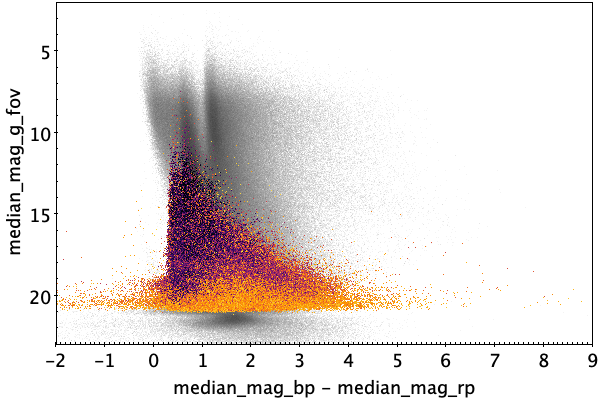}  
\hspace{2mm}
\stackinset{c}{8.8cm}{c}{3cm}{(c)}{} \includegraphics[width=0.45\hsize]{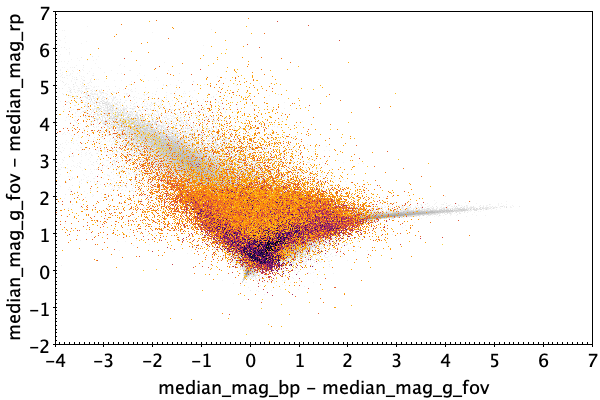} \\ 
\vspace{4mm}
\stackinset{c}{-0.3cm}{c}{3cm}{(d)}{} \includegraphics[width=0.45\hsize]{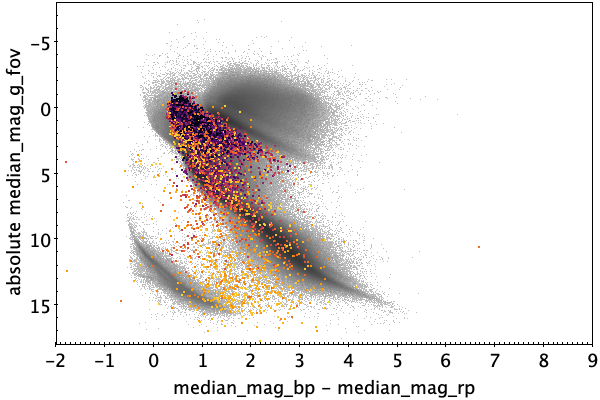}  
\hspace{2mm}
\stackinset{c}{8.8cm}{c}{3cm}{(e)}{} \includegraphics[width=0.45\hsize]{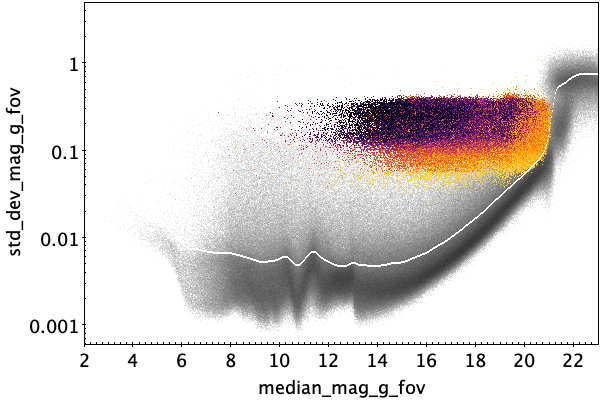} \\ 
\vspace{4mm}
\stackinset{c}{-0.3cm}{c}{3cm}{(f)}{} \includegraphics[width=0.45\hsize]{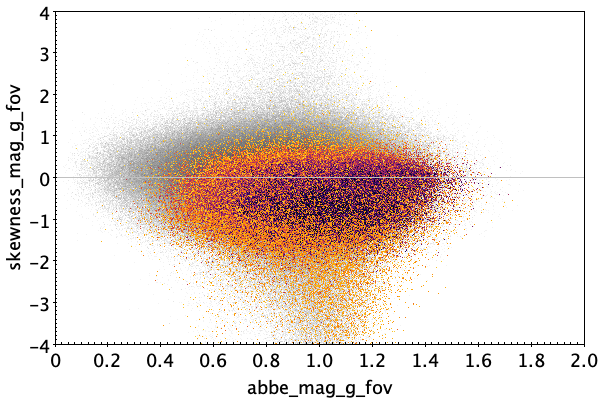}  
\hspace{2mm}
\stackinset{c}{8.8cm}{c}{3cm}{(g)}{} \includegraphics[width=0.45\hsize]{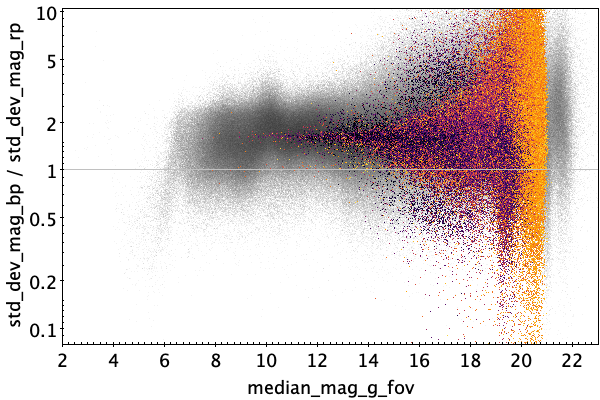}  \\ 
\vspace{4mm}
 \caption{RR: 297\,778 classified sources.}  
 \label{fig:app:RR}
\end{figure*}

\begin{figure*}
\centering
\stackinset{c}{-0.3cm}{c}{3cm}{(a)}{} \includegraphics[width=0.45\hsize]{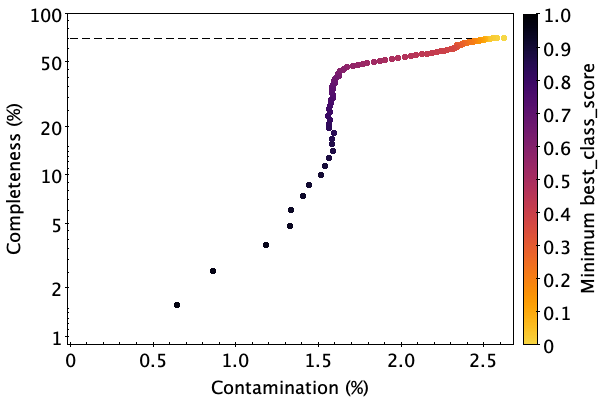}  
\hspace{2mm}
\stackinset{c}{8.8cm}{c}{3cm}{(b)}{} \includegraphics[width=0.45\hsize]{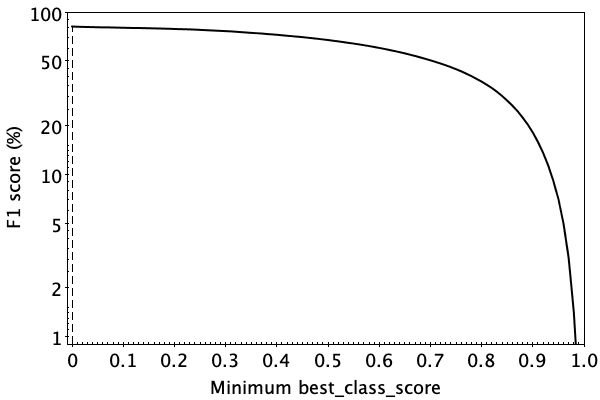} \\  
\vspace{4mm}
\stackinset{c}{-0.3cm}{c}{3cm}{(c)}{} \includegraphics[width=0.45\hsize]{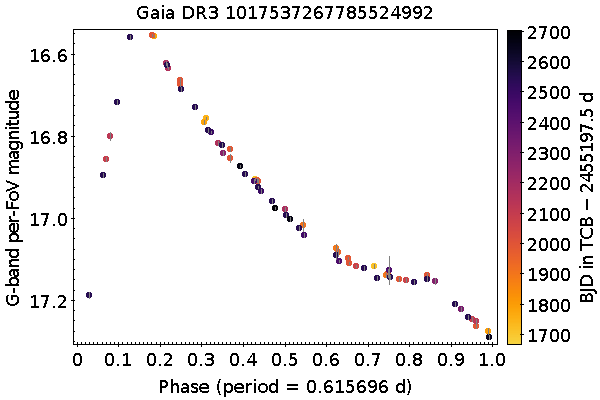}  
\hspace{2mm}
\stackinset{c}{8.8cm}{c}{3cm}{(d)}{} \includegraphics[width=0.45\hsize]{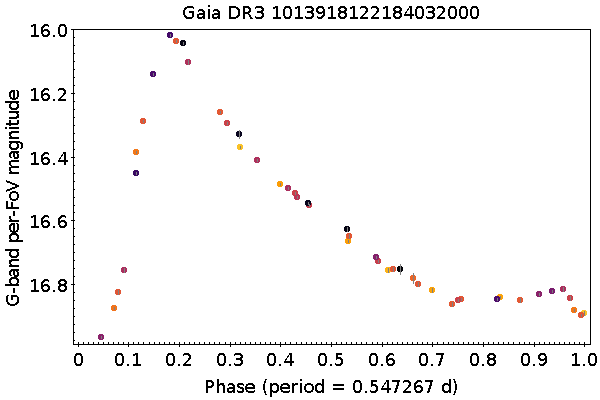} \\
\vspace{4mm}
\stackinset{c}{-0.3cm}{c}{3cm}{(e)}{} \includegraphics[width=0.45\hsize]{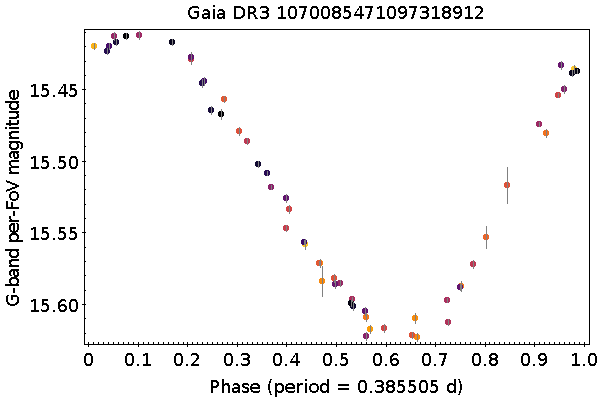}  
\hspace{2mm}
\stackinset{c}{8.8cm}{c}{3cm}{(f)}{} \includegraphics[width=0.45\hsize]{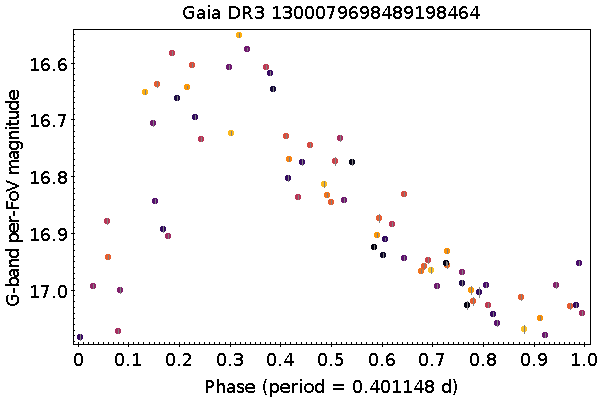} \\
\vspace{4mm}
 \caption{Same as Fig.~\ref{fig:app:ACV_cc}, but for RR:  (c,d) fundamental-mode, (e) first overtone, and (f) double-mode RR\,Lyrae stars.}
 \label{fig:app:RR_cc}
\end{figure*}

\begin{figure*}
\centering
\stackinset{c}{-0.7cm}{c}{2.7cm}{(a)}{} \includegraphics[width=0.6\hsize]{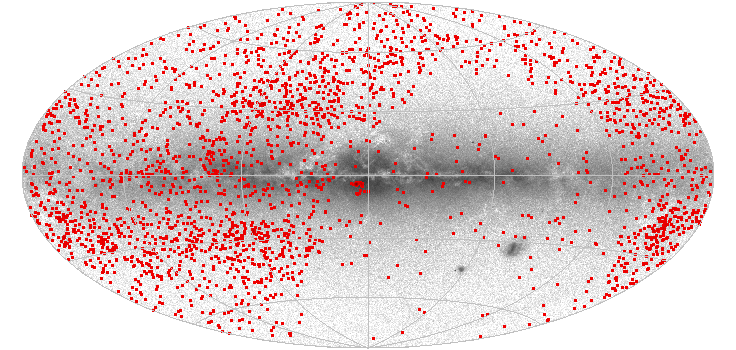} \\ 
\vspace{4mm}
\stackinset{c}{-0.3cm}{c}{3cm}{(b)}{} \includegraphics[width=0.45\hsize]{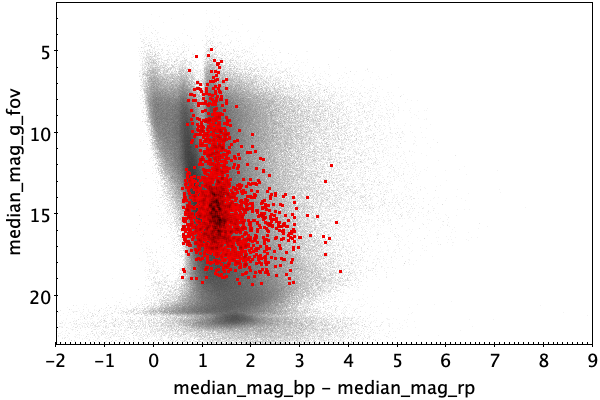}  
\hspace{2mm}
\stackinset{c}{8.8cm}{c}{3cm}{(c)}{} \includegraphics[width=0.45\hsize]{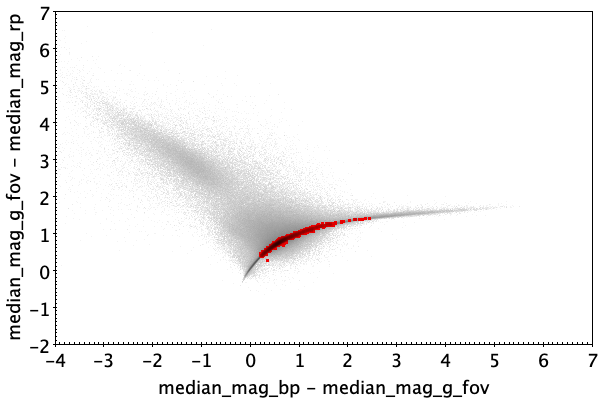} \\ 
\vspace{4mm}
\stackinset{c}{-0.3cm}{c}{3cm}{(d)}{} \includegraphics[width=0.45\hsize]{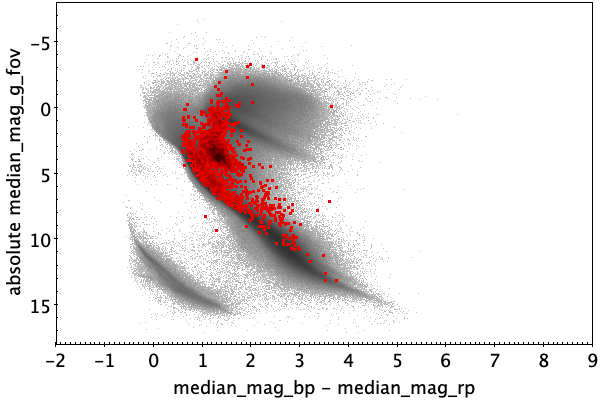}  
\hspace{2mm}
\stackinset{c}{8.8cm}{c}{3cm}{(e)}{} \includegraphics[width=0.45\hsize]{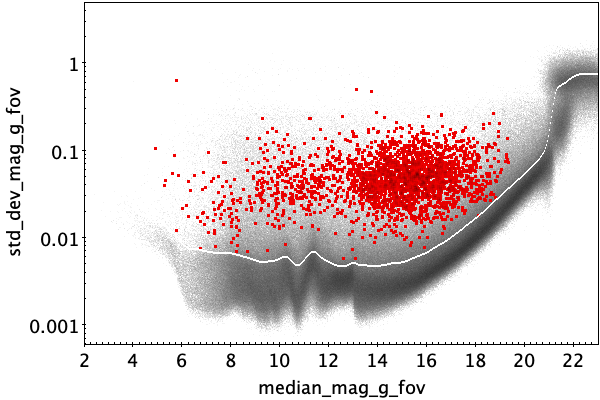} \\ 
\vspace{4mm}
\stackinset{c}{-0.3cm}{c}{3cm}{(f)}{} \includegraphics[width=0.45\hsize]{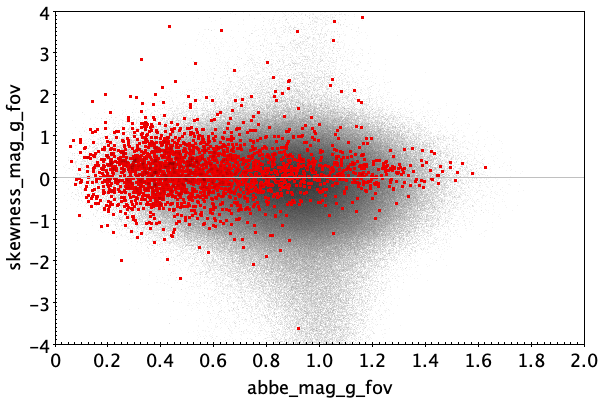}  
\hspace{2mm}
\stackinset{c}{8.8cm}{c}{3cm}{(g)}{} \includegraphics[width=0.45\hsize]{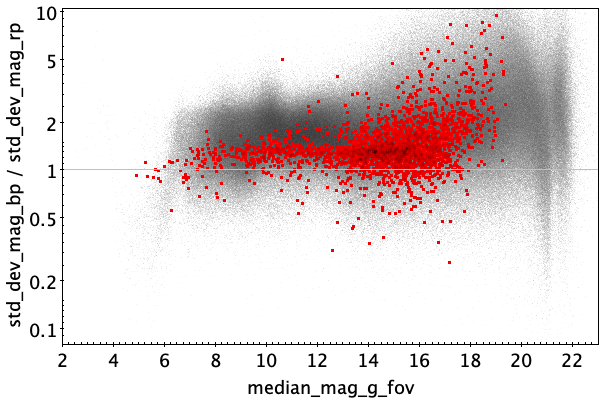}  \\ 
\vspace{4mm}
 \caption{RS: 2548 training sources.}  
 \label{fig:app:RS_trn}
\end{figure*}

\begin{figure*}
\centering
\stackinset{c}{-0.7cm}{c}{2.7cm}{(a)}{}
\includegraphics[width=0.6\hsize]{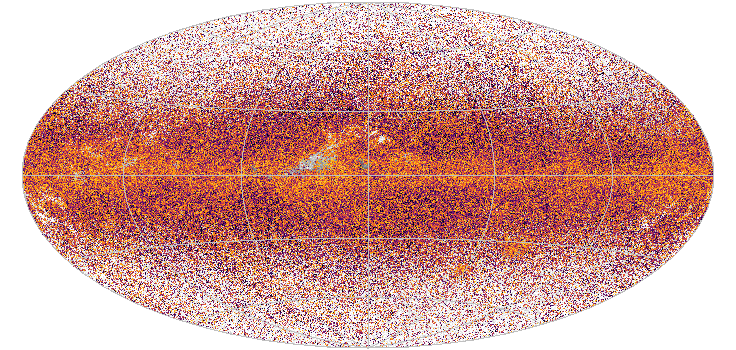} 
\stackinset{c}{2.2cm}{c}{2.7cm}{\includegraphics[height=5.5cm]{figures/appendix/vertical_best_class_score.png}}{} \\ 
\vspace{4mm}
\stackinset{c}{-0.3cm}{c}{3cm}{(b)}{} \includegraphics[width=0.45\hsize]{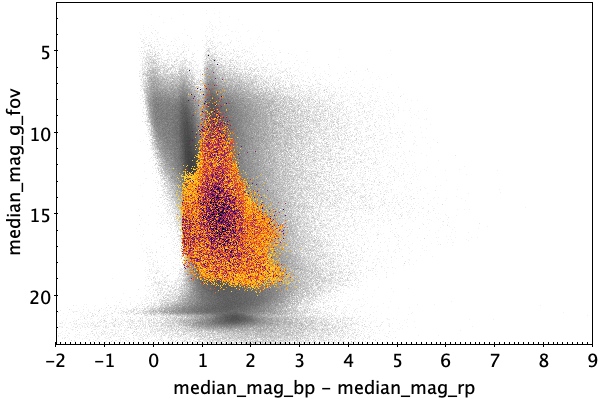}  
\hspace{2mm}
\stackinset{c}{8.8cm}{c}{3cm}{(c)}{} \includegraphics[width=0.45\hsize]{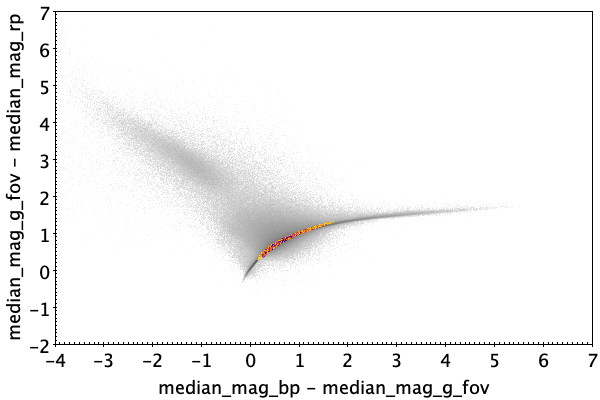} \\ 
\vspace{4mm}
\stackinset{c}{-0.3cm}{c}{3cm}{(d)}{} \includegraphics[width=0.45\hsize]{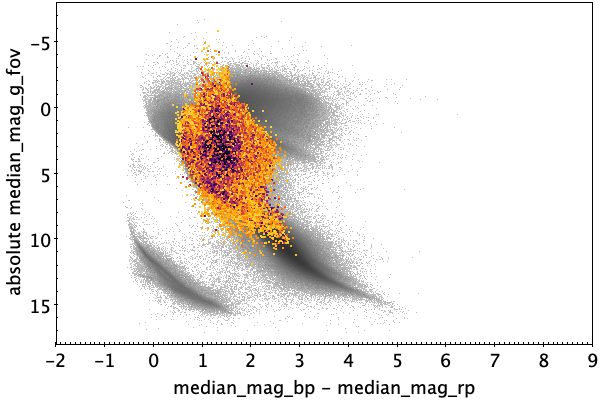}  
\hspace{2mm}
\stackinset{c}{8.8cm}{c}{3cm}{(e)}{} \includegraphics[width=0.45\hsize]{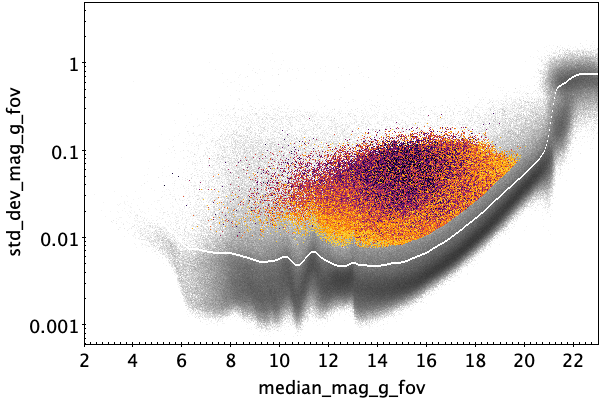} \\ 
\vspace{4mm}
\stackinset{c}{-0.3cm}{c}{3cm}{(f)}{} \includegraphics[width=0.45\hsize]{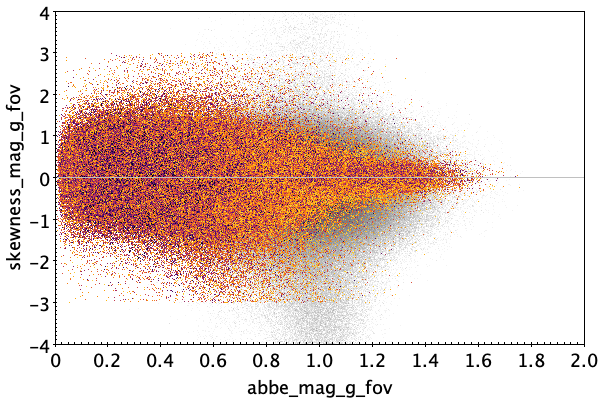}  
\hspace{2mm}
\stackinset{c}{8.8cm}{c}{3cm}{(g)}{} \includegraphics[width=0.45\hsize]{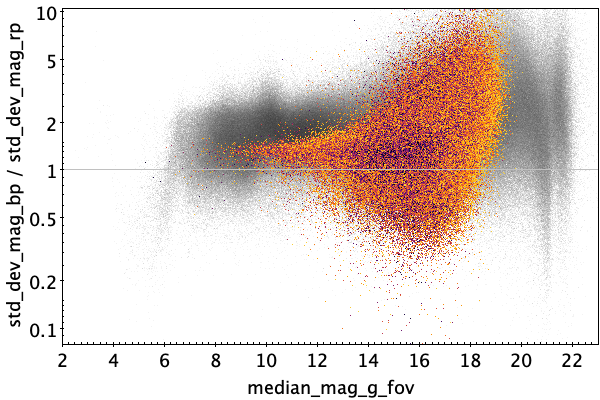}  \\ 
\vspace{4mm}
 \caption{RS: 742\,263 classified sources.}  
 \label{fig:app:RS}
\end{figure*}

\begin{figure*}
\centering
\stackinset{c}{-0.3cm}{c}{3cm}{(a)}{} \includegraphics[width=0.45\hsize]{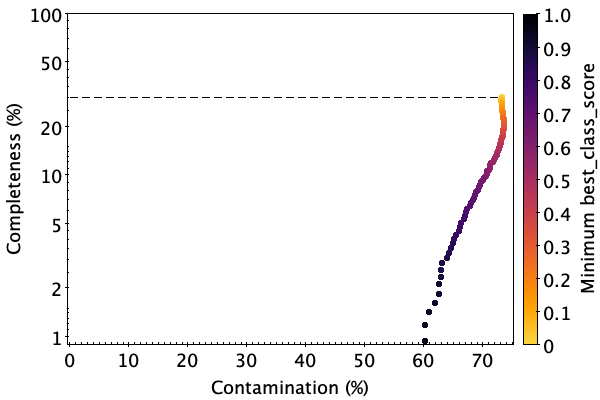}  
\hspace{2mm}
\stackinset{c}{8.8cm}{c}{3cm}{(b)}{} \includegraphics[width=0.45\hsize]{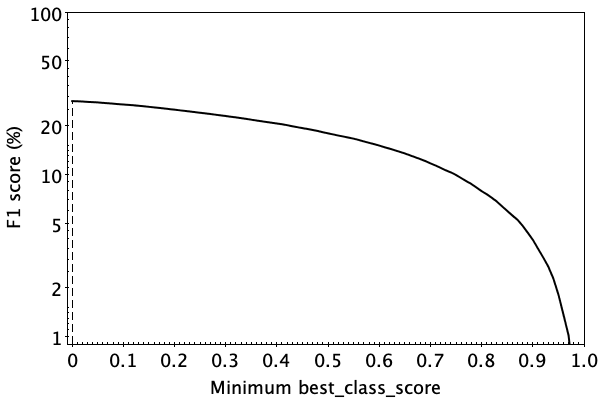} \\  
\vspace{4mm}
\stackinset{c}{-0.3cm}{c}{3cm}{(c)}{} \includegraphics[width=0.45\hsize]{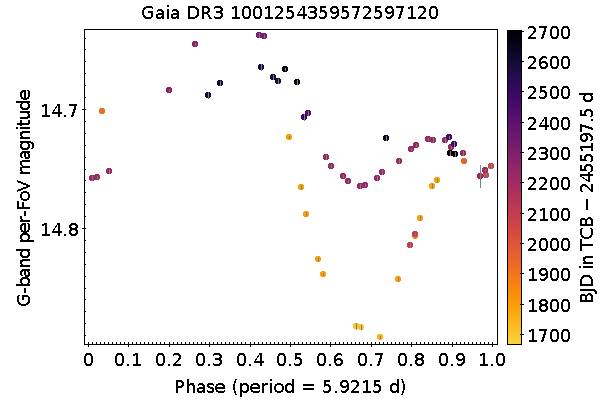}  
\hspace{2mm}
\stackinset{c}{8.8cm}{c}{3cm}{(d)}{} \includegraphics[width=0.45\hsize]{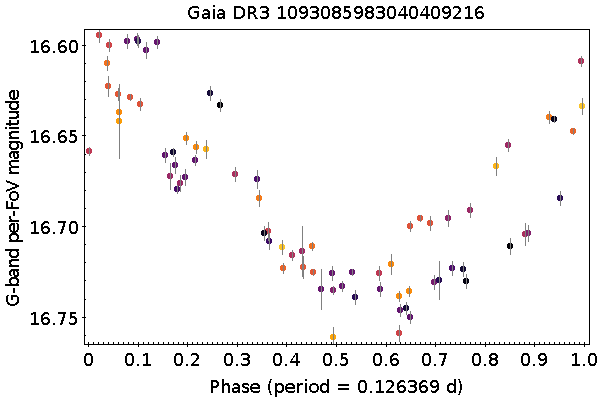} \\
\vspace{4mm}
\stackinset{c}{-0.3cm}{c}{3cm}{(e)}{} \includegraphics[width=0.45\hsize]{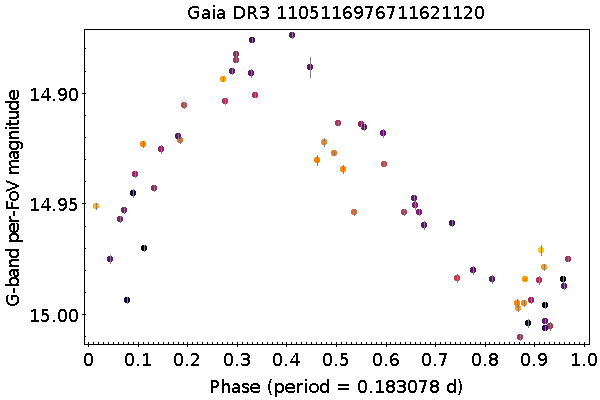}  
\hspace{2mm}
\stackinset{c}{8.8cm}{c}{3cm}{(f)}{} \includegraphics[width=0.45\hsize]{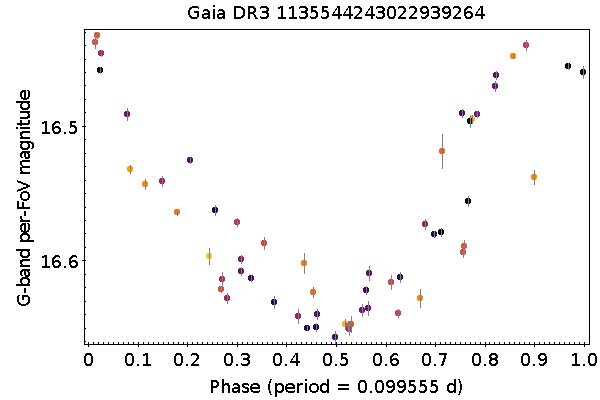} \\
\vspace{4mm}
 \caption{Same as Fig.~\ref{fig:app:ACV_cc}, but for RS.}
 \label{fig:app:RS_cc}
\end{figure*}

\begin{figure*}
\centering
\stackinset{c}{-0.7cm}{c}{2.7cm}{(a)}{} \includegraphics[width=0.6\hsize]{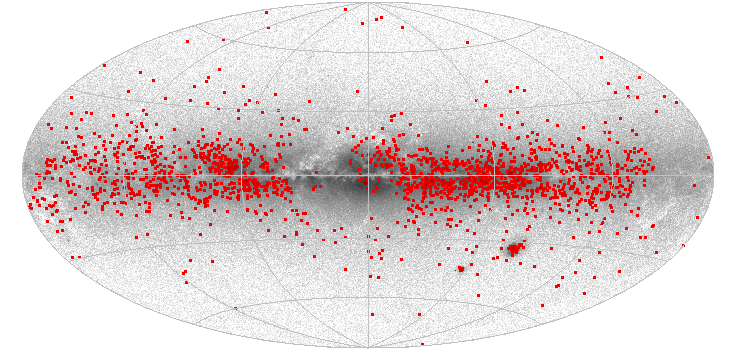} \\ 
\vspace{4mm}
\stackinset{c}{-0.3cm}{c}{3cm}{(b)}{} \includegraphics[width=0.45\hsize]{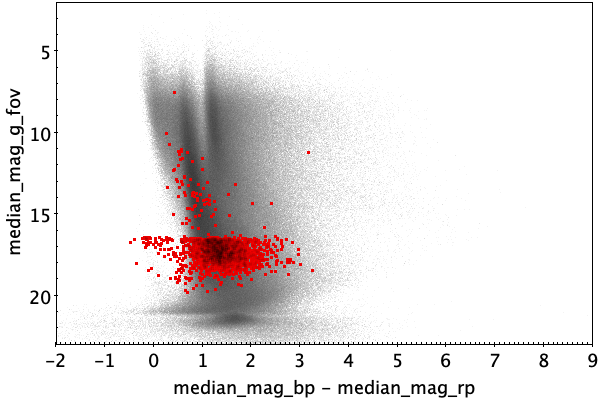}  
\hspace{2mm}
\stackinset{c}{8.8cm}{c}{3cm}{(c)}{} \includegraphics[width=0.45\hsize]{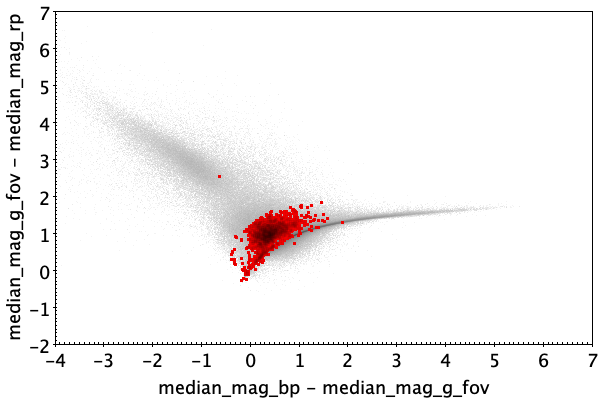} \\ 
\vspace{4mm}
\stackinset{c}{-0.3cm}{c}{3cm}{(d)}{} \includegraphics[width=0.45\hsize]{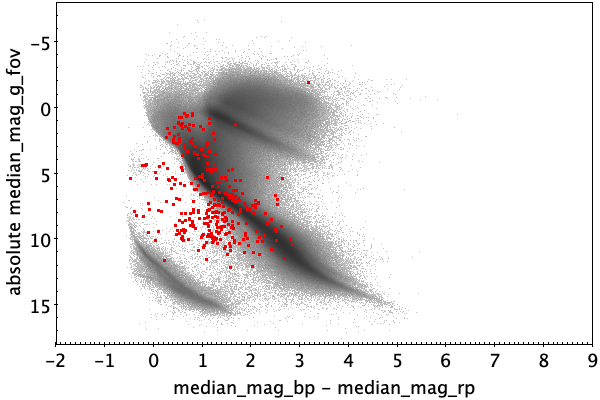}  
\hspace{2mm}
\stackinset{c}{8.8cm}{c}{3cm}{(e)}{} \includegraphics[width=0.45\hsize]{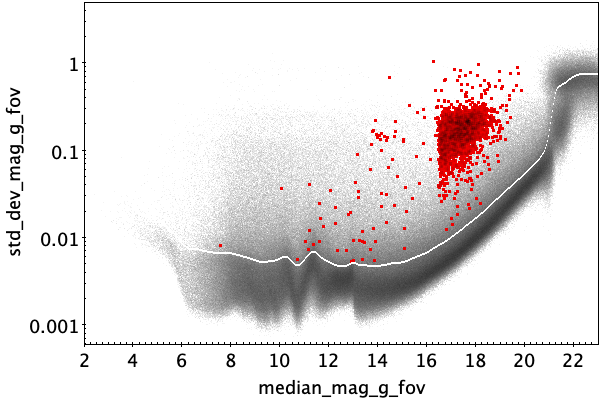} \\ 
\vspace{4mm}
\stackinset{c}{-0.3cm}{c}{3cm}{(f)}{} \includegraphics[width=0.45\hsize]{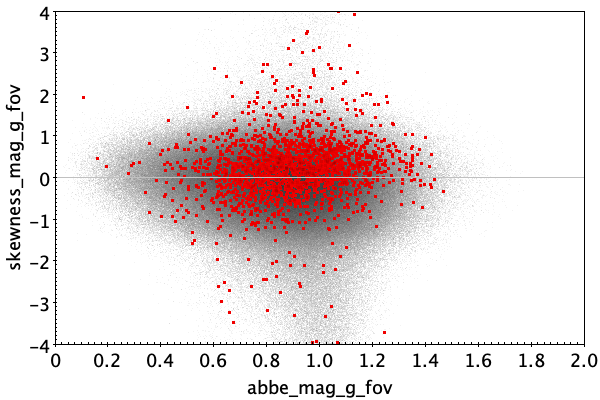}  
\hspace{2mm}
\stackinset{c}{8.8cm}{c}{3cm}{(g)}{} \includegraphics[width=0.45\hsize]{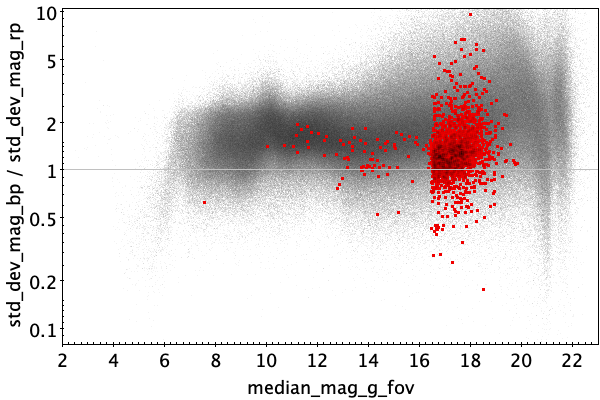}  \\ 
\vspace{4mm}
 \caption{S: 1965 training sources.}  
 \label{fig:app:S_trn}
\end{figure*}

\begin{figure*}
\centering
\stackinset{c}{-0.7cm}{c}{2.7cm}{(a)}{}
\includegraphics[width=0.6\hsize]{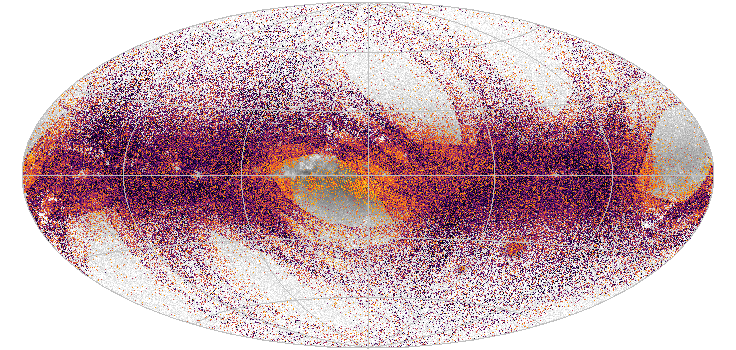} 
\stackinset{c}{2.2cm}{c}{2.7cm}{\includegraphics[height=5.5cm]{figures/appendix/vertical_best_class_score.png}}{} \\ 
\vspace{4mm}
\stackinset{c}{-0.3cm}{c}{3cm}{(b)}{} \includegraphics[width=0.45\hsize]{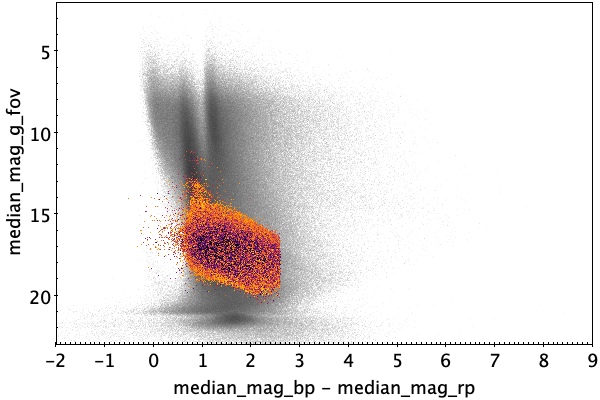}  
\hspace{2mm}
\stackinset{c}{8.8cm}{c}{3cm}{(c)}{} \includegraphics[width=0.45\hsize]{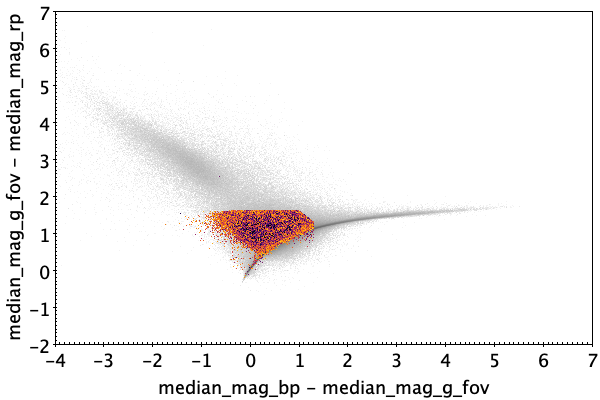} \\ 
\vspace{4mm}
\stackinset{c}{-0.3cm}{c}{3cm}{(d)}{} \includegraphics[width=0.45\hsize]{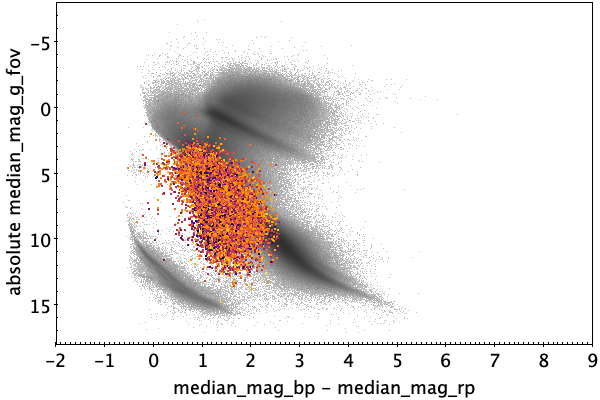}  
\hspace{2mm}
\stackinset{c}{8.8cm}{c}{3cm}{(e)}{} \includegraphics[width=0.45\hsize]{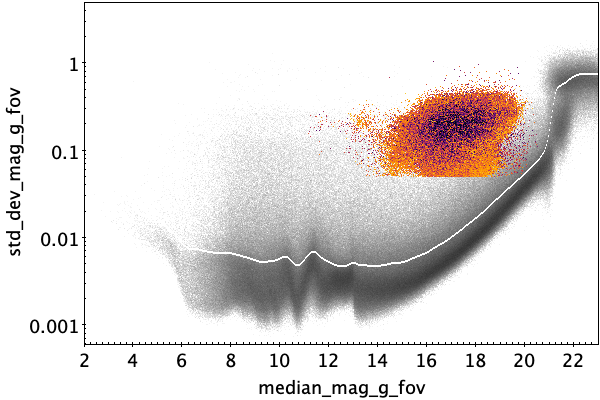} \\ 
\vspace{4mm}
\stackinset{c}{-0.3cm}{c}{3cm}{(f)}{} \includegraphics[width=0.45\hsize]{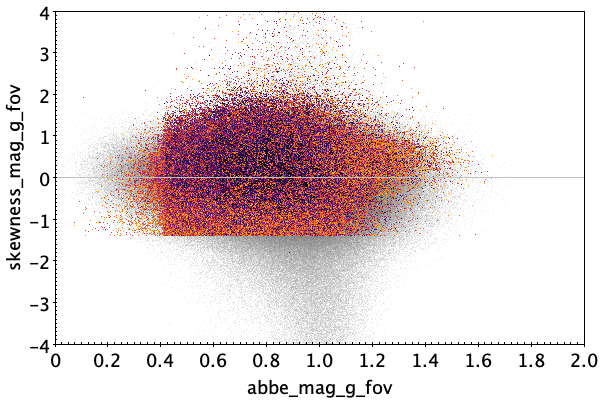}  
\hspace{2mm}
\stackinset{c}{8.8cm}{c}{3cm}{(g)}{} \includegraphics[width=0.45\hsize]{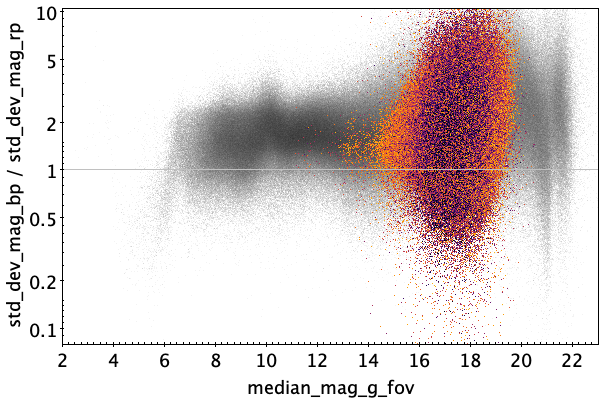}  \\ 
\vspace{4mm}
 \caption{S: 512\,005 classified sources.}  
 \label{fig:app:S}
\end{figure*}

\begin{figure*}
\centering
\stackinset{c}{-0.3cm}{c}{3cm}{(a)}{} \includegraphics[width=0.45\hsize]{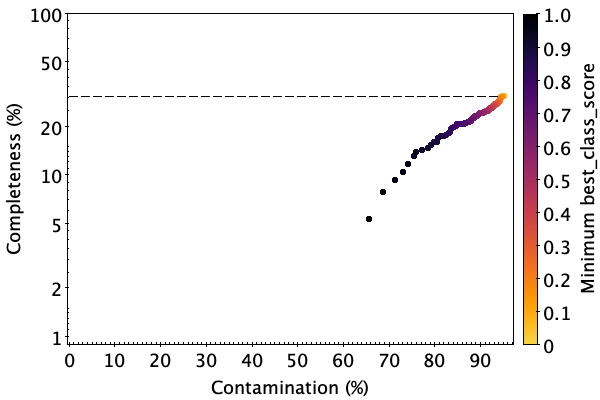}  
\hspace{2mm}
\stackinset{c}{8.8cm}{c}{3cm}{(b)}{} \includegraphics[width=0.45\hsize]{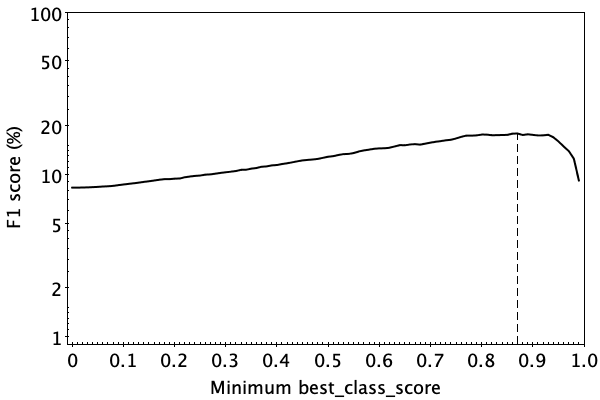} \\ 
\vspace{4mm}
\stackinset{c}{-0.3cm}{c}{3cm}{(c)}{} \includegraphics[width=0.45\hsize]{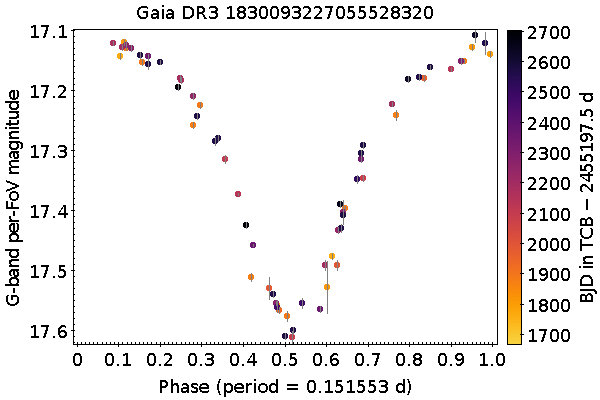}  
\hspace{2mm}
\stackinset{c}{8.8cm}{c}{3cm}{(d)}{} \includegraphics[width=0.45\hsize]{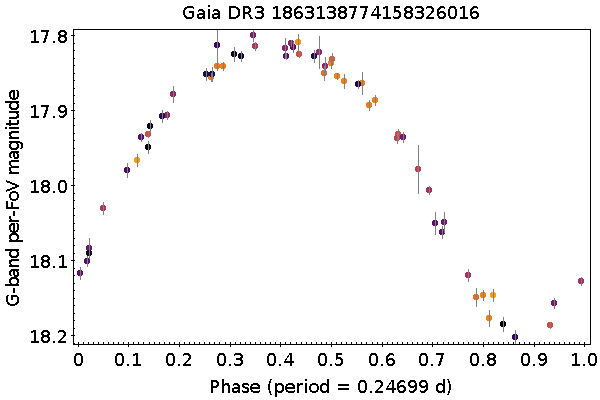} \\
\vspace{4mm}
\stackinset{c}{-0.3cm}{c}{3cm}{(e)}{} \includegraphics[width=0.45\hsize]{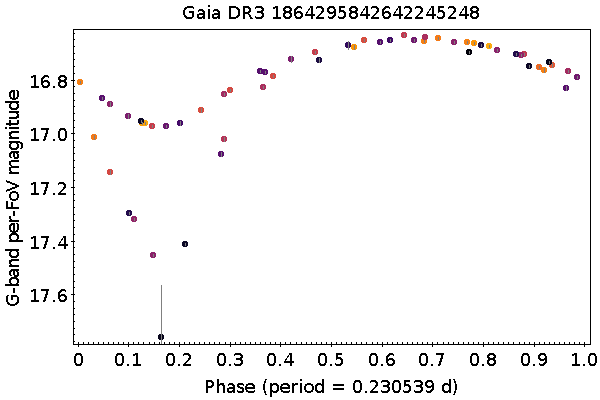}  
\hspace{2mm}
\stackinset{c}{8.8cm}{c}{3cm}{(f)}{} \includegraphics[width=0.45\hsize]{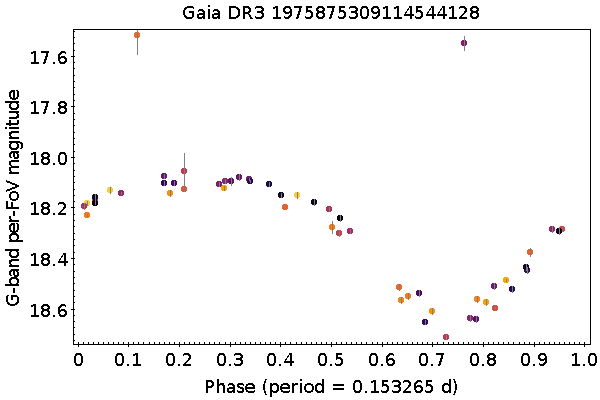} \\
\vspace{4mm}
 \caption{Same as Fig.~\ref{fig:app:ACV_cc}, but for S, including an eclipsing binary with halved period in panel (e).}
 \label{fig:app:S_cc}
\end{figure*}

\begin{figure*}
\centering
\stackinset{c}{-0.7cm}{c}{2.7cm}{(a)}{} \includegraphics[width=0.6\hsize]{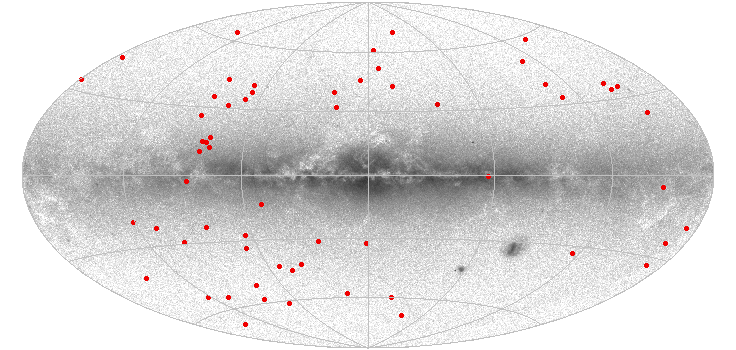} \\ 
\vspace{4mm}
\stackinset{c}{-0.3cm}{c}{3cm}{(b)}{} \includegraphics[width=0.45\hsize]{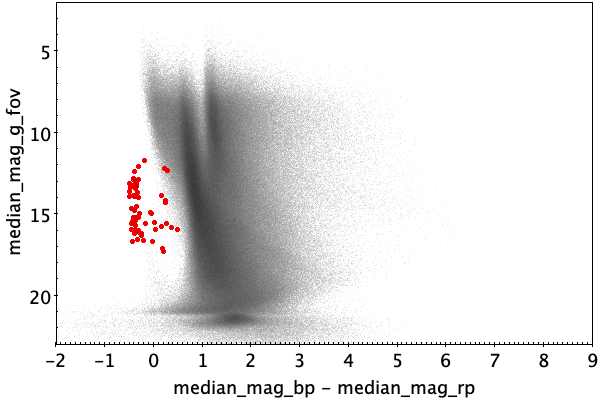}  
\hspace{2mm}
\stackinset{c}{8.8cm}{c}{3cm}{(c)}{} \includegraphics[width=0.45\hsize]{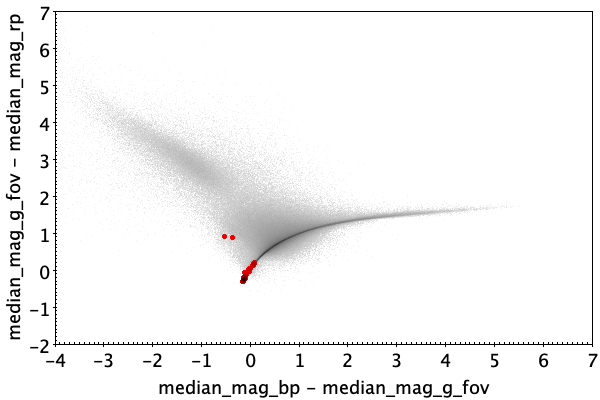} \\ 
\vspace{4mm}
\stackinset{c}{-0.3cm}{c}{3cm}{(d)}{} \includegraphics[width=0.45\hsize]{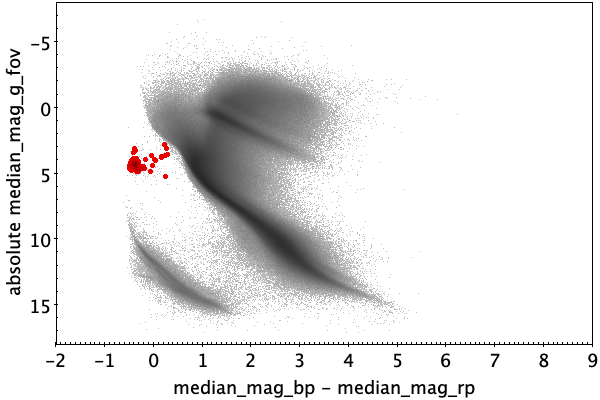}  
\hspace{2mm}
\stackinset{c}{8.8cm}{c}{3cm}{(e)}{} \includegraphics[width=0.45\hsize]{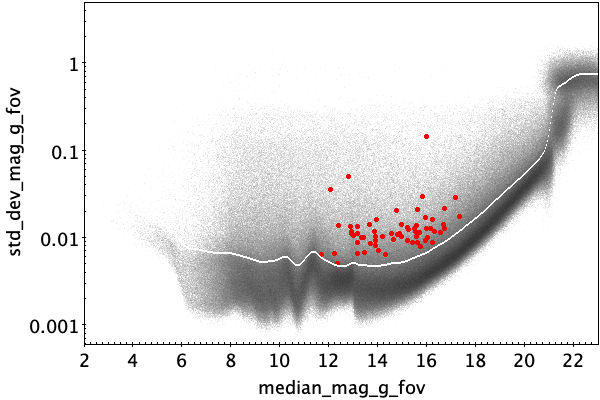} \\ 
\vspace{4mm}
\stackinset{c}{-0.3cm}{c}{3cm}{(f)}{} \includegraphics[width=0.45\hsize]{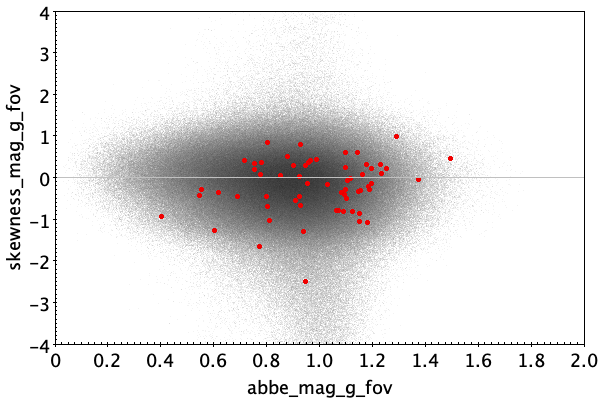}  
\hspace{2mm}
\stackinset{c}{8.8cm}{c}{3cm}{(g)}{} \includegraphics[width=0.45\hsize]{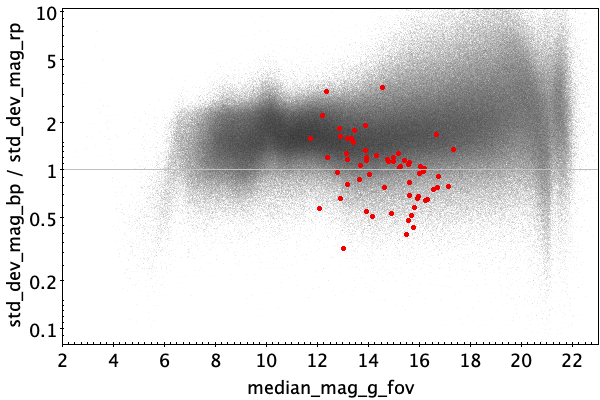}  \\ 
\vspace{4mm}
 \caption{SDB: 62 training sources.}  
 \label{fig:app:SDB_trn}
\end{figure*}

\begin{figure*}
\centering
\stackinset{c}{-0.7cm}{c}{2.7cm}{(a)}{}
\includegraphics[width=0.6\hsize]{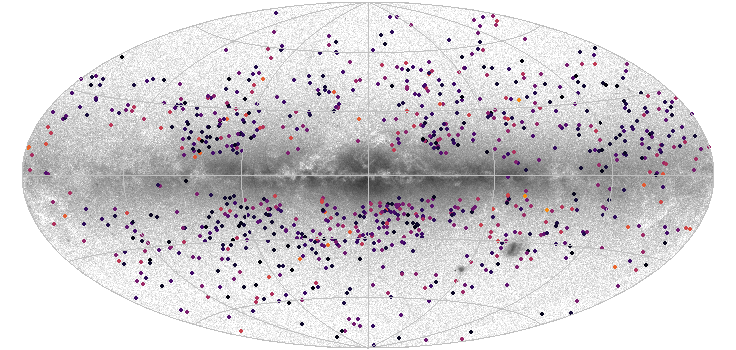} 
\stackinset{c}{2.2cm}{c}{2.7cm}{\includegraphics[height=5.5cm]{figures/appendix/vertical_best_class_score.png}}{} \\ 
\vspace{4mm}
\stackinset{c}{-0.3cm}{c}{3cm}{(b)}{} \includegraphics[width=0.45\hsize]{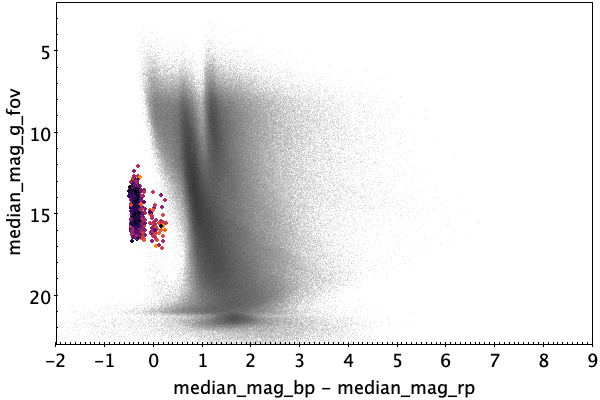}  
\hspace{2mm}
\stackinset{c}{8.8cm}{c}{3cm}{(c)}{} \includegraphics[width=0.45\hsize]{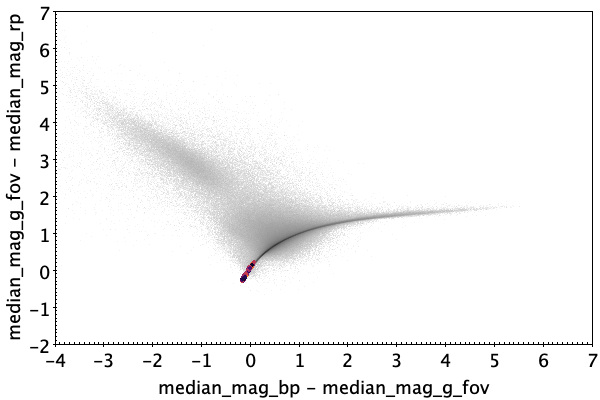} \\ 
\vspace{4mm}
\stackinset{c}{-0.3cm}{c}{3cm}{(d)}{} \includegraphics[width=0.45\hsize]{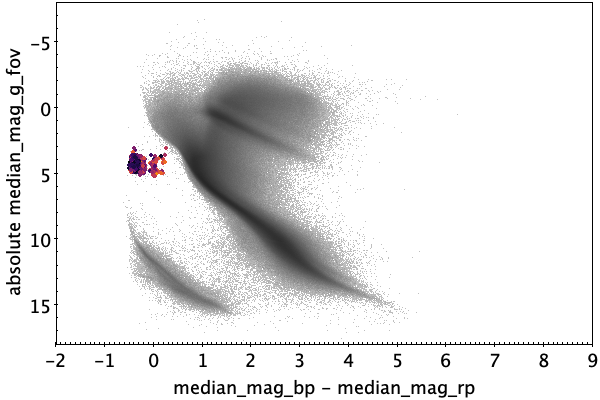}  
\hspace{2mm}
\stackinset{c}{8.8cm}{c}{3cm}{(e)}{} \includegraphics[width=0.45\hsize]{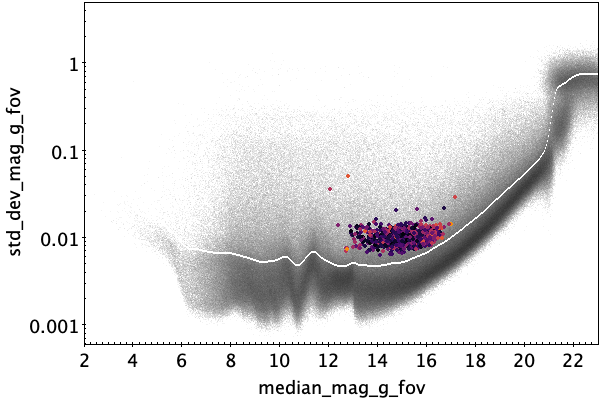} \\ 
\vspace{4mm}
\stackinset{c}{-0.3cm}{c}{3cm}{(f)}{} \includegraphics[width=0.45\hsize]{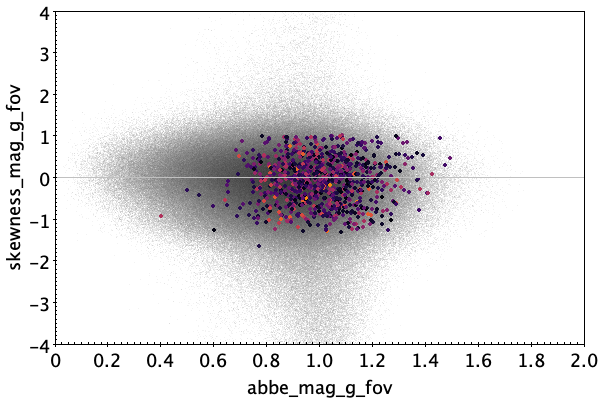}  
\hspace{2mm}
\stackinset{c}{8.8cm}{c}{3cm}{(g)}{} \includegraphics[width=0.45\hsize]{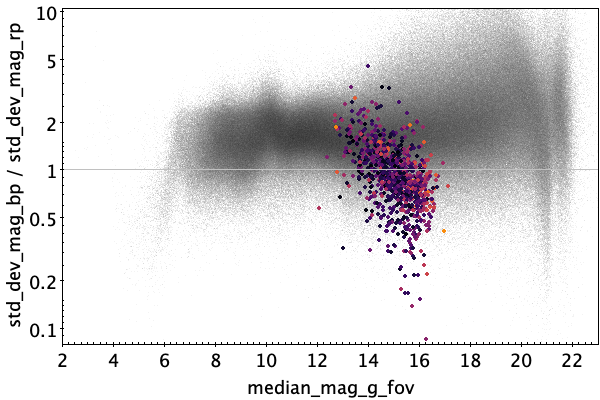}  \\ 
\vspace{4mm}
 \caption{SDB: 893 classified sources.}  
 \label{fig:app:SDB}
\end{figure*}

\begin{figure*}
\centering
\stackinset{c}{-0.3cm}{c}{3cm}{(a)}{} \includegraphics[width=0.45\hsize]{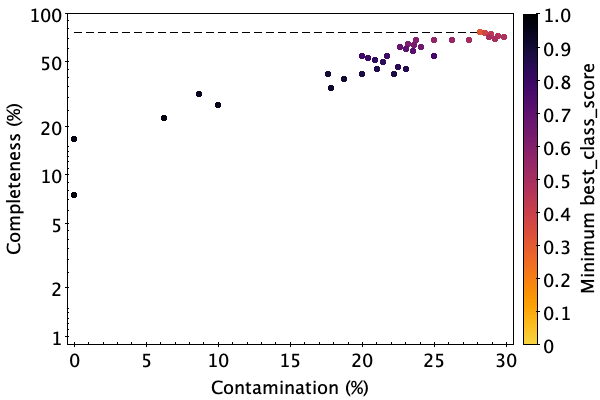}  
\hspace{2mm}
\stackinset{c}{8.8cm}{c}{3cm}{(b)}{} \includegraphics[width=0.45\hsize]{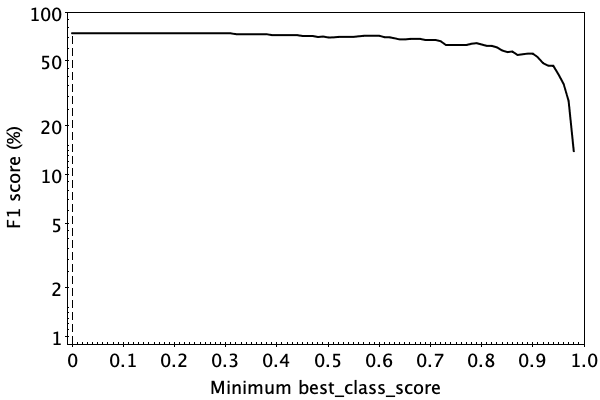} \\ 
\vspace{4mm}
\stackinset{c}{-0.3cm}{c}{3cm}{(c)}{} \includegraphics[width=0.45\hsize]{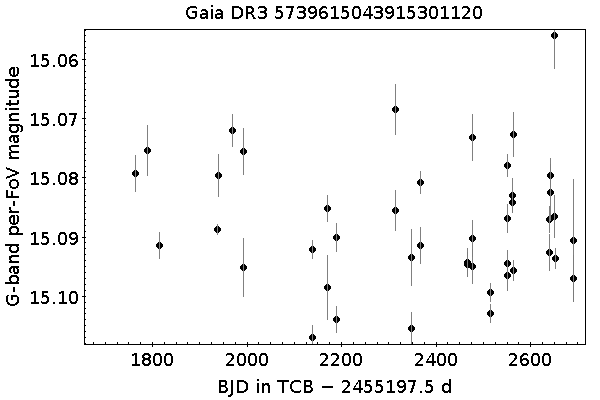}  
\hspace{2mm}
\stackinset{c}{8.8cm}{c}{3cm}{(d)}{} \includegraphics[width=0.45\hsize]{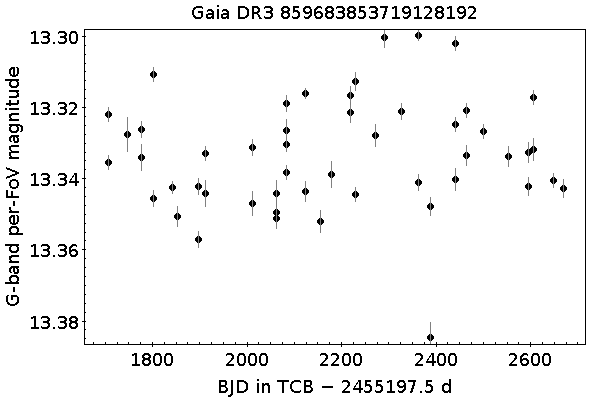} \\
\vspace{4mm}
\stackinset{c}{-0.3cm}{c}{3cm}{(e)}{} \includegraphics[width=0.45\hsize]{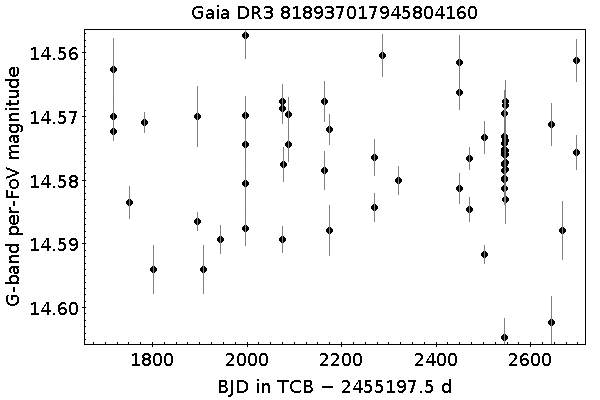}  
\hspace{2mm}
\stackinset{c}{8.8cm}{c}{3cm}{(f)}{} \includegraphics[width=0.45\hsize]{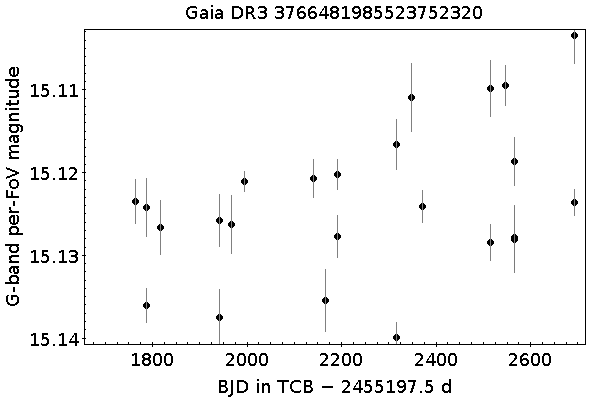} \\
\vspace{4mm}
 \caption{Same as Fig.~\ref{fig:app:ACV_cc}, but for SDB. Given the multiple significant periods of SDB stars, only time series are shown in panels (c)--(f).}
 \label{fig:app:SDB_cc}
\end{figure*}

\begin{figure*}
\centering
\stackinset{c}{-0.7cm}{c}{2.7cm}{(a)}{} \includegraphics[width=0.6\hsize]{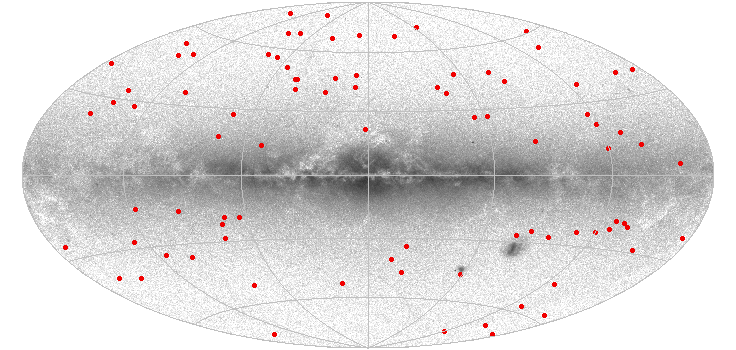} \\ 
\vspace{4mm}
\stackinset{c}{-0.3cm}{c}{3cm}{(b)}{} \includegraphics[width=0.45\hsize]{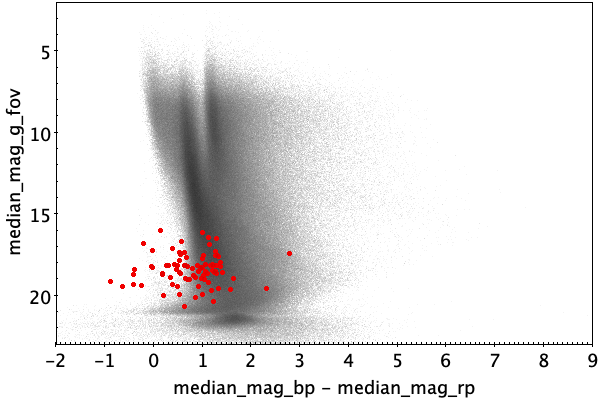}  
\hspace{2mm}
\stackinset{c}{8.8cm}{c}{3cm}{(c)}{} \includegraphics[width=0.45\hsize]{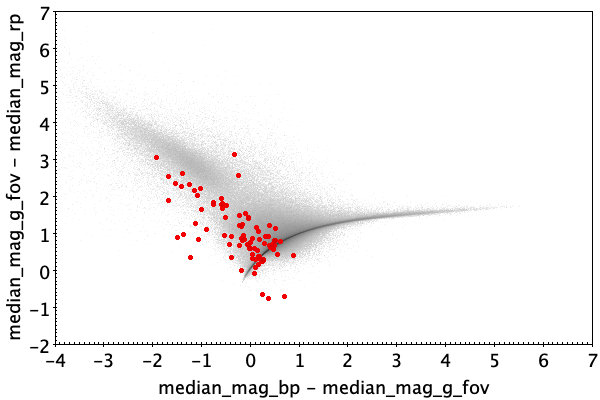} \\ 
\vspace{4mm}
\stackinset{c}{-0.3cm}{c}{3cm}{(d)}{} \includegraphics[width=0.45\hsize]{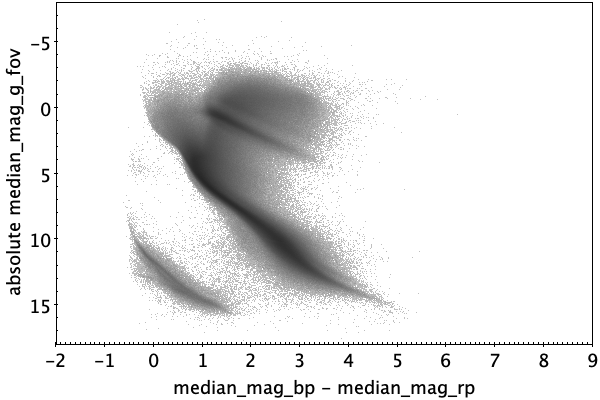}  
\hspace{2mm}
\stackinset{c}{8.8cm}{c}{3cm}{(e)}{} \includegraphics[width=0.45\hsize]{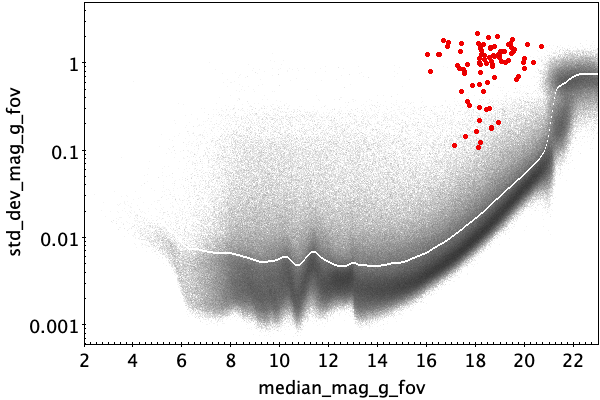} \\ 
\vspace{4mm}
\stackinset{c}{-0.3cm}{c}{3cm}{(f)}{} \includegraphics[width=0.45\hsize]{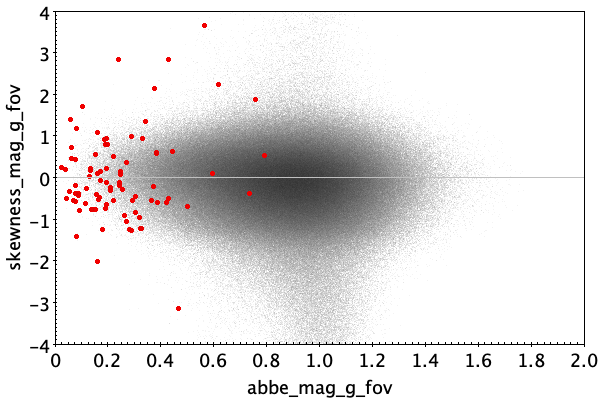}  
\hspace{2mm}
\stackinset{c}{8.8cm}{c}{3cm}{(g)}{} \includegraphics[width=0.45\hsize]{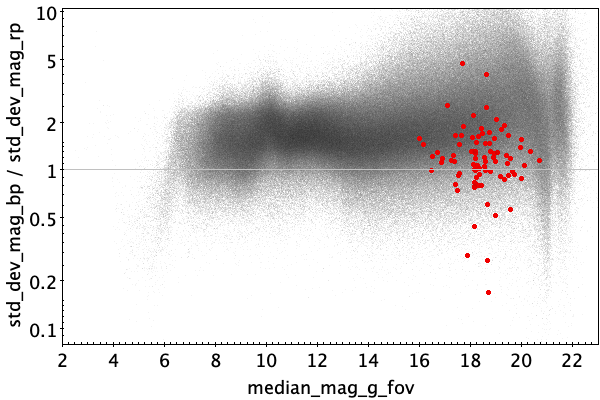}  \\ 
\vspace{4mm}
 \caption{SN: 86 training sources.}  
 \label{fig:app:SN_trn}
\end{figure*}

\begin{figure*}
\centering
\stackinset{c}{-0.7cm}{c}{2.7cm}{(a)}{}
\includegraphics[width=0.6\hsize]{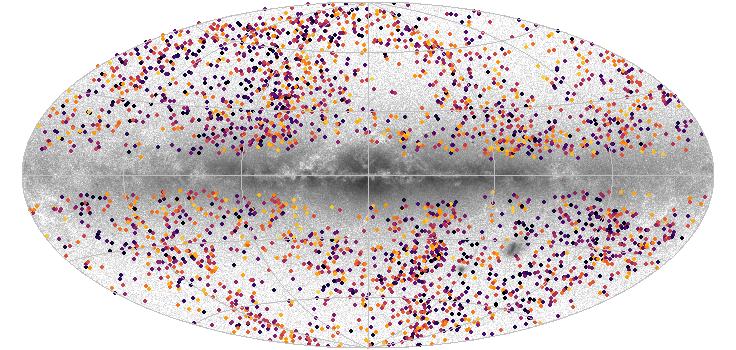} 
\stackinset{c}{2.2cm}{c}{2.7cm}{\includegraphics[height=5.5cm]{figures/appendix/vertical_best_class_score.png}}{} \\ 
\vspace{4mm}
\stackinset{c}{-0.3cm}{c}{3cm}{(b)}{} \includegraphics[width=0.45\hsize]{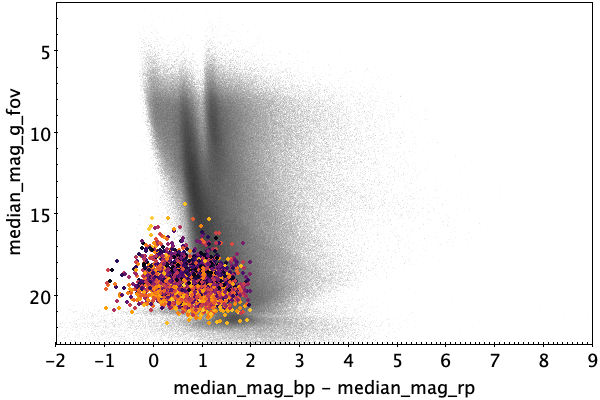}  
\hspace{2mm}
\stackinset{c}{8.8cm}{c}{3cm}{(c)}{} \includegraphics[width=0.45\hsize]{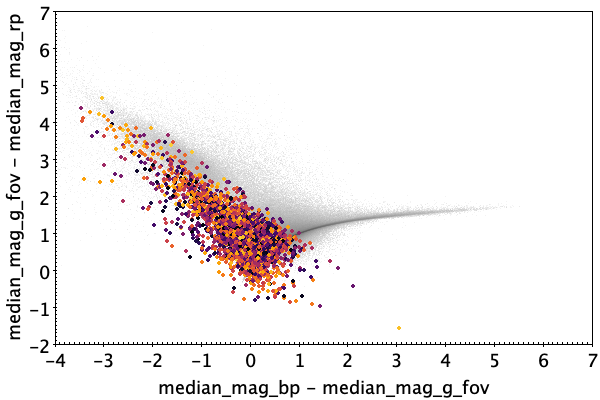} \\ 
\vspace{4mm}
\stackinset{c}{-0.3cm}{c}{3cm}{(d)}{} \includegraphics[width=0.45\hsize]{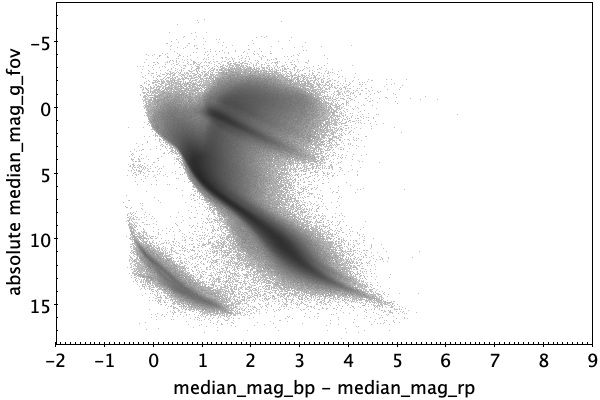}  
\hspace{2mm}
\stackinset{c}{8.8cm}{c}{3cm}{(e)}{} \includegraphics[width=0.45\hsize]{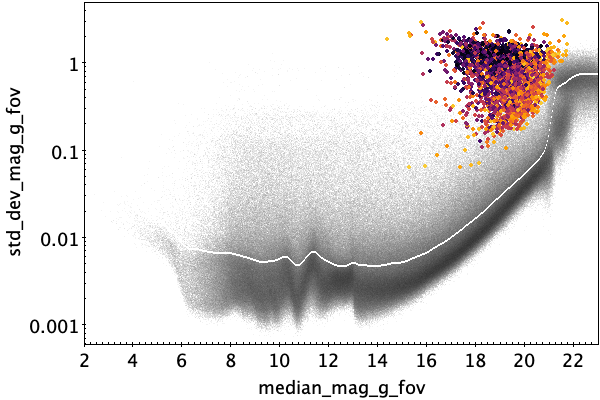} \\ 
\vspace{4mm}
\stackinset{c}{-0.3cm}{c}{3cm}{(f)}{} \includegraphics[width=0.45\hsize]{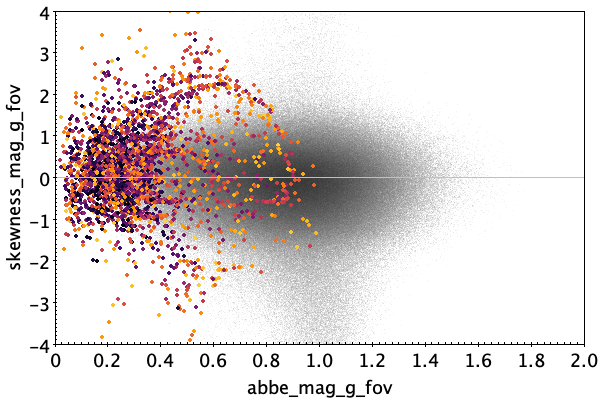}  
\hspace{2mm}
\stackinset{c}{8.8cm}{c}{3cm}{(g)}{} \includegraphics[width=0.45\hsize]{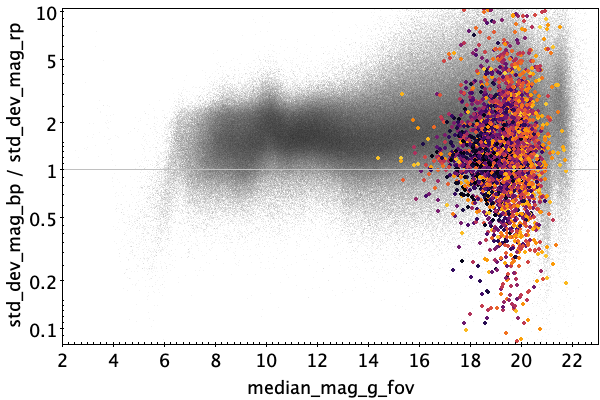}  \\ 
\vspace{4mm}
 \caption{SN: 3029 classified sources.}  
 \label{fig:app:SN}
\end{figure*}

\begin{figure*}
\centering
\stackinset{c}{-0.3cm}{c}{3cm}{(a)}{} \includegraphics[width=0.45\hsize]{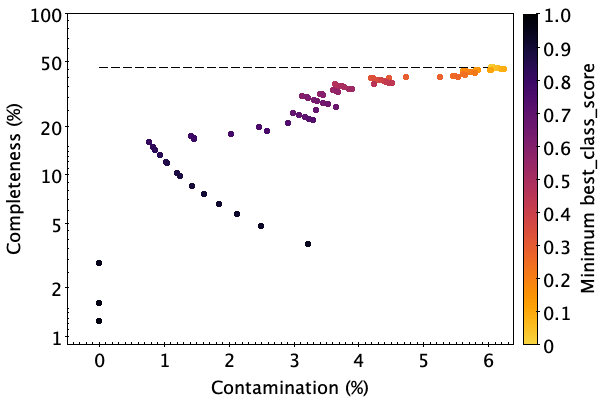}  
\hspace{2mm}
\stackinset{c}{8.8cm}{c}{3cm}{(b)}{} \includegraphics[width=0.45\hsize]{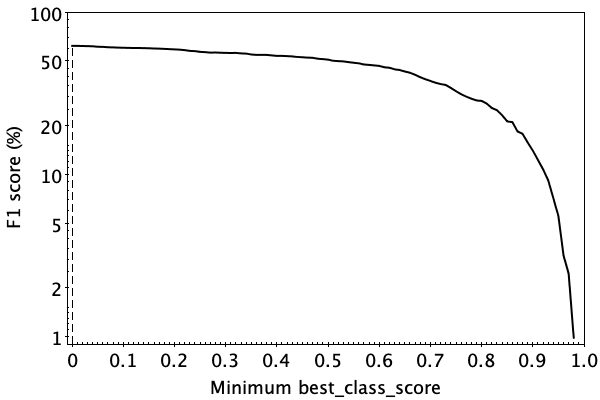} \\ 
\vspace{4mm}
\stackinset{c}{-0.3cm}{c}{3cm}{(c)}{} \includegraphics[width=0.45\hsize]{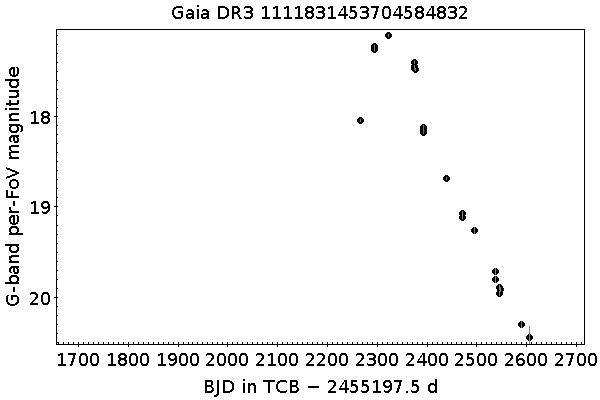}  
\hspace{2mm}
\stackinset{c}{8.8cm}{c}{3cm}{(d)}{} \includegraphics[width=0.45\hsize]{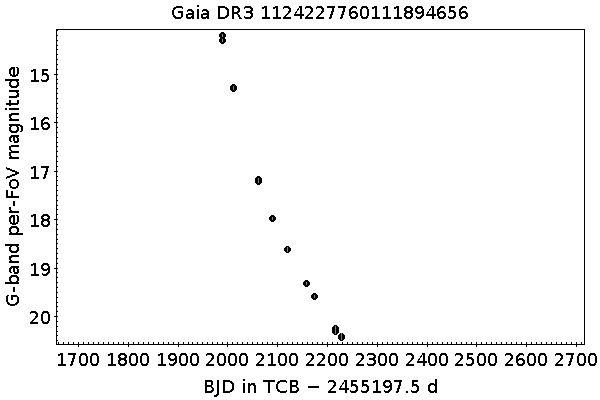} \\
\vspace{4mm}
\stackinset{c}{-0.3cm}{c}{3cm}{(e)}{} \includegraphics[width=0.45\hsize]{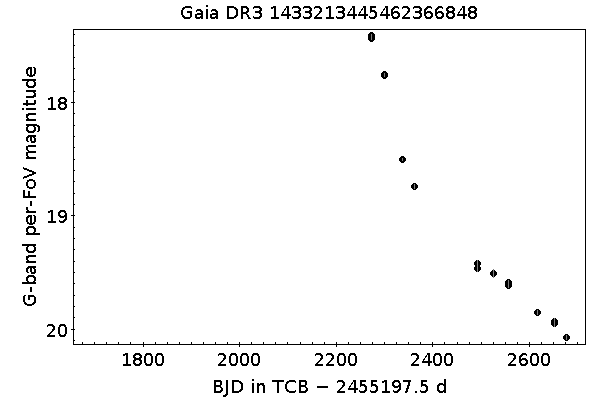}  
\hspace{2mm}
\stackinset{c}{8.8cm}{c}{3cm}{(f)}{} \includegraphics[width=0.45\hsize]{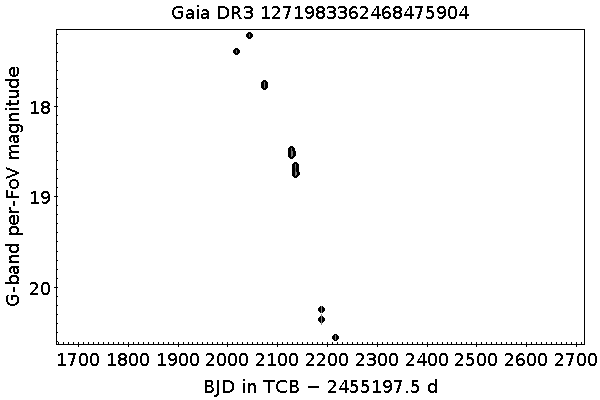} \\
\vspace{4mm}
 \caption{Same as Fig.~\ref{fig:app:ACV_cc}, but for SN.}
 \label{fig:app:SN_cc}
\end{figure*}

\begin{figure*}
\centering
\stackinset{c}{-0.7cm}{c}{2.7cm}{(a)}{} \includegraphics[width=0.6\hsize]{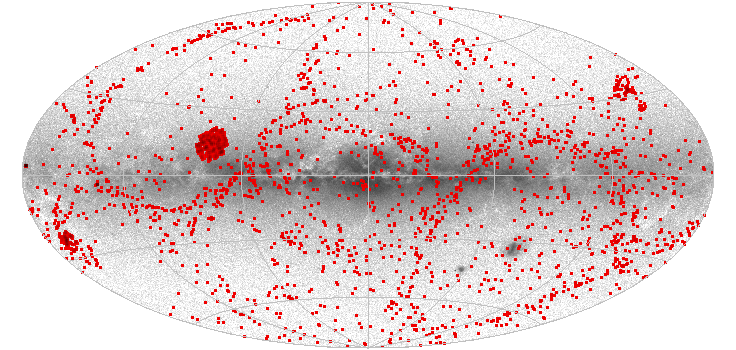} \\ 
\vspace{4mm}
\stackinset{c}{-0.3cm}{c}{3cm}{(b)}{} \includegraphics[width=0.45\hsize]{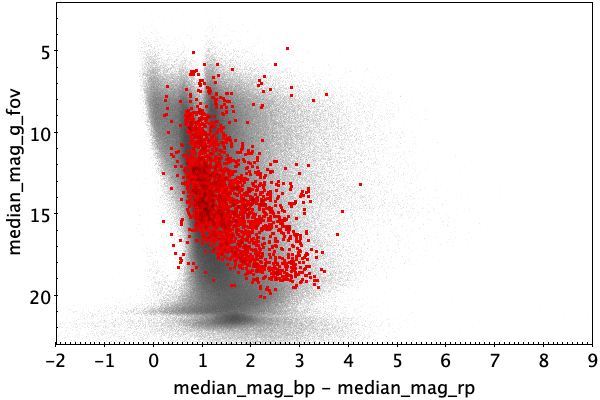}  
\hspace{2mm}
\stackinset{c}{8.8cm}{c}{3cm}{(c)}{} \includegraphics[width=0.45\hsize]{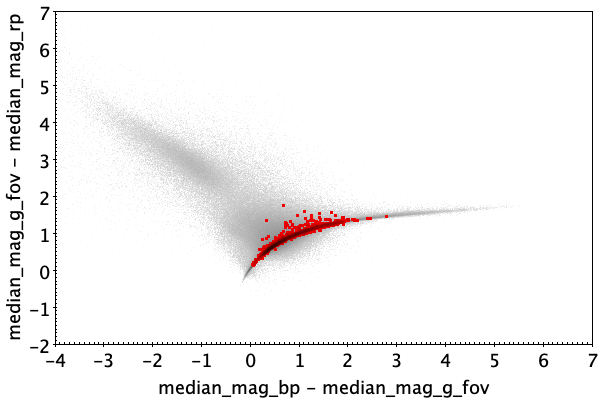} \\ 
\vspace{4mm}
\stackinset{c}{-0.3cm}{c}{3cm}{(d)}{} \includegraphics[width=0.45\hsize]{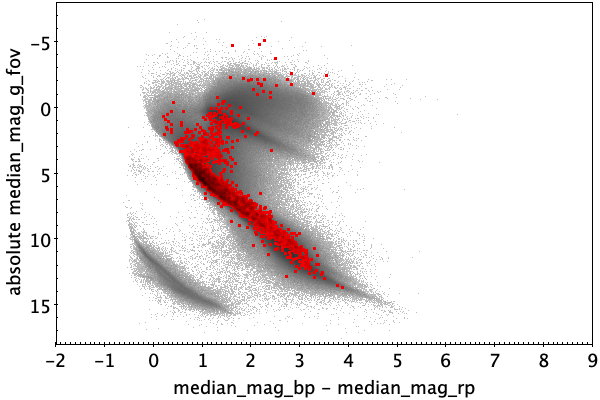}  
\hspace{2mm}
\stackinset{c}{8.8cm}{c}{3cm}{(e)}{} \includegraphics[width=0.45\hsize]{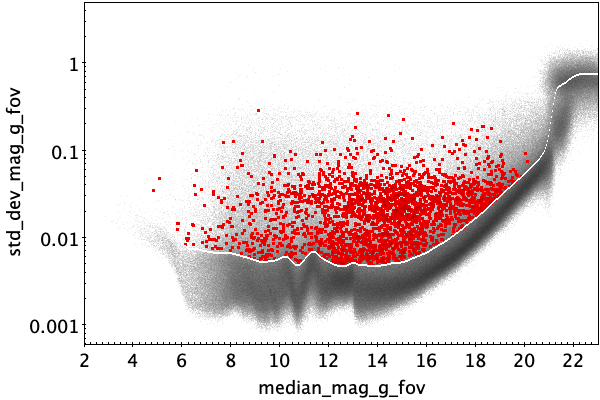} \\ 
\vspace{4mm}
\stackinset{c}{-0.3cm}{c}{3cm}{(f)}{} \includegraphics[width=0.45\hsize]{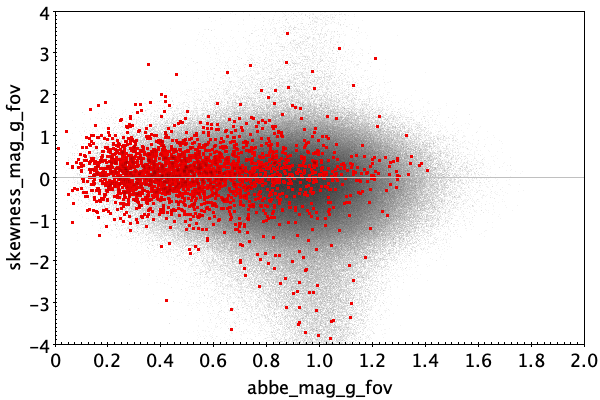}  
\hspace{2mm}
\stackinset{c}{8.8cm}{c}{3cm}{(g)}{} \includegraphics[width=0.45\hsize]{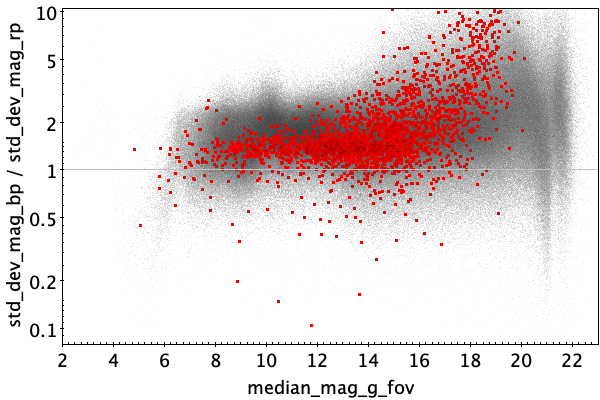}  \\ 
\vspace{4mm}
 \caption{SOLAR\_LIKE: 2628 training sources.}  
 \label{fig:app:SOLAR_trn}
\end{figure*}

\begin{figure*}
\centering
\stackinset{c}{-0.7cm}{c}{2.7cm}{(a)}{}
\includegraphics[width=0.6\hsize]{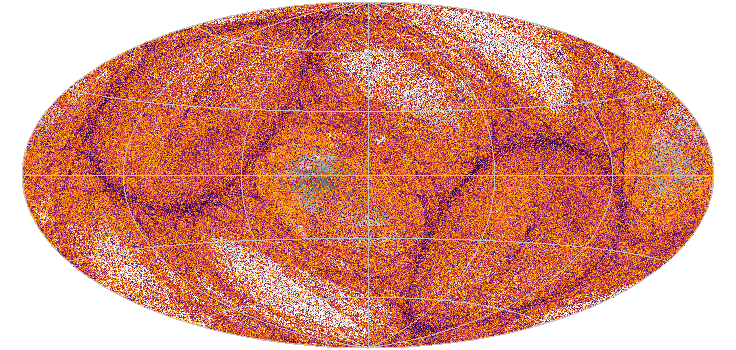} 
\stackinset{c}{2.2cm}{c}{2.7cm}{\includegraphics[height=5.5cm]{figures/appendix/vertical_best_class_score.png}}{} \\ 
\vspace{4mm}
\stackinset{c}{-0.3cm}{c}{3cm}{(b)}{} \includegraphics[width=0.45\hsize]{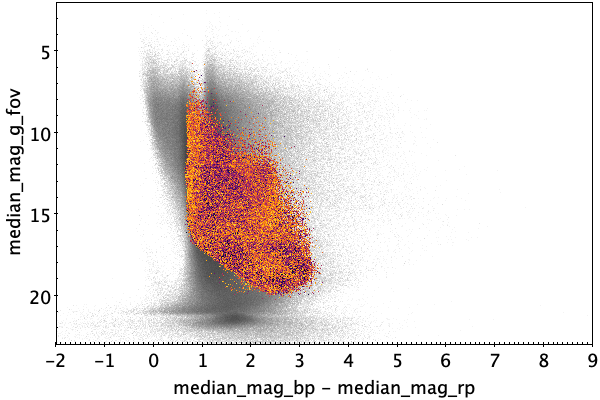}  
\hspace{2mm}
\stackinset{c}{8.8cm}{c}{3cm}{(c)}{} \includegraphics[width=0.45\hsize]{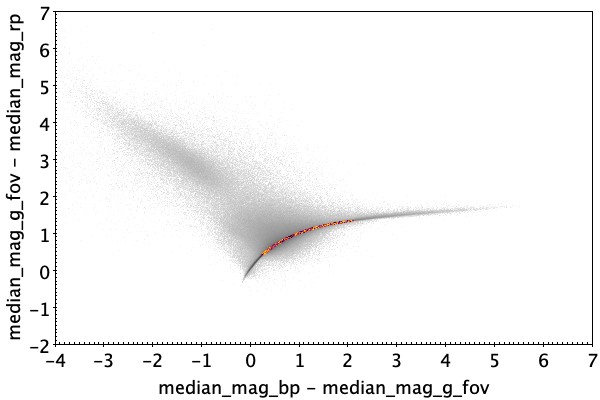} \\ 
\vspace{4mm}
\stackinset{c}{-0.3cm}{c}{3cm}{(d)}{} \includegraphics[width=0.45\hsize]{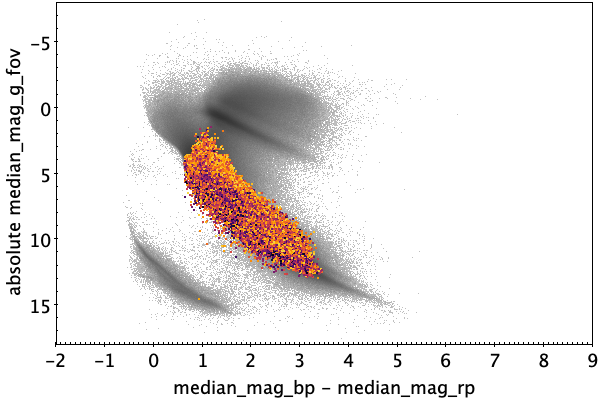}  
\hspace{2mm}
\stackinset{c}{8.8cm}{c}{3cm}{(e)}{} \includegraphics[width=0.45\hsize]{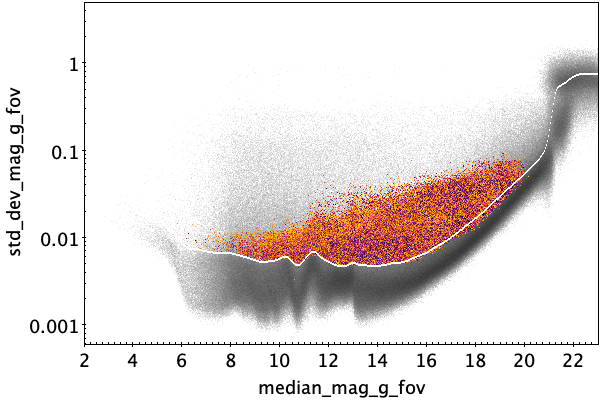} \\ 
\vspace{4mm}
\stackinset{c}{-0.3cm}{c}{3cm}{(f)}{} \includegraphics[width=0.45\hsize]{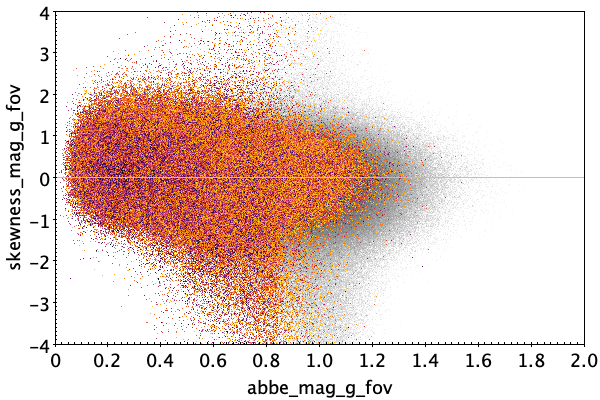}  
\hspace{2mm}
\stackinset{c}{8.8cm}{c}{3cm}{(g)}{} \includegraphics[width=0.45\hsize]{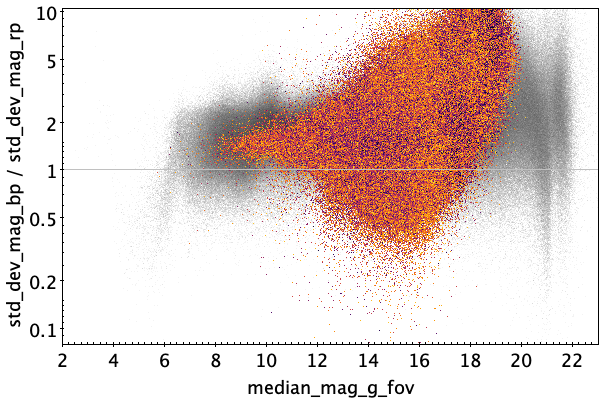}  \\ 
\vspace{4mm}
 \caption{SOLAR\_LIKE: 1\,934\,844 classified sources.}  
 \label{fig:app:SOLAR}
\end{figure*}

\begin{figure*}
\centering
\stackinset{c}{-0.3cm}{c}{3cm}{(a)}{} \includegraphics[width=0.45\hsize]{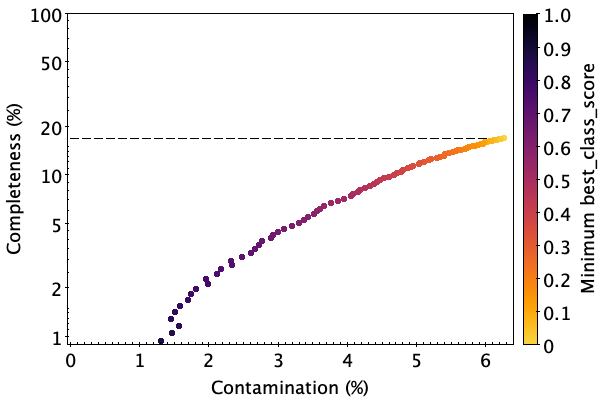}  
\hspace{2mm}
\stackinset{c}{8.8cm}{c}{3cm}{(b)}{} \includegraphics[width=0.45\hsize]{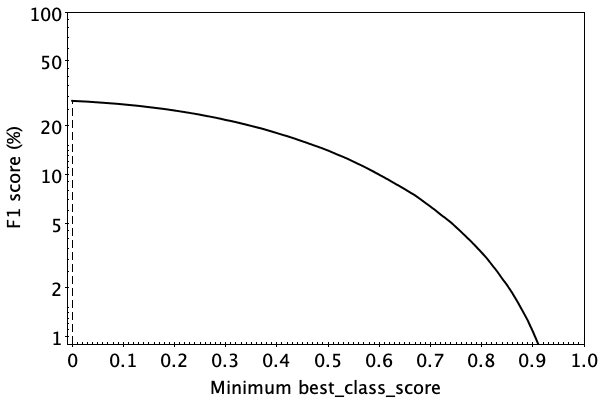} \\ 
\vspace{4mm}
\stackinset{c}{-0.3cm}{c}{3cm}{(c)}{} \includegraphics[width=0.45\hsize]{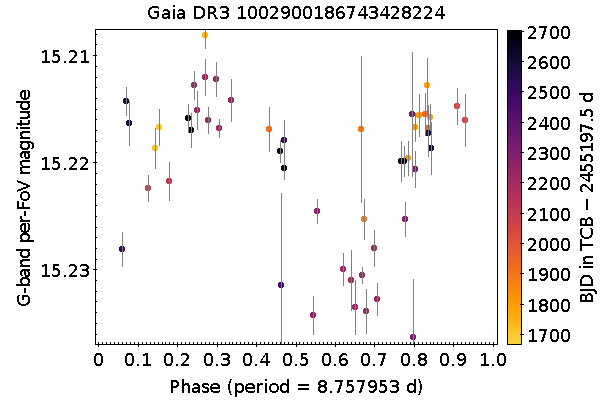}  
\hspace{2mm}
\stackinset{c}{8.8cm}{c}{3cm}{(d)}{} \includegraphics[width=0.45\hsize]{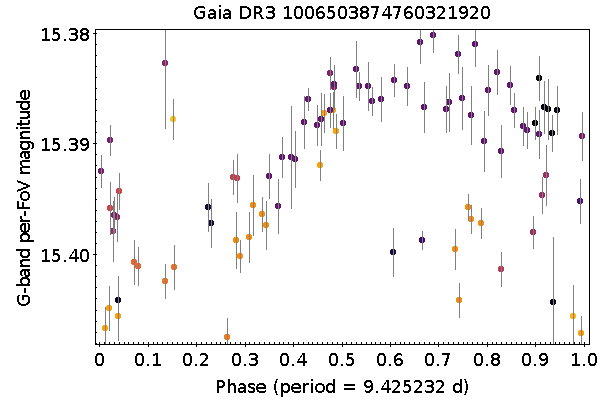} \\
\vspace{4mm}
\stackinset{c}{-0.3cm}{c}{3cm}{(e)}{} \includegraphics[width=0.45\hsize]{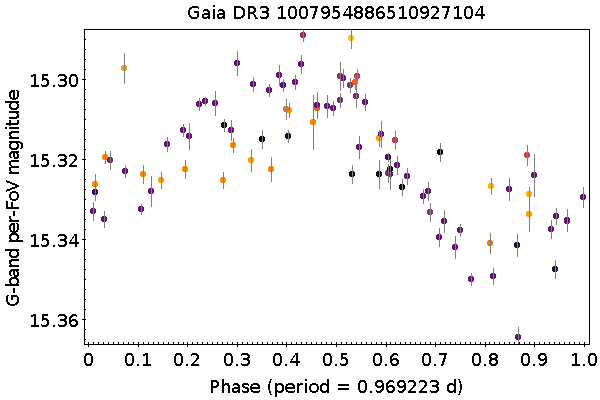}  
\hspace{2mm}
\stackinset{c}{8.8cm}{c}{3cm}{(f)}{} \includegraphics[width=0.45\hsize]{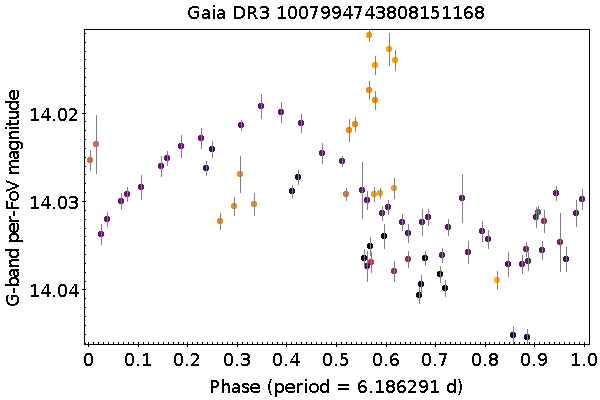} \\
\vspace{4mm}
 \caption{Same as Fig.~\ref{fig:app:ACV_cc}, but for SOLAR\_LIKE.}
 \label{fig:app:SOLAR_LIKE_cc}
\end{figure*}

\begin{figure*}
\centering
\stackinset{c}{-0.7cm}{c}{2.7cm}{(a)}{} \includegraphics[width=0.6\hsize]{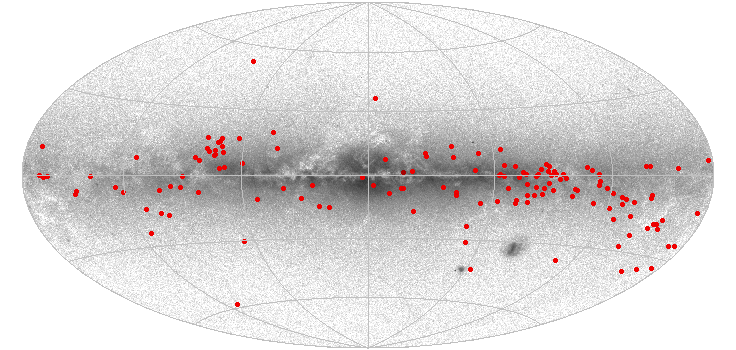} \\ 
\vspace{4mm}
\stackinset{c}{-0.3cm}{c}{3cm}{(b)}{} \includegraphics[width=0.45\hsize]{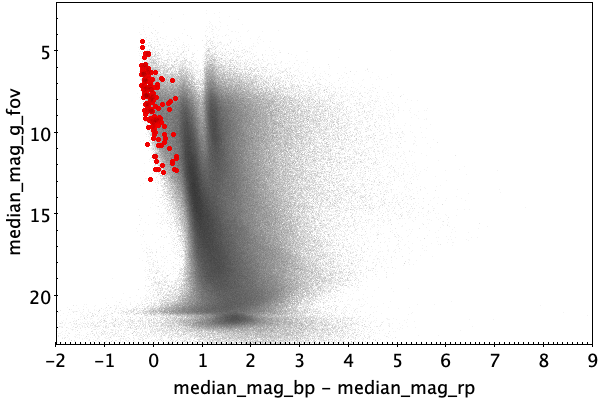}  
\hspace{2mm}
\stackinset{c}{8.8cm}{c}{3cm}{(c)}{} \includegraphics[width=0.45\hsize]{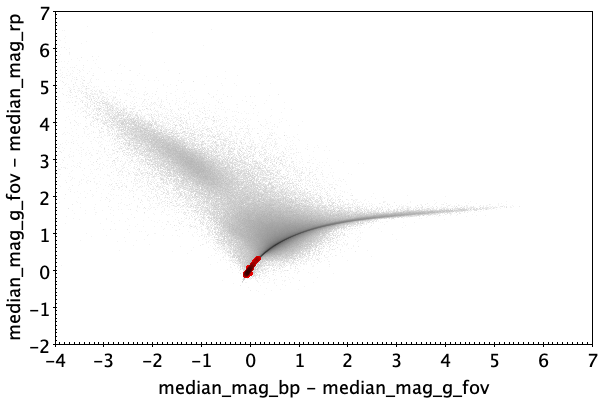} \\ 
\vspace{4mm}
\stackinset{c}{-0.3cm}{c}{3cm}{(d)}{} \includegraphics[width=0.45\hsize]{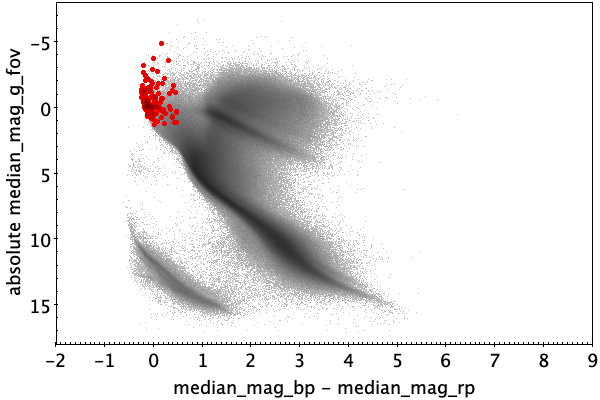}  
\hspace{2mm}
\stackinset{c}{8.8cm}{c}{3cm}{(e)}{} \includegraphics[width=0.45\hsize]{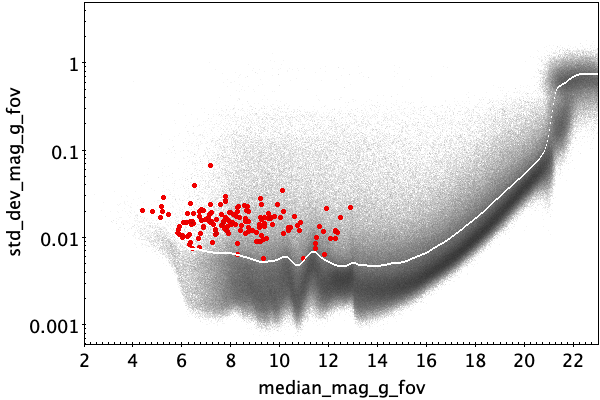} \\ 
\vspace{4mm}
\stackinset{c}{-0.3cm}{c}{3cm}{(f)}{} \includegraphics[width=0.45\hsize]{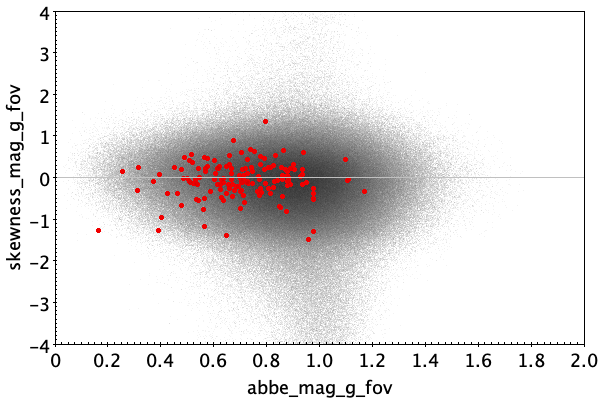}  
\hspace{2mm}
\stackinset{c}{8.8cm}{c}{3cm}{(g)}{} \includegraphics[width=0.45\hsize]{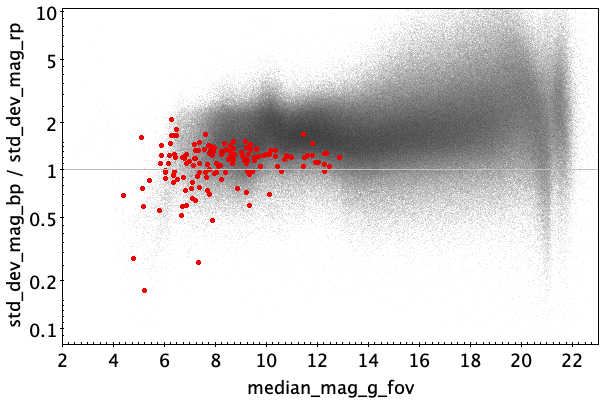}  \\ 
\vspace{4mm}
 \caption{SPB: 149 training sources.}  
 \label{fig:app:SPB_trn}
\end{figure*}

\begin{figure*}
\centering
\stackinset{c}{-0.7cm}{c}{2.7cm}{(a)}{}
\includegraphics[width=0.6\hsize]{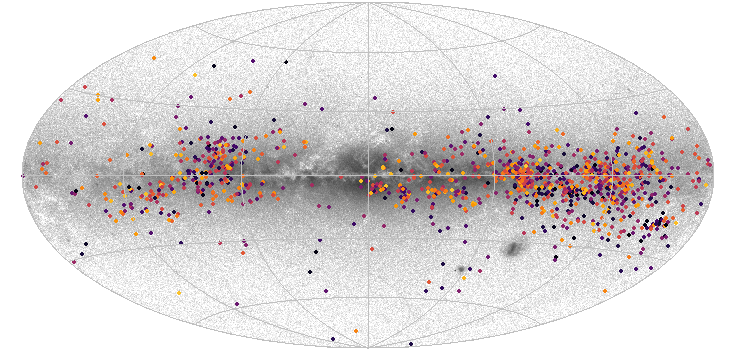} 
\stackinset{c}{2.2cm}{c}{2.7cm}{\includegraphics[height=5.5cm]{figures/appendix/vertical_best_class_score.png}}{} \\ 
\vspace{4mm}
\stackinset{c}{-0.3cm}{c}{3cm}{(b)}{} \includegraphics[width=0.45\hsize]{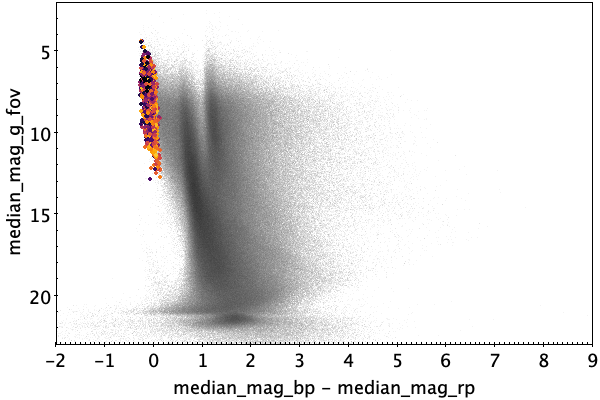}  
\hspace{2mm}
\stackinset{c}{8.8cm}{c}{3cm}{(c)}{} \includegraphics[width=0.45\hsize]{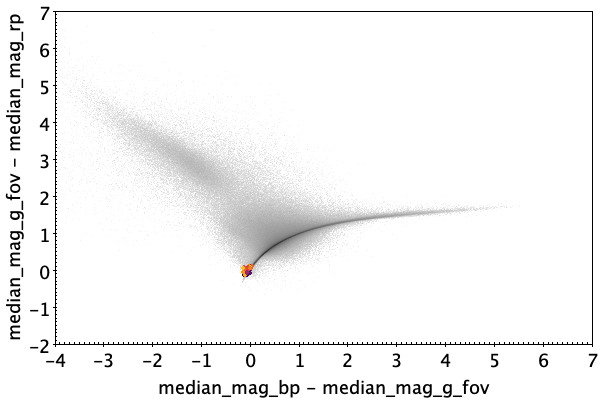} \\ 
\vspace{4mm}
\stackinset{c}{-0.3cm}{c}{3cm}{(d)}{} \includegraphics[width=0.45\hsize]{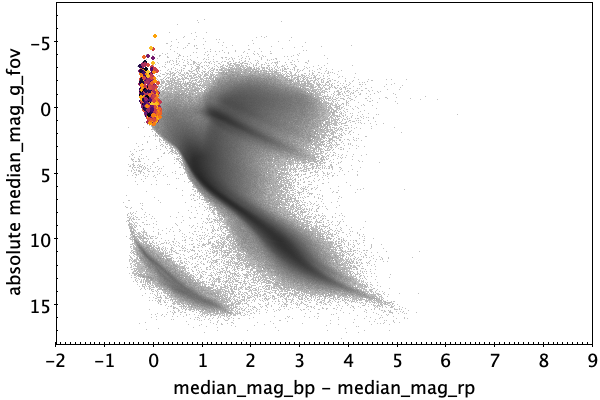}  
\hspace{2mm}
\stackinset{c}{8.8cm}{c}{3cm}{(e)}{} \includegraphics[width=0.45\hsize]{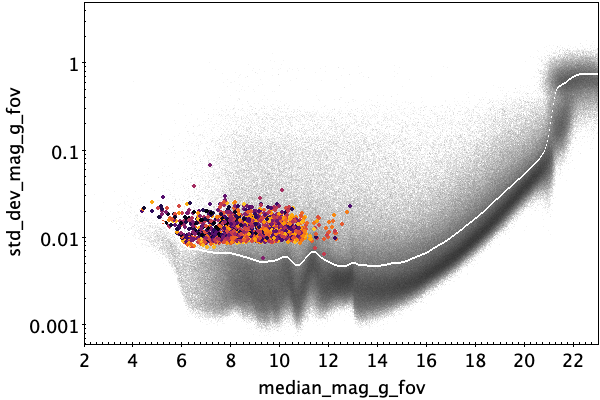} \\ 
\vspace{4mm}
\stackinset{c}{-0.3cm}{c}{3cm}{(f)}{} \includegraphics[width=0.45\hsize]{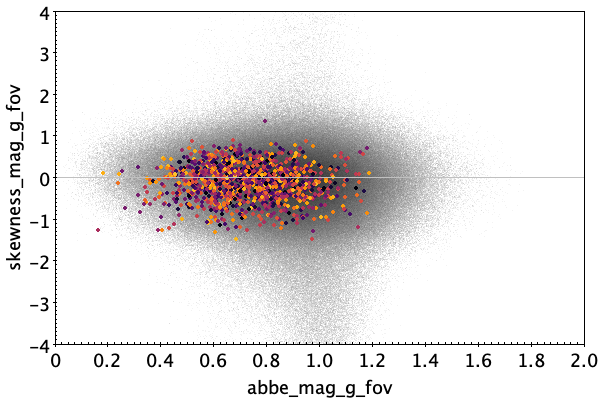}  
\hspace{2mm}
\stackinset{c}{8.8cm}{c}{3cm}{(g)}{} \includegraphics[width=0.45\hsize]{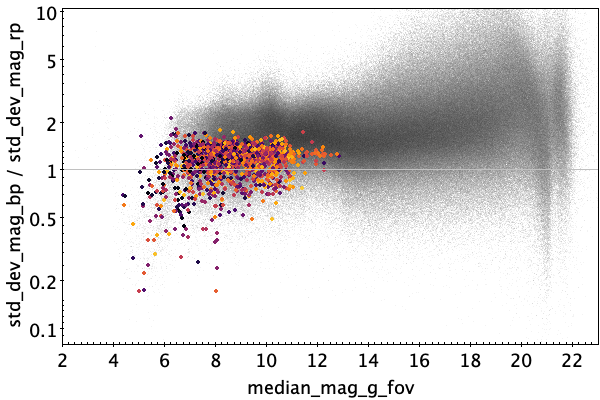}  \\ 
\vspace{4mm}
 \caption{SPB: 1228 classified sources.}  
 \label{fig:app:SPB}
\end{figure*}

\begin{figure*}
\centering
\stackinset{c}{-0.3cm}{c}{3cm}{(a)}{} \includegraphics[width=0.45\hsize]{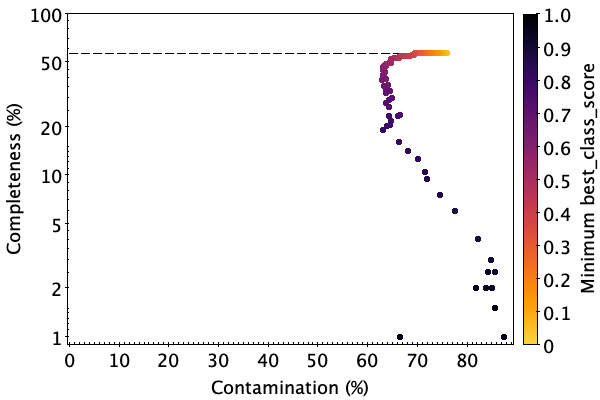}  
\hspace{2mm}
\stackinset{c}{8.8cm}{c}{3cm}{(b)}{} \includegraphics[width=0.45\hsize]{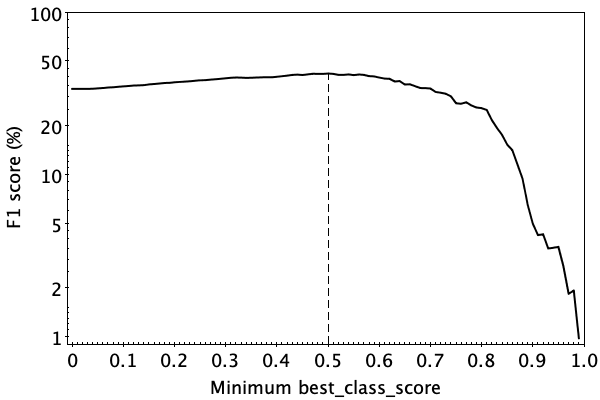} \\ 
\vspace{4mm}
\stackinset{c}{-0.3cm}{c}{3cm}{(c)}{} \includegraphics[width=0.45\hsize]{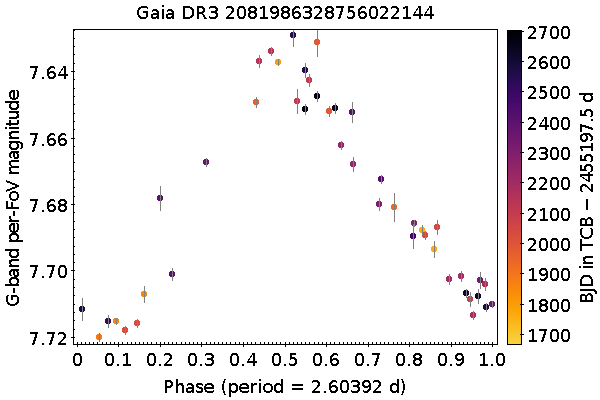}  
\hspace{2mm}
\stackinset{c}{8.8cm}{c}{3cm}{(d)}{} \includegraphics[width=0.45\hsize]{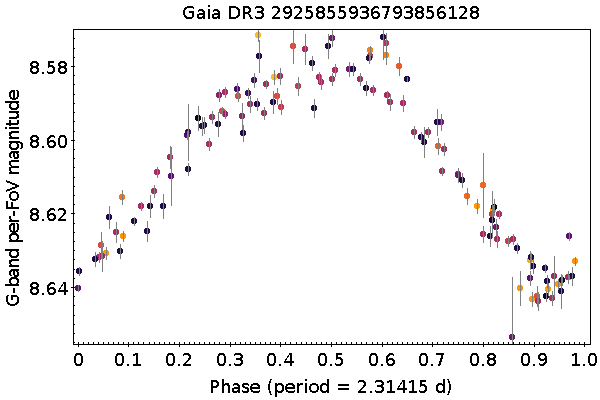} \\
\vspace{4mm}
\stackinset{c}{-0.3cm}{c}{3cm}{(e)}{} \includegraphics[width=0.45\hsize]{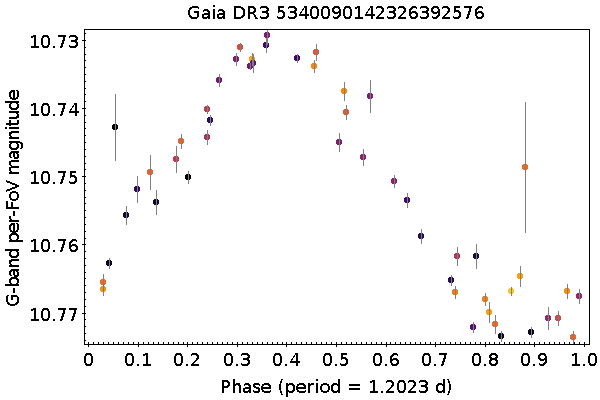}  
\hspace{2mm}
\stackinset{c}{8.8cm}{c}{3cm}{(f)}{} \includegraphics[width=0.45\hsize]{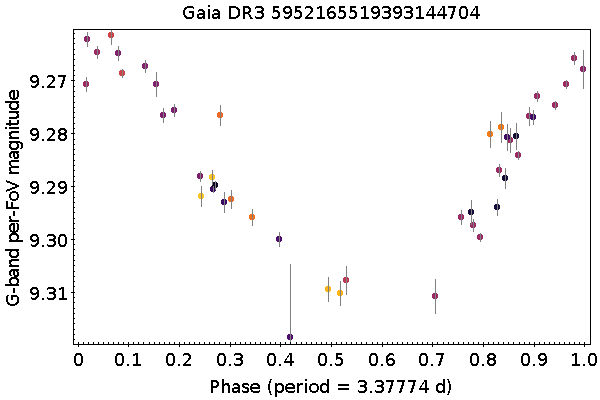} \\
\vspace{4mm}
 \caption{Same as Fig.~\ref{fig:app:ACV_cc}, but for SPB.}
 \label{fig:app:SPB_cc}
\end{figure*}

\begin{figure*}
\centering
\stackinset{c}{-0.7cm}{c}{2.7cm}{(a)}{} \includegraphics[width=0.6\hsize]{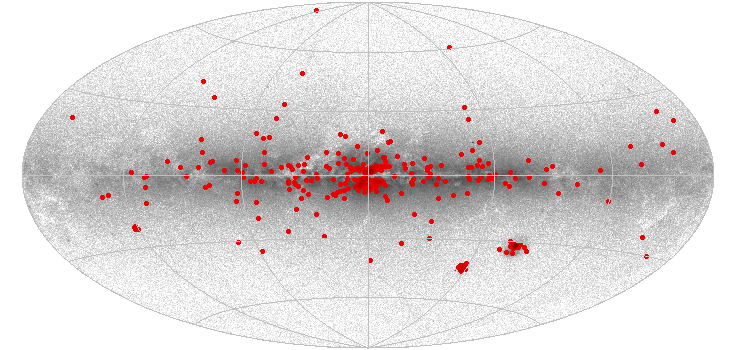} \\ 
\vspace{4mm}
\stackinset{c}{-0.3cm}{c}{3cm}{(b)}{} \includegraphics[width=0.45\hsize]{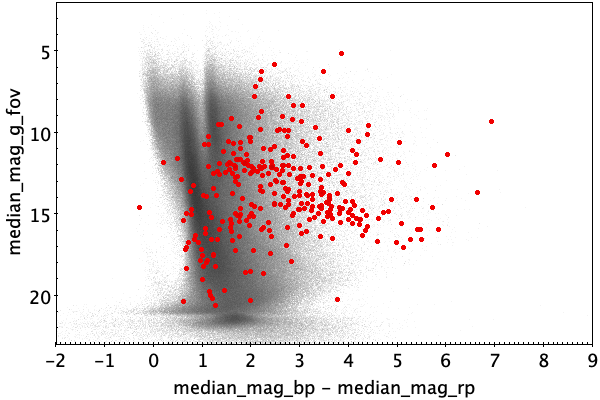}  
\hspace{2mm}
\stackinset{c}{8.8cm}{c}{3cm}{(c)}{} \includegraphics[width=0.45\hsize]{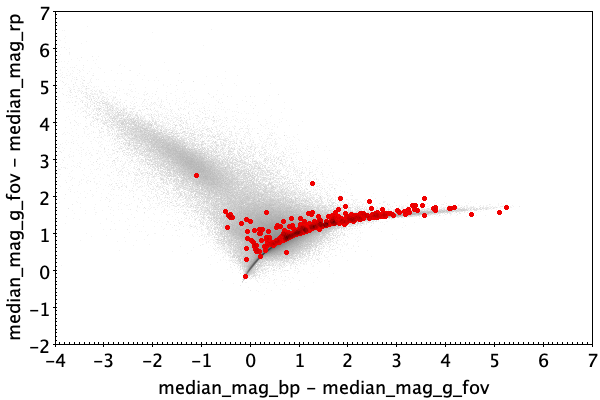} \\ 
\vspace{4mm}
\stackinset{c}{-0.3cm}{c}{3cm}{(d)}{} \includegraphics[width=0.45\hsize]{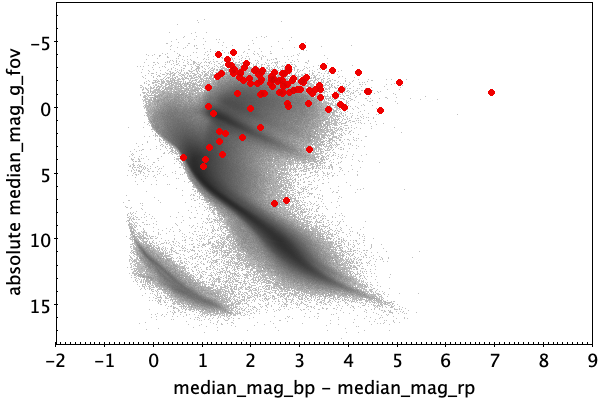}  
\hspace{2mm}
\stackinset{c}{8.8cm}{c}{3cm}{(e)}{} \includegraphics[width=0.45\hsize]{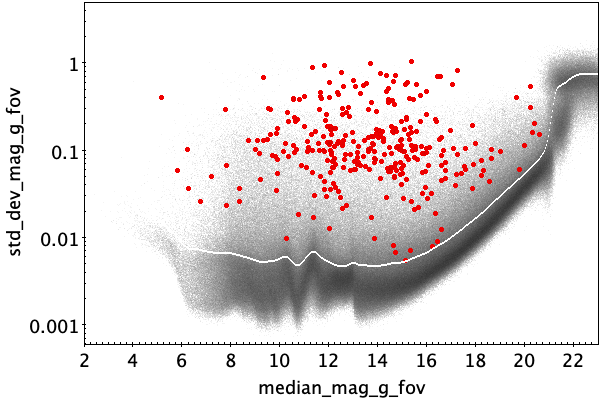} \\ 
\vspace{4mm}
\stackinset{c}{-0.3cm}{c}{3cm}{(f)}{} \includegraphics[width=0.45\hsize]{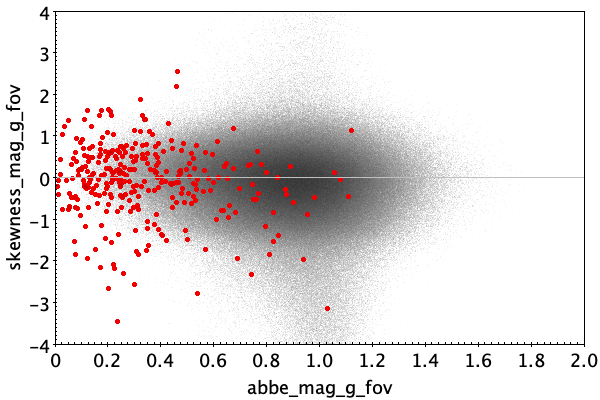}  
\hspace{2mm}
\stackinset{c}{8.8cm}{c}{3cm}{(g)}{} \includegraphics[width=0.45\hsize]{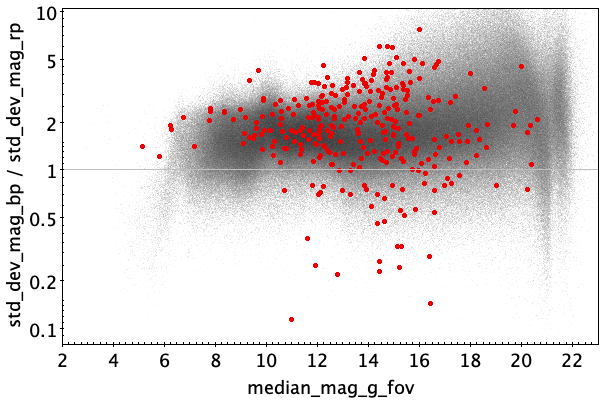}  \\ 
\vspace{4mm}
 \caption{SYST: 316 training sources.}  
 \label{fig:app:SYST_trn}
\end{figure*}

\begin{figure*}
\centering
\stackinset{c}{-0.7cm}{c}{2.7cm}{(a)}{}
\includegraphics[width=0.6\hsize]{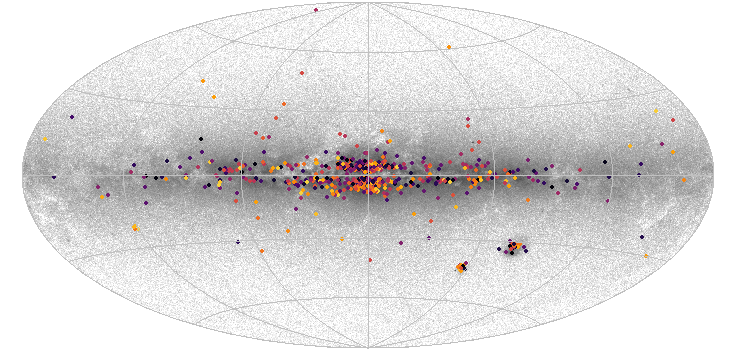} 
\stackinset{c}{2.2cm}{c}{2.7cm}{\includegraphics[height=5.5cm]{figures/appendix/vertical_best_class_score.png}}{} \\ 
\vspace{4mm}
\stackinset{c}{-0.3cm}{c}{3cm}{(b)}{} \includegraphics[width=0.45\hsize]{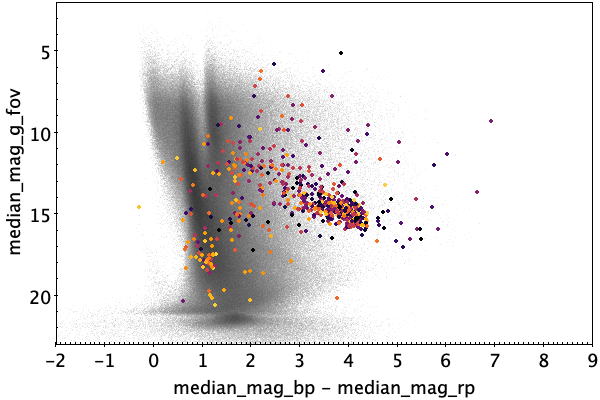}  
\hspace{2mm}
\stackinset{c}{8.8cm}{c}{3cm}{(c)}{} \includegraphics[width=0.45\hsize]{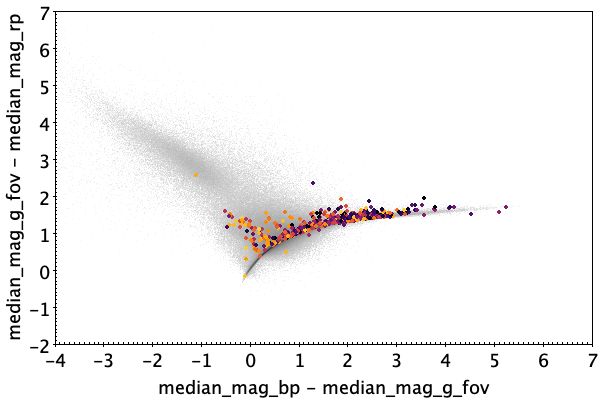} \\ 
\vspace{4mm}
\stackinset{c}{-0.3cm}{c}{3cm}{(d)}{} \includegraphics[width=0.45\hsize]{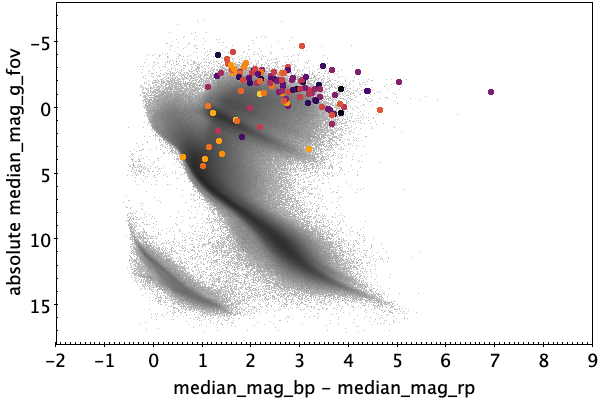}  
\hspace{2mm}
\stackinset{c}{8.8cm}{c}{3cm}{(e)}{} \includegraphics[width=0.45\hsize]{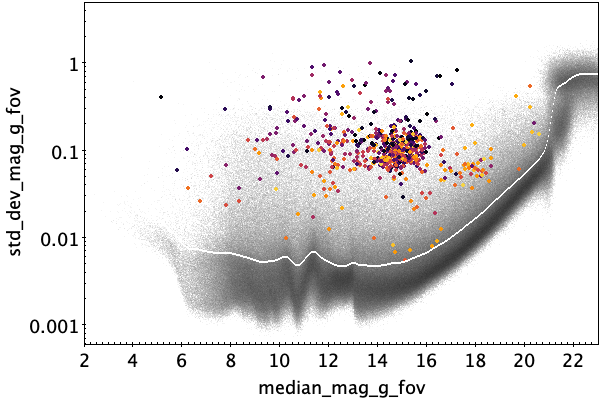} \\ 
\vspace{4mm}
\stackinset{c}{-0.3cm}{c}{3cm}{(f)}{} \includegraphics[width=0.45\hsize]{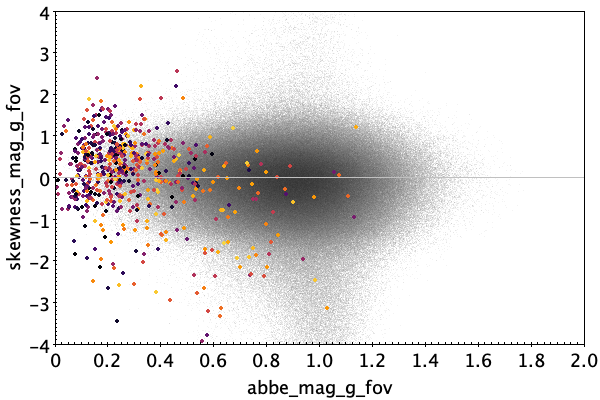}  
\hspace{2mm}
\stackinset{c}{8.8cm}{c}{3cm}{(g)}{} \includegraphics[width=0.45\hsize]{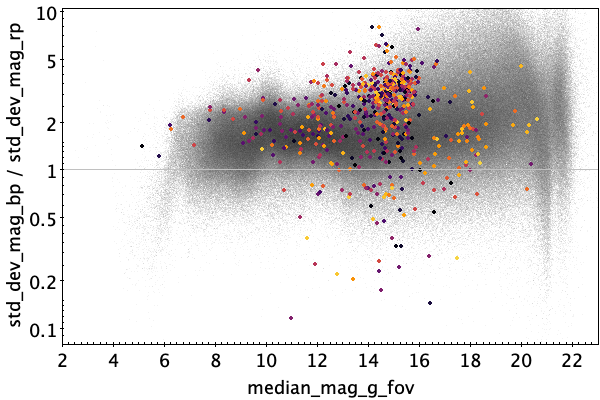}  \\ 
\vspace{4mm}
 \caption{SYST: 649 classified sources.}  
 \label{fig:app:SYST}
\end{figure*}

\begin{figure*}
\centering
\stackinset{c}{-0.3cm}{c}{3cm}{(a)}{} \includegraphics[width=0.45\hsize]{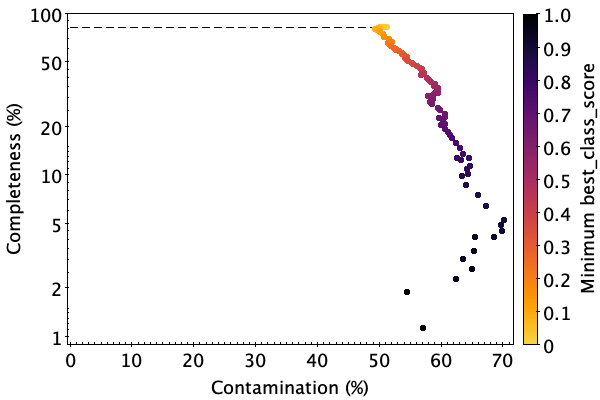}  
\hspace{2mm}
\stackinset{c}{8.8cm}{c}{3cm}{(b)}{} \includegraphics[width=0.45\hsize]{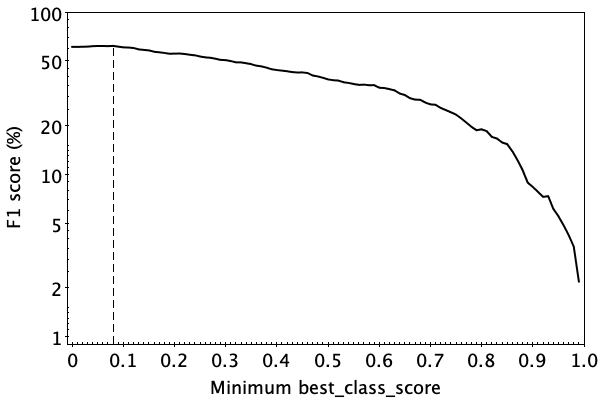} \\ 
\vspace{4mm}
\stackinset{c}{-0.3cm}{c}{3cm}{(c)}{} \includegraphics[width=0.45\hsize]{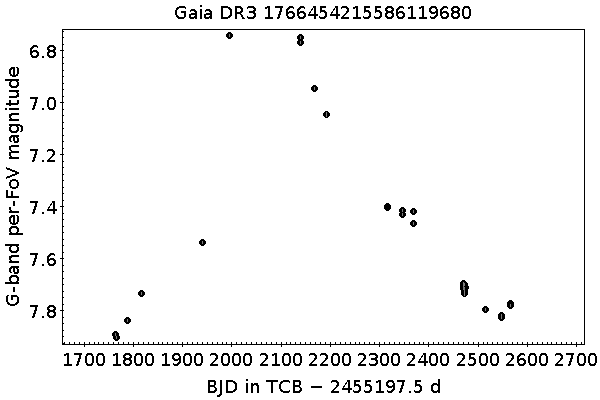}  
\hspace{2mm}
\stackinset{c}{8.8cm}{c}{3cm}{(d)}{} \includegraphics[width=0.45\hsize]{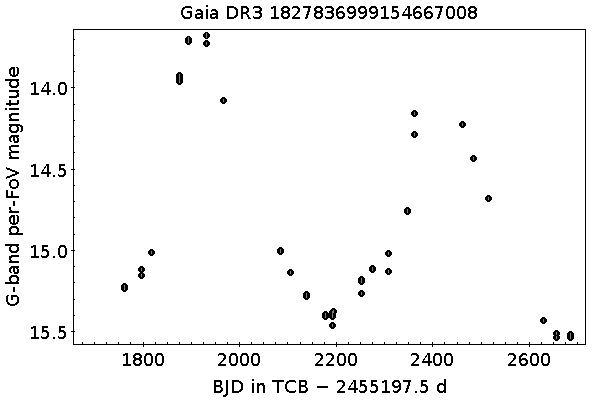} \\
\vspace{4mm}
\stackinset{c}{-0.3cm}{c}{3cm}{(e)}{} \includegraphics[width=0.45\hsize]{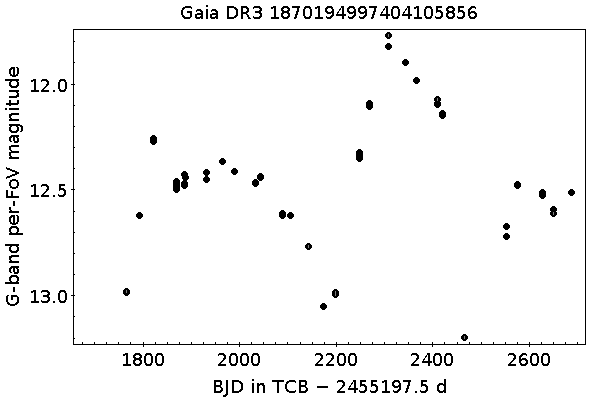}  
\hspace{2mm}
\stackinset{c}{8.8cm}{c}{3cm}{(f)}{} \includegraphics[width=0.45\hsize]{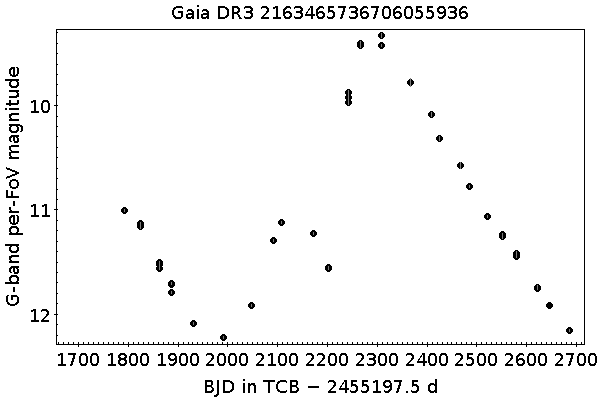} \\
\vspace{4mm}
 \caption{Same as Fig.~\ref{fig:app:ACV_cc}, but for SYST.}
 \label{fig:app:SYST_cc}
\end{figure*}

\begin{figure*}
\centering
\stackinset{c}{-0.7cm}{c}{2.7cm}{(a)}{} \includegraphics[width=0.6\hsize]{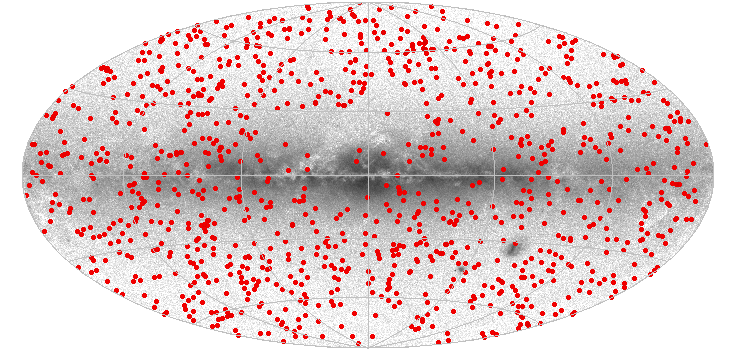} \\ 
\vspace{4mm}
\stackinset{c}{-0.3cm}{c}{3cm}{(b)}{} \includegraphics[width=0.45\hsize]{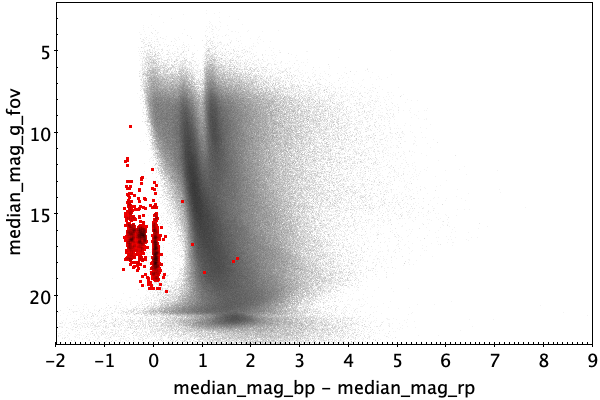}  
\hspace{2mm}
\stackinset{c}{8.8cm}{c}{3cm}{(c)}{} \includegraphics[width=0.45\hsize]{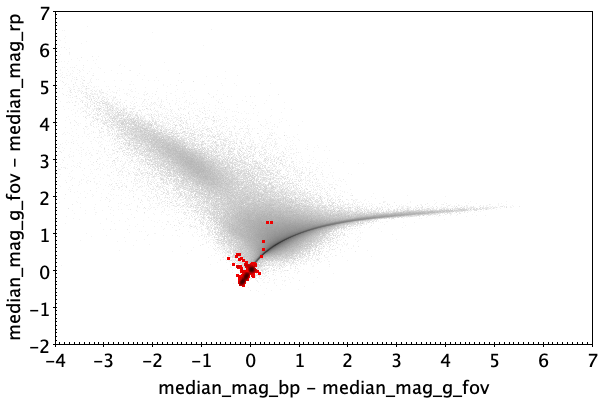} \\ 
\vspace{4mm}
\stackinset{c}{-0.3cm}{c}{3cm}{(d)}{} \includegraphics[width=0.45\hsize]{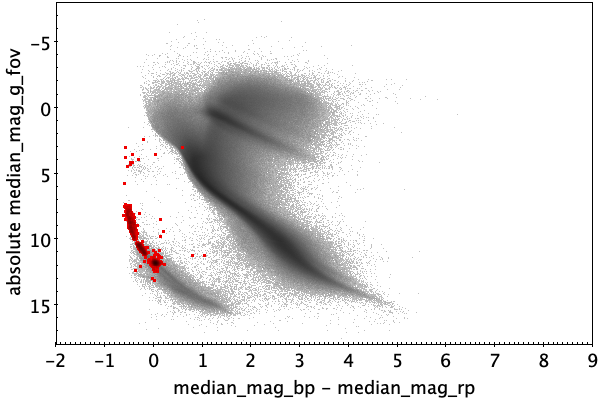}  
\hspace{2mm}
\stackinset{c}{8.8cm}{c}{3cm}{(e)}{} \includegraphics[width=0.45\hsize]{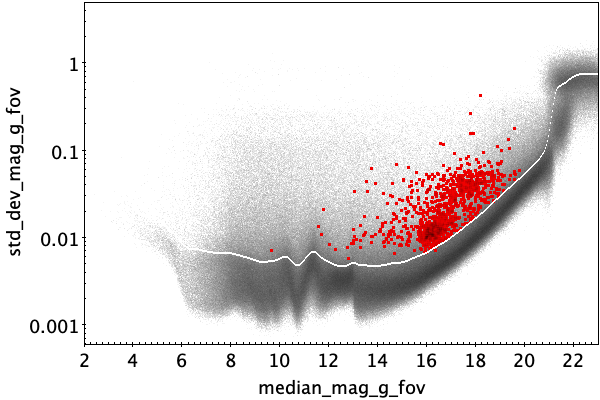} \\ 
\vspace{4mm}
\stackinset{c}{-0.3cm}{c}{3cm}{(f)}{} \includegraphics[width=0.45\hsize]{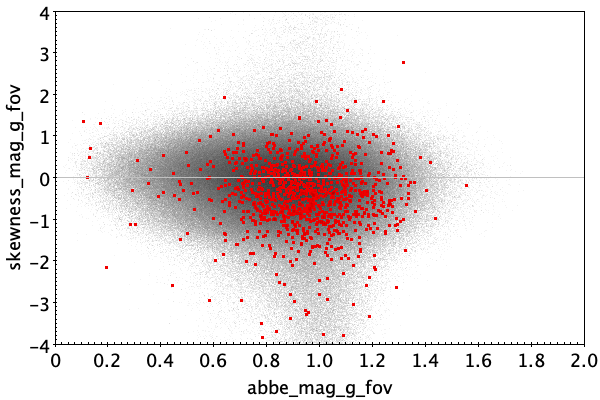}  
\hspace{2mm}
\stackinset{c}{8.8cm}{c}{3cm}{(g)}{} \includegraphics[width=0.45\hsize]{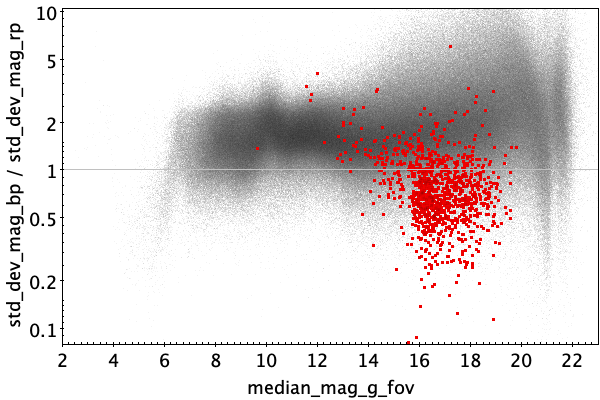}  \\ 
\vspace{4mm}
 \caption{WD: 1075 training sources.}  
 \label{fig:app:WD_trn}
\end{figure*}

\begin{figure*}
\centering
\stackinset{c}{-0.7cm}{c}{2.7cm}{(a)}{}
\includegraphics[width=0.6\hsize]{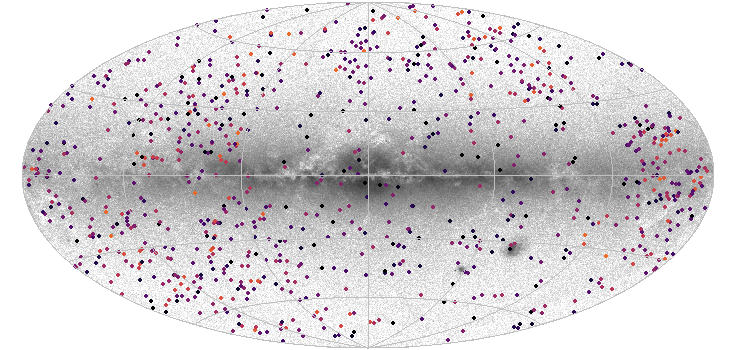} 
\stackinset{c}{2.2cm}{c}{2.7cm}{\includegraphics[height=5.5cm]{figures/appendix/vertical_best_class_score.png}}{} \\ 
\vspace{4mm}
\stackinset{c}{-0.3cm}{c}{3cm}{(b)}{} \includegraphics[width=0.45\hsize]{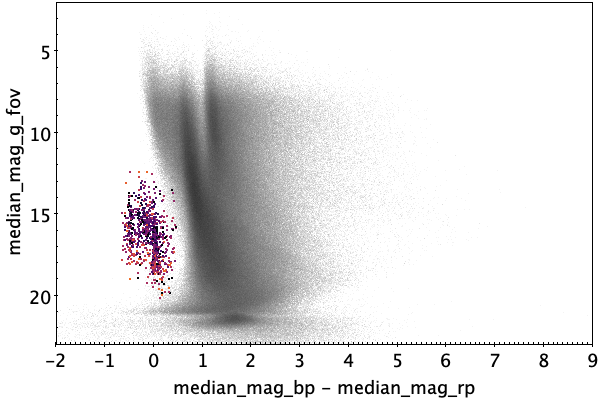}  
\hspace{2mm}
\stackinset{c}{8.8cm}{c}{3cm}{(c)}{} \includegraphics[width=0.45\hsize]{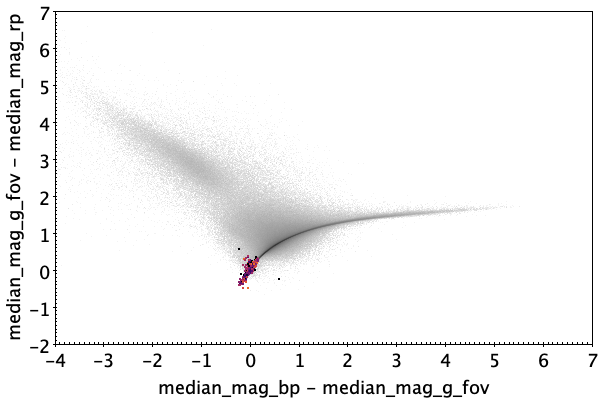} \\ 
\vspace{4mm}
\stackinset{c}{-0.3cm}{c}{3cm}{(d)}{} \includegraphics[width=0.45\hsize]{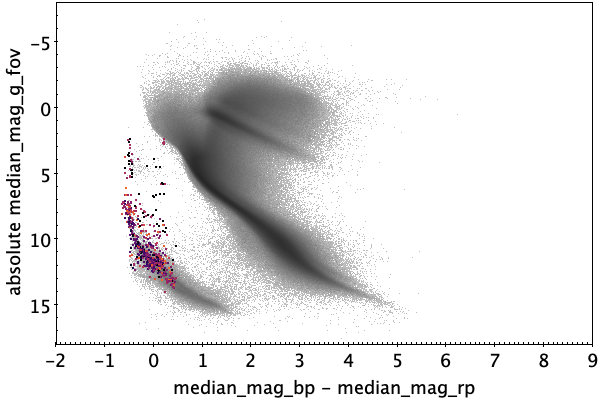}  
\hspace{2mm}
\stackinset{c}{8.8cm}{c}{3cm}{(e)}{} \includegraphics[width=0.45\hsize]{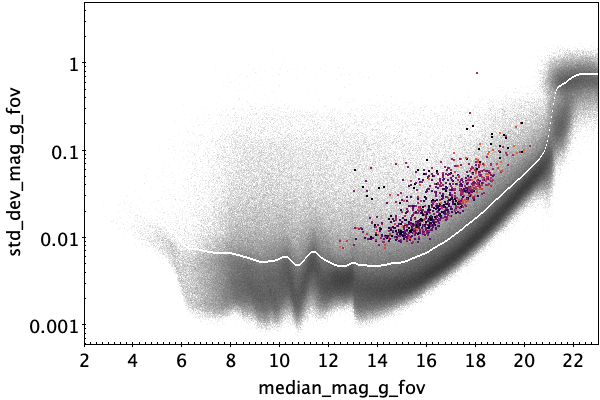} \\ 
\vspace{4mm}
\stackinset{c}{-0.3cm}{c}{3cm}{(f)}{} \includegraphics[width=0.45\hsize]{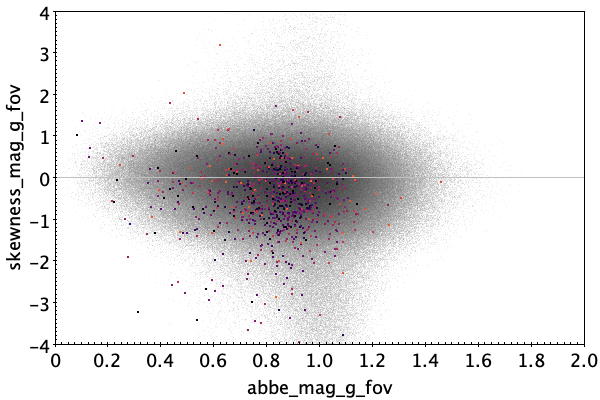}  
\hspace{2mm}
\stackinset{c}{8.8cm}{c}{3cm}{(g)}{} \includegraphics[width=0.45\hsize]{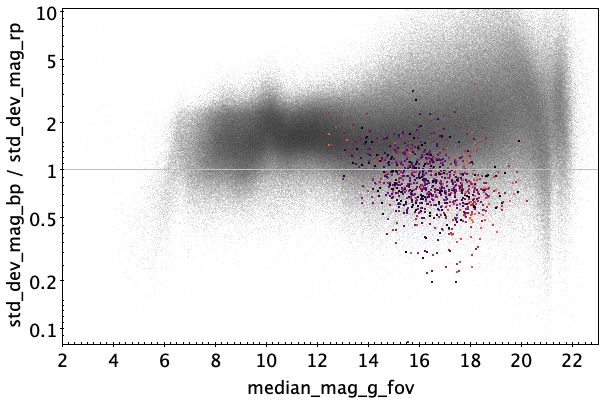}  \\ 
\vspace{4mm}
 \caption{WD: 910 classified sources.}  
 \label{fig:app:WD}
\end{figure*}

\begin{figure*}
\centering
\stackinset{c}{-0.3cm}{c}{3cm}{(a)}{} \includegraphics[width=0.45\hsize]{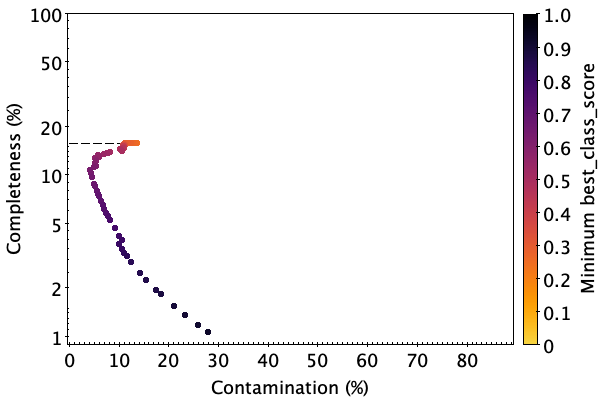}  
\hspace{2mm}
\stackinset{c}{8.8cm}{c}{3cm}{(b)}{} \includegraphics[width=0.45\hsize]{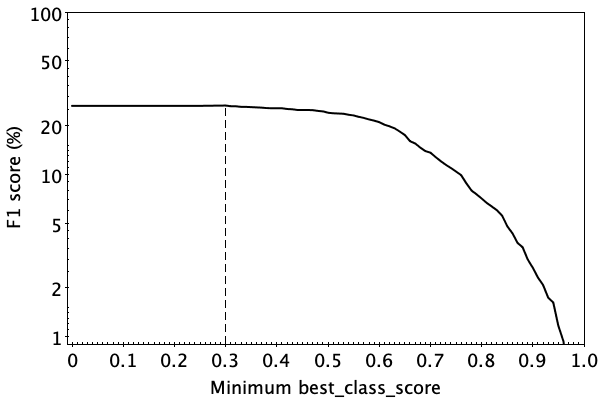} \\ 
\vspace{4mm}
\stackinset{c}{-0.3cm}{c}{3cm}{(c)}{} \includegraphics[width=0.45\hsize]{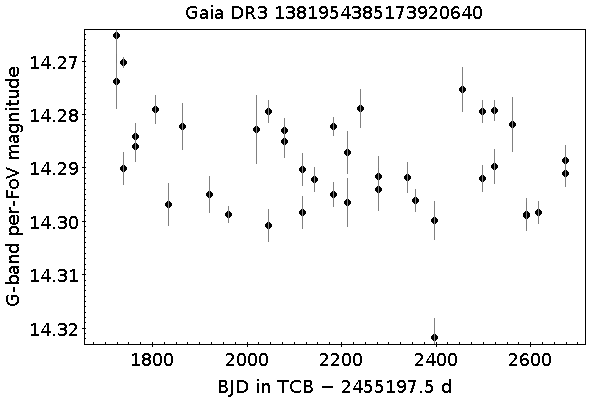}  
\hspace{2mm}
\stackinset{c}{8.8cm}{c}{3cm}{(d)}{} \includegraphics[width=0.45\hsize]{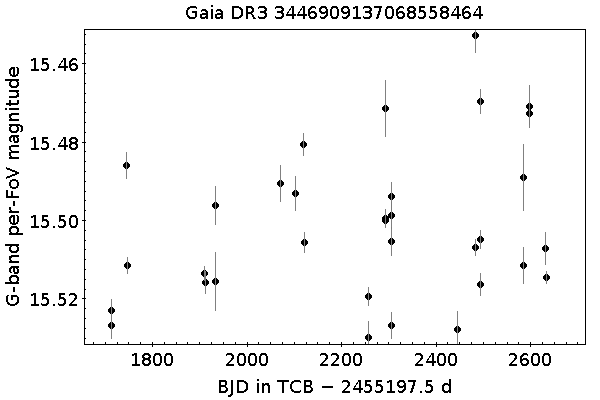} \\
\vspace{4mm}
\stackinset{c}{-0.3cm}{c}{3cm}{(e)}{} \includegraphics[width=0.45\hsize]{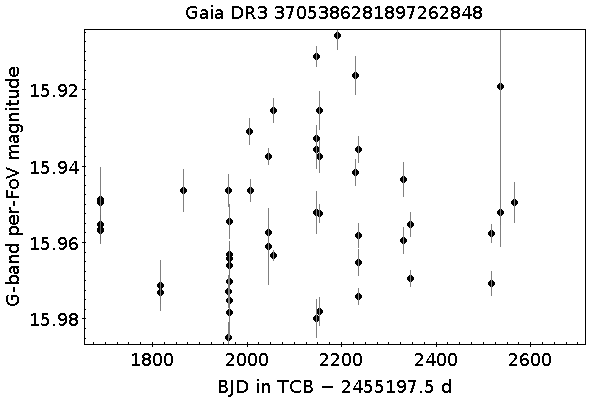}  
\hspace{2mm}
\stackinset{c}{8.8cm}{c}{3cm}{(f)}{} \includegraphics[width=0.45\hsize]{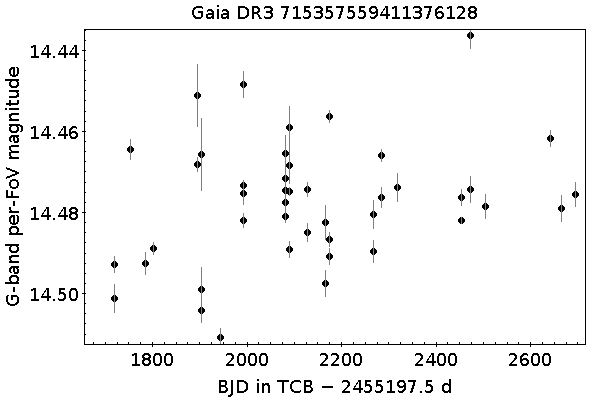} \\
\vspace{4mm}
 \caption{Same as Fig.~\ref{fig:app:ACV_cc}, but for WD. Given the multiple significant periods of WD variables, only time series are shown in panels (c)--(f).}
 \label{fig:app:WD_cc}
\end{figure*}

\begin{figure*}
\centering
\stackinset{c}{-0.7cm}{c}{2.7cm}{(a)}{} \includegraphics[width=0.6\hsize]{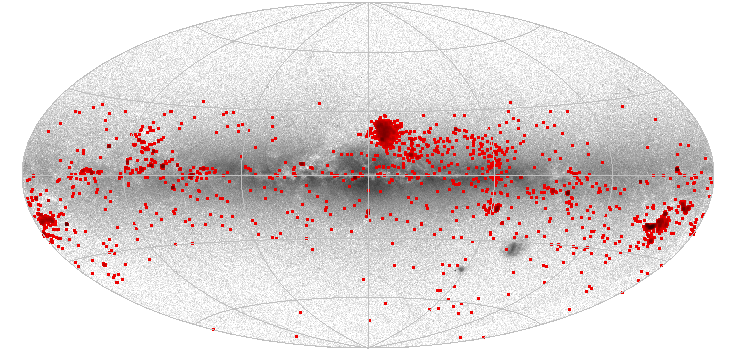} \\ 
\vspace{4mm}
\stackinset{c}{-0.3cm}{c}{3cm}{(b)}{} \includegraphics[width=0.45\hsize]{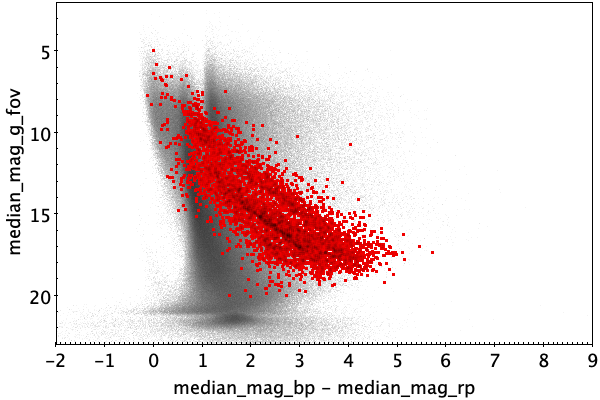}  
\hspace{2mm}
\stackinset{c}{8.8cm}{c}{3cm}{(c)}{} \includegraphics[width=0.45\hsize]{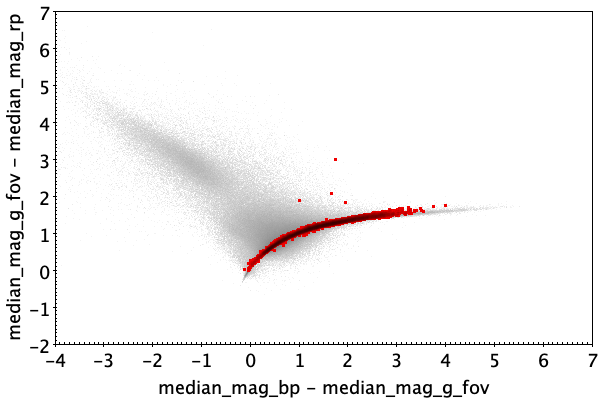} \\ 
\vspace{4mm}
\stackinset{c}{-0.3cm}{c}{3cm}{(d)}{} \includegraphics[width=0.45\hsize]{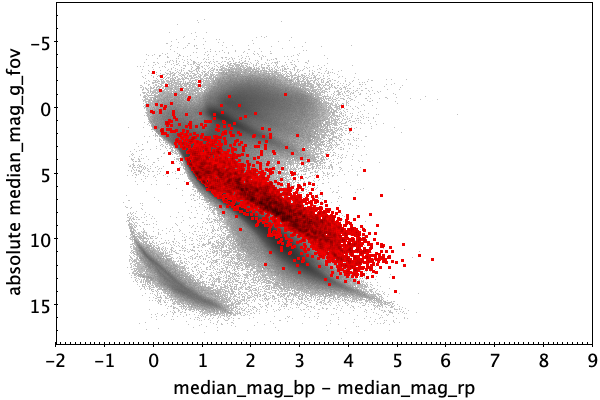}  
\hspace{2mm}
\stackinset{c}{8.8cm}{c}{3cm}{(e)}{} \includegraphics[width=0.45\hsize]{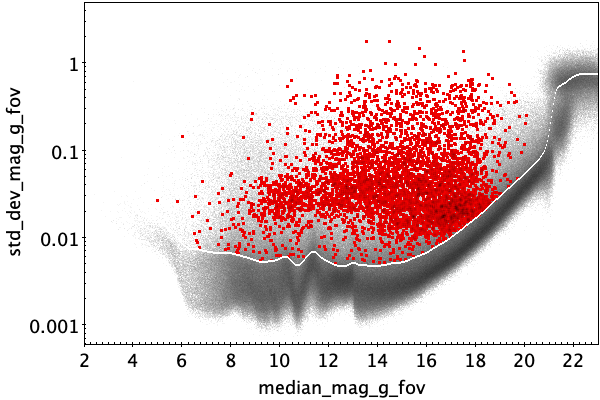} \\ 
\vspace{4mm}
\stackinset{c}{-0.3cm}{c}{3cm}{(f)}{} \includegraphics[width=0.45\hsize]{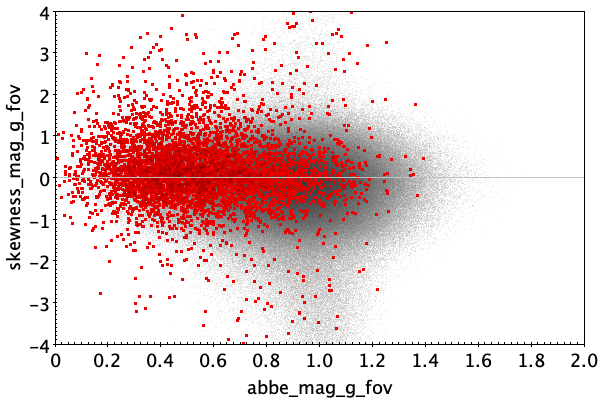}  
\hspace{2mm}
\stackinset{c}{8.8cm}{c}{3cm}{(g)}{} \includegraphics[width=0.45\hsize]{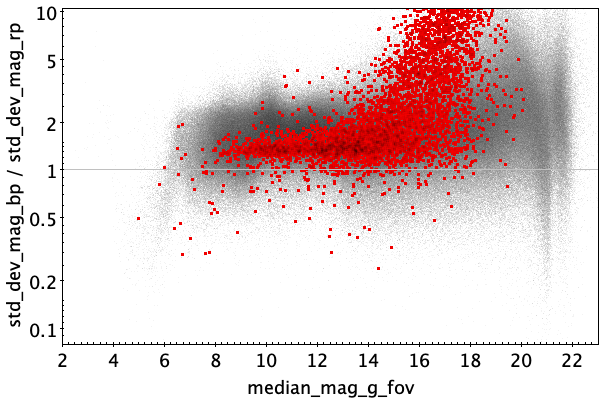}  \\ 
\vspace{4mm}
 \caption{YSO: 5148 training sources.}  
 \label{fig:app:YSO_trn}
\end{figure*}

\begin{figure*}
\centering
\stackinset{c}{-0.7cm}{c}{2.7cm}{(a)}{}
\includegraphics[width=0.6\hsize]{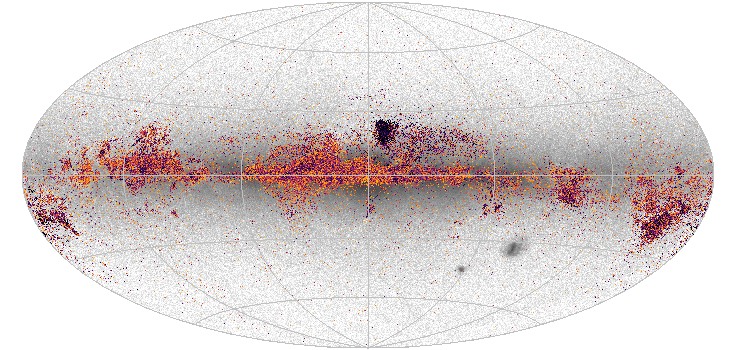} 
\stackinset{c}{2.2cm}{c}{2.7cm}{\includegraphics[height=5.5cm]{figures/appendix/vertical_best_class_score.png}}{} \\ 
\vspace{4mm}
\stackinset{c}{-0.3cm}{c}{3cm}{(b)}{} \includegraphics[width=0.45\hsize]{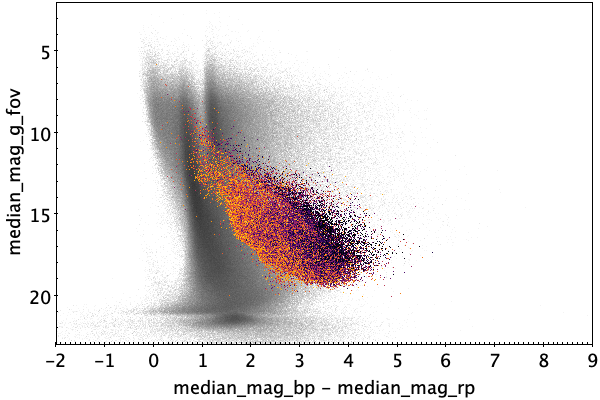}  
\hspace{2mm}
\stackinset{c}{8.8cm}{c}{3cm}{(c)}{} \includegraphics[width=0.45\hsize]{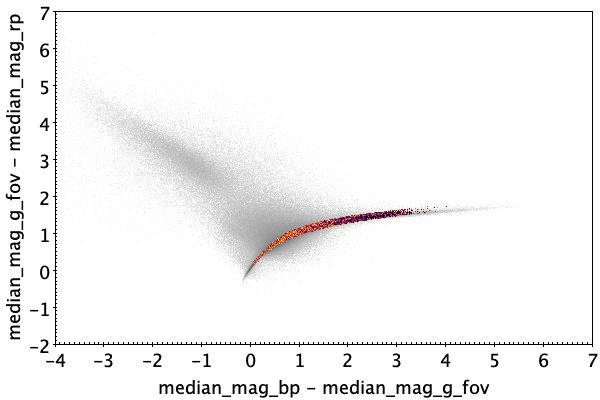} \\ 
\vspace{4mm}
\stackinset{c}{-0.3cm}{c}{3cm}{(d)}{} \includegraphics[width=0.45\hsize]{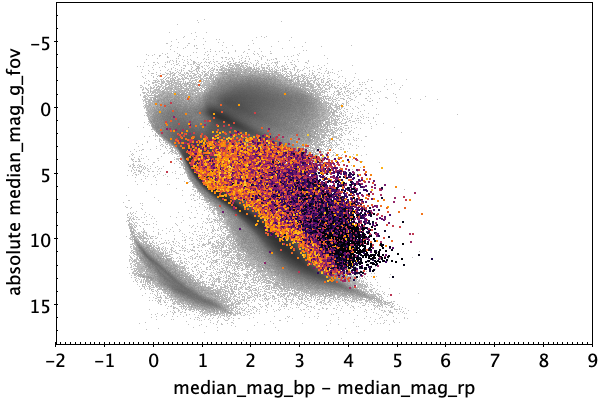}  
\hspace{2mm}
\stackinset{c}{8.8cm}{c}{3cm}{(e)}{} \includegraphics[width=0.45\hsize]{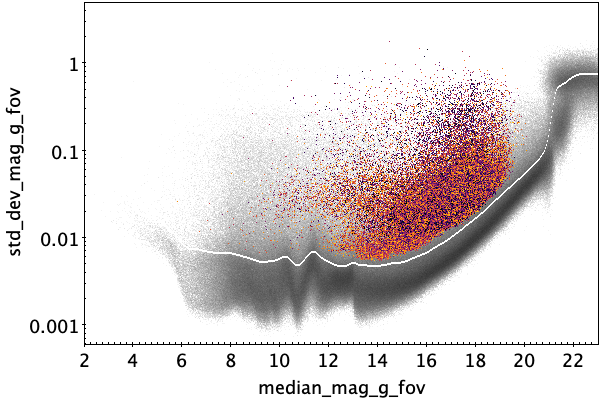} \\ 
\vspace{4mm}
\stackinset{c}{-0.3cm}{c}{3cm}{(f)}{} \includegraphics[width=0.45\hsize]{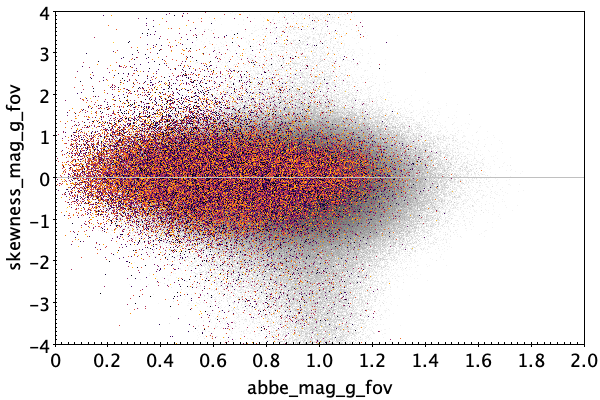}  
\hspace{2mm}
\stackinset{c}{8.8cm}{c}{3cm}{(g)}{} \includegraphics[width=0.45\hsize]{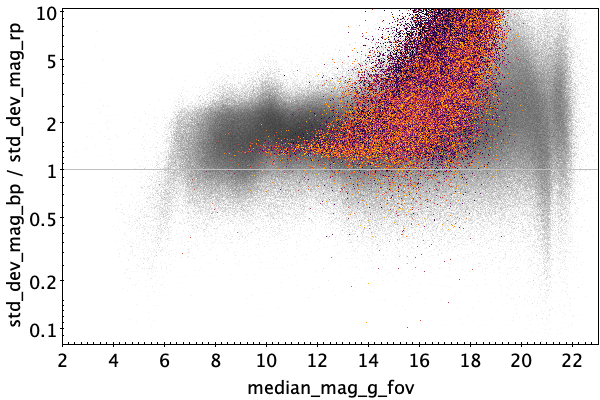}  \\ 
\vspace{4mm}
 \caption{YSO: 79\,375 classified sources.}  
 \label{fig:app:YSO}
\end{figure*}

\begin{figure*}
\centering
\stackinset{c}{-0.3cm}{c}{3cm}{(a)}{} \includegraphics[width=0.45\hsize]{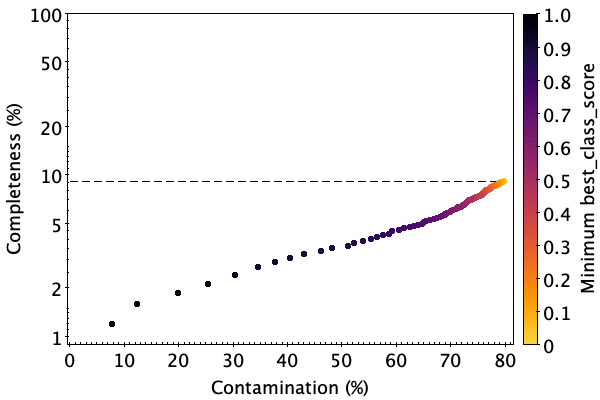}  
\hspace{2mm}
\stackinset{c}{8.8cm}{c}{3cm}{(b)}{} \includegraphics[width=0.45\hsize]{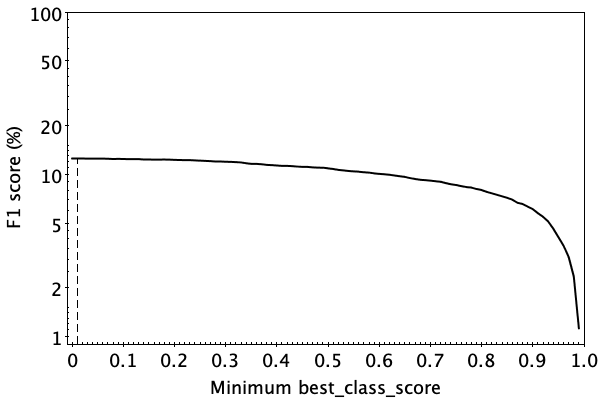} \\ 
\vspace{4mm}
\stackinset{c}{-0.3cm}{c}{3cm}{(c)}{} \includegraphics[width=0.45\hsize]{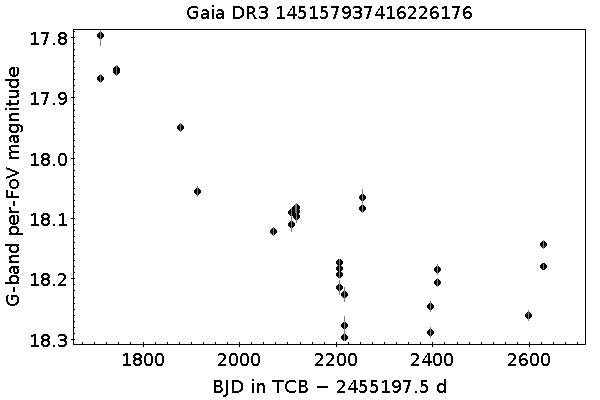}  
\hspace{2mm}
\stackinset{c}{8.8cm}{c}{3cm}{(d)}{} \includegraphics[width=0.45\hsize]{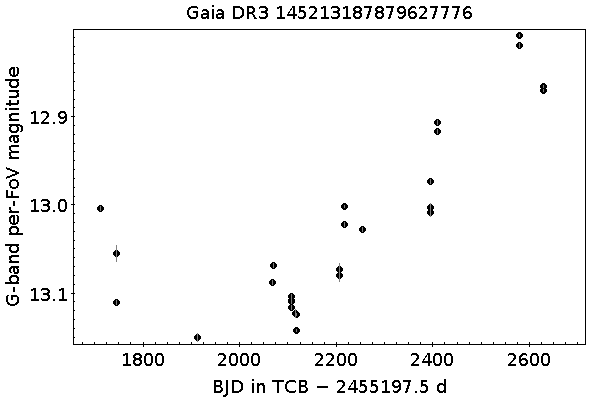} \\
\vspace{4mm}
\stackinset{c}{-0.3cm}{c}{3cm}{(e)}{} \includegraphics[width=0.45\hsize]{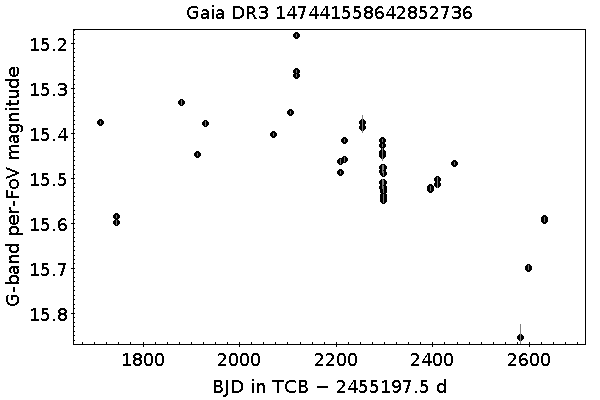}  
\hspace{2mm}
\stackinset{c}{8.8cm}{c}{3cm}{(f)}{} \includegraphics[width=0.45\hsize]{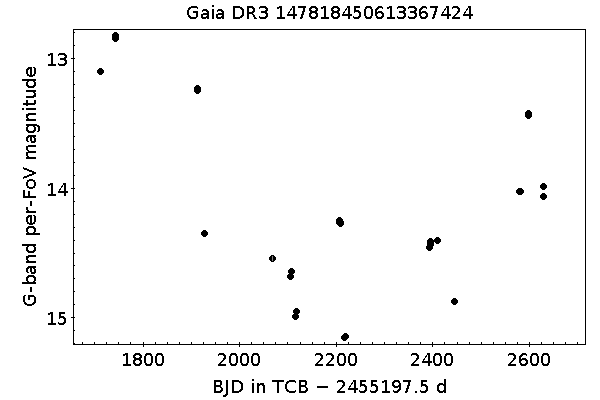} \\
\vspace{4mm}
 \caption{Same as Fig.~\ref{fig:app:ACV_cc}, but for YSO.}
 \label{fig:app:YSO_cc}
\end{figure*}

\begin{figure*}
\centering
\stackinset{c}{-0.7cm}{c}{2.7cm}{(a)}{} \includegraphics[width=0.6\hsize]{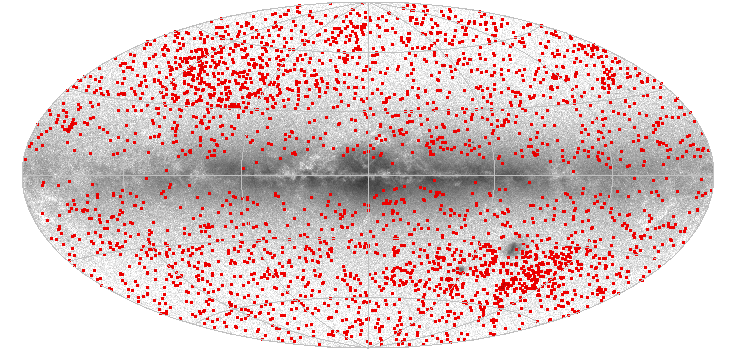} \\ 
\vspace{4mm}
\stackinset{c}{-0.3cm}{c}{3cm}{(b)}{} \includegraphics[width=0.45\hsize]{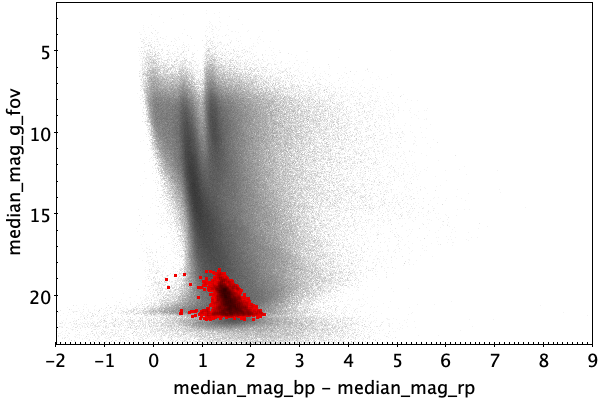}  
\hspace{2mm}
\stackinset{c}{8.8cm}{c}{3cm}{(c)}{} \includegraphics[width=0.45\hsize]{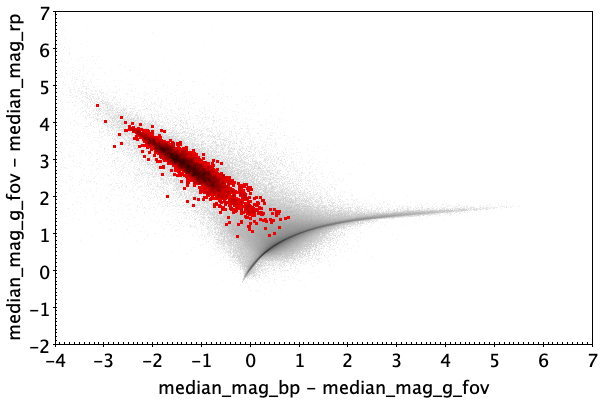} \\ 
\vspace{4mm}
\stackinset{c}{-0.3cm}{c}{3cm}{(d)}{} \includegraphics[width=0.45\hsize]{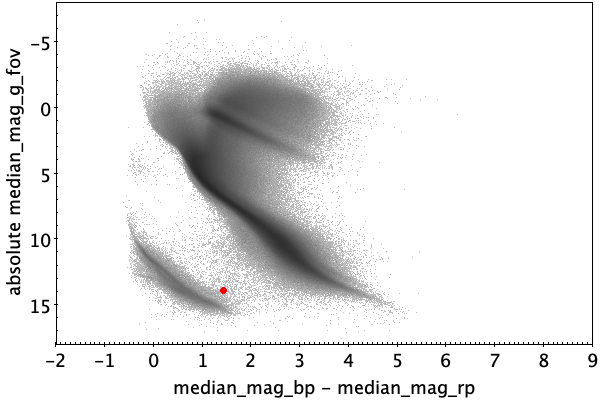}  
\hspace{2mm}
\stackinset{c}{8.8cm}{c}{3cm}{(e)}{} \includegraphics[width=0.45\hsize]{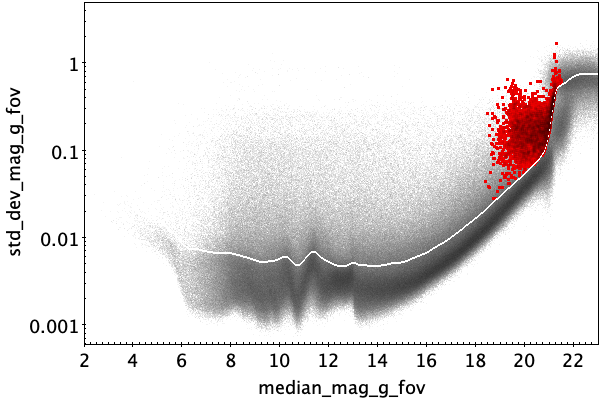} \\ 
\vspace{4mm}
\stackinset{c}{-0.3cm}{c}{3cm}{(f)}{} \includegraphics[width=0.45\hsize]{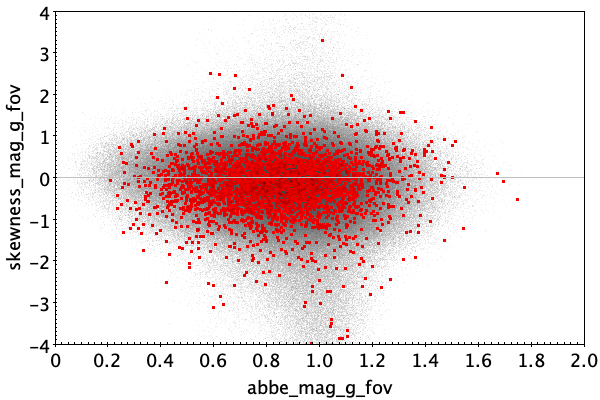}  
\hspace{2mm}
\stackinset{c}{8.8cm}{c}{3cm}{(g)}{} \includegraphics[width=0.45\hsize]{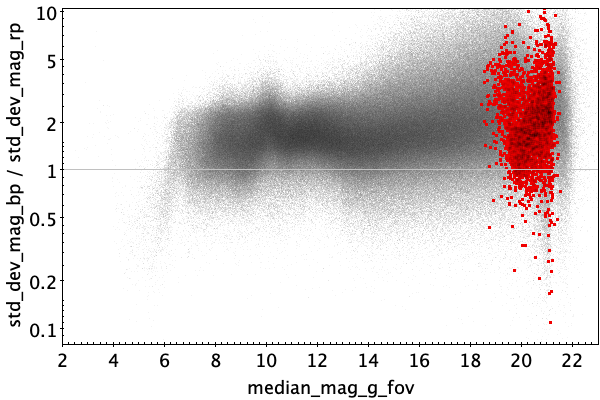}  \\ 
\vspace{4mm}
 \caption{GALAXY: 3116 training sources.}  
 \label{fig:app:GALAXY_trn}
\end{figure*}

\begin{figure*}
\centering
\stackinset{c}{-0.7cm}{c}{2.7cm}{(a)}{}
\includegraphics[width=0.6\hsize]{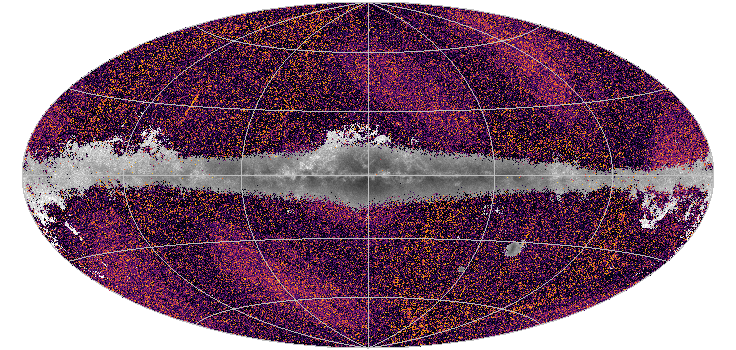} 
\stackinset{c}{2.2cm}{c}{2.7cm}{\includegraphics[height=5.5cm]{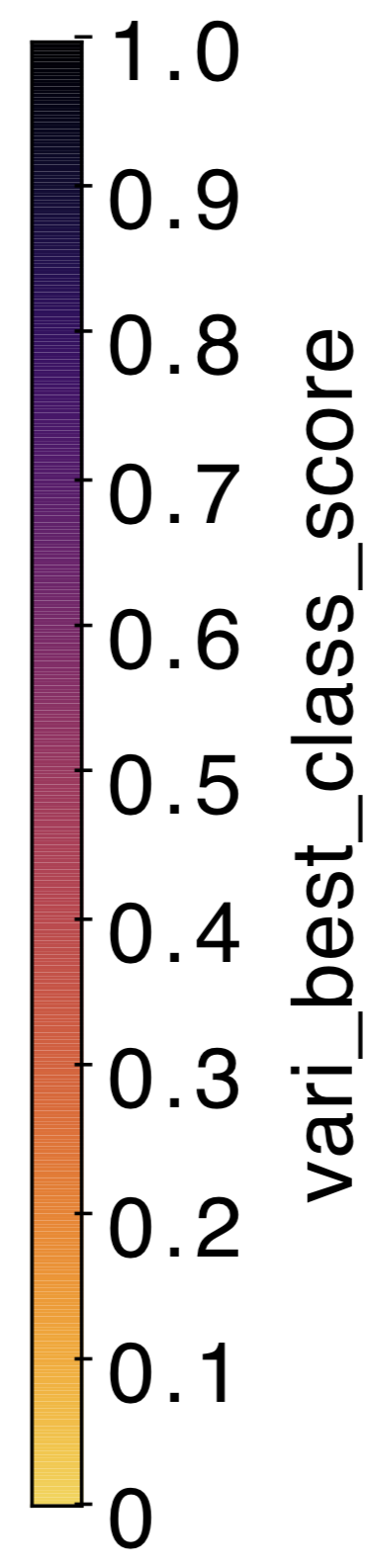}}{} \\ 
\vspace{4mm}
\stackinset{c}{-0.3cm}{c}{3cm}{(b)}{} \includegraphics[width=0.45\hsize]{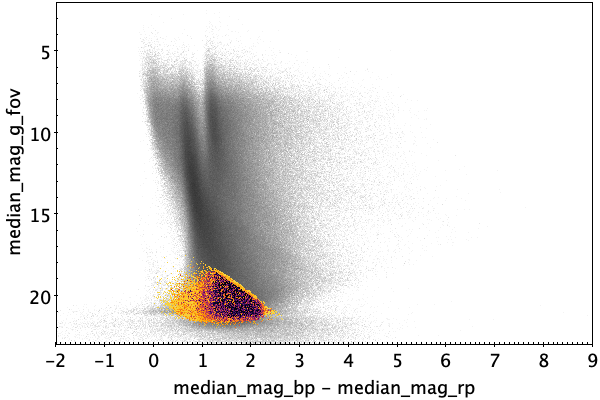}  
\hspace{2mm}
\stackinset{c}{8.8cm}{c}{3cm}{(c)}{} \includegraphics[width=0.45\hsize]{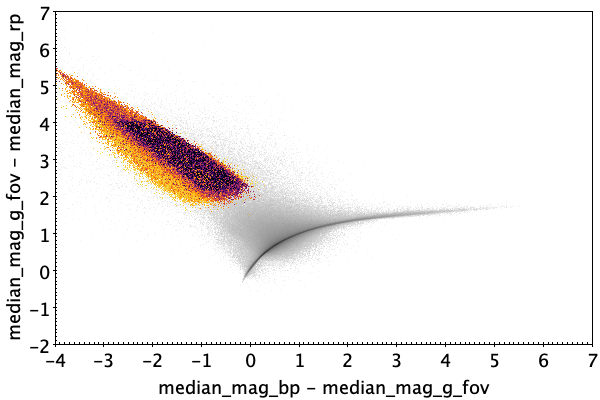} \\ 
\vspace{4mm}
\stackinset{c}{-0.3cm}{c}{3cm}{(d)}{} \includegraphics[width=0.45\hsize]{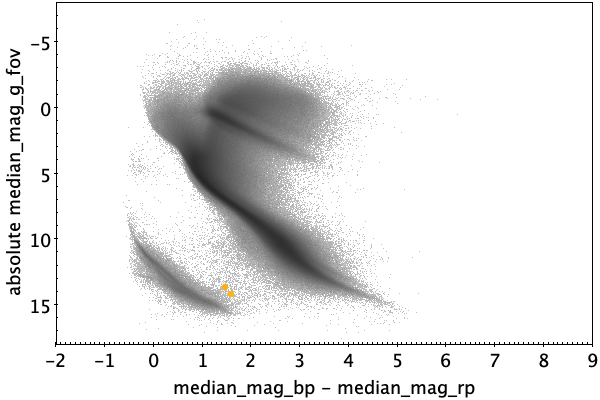}  
\hspace{2mm}
\stackinset{c}{8.8cm}{c}{3cm}{(e)}{} \includegraphics[width=0.45\hsize]{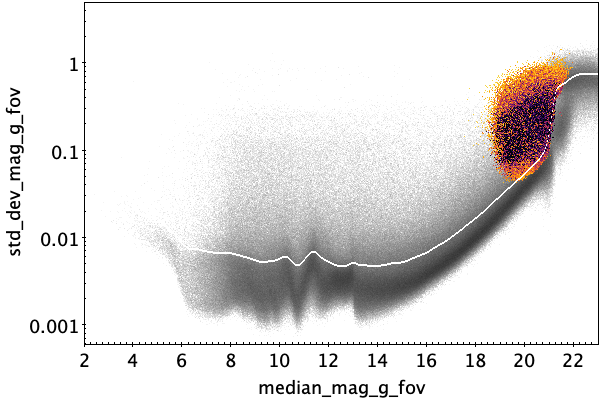} \\ 
\vspace{4mm}
\stackinset{c}{-0.3cm}{c}{3cm}{(f)}{} \includegraphics[width=0.45\hsize]{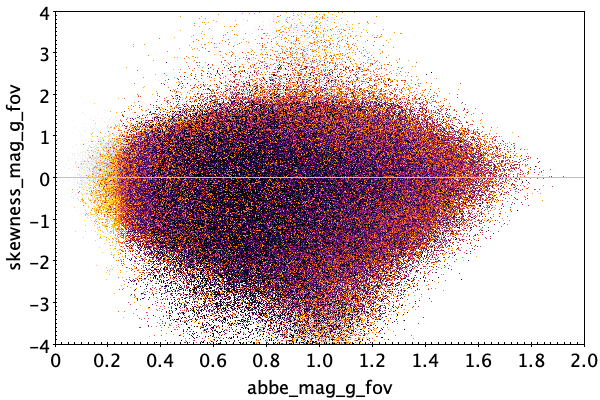}  
\hspace{2mm}
\stackinset{c}{8.8cm}{c}{3cm}{(g)}{} \includegraphics[width=0.45\hsize]{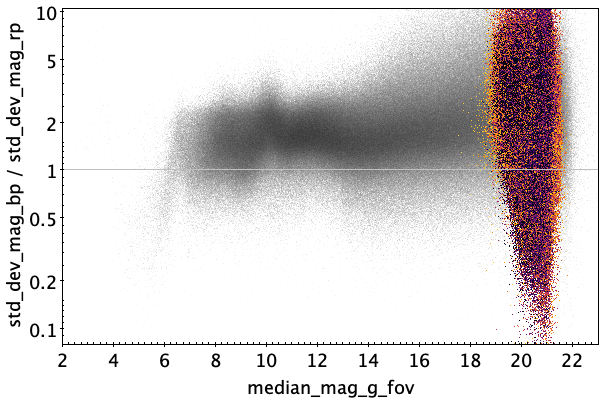}  \\ 
\vspace{4mm}
 \caption{GALAXY: 2\,451\,364 classified sources.}  
 \label{fig:app:GALAXY_ALL}
\end{figure*}

\begin{figure*}
\centering
\stackinset{c}{-0.3cm}{c}{3cm}{(a)}{} \includegraphics[width=0.45\hsize]{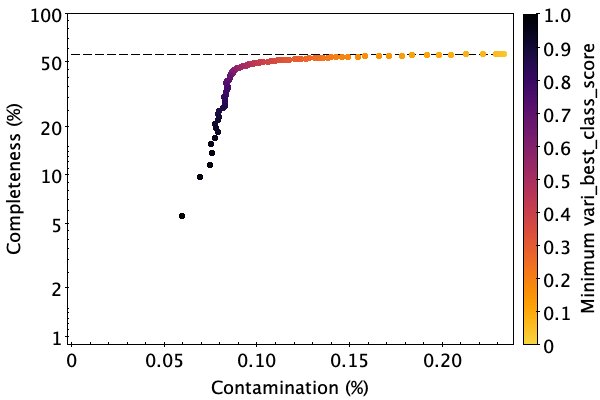}  
\hspace{2mm}
\stackinset{c}{8.8cm}{c}{3cm}{(b)}{} \includegraphics[width=0.45\hsize]{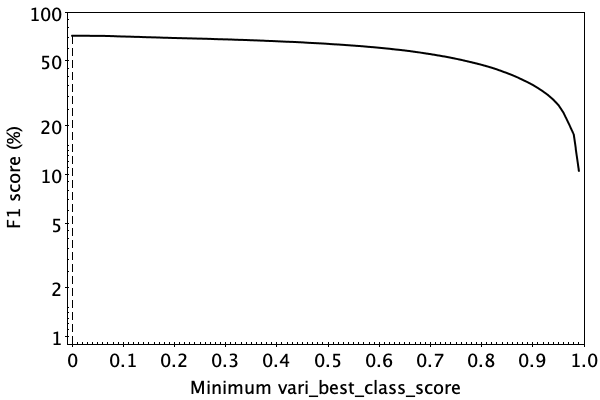} \\ 
\vspace{4mm}
\stackinset{c}{-0.3cm}{c}{3cm}{(c)}{} \includegraphics[width=0.45\hsize]{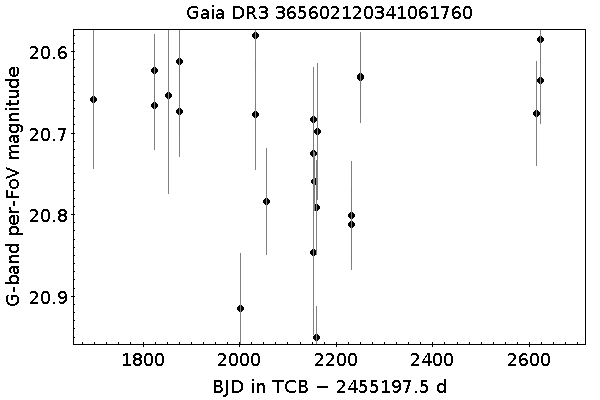}  
\hspace{2mm}
\stackinset{c}{8.8cm}{c}{3cm}{(d)}{} \includegraphics[width=0.45\hsize]{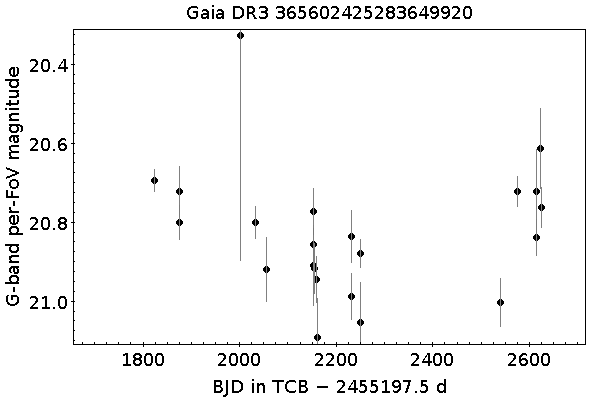} \\
\vspace{4mm}
\stackinset{c}{-0.3cm}{c}{3cm}{(e)}{} \includegraphics[width=0.45\hsize]{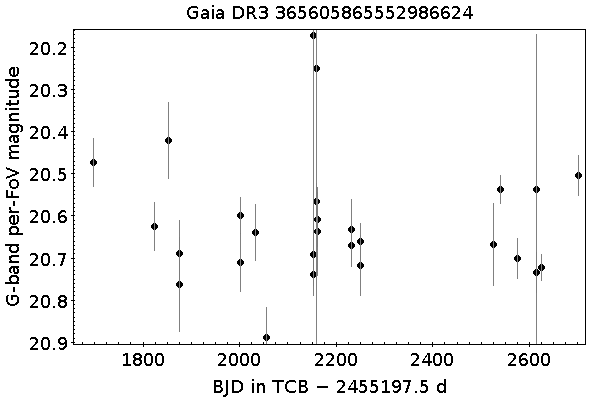}  
\hspace{2mm}
\stackinset{c}{8.8cm}{c}{3cm}{(f)}{} \includegraphics[width=0.45\hsize]{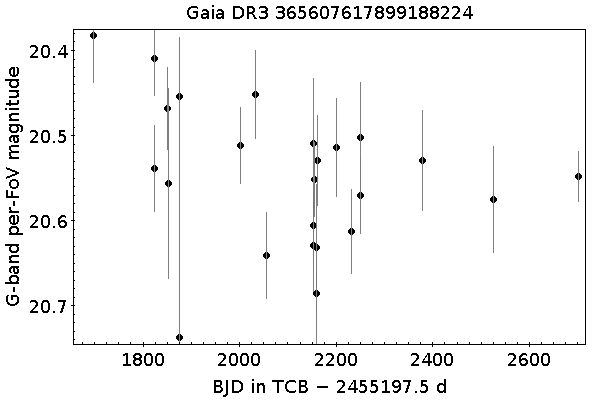} \\
\vspace{4mm}
 \caption{Same as Fig.~\ref{fig:app:ACV_cc}, but for GALAXY  \citep[light curves are limited to the subset with published photometric time series in the \gaia Andromeda Photometric Survey;][]{DR3-DPACP-142}.}
 \label{fig:app:GALAXY_ALL_cc}
\end{figure*}

\begin{figure*}
\centering
\stackinset{c}{-0.7cm}{c}{2.7cm}{(a)}{}
\includegraphics[width=0.6\hsize]{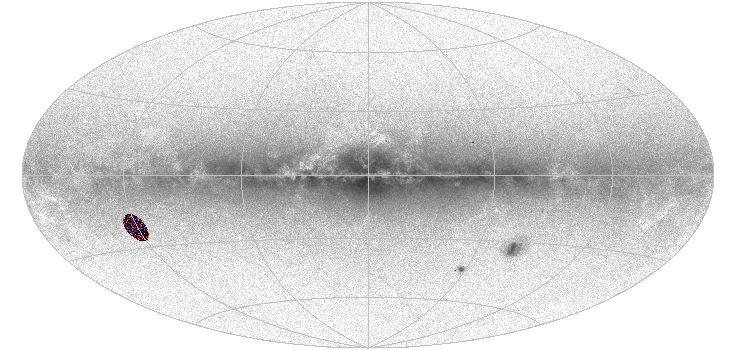} 
\stackinset{c}{2.2cm}{c}{2.7cm}{\includegraphics[height=5.5cm]{figures/appendix/vertical_vari_best_class_score.png}}{} \\ 
\vspace{4mm}
\stackinset{c}{-0.3cm}{c}{3cm}{(b)}{} \includegraphics[width=0.45\hsize]{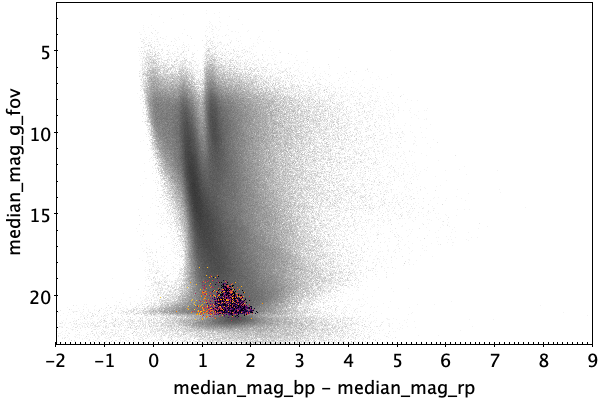}  
\hspace{2mm}
\stackinset{c}{8.8cm}{c}{3cm}{(c)}{} \includegraphics[width=0.45\hsize]{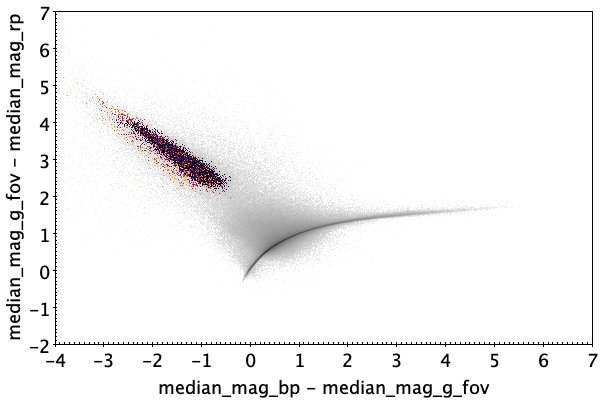} \\ 
\vspace{4mm}
\stackinset{c}{-0.3cm}{c}{3cm}{(d)}{} \includegraphics[width=0.45\hsize]{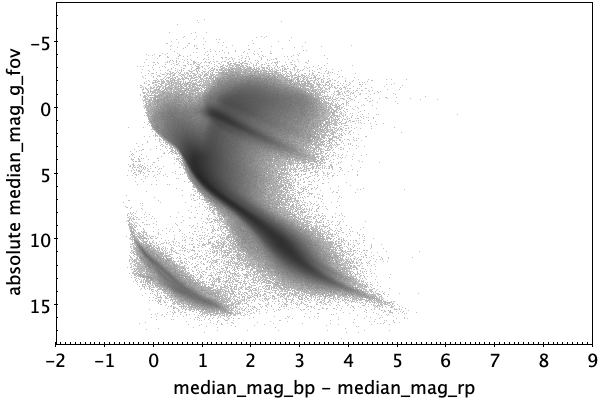}  
\hspace{2mm}
\stackinset{c}{8.8cm}{c}{3cm}{(e)}{} \includegraphics[width=0.45\hsize]{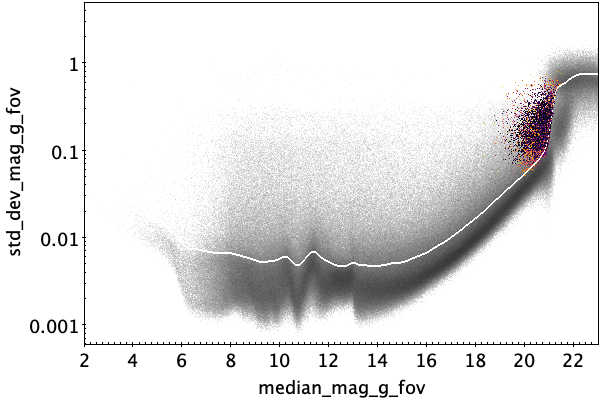} \\ 
\vspace{4mm}
\stackinset{c}{-0.3cm}{c}{3cm}{(f)}{} \includegraphics[width=0.45\hsize]{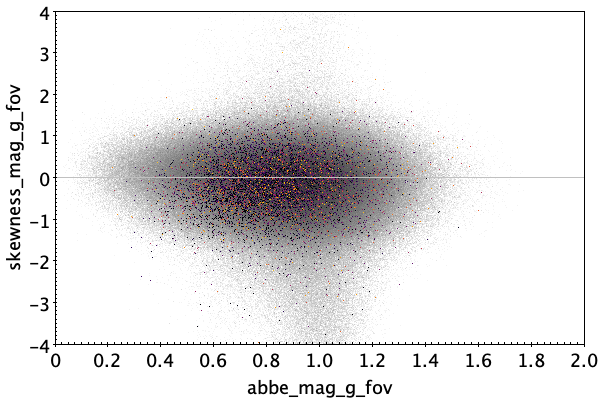}  
\hspace{2mm}
\stackinset{c}{8.8cm}{c}{3cm}{(g)}{} \includegraphics[width=0.45\hsize]{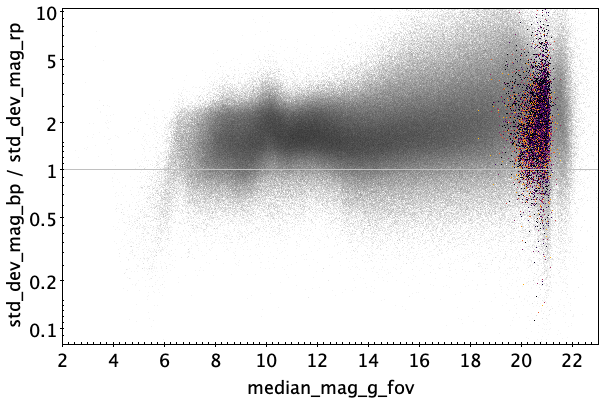}  \\ 
\vspace{4mm}
 \caption{GALAXY \citep[subset of candidates in the \gaia Andromeda Photometric Survey;][]{DR3-DPACP-142}: 7579 classified sources.}  
 \label{fig:app:GALAXY}
\end{figure*}

\end{appendix}

\end{document}